\input amstexl


\catcode`\@=11
\ifx\amstexloaded@\relax\else
 \errmessage{AmS-TeX must be loaded before LamS-TeX}\fi
\ifx\laxread@\undefined\else\catcode`\@=\active \fi
\def\err@#1{\errmessage{LamS-TeX error: #1}}
\def^^L{\par}
\let\+\tabalign
\def\newcount{\alloc@0\count\countdef\insc@unt}
\def\newdimen{\alloc@1\dimen\dimendef\insc@unt}
\def\newskip{\alloc@2\skip\skipdef\insc@unt}
\def\newmuskip{\alloc@3\muskip\muskipdef\@cclvi}
\def\newbox{\alloc@4\box\chardef\insc@unt}
\let\newtoks\relax
\def\newhelp#1#2{\newtoks#1#1\expandafter{\csname#2\endcsname}}
\def\newtoks{\alloc@5\toks\toksdef\@cclvi}
\def\newread{\alloc@6\read\chardef\sixt@@n}
\def\newwrite{\alloc@7\write\chardef\sixt@@n}
\def\newfam{\alloc@8\fam\chardef\sixt@@n}
\def\newlanguage{\alloc@9\language\chardef\@cclvi}
\def\newinsert#1{\global\advance\insc@unt by\m@ne
  \ch@ck0\insc@unt\count
  \ch@ck1\insc@unt\dimen
  \ch@ck2\insc@unt\skip
  \ch@ck4\insc@unt\box
  \allocationnumber=\insc@unt
  \global\chardef#1=\allocationnumber
  \wlog{\string#1=\string\insert\the\allocationnumber}}
\def\newif#1{\count@\escapechar \escapechar\m@ne
  \expandafter\expandafter\expandafter
   \edef\@if#1{true}{\let\noexpand#1=\noexpand\iftrue}%
  \expandafter\expandafter\expandafter
   \edef\@if#1{false}{\let\noexpand#1=\noexpand\iffalse}%
  \@if#1{false}\escapechar\count@}

\def\Err@#1{\errhelp\defaulthelp@\err@{#1}}
{\catcode`\@=\active
 \edef\next{\gdef\noexpand@{\futurelet\noexpand\next
  \csname at\string@\endcsname}}
 \next
}
\def\at@{\ifcat\noexpand\next a\let\next@\at@@\else
 \ifcat\noexpand\next0\let\next@\at@@\else
 \ifcat\noexpand\next\relax\let\next@\at@@\else
 \let\next@\at@@@\fi\fi\fi\next@}
\def\at@@@{\errhelp\athelp@\err@{Invalid use of @}}
\def\at@@#1{\expandafter
 \ifx\csname\string#1@at\endcsname\relax\let\next@\at@@@\else
 \DN@{\csname\string#1@at\endcsname}\fi\next@}
\def\atdef@#1{\expandafter\def\csname\string#1@at\endcsname}
\newif\iftest@
\def\tagin@#1{\tagin@false
 \DN@##1\tag##2##3\next@{\test@true\ifx\tagin@##2\test@false\fi}%
 \next@#1\tag\tagin@\next@\tagin@false\iftest@\tagin@true\fi}
\let\lkerns@\relax
\def\nolinebreak{\RIfM@\mathmodeerr@\nolinebreak\else
 \ifhmode\saveskip@\lastskip\unskip
 \nobreak\ifdim\saveskip@>\z@\hskip\saveskip@\fi\lkerns@
 \else\vmodeerr@\nolinebreak\fi\fi}
\def\allowlinebreak{\RIfM@\mathmodeerr@\allowlinebreak\else
 \ifhmode\saveskip@\lastskip\unskip
 \allowbreak\ifdim\saveskip@>\z@\hskip\saveskip@\fi\lkerns@
 \else\vmodeerr@\allowlinebreak\fi\fi}
\def\linebreak{\RIfM@\mathmodeerr@\linebreak\else
 \ifhmode\unskip\unkern\break\lkerns@
 \else\vmodeerr@\linebreak\fi\fi}
\let\nkerns@\relax
\def\newline{\RIfM@\mathmodeerr@\newline\else
 \ifhmode\unskip\unkern\null\hfill\break\nkerns@
 \else\vmodeerr@\newline\fi\fi}%
\def\newbox@{\alloc@@4\box\chardef\insc@unt}
\def\newcount@{\alloc@@0\count\countdef\insc@unt}
\def\accentedsymbol#1#2{\expandafter\newbox@\csname\exstring@#1@box\endcsname
 \setbox\csname\exstring@#1@box\endcsname\hbox{$\m@th#2$}%
 \define#1{\copy\csname\exstring@#1@box\endcsname{}}}
\def\rightadd@#1\to#2{\toks@{\\#1}\toks@@\expandafter{#2}\xdef#2{\the\toks@@
 \the\toks@}\toks@{}\toks@@{}}
\def\fontlist@{\\\tenrm\\\sevenrm\\\fiverm\\\teni\\\seveni\\\fivei
 \\\tensy\\\sevensy\\\fivesy\\\tenex\\\tenbf\\\sevenbf\\\fivebf
 \\\tensl\\\tenit}
\def\font@#1=#2 {\rightadd@#1\to\fontlist@\font#1=#2 }
\def\ismember@#1#2{\global\let\Next@ F\let\next@= #2%
 {\def\\##1{\let\nextii@##1\ifx\nextii@\next@\global\let\Next@ T\fi}#1}%
 \test@false\ifx\Next@ T\test@true\fi\let\next@\relax}
\def\FNSS@#1{\let\FNSS@@#1\FN@\FNSS@@@}
\def\FNSS@@@{\ifx\next\space@\def\FNSS@@@@. {\FN@\FNSS@@@}\else
 \def\FNSS@@@@.{\FNSS@@}\fi\FNSS@@@@.}
\atdef@"{\unskip
 \DN@{\ifx\next`\DN@`{\FN@\nextii@}%
  \else\ifx\next\lq\DN@\lq{\FN@\nextii@}%
  \else\DN@####1{\FN@\nextiii@}\fi\fi
  \next@}%
 \DNii@{\ifx\next`\DN@`{\sldl@``}%
  \else\ifx\next\lq\DN@\lq{\sldl@``}%
  \else\DN@{\dlsl@`}\fi\fi\next@}%
 \def\nextiii@{\ifx\next'\DN@'{\srdr@''}%
  \else\ifx\next\rq\DN@\rq{\srdr@''}%
  \else\DN@{\drsr@'}\fi\fi\next@}%
 \FNSS@\next@}
\def\root{%
  \DN@{\ifx\next\uproot\let\next@\nextii@\else
   \ifx\next\leftroot\let\next@\nextiii@\else
   \let\next@\plainroot@\fi\fi\next@}%
  \DNii@\uproot##1{\uproot@##1\relax\FNSS@\nextiv@}%
  \def\nextiv@{\ifx\next\leftroot\let\next@\nextv@\else
   \let\next@\plainroot@\fi\next@}%
  \def\nextv@\leftroot##1{\leftroot@##1\relax\plainroot@}%
  \def\nextiii@\leftroot##1{\leftroot@##1\relax\FNSS@\nextvi@}%
  \def\nextvi@{\ifx\next\uproot\let\next@\nextvii@\else
   \let\next@\plainroot@\fi\next@}%
  \def\nextvii@\uproot##1{\uproot@##1\relax\plainroot@}%
  \bgroup\uproot@\z@\leftroot@\z@
 \FNSS@\next@}
\def\loop#1\repeat{\def\iterate{#1\relax\expandafter\iterate\fi}%
 \iterate\let\iterate\relax}
\def\gloop@#1\repeat{\gdef\iterate@{#1\relax\expandafter\iterate@\fi}%
 \iterate@\global\let\iterate@\relax}
\def\printoptions{\W@{Do you want S(yntax check),
  G(alleys) or P(ages)?^^JType S, G or P, follow by <return>: }\loop
 \read\m@ne to\ans@
 \edef\next@{\def\noexpand\Ans@{\ans@}}\uppercase\expandafter{\next@}%
 \ifx\Ans@\S@\test@true\syntax\else
 \ifx\Ans@\G@\test@true\galleys\else
 \ifx\Ans@\P@\test@true\else
 \test@false\fi\fi\fi
 \iftest@\else\W@{Type S, G or P, follow by <return>: }%
 \repeat}
\expandafter\let\csname A@;\endcsname;
\expandafter\let\csname A@:\endcsname:
\expandafter\let\csname A@?\endcsname?
\expandafter\let\csname A@!\endcsname!
\def\APdef#1{\def\next@{\expandafter\let\csname A@\string#1\endcsname#1}%
 \afterassignment\next@\def#1}
\let\fextra@\,
\def\tdots@{\unskip
 \DN@{$\m@th\mathinner{\ldotp\ldotp\ldotp}\,
   \ifx\next,\,$\else\ifx\next.\,$\else
   \ifx\next;\,$\else
   \expandafter\ifx\csname A@\string;\endcsname\next\fextra@$\else
   \ifx\next:\,$\else
   \expandafter\ifx\csname A@\string:\endcsname\next\fextra@$\else
   \ifx\next?\,$\else
   \expandafter\ifx\csname A@\string?\endcsname\next\fextra@$\else
   \ifx\next!\,$\else
   \expandafter\ifx\csname A@\string!\endcsname\next\fextra@$\else
   $ \fi\fi\fi\fi\fi\fi\fi\fi\fi\fi}%
 \ \FN@\next@}
\def\extrap@#1{%
 \ifx\next,\DN@{#1\,}\else
 \ifx\next;\DN@{#1\,}\else
 \expandafter\ifx\csname A@\string;\endcsname\next\DN@{#1\fextra@}\else
 \ifx\next.\DN@{#1\,}\else\extra@
 \ifextra@\DN@{#1\,}\else
 \let\next@#1\fi\fi\fi\fi\fi\next@}
\def\dotsc{\DN@{\ifx\next;\plainldots@\,\else
 \expandafter\ifx\csname A@\string;\endcsname\next\plainldots@\fextra@\else
 \ifx\next.\plainldots@\,\else\extra@\plainldots@
 \ifextra@\,\fi\fi\fi\fi}%
 \FN@\next@}
\def\keybin@{\keybin@true
 \ifx\next+\else\ifx\next=\else\ifx\next<\else\ifx\next>\else\ifx\next-\else
 \ifx\next*\else\ifx\next:\else
 \expandafter\ifx\csname A@\string;\endcsname\next\else
 \keybin@false\fi\fi\fi\fi\fi\fi\fi\fi}
\def\boldkey#1{\ifcat\noexpand#1A%
  \ifcmmibloaded@{\fam\cmmibfam#1}\else
   \Err@{First bold symbol font not loaded}\fi
 \else
 \let\next=#1%
 \ifx#1!\mathchar"5\bffam@21 \else
 \expandafter\ifx\csname A@\string!\endcsname\next\mathchar"5\bffam@21 \else
 \ifx#1(\mathchar"4\bffam@28 \else\ifx#1)\mathchar"5\bffam@29 \else
 \ifx#1+\mathchar"2\bffam@2B \else\ifx#1:\mathchar"3\bffam@3A \else
 \expandafter\ifx\csname A@\string:\endcsname\next\mathchar"3\bffam@3A \else
 \ifx#1;\mathchar"6\bffam@3B \else
 \expandafter\ifx\csname A@\string;\endcsname\next\mathchar"6\bffam@3B \else
 \ifx#1=\mathchar"3\bffam@3D \else
 \ifx#1?\mathchar"5\bffam@3F \else
 \expandafter\ifx\csname A@\string?\endcsname\next\mathchar"5\bffam@3F \else
 \ifx#1[\mathchar"4\bffam@5B \else
 \ifx#1]\mathchar"5\bffam@5D \else
 \ifx#1,\mathchari@63B \else
 \ifx#1-\mathcharii@200 \else
 \ifx#1.\mathchari@03A \else
 \ifx#1/\mathchari@03D \else
 \ifx#1<\mathchari@33C \else
 \ifx#1>\mathchari@33E \else
 \ifx#1*\mathcharii@203 \else
 \ifx#1|\mathcharii@06A \else
 \ifx#10\bold0\else\ifx#11\bold1\else\ifx#12\bold2\else\ifx#13\bold3\else
 \ifx#14\bold4\else\ifx#15\bold5\else\ifx#16\bold6\else\ifx#17\bold7\else
 \ifx#18\bold8\else\ifx#19\bold9\else
  \Err@{\noexpand\boldkey can't be used with #1}%
 \fi\fi\fi\fi\fi\fi\fi\fi\fi\fi\fi\fi\fi\fi\fi
 \fi\fi\fi\fi\fi\fi\fi\fi\fi\fi\fi\fi\fi\fi\fi\fi\fi\fi}
\def\arabic#1{#1}
\def\alph#1{\count@#1\relax\advance\count@96 \ifnum\count@>122
 \Err@{\noexpand\alph invalid for numbers > 26}\else\char\count@\fi}
\def\Alph#1{\count@#1\relax\advance\count@64 \ifnum\count@>90
 \Err@{\noexpand\Alph invalid for numbers > 26}\else\char\count@\fi}

\def\Roman#1{\uppercase\expandafter{\romannumeral#1}}
\def\fnsymbol#1{\count@#1\relax
 \count@@\count@
 \advance\count@\m@ne\divide\count@7
 \count@@@\count@\advance\count@@@\@ne
 \multiply\count@7 \advance\count@@-\count@
 \count@\count@@@
 {\loop
  \ifcase\count@@\or*\or\dag\or\ddag\or\P\or\S\or\text{$\|$}\or\#\fi
  \advance\count@\m@ne\ifnum\count@>\z@\repeat}}
\def\cardnine@#1{\ifcase#1\or one\or two\or three\or four\or five\or
 six\or seven\or eight\or nine\fi}
\let\alloc@\alloc@@
\newcount\ten@
\ten@10
\def\cardinal#1{\count@#1\relax
 \ifnum\count@>99 \number\count@
 \else
  \ifnum\count@=\z@ zero%
  \else
   \ifnum\count@<\ten@\cardnine@\count@
   \else
    \ifnum\count@<20
     \advance\count@-\ten@
     \ifcase\count@ ten\or eleven\or twelve\or thirteen\or fourteen\or
      fifteen\or sixteen\or seventeen\or eighteen\or nineteen\fi
    \else
     \count@@\count@\count@@@\count@@
     \divide\count@\ten@\multiply\count@\ten@
     \advance\count@@@-\count@\divide\count@\ten@
     \ifcase\count@\or\or twenty\or thirty\or forty\or fifty\or sixty\or
      seventy\or eighty\or ninety\fi
     \ifnum\count@@@=\z@\else-\cardnine@\count@@@\fi
    \fi
   \fi
  \fi
 \fi}
\def\ordnine@#1{\ifcase#1\or first\or second\or third\or fourth\or fifth\or
 sixth\or seventh\or eighth\or ninth\fi}
\newcount\count@@@@
\def\ordsuffix@{\count@@@@\count@
 \divide\count@\ten@
 \count@@@\count@\count@@\count@
 \divide\count@@\ten@\multiply\count@@\ten@
 \advance\count@@@-\count@@
 \ifnum\count@@@=\@ne th%
 \else
  \count@@@\count@@@@
  \count@@\count@@@@
  \divide\count@@\ten@\multiply\count@@\ten@
  \advance\count@@@-\count@@
  \ifcase\count@@@ th\or st\or nd\or rd\else th\fi
 \fi}
\def\nordinal#1{\count@#1\relax\number\count@\ordsuffix@}
\def\spordinal#1{\count@#1\relax\number\count@$^{\text{\ordsuffix@}}$}
\def\ordinal#1{\count@#1\relax
 \ifnum\count@>99 \number\count@\ordsuffix@
 \else
   \ifnum\count@=\z@ zeroth%
  \else
    \ifnum\count@<\ten@\ordnine@\count@
    \else
     \ifnum\count@<20 \advance\count@-\ten@
      \ifcase\count@ tenth\or eleventh\or twelfth\or thirteenth\or
       fourteenth\or fifteenth\or sixteenth\or seventeenth\or eighteenth\or
       nineteenth\fi
     \else
      \count@@\count@
      \divide\count@\ten@\multiply\count@\ten@
      \count@@@\count@@\advance\count@@@-\count@
      \divide\count@\ten@
      \ifcase\count@\or\or twent\or thirt\or fort\or fift\or sixt\or sevent\or
       eight\or ninet\fi
      \ifnum\count@@@=\z@ ieth\else y-\ordnine@\count@@@\fi
     \fi
    \fi
  \fi
 \fi}
\font@\tensmc=cmcsc10
\textonlyfont@\smc\tensmc
\newtoks\noexpandtoks@
\noexpandtoks@{\let\arabic\relax\let\alph\relax\let\Alph\relax
 \let\Roman\relax\let\fnsymbol\relax\let\rm\relax
 \let\it\relax\let\bf\relax\let\sl\relax\let\smc\relax
 \let\/\relax\let\null\relax}
\def\noexpands@{\the\noexpandtoks@}
\def\Nonexpanding#1{\global\noexpandtoks@
 \expandafter{\the\noexpandtoks@\let#1\relax}}
\def\prevanish@{\saveskip@\z@\ifhmode\saveskip@\lastskip\unskip\fi}
\def\postvanish@{\ifdim\saveskip@>\z@\hskip\saveskip@\fi\FN@\postvanish@@}
\def\postvanish@@{\DN@.{}%
 \ifx\next\space@\ifdim\saveskip@>\z@\DN@. {}\fi\fi\next@.}
\def\invisible#1{\prevanish@\ignorespaces#1\unskip\postvanish@}
\def\vanishlist@{\\\invisible}
\let\noindent@\noindent
\def\noindent{\par\noindent@\FN@\pretendspace@}
\def\pretendspace@{\ismember@\vanishlist@\next
 \iftest@\nobreak\hskip-\p@\hskip\p@\fi}
\let\flushpar\noindent
\newtoks\everypartoks@
\def\noindent@@{\par\everypartoks@\expandafter{\the\everypar}\everypar{}%
 \noindent@\everypar\expandafter{\the\everypartoks@}}
\def\page{\Err@{\noexpand\page has no meaning by itself}}
\let\page@C\pageno
\let\page@P\empty
\let\page@Q\empty
\def\page@S#1{#1\/}
\def\page@F{\rm}
\def\page@N{\arabic}   
\newif\ifindexing@
\def\indexfile{\ifindexing@\else
 \alloc@@7\write\chardef\sixt@@n\ndx@
 \immediate\openout\ndx@=\jobname.ndx
 \global\indexing@true\fi}
\global\advance\insc@unt\m@ne
\ch@ck0\insc@unt\count
\ch@ck1\insc@unt\dimen
\ch@ck2\insc@unt\skip
\ch@ck4\insc@unt\box
\allocationnumber\insc@unt
\global\chardef\margin@\allocationnumber
\dimen\margin@\maxdimen
\count\margin@\z@
\skip\margin@\z@
\newif\ifindexproofing@
\def\indexproofing{\indexproofing@true}
\def\noindexproofing{\indexproofing@false}
\def\unmacro@#1:#2->#3\unmacro@{\def\macpar@{#2}\def\macdef@{#3}}
\def\starparts@#1{\def\stari@{#1}\def\starii@{#1}\let\stariii@\empty
 \test@false
 \DN@##1*##2##3\next@{\ifx\starparts@##2\test@false\else\test@true\fi}%
 \next@#1*\starparts@\next@
 \iftest@\DN@{\starparts@@#1\starparts@@}\else\let\next@\relax\fi\next@}
\def\starparts@@#1*#2\starparts@@{\def\starii@{#1}\def\stariii@{*#2}}
\def\windex@{\ifindexing@
 \expandafter\unmacro@\meaning\stari@\unmacro@
 \edef\macdef@{\string"\macdef@\string"}%
 \edef\next@{\write\ndx@{\macdef@}}\next@
 \write\ndx@{{\number\pageno}{\page@N}{\page@P}{\page@Q}}%
 \fi
 \ifindexproofing@
  \ifx\stariii@\empty\else
   \expandafter\unmacro@\meaning\stariii@\unmacro@\fi
  \insert\margin@{\hbox{\rm\vrule\height9\p@\depth2\p@\width\z@\starii@
  \ifx\stariii@\empty\else\tt\macdef@\fi}}\fi}
\catcode`\"=\active
\def"{\FN@\quote@}
\def\quote@{\ifx\next"\expandafter\quote@@\else\expandafter\quote@@@\fi}
\def\quote@@@#1"{\starparts@{#1}\starii@\windex@}
\def\quote@@"#1"{\prevanish@\starparts@{#1}\windex@\FN@\quote@@@@}
\def\quote@@@@{\ifx\next"\DN@"{\postvanish@}\else
 \let\next@\postvanish@\fi\next@}
\rightadd@"\to\vanishlist@
\def\idefine#1{\DN@{#1}\DNii@{\noexpand#1}%
 \afterassignment\idefine@\def\nextiii@}
\def\idefine@{\ifindexing@
 \expandafter\let\next@\nextiii@
 \expandafter\unmacro@\meaning\nextiii@\unmacro@
 \immediate\write\ndx@{\noexpand\define\nextii@\macpar@{\macdef@}}\fi}
\def\iabbrev*#1#2{\ifindexing@\toks@{#2}%
 \immediate\write\ndx@{\noexpand\abbrev*\noexpand#1{\the\toks@}}\fi}
\newread\laxread@
\newwrite\laxwrite@
\let\fnpages@\empty
\def\Finit@#1#2\Finit@{\let\nextii@#1\def\nextiii@{#2}}
\catcode`\~=11
\def\getparts@ @#1~#2~#3~#4~#5~#6{\def\nextiv@{#1}%
 \def\nextiii@{#2~#3~#4~#5~}\count@#6\relax}
\newif\ifdocument@
\def\document{\ifdocument@\else\global\document@true
 \let\fontlist@\empty
 \immediate\openin\laxread@=\jobname.lax\relax
 {\endlinechar\m@ne\noexpands@\catcode`\@=11 \catcode`\~=11
  \loop\ifeof\laxread@\else
   \read\laxread@ to\next@
   \ifx\next@\empty
   \else
    \expandafter\Finit@\next@\Finit@
    \if\nextii@ F%
     \expandafter\rightadd@\nextiii@\to\fnpages@
    \else
     \expandafter\getparts@\next@
     \edef\next@{\gdef\csname\nextiv@ @L\endcsname{\nextiii@\number\count@}}%
     \next@
    \fi
   \fi
  \repeat}%
 \immediate\closein\laxread@
 \immediate\openout\laxwrite@=\jobname.lax\relax\fi}
\let\thelabel@\relax
\def\thelabels@{\thelabel@ ~\thelabel@@ ~\thelabel@@@ ~\thelabel@@@@ ~}
\def\label#1{\prevanish@
 \ifx\thelabel@\relax
  \Err@{There's nothing here to be labelled}%
 \else
  {\noexpands@
  \expandafter\ifx\csname#1@L\endcsname\relax
   \expandafter\xdef\csname#1@L\endcsname{\thelabels@0}%
   \immediate\write\laxwrite@{@#1~\thelabels@1}%
  \else
   \edef\next@{@~\csname#1@L\endcsname}%
    \expandafter\getparts@\next@
    \ifodd\count@
    \expandafter\xdef\csname#1@L\endcsname{\thelabels@0}%
    \immediate\write\laxwrite@{@#1~\thelabels@1}%
   \else
    \Err@{Label #1 already used}%
   \fi
  \fi
  }%
 \fi
 \postvanish@}
\rightadd@\label\to\vanishlist@
\def\thepages@{\page@N{\number\page@C}~%
 \page@S{\page@P\page@N{\number\page@C}\page@Q}~%
 \number\page@C ~\page@P\page@N{\number\page@C}\page@Q ~}
\def\pagelabel#1{\prevanish@
 \expandafter\ifx\csname#1@L\endcsname\relax
  {\noexpands@
  \expandafter\xdef\csname#1@L\endcsname{\thepages@2}}%
  \write\laxwrite@{@#1~\thepages@3}%
 \else
  {\noexpands@
  \edef\next@{@~\csname#1@L\endcsname}%
  \expandafter\getparts@\next@
  \ifodd\count@
   \ifnum\count@=\@ne
    \expandafter\xdef\csname#1@L\endcsname{\thelabels@2}%
   \fi
   \write\laxwrite@{@#1~\thepages@3}%
  \else
   \Err@{Label #1 already used}%
  \fi
  }%
 \fi
 \postvanish@}
\rightadd@\pagelabel\to\vanishlist@
\newif\ifreferr@
\referr@true
\def\RefErrors{\global\referr@true}
\def\RefWarnings{\global\referr@false}
\setbox\z@\hbox{\global\count@=`^^30}
\ifnum\count@=48 \let\versionthree@\relax\fi
\def\nolabel@#1#2#3{\expandafter\ifx\csname#2@L\endcsname\relax
 \ifreferr@\Err@{No \noexpand\label found for #2}\else
 \W@{Warning: No \noexpand\label found for #2.}%
 \ifx\versionthree@\relax\W@{l.\number\inputlineno\space ... \string#1{#2}}\fi
 \fi#3\else}
\def\csL@#1{{\noexpands@\xdef\Next@{\csname#1@L\endcsname}}}
\def\ref#1{\nolabel@\ref{#1}\relax
 \DNii@##1~##2\nextii@{##1}%
 \csL@{#1}\expandafter\nextii@\Next@\nextii@\fi}
\def\Ref#1{\nolabel@\Ref{#1}\relax
 \DNii@##1~##2~##3\nextii@{##2}%
 \csL@{#1}\expandafter\nextii@\Next@\nextii@\fi}
\def\nref#1{\nolabel@\nref{#1}\relax
 \DNii@##1~##2~##3~##4\nextii@{##3}%
 \csL@{#1}\expandafter\nextii@\Next@\nextii@\fi}
\def\pref#1{\nolabel@\pref{#1}\relax
 \DNii@##1~##2~##3~##4~##5\nextii@{##4}%
 \csL@{#1}\expandafter\nextii@\Next@\nextii@\fi}
\let\pref@\pref
\def\Evaluatenref#1{\nolabel@\Evaluatenref{#1}{\gdef\Nref{-10000 }}%
 \DNii@##1~##2~##3~##4\nextii@{\DNii@{##3}}%
 \csL@{#1}\expandafter\nextii@\Next@\nextii@
 \xdef\Nref{\nextii@}\fi}
\def\Evaluatepref#1{\nolabel@\Evaluatepref{#1}{\global\let\Pref\empty}%
 \DNii@##1~##2~##3~##4~##5\nextii@{\DNii@{##4}}%
 \csL@{#1}\expandafter\nextii@\Next@\nextii@
 \xdef\Pref{\nextii@}\fi}
\def\readlax#1{\immediate\openin\laxread@=#1.lax\relax
 \ifeof\laxread@\W@{}\W@{File #1.lax not found.}\W@{}\fi
 {\endlinechar\m@ne\noexpands@\catcode`\@=11 \catcode`\~=11
  \loop\ifeof\laxread@\else
   \read\laxread@ to\nextv@
   \ifx\nextv@\empty
   \else
    \expandafter\Finit@\nextv@\Finit@
    \ifx\nextii@ F%
    \else
     \expandafter\getparts@\nextv@
     \expandafter\ifx\csname\nextiv@ @L\endcsname\relax
      \edef\next@{\gdef\csname\nextiv@ @L\endcsname
       {\nextiii@\ifnum\count@=\@ne0\else2\fi}}%
      \next@
     \else
      \Err@{Label \nextiv@\space in #1.lax already used}%
     \fi
    \fi
   \fi
  \repeat}%
 \immediate\closein\laxread@}
\catcode`\~=\active

\def\FNSSP@{\FNSS@\pretendspace@}
\everydisplay{\csname displaymath \endcsname}
\expandafter\def\csname displaymath \endcsname#1$${#1$$\FNSSP@}
\def\locallabel@{\let\thelabel@\Thelabel@\let\thelabel@@\Thelabel@@
 \let\thelabel@@@\Thelabel@@@\let\thelabel@@@@\Thelabel@@@@}
\newcount\tag@C
\tag@C\z@
\let\tag@P\empty
\let\tag@Q\empty
\def\tag@S#1{{\rm(}{#1\/}{\rm)}}
\let\tag@N\arabic
\def\tag@F{\rm}
\def\maketag@{\FN@\maketag@@}
\def\maketag@@{\ifx\next\relax\DN@\relax{\FN@\maketag@@}\else
 \ifx\next"\let\next@\maketag@@@\else
 \let\next@\maketag@@@@\fi\fi\next@}
\def\xdefThelabel@#1{\xdef\Thelabel@{#1{\Thelabel@@@}}}
\def\xdefThelabel@@#1{\xdef\Thelabel@@{#1{\Thelabel@@@@}}}
\def\maketag@@@@#1\maketag@{\global\advance\tag@C\@ne
 {\noexpands@
  \xdef\Thelabel@@@{\number\tag@C}%
  \xdefThelabel@\tag@N
  \xdef\Thelabel@@@@{\ifmathtags@$\tag@P\Thelabel@\tag@Q$\else
   \tag@P\Thelabel@\tag@Q\fi}%
  \xdefThelabel@@\tag@S
  }%
 \locallabel@
 \hbox{\tag@F\thelabel@@}%
 #1}
\def\Qlabel@#1{{\noexpands@\xdef\Thelabel@@{#1}%
 \let\style\empty\xdef\Thelabel@@@@{#1}%
 \let\pre\empty\let\post\empty\xdef\Thelabel@{#1}%
 \let\numstyle\empty\xdef\Thelabel@@@{#1}}}
\def\maketag@@@"#1"#2\maketag@{%
 {\let\pre\tag@P\let\post\tag@Q\let\style\tag@S\let\numstyle\tag@N
  \hbox{\tag@F#1}%
  \noexpands@
  \Qlabel@{#1}%
  }%
 \locallabel@
 #2}
\def\align@{\inalign@true\inany@true
 \vspace@\allowdisplaybreak@\displaybreak@\intertext@
 \def\tag{\global\tag@true\ifnum\and@=\z@
  \DN@{&\omit\global\rwidth@\z@&\relax}\else
  \DN@{&\relax}\fi\next@}%
 \iftagsleft@\DN@{\csname align \endcsname}\else
  \DN@{\csname align \space\endcsname}\fi\next@}
\def\noset@{\def\Offset##1##2{\prevanish@\postvanish@}%
 \def\Reset##1##2{\prevanish@\postvanish@}}
\def\measure@#1\endalign{\global\lwidth@\z@\global\rwidth@\z@
 \global\maxlwidth@\z@\global\maxrwidth@\z@
 \global\and@\z@
 \setbox\z@\vbox
  {\noset@\everycr{\noalign{\global\tag@false\global\and@\z@}}\Let@
  \halign{\setboxz@h{$\m@th\displaystyle{\@lign##}$}%
   \global\lwidth@\wdz@
   \ifdim\lwidth@>\maxlwidth@\global\maxlwidth@\lwidth@\fi
   \global\advance\and@\@ne
   &\setboxz@h{$\m@th\displaystyle{{}\@lign##}$}\global\rwidth@\wdz@
   \ifdim\rwidth@>\maxrwidth@\global\maxrwidth@\rwidth@\fi
   \global\advance\and@\@ne
   &\Tag@\eat@{##}\crcr#1\crcr}}%
 \totwidth@\maxlwidth@\advance\totwidth@\maxrwidth@}
\def\prepost@{\global\let\tag@P@\tag@P\global\let\tag@Q@\tag@Q}
\def\reprepost@{\let\tag@P\tag@P@\let\tag@Q\tag@Q@}
\expandafter\def\csname align \space\endcsname#1\endalign
 {\measure@#1\endalign\global\and@\z@
 \ifingather@\everycr{\noalign{\global\and@\z@}}\else\displ@y@\fi
 \Let@\tabskip\centering@
 \halign to\displaywidth
  {\hfil\strut@\setboxz@h{$\m@th\displaystyle{\@lign##\prepost@}$}%
  \boxz@\global\advance\and@\@ne
  \tabskip\z@skip
  &\setboxz@h{$\m@th\displaystyle{{}\@lign##\prepost@}$}%
  \global\rwidth@\wdz@\boxz@\hfil\global\advance\and@\@ne
  \tabskip\centering@
  &\setboxz@h{\@lign\strut@\reprepost@\maketag@##\maketag@}%
  \dimen@\displaywidth\advance\dimen@-\totwidth@
  \divide\dimen@\tw@\advance\dimen@\maxrwidth@\advance\dimen@-\rwidth@
  \ifdim\dimen@<\tw@\wdz@\llap{\vtop{\normalbaselines\null\boxz@}}%
  \else\llap{\boxz@}\fi
  \tabskip\z@skip
  \crcr#1\crcr
  \black@\totwidth@}}
\expandafter\def\csname align \endcsname#1\endalign{\measure@#1\endalign
 \global\and@\z@
 \ifdim\totwidth@>\displaywidth\let\displaywidth@\totwidth@\else
  \let\displaywidth@\displaywidth\fi
 \ifingather@\everycr{\noalign{\global\and@\z@}}\else\displ@y@\fi
 \Let@\tabskip\centering@\halign to\displaywidth
  {\hfil\strut@\setboxz@h{$\m@th\displaystyle{\@lign##\prepost@}$}%
  \global\lwidth@\wdz@\global\lineht@\ht\z@
  \boxz@\global\advance\and@\@ne
  \tabskip\z@skip&\setboxz@h{$\m@th\displaystyle{{}\@lign##\prepost@}$}%
  \ifdim\ht\z@>\lineht@\global\lineht@\ht\z@\fi
  \boxz@\hfil\global\advance\and@\@ne
  \tabskip\centering@&\kern-\displaywidth@
  \setboxz@h{\@lign\strut@\reprepost@\maketag@##\maketag@}%
  \dimen@\displaywidth\advance\dimen@-\totwidth@
  \divide\dimen@\tw@\advance\dimen@\maxlwidth@\advance\dimen@-\lwidth@
  \ifdim\dimen@<\tw@\wdz@
   \rlap{\vbox{\normalbaselines\boxz@\vbox to\lineht@{}}}\else
   \rlap{\boxz@}\fi
  \tabskip\displaywidth@\crcr#1\crcr\black@\totwidth@}}
\def\attag@#1{\let\Maketag@\maketag@\let\TAG@\Tag@
 \let\Prepost@\prepost@\let\Reprepost@\reprepost@
 \let\Tag@\relax\let\maketag@\relax
 \let\prepost@\relax\let\reprepost@\relax
 \ifmeasuring@
  \def\llap@##1{\setboxz@h{##1}\hbox to\tw@\wdz@{}}%
  \def\rlap@##1{\setboxz@h{##1}\hbox to\tw@\wdz@{}}%
 \else\let\llap@\llap\let\rlap@\rlap\fi
 \toks@{\hfil\strut@
  $\m@th\displaystyle{\@lign\the\hashtoks@\prepost@}$%
  \tabskip\z@skip\global\advance\and@\@ne&
  $\m@th\displaystyle{{}\@lign\the\hashtoks@\prepost@}$\hfil
  \ifxat@\tabskip\centering@\fi\global\advance\and@\@ne}%
 \iftagsleft@
  \toks@@{\tabskip\centering@&\Tag@\kern-\displaywidth
   \rlap@{\@lign\reprepost@\maketag@\the\hashtoks@\maketag@}%
   \global\advance\and@\@ne\tabskip\displaywidth}\else
  \toks@@{\tabskip\centering@&\Tag@\llap@{\@lign\reprepost@\maketag@
   \the\hashtoks@\maketag@}\global\advance\and@\@ne\tabskip\z@skip}\fi
 \atcount@#1\relax\advance\atcount@\m@ne
 \loop\ifnum\atcount@>\z@
  \toks@\expandafter{\the\toks@&\hfil$\m@th\displaystyle{\@lign
  \the\hashtoks@\prepost@}$\global\advance\and@\@ne
  \tabskip\z@skip
  &$\m@th\displaystyle{{}\@lign\the\hashtoks@\prepost@}$\hfil\ifxat@
  \tabskip\centering@\fi\global\advance\and@\@ne}\advance\atcount@\m@ne
 \repeat
 \edef\preamble@{\the\toks@\the\toks@@}%
 \edef\preamble@@{\preamble@}%
 \let\maketag@\Maketag@\let\Tag@\TAG@
 \let\prepost@\Prepost@\let\reprepost@\Reprepost@}
\def\unlabel@{\def\label##1{\prevanish@\postvanish@}%
 \def\pagelabel##1{\prevanish@\postvanish@}}
\newcount\tag@CC
\expandafter\def\csname alignat \endcsname#1#2\endalignat
 {\inany@true\xat@false
 \def\tag{\global\tag@true
  \count@#1\relax\multiply\count@\tw@\advance\count@\m@ne
  \gdef\tag@{&}%
  \loop\ifnum\count@>\and@\xdef\tag@{&\omit\tag@}%
  \advance\count@\m@ne\repeat
  \tag@\relax}%
 \vspace@\allowdisplaybreak@\displaybreak@\intertext@
 \displ@y@\measuring@true\tag@CC\tag@C
 \setbox\savealignat@\hbox{\noset@\unlabel@$\m@th\displaystyle\Let@
  \attag@{#1}\vbox{\halign{\span\preamble@@\crcr#2\crcr}}$}%
 \measuring@false
 \Let@\attag@{#1}\tag@C\tag@CC
 \tabskip\centering@\halign to\displaywidth
  {\span\preamble@@\crcr#2\crcr\black@{\wd\savealignat@}}}
\expandafter\def\csname xalignat \endcsname#1#2\endxalignat
 {\inany@true\xat@true
 \def\tag{\global\tag@true
  \count@#1\relax\multiply\count@\tw@\advance\count@\m@ne
  \gdef\tag@{&}%
  \loop\ifnum\count@>\and@\xdef\tag@{&\omit\tag@}%
  \advance\count@\m@ne\repeat
  \tag@\relax}%
 \vspace@\allowdisplaybreak@\displaybreak@\intertext@
 \displ@y@\measuring@true\tag@CC\tag@C
 \setbox\savealignat@\hbox{\noset@\unlabel@$\m@th\displaystyle\Let@
  \attag@{#1}\vbox{\halign{\span\preamble@@\crcr#2\crcr}}$}%
 \measuring@false\Let@\attag@{#1}\tag@C\tag@CC
 \tabskip\centering@\halign to\displaywidth
 {\span\preamble@@\crcr#2\crcr\black@{\wd\savealignat@}}}
\def\gather{\RIfMIfI@\DN@{\onlydmatherr@\gather}\else
 \ingather@true\inany@true\def\tag{&\relax}%
 \vspace@\allowdisplaybreak@\displaybreak@\intertext@
 \displ@y\Let@
 \iftagsleft@\DN@{\csname gather \endcsname}\else
  \DN@{\csname gather \space\endcsname}\fi\fi
 \else\DN@{\onlydmatherr@\gather}\fi\next@}
\def\exstring@{\expandafter\eat@\string}
\def\newcounter#1{\define#1{}%
 \edef\next@{\def\noexpand#1{\futurelet\noexpand\next
  \csname\exstring@#1@Z\endcsname}}\next@
 \edef\next@{\def\csname\exstring@#1@Z\endcsname
  {\global\advance\csname\exstring@#1@C\endcsname\@ne
  {\csname\exstring@#1@F\endcsname\csname\exstring@#1@S\endcsname
   {\csname\exstring@#1@P\endcsname\csname\exstring@#1@N\endcsname
   {\noexpand\number\csname\exstring@#1@C\endcsname}%
   \csname\exstring@#1@Q\endcsname}}%
  \noexpand\ifx\noexpand\next\noexpand\label
   \def\noexpand\next@\noexpand\label########1{{\noexpand\noexpands@
    \xdef\noexpand\Thelabel@{\csname\exstring@#1@N\endcsname
     {\noexpand\number\csname\exstring@#1@C\endcsname}}%
    \xdef\noexpand\Thelabel@@@{\noexpand\number
     \csname\exstring@#1@C\endcsname}%
    \xdef\noexpand\Thelabel@@{\csname\exstring@#1@S\endcsname
     {\csname\exstring@#1@P\endcsname
     \csname\exstring@#1@N\endcsname
     {\noexpand\number\csname\exstring@#1@C\endcsname}%
     \csname\exstring@#1@Q\endcsname}}%
    \xdef\noexpand\Thelabel@@@@{\csname\exstring@#1@P\endcsname
     \csname\exstring@#1@N\endcsname
     {\noexpand\number\csname\exstring@#1@C\endcsname}%
     \csname\exstring@#1@Q\endcsname}}%
    {\noexpand\locallabel@\noexpand\label{########1}}}%
   \noexpand\else\let\noexpand\next@\relax\noexpand\fi\noexpand\next@}}\next@
 \expandafter\newcount@\csname\exstring@#1@C\endcsname
 \expandafter\let\csname\exstring@#1@N\endcsname\arabic
 \expandafter\def\csname\exstring@#1@S\endcsname##1{##1\/}%
 \expandafter\let\csname\exstring@#1@P\endcsname\empty
 \expandafter\let\csname\exstring@#1@Q\endcsname\empty
 \expandafter\def\csname\exstring@#1@F\endcsname{\rm}%
 }
\def\HASH@#1#2{\ifnum#2=\z@\else
 \edef\next@{\toks@{\the\toks@\the\hashtoks@#2}%
 \toks@@{\the\toks@@{\the\hashtoks@#2}}}\next@\expandafter\HASH@\fi}
\def\HASH@@{\toks@{}\toks@@{}\expandafter\HASH@\macpar@00}
\def\usecounter#1#2{\expandafter\ifx\csname\exstring@#1@Z\endcsname
 \relax\Err@{\noexpand#1not created with \string\newcounter}\fi
 \expandafter\let\csname\exstring@#1@@Z\endcsname\relax
 \expandafter\let\csname\exstring@#1@@Z@\endcsname\relax
 \expandafter\let\csname\exstring@#1@@Z@@\endcsname\relax
 \edef\next@{\def\noexpand#2{\futurelet\noexpand\next
  \csname\exstring@#1@@Z\endcsname}}\next@
 \edef\next@{\def\csname\exstring@#1@@Z\endcsname{\noexpand\ifx
  \noexpand\next\noexpand\label\def\noexpand\next@\noexpand\label
   ########1{\csname\exstring@#1@@Z@\endcsname
   {\noexpand#1\noexpand\label{########1}}}%
   \noexpand\else\noexpand\ifx\noexpand\next
   \noexpand"\def\noexpand\next@\noexpand"########1\noexpand"%
   {\csname\exstring@#1@@Z@\endcsname{{\expandafter\noexpand
   \csname\exstring@#1@F\endcsname
   \let\noexpand\pre\expandafter\noexpand\csname\exstring@#1@P\endcsname
   \let\noexpand\post\expandafter\noexpand\csname\exstring@#1@Q\endcsname
   \let\noexpand\style\expandafter\noexpand\csname\exstring@#1@S\endcsname
   \let\noexpand\numstyle\expandafter\noexpand\csname\exstring@#1@N\endcsname
   ########1}}}\noexpand\else
   \def\noexpand\next@{\csname\exstring@#1@@Z@\endcsname{\noexpand#1}}%
   \noexpand\fi\noexpand\fi\noexpand\next@}}\next@
 \def\next@{\expandafter\expandafter\expandafter\unmacro@\expandafter
  \meaning\csname\exstring@#1@@Z@@\endcsname\unmacro@
  \HASH@@
  \edef\next@{\def\csname\exstring@#1@@Z@\endcsname\the\toks@{%
   \expandafter\noexpand\csname\exstring@#1@@Z@@\endcsname\the\toks@@
   \noexpand\FNSSP@}}\next@}%
 \afterassignment\next@
 \expandafter\def\csname\exstring@#1@@Z@@\endcsname}
\def\listbi@{\penalty50 \medskip}
\def\listbii@{\penalty100 \smallskip}
\let\listbiii@\relax
\let\listbiv@\relax
\let\listbv@\relax
\def\listmi@{\advance\leftskip30\p@\relax}
\let\listmii@\listmi@
\let\listmiii@\listmi@
\let\listmiv@\listmi@
\let\listmv@\listmi@
\def\itemi@#1{\noindent@@\llap{#1\hskip5\p@}}
\let\itemii@\itemi@
\let\itemiii@\itemi@
\let\itemiv@\itemi@
\let\itemv@\itemi@
\def\liste@{\penalty-50 \medskip}
\def\listei@{\penalty-100 \smallskip}
\let\listeii@\relax
\let\listeiii@\relax
\let\listeiv@\relax
\expandafter\newcount\csname list@C1\endcsname
\csname list@C1\endcsname\z@
\expandafter\newcount\csname list@C2\endcsname
\csname list@C2\endcsname\z@
\expandafter\newcount\csname list@C3\endcsname
\csname list@C3\endcsname\z@
\expandafter\newcount\csname list@C4\endcsname
\csname list@C4\endcsname\z@
\expandafter\newcount\csname list@C5\endcsname
\csname list@C5\endcsname\z@
\expandafter\let\csname list@P1\endcsname\empty
\expandafter\let\csname list@P2\endcsname\empty
\expandafter\let\csname list@P3\endcsname\empty
\expandafter\let\csname list@P4\endcsname\empty
\expandafter\let\csname list@P5\endcsname\empty
\expandafter\let\csname list@Q1\endcsname\empty
\expandafter\let\csname list@Q2\endcsname\empty
\expandafter\let\csname list@Q3\endcsname\empty
\expandafter\let\csname list@Q4\endcsname\empty
\expandafter\let\csname list@Q5\endcsname\empty
\expandafter\def\csname list@S1\endcsname#1{{\rm(}{#1\/}{\rm)}}
\expandafter\def\csname list@S2\endcsname#1{{\rm(}{#1\/}{\rm)}}
\expandafter\def\csname list@S3\endcsname#1{{\rm(}{#1\/}{\rm)}}
\expandafter\def\csname list@S4\endcsname#1{{\rm(}{#1\/}{\rm)}}
\expandafter\def\csname list@S5\endcsname#1{{\rm(}{#1\/}{\rm)}}
\expandafter\let\csname list@N1\endcsname\arabic
\expandafter\let\csname list@N2\endcsname\arabic
\expandafter\let\csname list@N3\endcsname\arabic
\expandafter\let\csname list@N4\endcsname\arabic
\expandafter\let\csname list@N5\endcsname\arabic
\expandafter\def\csname list@F1\endcsname{\rm}
\expandafter\def\csname list@F2\endcsname{\rm}
\expandafter\def\csname list@F3\endcsname{\rm}
\expandafter\def\csname list@F4\endcsname{\rm}
\expandafter\def\csname list@F5\endcsname{\rm}
\newcount\listlevel@
\listlevel@\z@
\def\list@@C{\csname list@C\number\listlevel@\endcsname}
\def\list@@P{\csname list@P\number\listlevel@\endcsname}
\def\list@@Q{\csname list@Q\number\listlevel@\endcsname}
\def\list@@S{\csname list@S\number\listlevel@\endcsname}
\def\list@@N{\csname list@N\number\listlevel@\endcsname}
\def\list@@F{\csname list@F\number\listlevel@\endcsname}
\newif\iffirstitemi@
\newif\iffirstitemii@
\newif\iffirstitemiii@
\newif\iffirstitemiv@
\newif\iffirstitemv@
\def\Firstitem@true{\csname firstitem\romannumeral\listlevel@
 @true\endcsname}
\def\Firstitem@false{\csname firstitem\romannumeral\listlevel@
 @false\endcsname}
\def\Listm@{\csname listm\romannumeral\listlevel@ @\endcsname}
\def\Item@{\csname item\romannumeral\listlevel@ @\endcsname}
\def\Liste@{\csname liste\romannumeral\listlevel@ @\endcsname}
\newif\iflistcontinue@
\def\keepitem{\listcontinue@true}
\newcount\list@C@
\def\list{%
 \iflistcontinue@\csname list@C1\endcsname\csname list@C@\endcsname\fi
 \global\csname list@C2\endcsname\z@
 \global\csname list@C3\endcsname\z@
 \global\csname list@C4\endcsname\z@
 \global\csname list@C5\endcsname\z@
 \begingroup
 \firstitemi@true
 \listlevel@\@ne
 \def\item{\FN@\item@}%
 \FN@\list@}
\Invalid@\runinitem
\def\list@{\ifx\next\par
 \DN@\par{\FN@\list@}\else
 \ifx\next\runinitem
  \DN@\runinitem{\FN@\runinitem@}\else
  \DN@{\par\dimen@\parskip\parskip\dimen@}\fi\fi\next@}
\newif\ifoutlevel@
\newif\ifrunin@
\def\item@{%
 \ifoutlevel@\Liste@\outlevel@false\fi
 \ifrunin@\runin@false\par
  \dimen@\parskip\parskip\dimen@
  \Listm@\fi
 \iffirstitemi@\listbi@\listmi@\firstitemi@false\else\par\fi
 \iffirstitemii@\listbii@\listmii@\firstitemii@false\else\par\fi
 \iffirstitemiii@\listbiii@\listmiii@\firstitemiii@false\else\par\fi
 \iffirstitemiv@\listbiv@\listmiv@\firstitemiv@false\else\par\fi
 \iffirstitemv@\listbv@\listmv@\firstitemv@false\else\par\fi
 \DN@"##1"{{\let\pre\list@@P\let\post\list@@Q
  \let\style\list@@S\let\numstyle\list@@N
  \vskip-\parskip
  \Item@{\list@@F##1}%
  \noexpands@
  \Qlabel@{##1}}%
  \locallabel@
  \FNSSP@}%
 \DNii@{\global\advance\list@@C\@ne
  {\noexpands@
   \xdef\Thelabel@@@{\number\list@@C}%
   \xdefThelabel@\list@@N
   \xdef\Thelabel@@@@{\list@@P\Thelabel@\list@@Q}%
   \xdefThelabel@@\list@@S
  }%
  \locallabel@
  \vskip-\parskip
  \Item@{\list@@F\thelabel@@}%
  \FN@\pretendspace@}%
 \ifx\next"\expandafter\next@\else\expandafter\nextii@\fi}
\def\runinitem@{%
  \runin@true
  \Firstitem@false
  \DN@"##1"{{\let\pre\list@@P\let\post\list@@Q
   \let\style\list@@S\let\numstyle\list@@N
   \unskip\space{\list@@F##1} %
   \noexpands@
   \Qlabel@{##1}}%
   \locallabel@
   \ignorespaces}%
  \DNii@{\global\advance\list@@C\@ne
   {\noexpands@
    \xdef\Thelabel@@@{\number\list@@C}%
    \xdefThelabel@\list@@N
    \xdef\Thelabel@@@@{\list@@P\Thelabel@\list@@Q}%
    \xdefThelabel@@\list@@S
   }%
   \locallabel@
   \unskip\space{\list@@F\thelabel@@} }%
  \ifx\next"\expandafter\next@\else\expandafter\nextii@\fi}
\def\inlevel{\ifnum\listlevel@=5
 \DN@{\Err@{Already 5 levels down}}\else
 \DN@{\begingroup\advance\listlevel@\@ne
 \Firstitem@true\FN@\inlevel@}\fi\next@}
\def\inlevel@{\ifx\next\par
 \DN@\par{\FN@\inlevel@}\else
 \ifx\next\runinitem
  \DN@\runinitem{\FN@\runinitem@}\else
  \let\next@\relax\fi\fi\next@}
\def\outlevel{\ifnum\listlevel@=\@ne
 \Err@{At top level}\else
 \par\global\list@@C\z@\endgroup\outlevel@true\fi}
\def\endlist{%
 \expandafter\global\csname list@C@\endcsname\csname list@C1\endcsname
 \par
 \global\toks\@ne{}\count@\listlevel@
 {\loop
  \ifnum\count@>\z@\global\toks\@ne\expandafter{\the\toks\@ne\endgroup}%
  \advance\count@\m@ne
  \repeat}%
 \the\toks\@ne
 \liste@
 \listcontinue@false\global\csname list@C1\endcsname\z@
 \vskip-\parskip
 \noindent@@
 \FN@\pretendspace@}
\newif\iffirstdescribe@
\def\describe{\par
 \begingroup\firstdescribe@true
 \def\item##1{%
  \iffirstdescribe@\penalty50 \medskip\vskip-\parskip
  \firstdescribe@false\else\par\fi
  \noindent@@\hangindent2pc\hangafter\@ne
  {\bf##1}\hskip.5em}}

\Invalid@\pullin
\Invalid@\pullinmore
\newif\iffirstpull@
\def\margins{\par\begingroup\firstpull@true
 \def\pullin##1##2{\par
  \iffirstpull@\firstpull@false\else\endgroup\fi
  \begingroup\DN@{##1}%
  \ifx\next@\empty\leftskip\z@\else\ifx\next@\space\leftskip\z@
  \else\leftskip##1\fi\fi
  \DN@{##2}\ifx\next@\empty\rightskip\z@\else\ifx\next@\space
  \rightskip\z@\else\rightskip##2\fi\fi\ignorespaces}%
 \def\pullinmore##1##2{\par
  \xdef\Next@{\leftskip\the\leftskip\relax\rightskip\the\rightskip\relax}%
  \iffirstpull@\firstpull@false\else\endgroup\fi
  \begingroup\Next@
  \DN@{##1}%
  \ifx\next@\empty\else\ifx\next@\space\else\advance\leftskip##1\fi\fi
  \DN@{##2}\ifx\next@\empty\else\ifx\next@\space\else
  \advance\rightskip##2\fi\fi\ignorespaces}}

\newif\ifnopunct@
\newif\ifnospace@
\newif\ifoverlong@
\let\nofrillslist@\empty
\let\overlonglist@\empty
\def\nopunct{\nopunct@true\FN@\nopunct@}
\def\nospace{\nospace@true\FN@\nospace@}
\def\overlong{\overlong@true\FN@\overlong@}
\def\nopunct@{\ifx\next\nospace
 \DN@\nospace{\nospace@true\FN@\nopnos@}\else\ifx\next\overlong
 \DN@\overlong{\overlong@true\FN@\nopol@}\else
 \let\next@\nopunct@@\fi\fi\next@}
\def\nopunct@@#1{\ismember@\nofrillslist@#1%
 \iftest@\let\next@#1\else
 \DN@{\nopunct@false\Err@{\noexpand\nopunct can't be used with
 \string#1}#1}\fi\next@}
\def\nospace@{\ifx\next\nopunct
 \DN@\nopunct{\nopunct@true\FN@\nopnos@}\else\ifx\next\overlong
 \DN@\overlong{\overlong@true\FN@\nosol@}\else
 \let\next@\nospace@@\fi\fi\next@}
\def\nospace@@#1{\ismember@\nofrillslist@#1%
 \iftest@\let\next@#1\else
 \DN@{\nospace@false\Err@{\noexpand\nospace can't be used with
 \string#1}#1}\fi\next@}
\def\overlong@{\ifx\next\nopunct
 \DN@\nopunct{\nopunct@true\FN@\nopol@}\else\ifx\next\nospace
 \DN@\nospace{\nospace@true\FN@\nosol@}\else
 \let\next@\overlong@@\fi\fi\next@}
\def\overlong@@#1{\ismember@\overlonglist@#1%
 \iftest@\let\next@#1\else
 \DN@{\overlong@false\Err@{\noexpand\overlong can't be used with
 \string#1}#1}\fi\next@}
\def\nopnos@{\ifx\next\overlong
 \DN@\overlong{\overlong@true\nopnosol@}\else
 \let\next@\nopnos@@\fi\next@}
\def\nopol@{\ifx\next\nospace
 \DN@\nospace{\nospace@true\nopnosol@}\else
 \let\next@\nopol@@\fi\next@}
\def\nosol@{\ifx\next\nopunct
 \DN@\nopunct{\nopunct@true\nopnosol@}\else
 \let\next@\nosol@@\fi\next@}
\def\nopnos@@#1{\ismember@\nofrillslist@#1%
 \iftest@\let\next@#1\else
 \DN@{\nopunct@false\nospace@false
  \Err@{\noexpand\nopunct\noexpand\nospace
   can't be used with \string#1}#1}\fi\next@}
\def\testii@#1{\ismember@\nofrillslist@#1%
 \iftest@\let\nextiii@ T\else\let\nextiii@ F\fi
 \ismember@\overlonglist@#1%
 \iftest@\let\nextiv@ T\else\let\nextiv@ F\fi
 \test@false\if\nextiii@ T\if\nextiv@ T\test@true\fi\fi}
\def\nopol@@#1{\testii@{#1}%
 \iftest@\let\next@#1%
 \else\DN@{\if\nextiii@ T\else\nopunct@false\fi
  \if\nextiv@ T\else\overlong@false\fi
  \Err@{\if\nextiii@ T\else\noexpand\nopunct\fi
  \if\nextiv@ T\else\noexpand\overlong\fi can't be used
  with \string#1}#1}\fi\next@}
\def\nosol@@#1{\testii@{#1}%
 \iftest@\let\next@#1%
 \else\DN@{\if\nextiii@ T\else\nospace@false\fi
  \if\nextiv@ T\else\overlong@false\fi
  \Err@{\if\nextiii@ T\else\noexpand\nospace\fi
  \if\nextiv@ T\else\noexpand\overlong\fi can't be used
  with \string#1}#1}\fi\next@}
\def\nopnosol@#1{\testii@{#1}%
 \iftest@\let\next@#1%
 \else\DN@{\if\nextiii@ T\else\nopunct@false\nospace@false\fi
  \if\nextiv@ T\else\overlong@false\fi
  \Err@{\if\nextiii@ T\else\noexpand\nopunct\noexpand\nospace\fi
  \if\nextiv@ T\else\noexpand\overlong\fi can't be used
  with \string#1}#1}\fi\next@}
\def\punct@#1{\ifnopunct@\else#1\fi}
\def\addspace@#1{\ifnospace@\else#1\fi}
\def\hss@{\ifoverlong@\z@ plus\@m\p@ minus\@m\p@
 \else \z@ plus\@m\p@\fi}
\rightadd@\demo\to\nofrillslist@
\newif\ifclaim@
\def\exxx@{\expandafter\expandafter\expandafter\eat@\expandafter\string}
\let\colon@:
\def\demo#1{\ifclaim@
 \Err@{Previous \expandafter\noexpand\claimtype@ has
  no matching \string\end\exxx@\claimtype@}%
 \let\next@\relax
 \else
  \par
  \ifdim\lastskip<\smallskipamount\removelastskip\smallskip\fi
  \begingroup
  \noindent@@{\smc\ignorespaces#1\unskip
   \punct@{\null\colon@}\addspace@\enspace}%
  \nopunct@false\nospace@false
  \rm
  \DN@{\FNSSP@}%
 \fi
 \next@}
\def\enddemo{\par\endgroup\nopunct@false\nospace@false\smallskip}
\rightadd@\claim\to\nofrillslist@
\def\claim@F{\smc}
\def\claim@@@F{\csname\exxx@\claimtype@ @F\endcsname}
\def\claimformat@#1#2#3{%
 \medbreak\noindent@@{\smc#1 {\claim@@@F#2} #3%
 \punct@{\null.}\addspace@\enspace}\sl}
\def\claimformat@@#1#2{\claimformat@{\ignorespaces#1\unskip}%
 {\ifx\thelabel@@\empty\unskip\else\thelabel@@\fi}%
 {\ignorespaces#2\unskip}%
 \let\Claimformat@@\claimformat@@\FNSSP@}
\let\Claimformat@@\claimformat@@
\def\claim@@@P{\csname\exxx@\claimtype@ @P\endcsname}
\def\claim@@@Q{\csname\exxx@\claimtype@ @Q\endcsname}
\def\claim@@@S{\csname\exxx@\claimtype@ @S\endcsname}
\def\claim@@@N{\csname\exxx@\claimtype@ @N\endcsname}
\def\claim@@@C{\csname claim@C\claimclass@\endcsname}
\newcount\claim@C
\claim@C\z@
\let\claim@P\empty
\let\claim@Q\empty
\def\claim@S#1{#1\/}
\let\claim@N\arabic
\def\claim{\claim@true\let\claimclass@\empty
 \def\claimtype@{\claim}\FN@\claim@}
\def\claim@{%
 \ifx\next\c
  \let\next@\claim@c
 \else
  \ifx\next"%
   \let\next@\claim@q
  \else
   \begingroup\global\advance\claim@C\@ne
   {\noexpands@
    \xdef\Thelabel@@@{\number\claim@C}%
    \xdefThelabel@\claim@N
    \xdef\Thelabel@@@@{\claim@P\Thelabel@\claim@Q}%
    \xdefThelabel@@\claim@S
   }%
   \locallabel@
   \let\next@\Claimformat@@
  \fi
 \fi
 \next@}
\def\claim@c\c#1{\claim@true\begingroup
 \expandafter
 \ifx\csname claim@C#1\endcsname\relax
  \expandafter\newcount@\csname claim@C#1\endcsname
  \global\csname claim@C#1\endcsname\@ne
 \else
  \global\advance\csname claim@C#1\endcsname\@ne
 \fi
 \def\claimclass@{#1}%
 {\noexpands@
  \xdef\Thelabel@@@{\number\claim@@@C}%
  \xdefThelabel@\claim@@@N
  \xdef\Thelabel@@@@{\claim@@@P\Thelabel@\claim@@@Q}%
  \xdefThelabel@@\claim@@@S
 }%
 \locallabel@
 \FNSS@\claim@c@}
\def\claim@q"#1"{\begingroup
 {\let\pre\claim@@@P\let\post\claim@@@Q
  \let\style\claim@@@S\let\numstyle\claim@@@N
  \noexpands@
  \Qlabel@{#1}}%
 \locallabel@
 \FNSS@\claim@q@}
\def\claim@c@{\ifx\next"%
 \global\advance\claim@@@C\m@ne\let\next@\claim@cq
 \else\let\next@\Claimformat@@\fi\next@}
\def\claim@cq"#1"{{\let\pre\claim@@@P\let\post\claim@@@Q
 \let\style\claim@@@S\let\numstyle\claim@@@N
 \noexpands@
 \Qlabel@{#1}}%
 \locallabel@
 \FNSS@\Claimformat@@}
\def\claim@q@{\ifx\next\c\expandafter\claim@qc
 \else\expandafter\Claimformat@@\fi}
\def\claim@qc\c#1{\expandafter\ifx\csname claim@C#1\endcsname\relax
 \expandafter\newcount@\csname claim@C#1\endcsname
 \global\csname claim@C#1\endcsname\z@\fi
 \FNSS@\Claimformat@@}
\def\endclaim{\endgroup\claim@false\nopunct@false\nospace@false
 \let\Claimformat@@\claimformat@@\medbreak}
\Invalid@\claimclause
\def\newclaim{\FN@\newclaim@}
\def\newclaim@{\ifx\next\claimclause
 \DN@\claimclause##1{\newclaim@@{##1}}\else
 \DN@{\newclaim@@\relax}\fi\next@}
\def\claimlist@{\\\claim}
\newtoks\claim@i
\newtoks\claim@v
\let\noclaimclause@=F
\def\newclaim@@#1#2#3\c#4#5{\define#2{}%
 \rightadd@#2\to\claimlist@\rightadd@#2\to\nofrillslist@%
 \expandafter\def\csname\exstring@#2@P\endcsname{\claim@P}%
 \expandafter\def\csname\exstring@#2@Q\endcsname{\claim@Q}%
 \expandafter\def\csname\exstring@#2@S\endcsname{\claim@S}%
 \expandafter\def\csname\exstring@#2@N\endcsname{\claim@N}%
 \expandafter\def\csname\exstring@#2@F\endcsname{\claim@F}%
 \expandafter\def\csname end\exstring@#2\endcsname{\endclaim}%
 \expandafter\ifx\csname claim@C#4\endcsname\relax
  \expandafter\newcount@\csname claim@C#4\endcsname
  \global\csname claim@C#4\endcsname\z@\fi
 \edef\next@{\let\csname\exstring@#2@C\endcsname
   \csname claim@C#4\endcsname}\next@
 \def#2{\ifx\noclaimclause@ T\else#1\fi
  \global\claim@i{#1}\gdef\claim@iv{#4}\global\claim@v{#5}%
  \def\claimtype@{#2}\def\Claimformat@@{\claimformat@@{#5}}\claim@c\c{#4}}}
\def\shortenclaim#1#2{\define#2{}%
 \ismember@\claimlist@#1%
 \iftest@
  \rightadd@#2\to\nofrillslist@%
  \expandafter\def\csname\exstring@#2@P\endcsname
   {\csname\exstring@#1@P\endcsname}%
  \expandafter\def\csname\exstring@#2@Q\endcsname
   {\csname\exstring@#1@Q\endcsname}%
  \expandafter\def\csname\exstring@#2@S\endcsname
   {\csname\exstring@#1@S\endcsname}%
  \expandafter\def\csname\exstring@#2@N\endcsname
   {\csname\exstring@#1@N\endcsname}%
  \expandafter\def\csname\exstring@#2@F\endcsname
   {\csname\exstring@#1@F\endcsname}%
  \expandafter\def\csname end\exstring@#2\endcsname{\endclaim}%
  \edef\next@{\let\csname\exstring@#2@C\endcsname
    \csname claim\exstring@#1C\endcsname}\next@
  \setbox\z@\vbox{\let\noclaimclause@ T#1""\relax\endgroup}%
  \edef#2{\the\claim@i
   \def\noexpand\claimtype@{\noexpand#2}%
   \def\noexpand\Claimformat@@{\noexpand\claimformat@@{\the\claim@v}\relax}%
   \noexpand\claim@c\noexpand\c{\claim@iv}}%
 \else
  \Err@{\noexpand#1not yet created by \string\newclaim}%
 \fi}
\def\classtest@#1{\DN@{#1}\ifx\next@\claimclass@
 \test@true\else\test@false\fi}
\def\typetest@#1{\DN@{#1}\ifx\next@\claimtype@\test@true\else
  \test@false\fi}
\newif\iftoc@
\def\tocfile{\iftoc@\else\alloc@@7\write\chardef\sixt@@n\toc@
 \immediate\openout\toc@=\jobname.toc
 \alloc@@7\write\chardef\sixt@@n\tic@
 \immediate\openout\tic@=\jobname.tic
 \global\toc@true\fi}
\rightadd@\hl\to\nofrillslist@
\rightadd@\HL\to\overlonglist@
\def\HL@@C{\csname HL@C\HLlevel@\endcsname}
\def\HL@@P{\csname HL@P\HLlevel@\endcsname}
\def\HL@@Q{\csname HL@Q\HLlevel@\endcsname}
\def\HL@@S{\csname HL@S\HLlevel@\endcsname}
\def\HL@@N{\csname HL@N\HLlevel@\endcsname}
\def\HL@@F{\csname HL@F\HLlevel@\endcsname}
\def\HL@@@C{\csname\exxx@\HLtype@ @C\endcsname}
\def\HL@@@P{\csname\exxx@\HLtype@ @P\endcsname}
\def\HL@@@Q{\csname\exxx@\HLtype@ @Q\endcsname}
\def\HL@@@S{\csname\exxx@\HLtype@ @S\endcsname}
\def\HL@@@N{\csname\exxx@\HLtype@ @N\endcsname}
\def\HL#1{\expandafter
 \ifx\csname HL@C#1\endcsname\relax
  \DN@{\Err@{\string\HL#1 not defined in this style}}%
 \else
  \DN@{\gdef\HLlevel@{#1}\def\HLname@{\HL{#1}}\let\HLtype@\relax\FNSS@\HL@}%
 \fi
 \next@}%
\newif\ifquoted@
\let\aftertoc@\relax
\def\HL@{%
 \DN@"##1"##2\endHL{\def\entry@{##2}\quoted@true
  {\noexpands@
  \ifx\HLtype@\relax
   \let\pre\HL@@P\let\post\HL@@Q\let\style\HL@@S\let\numstyle\HL@@N
  \else
   \let\pre\HL@@@P\let\post\HL@@@Q\let\style\HL@@@S\let\numstyle\HL@@@N
  \fi
  \Qlabel@{##1}\let\style\relax\xdef\Qlabel@@@@{##1}%
  \xdef\Thepref@{\Thelabel@@@@}}%
  \csname HL@\HLlevel@\endcsname##2\endHL
  \let\pref\Thepref@
  \csname HL@I\HLlevel@\endcsname
  \csname HL@J\HLlevel@\endcsname
  \let\pref\pref@
  \HLtoc@	
  \aftertoc@
  \let\aftertoc@\relax\overlong@false}%
 \DNii@##1\endHL{\def\entry@{##1}\quoted@false
  {\noexpands@
  \ifx\HLtype@\relax
   \global\advance\HL@@C\@ne
   \xdef\Thelabel@@@{\number\HL@@C}%
   \xdefThelabel@{\HL@@N}%
   \xdef\Thelabel@@@@{\HL@@P\Thelabel@\HL@@Q}%
   \xdefThelabel@@{\HL@@S}%
  \else
   \global\advance\HL@@@C\@ne
   \xdef\Thelabel@@@{\number\HL@@@C}%
   \xdefThelabel@{\HL@@@N}%
   \xdef\Thelabel@@@@{\HL@@@P\Thelabel@\HL@@@Q}%
   \xdefThelabel@@{\HL@@@S}%
  \fi
  \xdef\Thepref@{\Thelabel@@@@}}%
  \csname HL@\HLlevel@\endcsname##1\endHL
  \let\pref\Thepref@
  \csname HL@I\HLlevel@\endcsname
  \csname HL@J\HLlevel@\endcsname
  \let\pref\pref@
  \HLtoc@
  \aftertoc@
  \let\aftertoc@\relax\overlong@false}%
 \ifx\next"\expandafter\next@\else\expandafter\nextii@\fi}%
\Invalid@\endHL
\def\hl@@C{\csname hl@C\hllevel@\endcsname}
\def\hl@@P{\csname hl@P\hllevel@\endcsname}
\def\hl@@Q{\csname hl@Q\hllevel@\endcsname}
\def\hl@@S{\csname hl@S\hllevel@\endcsname}
\def\hl@@N{\csname hl@N\hllevel@\endcsname}
\def\hl@@F{\csname hl@F\hllevel@\endcsname}
\def\hl@@@C{\csname\exxx@\hltype@ @C\endcsname}
\def\hl@@@P{\csname\exxx@\hltype@ @P\endcsname}
\def\hl@@@Q{\csname\exxx@\hltype@ @Q\endcsname}
\def\hl@@@S{\csname\exxx@\hltype@ @S\endcsname}
\def\hl@@@N{\csname\exxx@\hltype@ @N\endcsname}
\def\hl#1{\expandafter
 \ifx\csname hl@C#1\endcsname\relax
  \DN@{\Err@{\string\hl#1 not defined in this style}}%
 \else
  \DN@{\gdef\hllevel@{#1}\def\hlname@{\hl{#1}}\let\hltype@\relax\FNSS@\hl@}%
 \fi
 \next@}
\def\hl@{%
 \DN@"##1"##2{\def\entry@{##2}\quoted@true
  {\noexpands@
  \ifx\hltype@\relax
   \let\pre\hl@@P\let\post\hl@@Q\let\style\hl@@S\let\numstyle\hl@@N
  \else
   \let\pre\hl@@@P\let\post\hl@@@Q\let\style\hl@@@S\let\numstyle\hl@@@N
  \fi
  \Qlabel@{##1}\let\style\relax\xdef\Qlabel@@@@{##1}%
  \xdef\Thepref@{\Thelabel@@@@}}%
  \csname hl@\hllevel@\endcsname{##2}%
  \let\pref\Thepref@
  \csname hl@I\hllevel@\endcsname
  \csname hl@J\hllevel@\endcsname
  \let\pref\pref@
  \hltoc@
  \aftertoc@
  \let\aftertoc@\relax\nopunct@false\nospace@false\FNSSP@}%
 \DNii@##1{\def\entry@{##1}\quoted@false
  {\noexpands@
  \ifx\hltype@\relax
   \global\advance\hl@@C\@ne
   \xdef\Thelabel@@@{\number\hl@@C}%
   \xdefThelabel@{\hl@@N}%
   \xdef\Thelabel@@@@{\hl@@P\Thelabel@\hl@@Q}%
   \xdefThelabel@@{\hl@@S}%
  \else
   \global\advance\hl@@@C\@ne
   \xdef\Thelabel@@@{\number\hl@@@C}%
   \xdefThelabel@{\hl@@@N}%
   \xdef\Thelabel@@@@{\hl@@@P\Thelabel@\hl@@@Q}%
   \xdefThelabel@@{\hl@@@S}%
  \fi
  \xdef\Thepref@{\Thelabel@@@@}}%
  \csname hl@\hllevel@\endcsname{##1}%
  \let\pref\Thepref@
  \csname hl@I\hllevel@\endcsname
  \csname hl@J\hllevel@\endcsname
  \let\pref\pref@
  \hltoc@
  \aftertoc@
  \let\aftertoc@\relax\nopunct@false\nospace@false\FNSSP@}%
 \ifx\next"\expandafter\next@\else\expandafter\nextii@\fi}%
\def\six@#1#2 #3 #4 #5 #6 #7 {\DN@{#2}\ifx\next@\empty
 \DN@##1\six@{}\else
 \write#1{ #2 #3 #4 #5 #6 #7}\DN@{\six@#1}\fi
 \next@}
\def\Sixtoc@{\ifx\macdef@\empty\else
 \DN@##1##2\next@{\def\macdef@{##1##2}}%
 \expandafter\next@\macdef@\next@
 \edef\next@
  {\noexpand\six@\toc@\macdef@
  \space\space\space\space\space\space\space\space\space\space\space\space
  \noexpand\six@}%
 \next@\let\macdef@\relax\fi}
\def\QorThelabel@@@@{\ifquoted@
 \noexpand\noexpand\noexpand"\Qlabel@@@@\noexpand\noexpand\noexpand"\else
 \Thelabel@@@@\fi}
\def\HLtoc@{%
 \iftoc@
 \expandafter\expandafter\expandafter\unmacro@
  \expandafter\meaning\csname HL@W\HLlevel@\endcsname\unmacro@
  {\noexpands@\let\style\relax
   \edef\next@{\write\toc@{\noexpand\noexpand\expandafter\noexpand\HLname@
   {\macdef@}{\QorThelabel@@@@}}}%
  \next@}%
  \expandafter\unmacro@\meaning\entry@\unmacro@
  \Sixtoc@
  \write\toc@{\noexpand\Page{\number\pageno}{\page@N}%
   {\page@P}{\page@Q}^^J}%
 \fi}
\def\hltoc@{%
 \iftoc@
 \expandafter\expandafter\expandafter\unmacro@
  \expandafter\meaning\csname hl@W\hllevel@\endcsname\unmacro@
  {\noexpands@\let\style\relax
  \edef\next@{\write\toc@{%
   \ifnopunct@\noexpand\noexpand\noexpand\nopunct\fi
   \ifnospace@\noexpand\noexpand\noexpand\nospace\fi
   \noexpand\noexpand\expandafter\noexpand\hlname@
   {\macdef@}{\QorThelabel@@@@}}}%
  \next@}%
  \expandafter\unmacro@\meaning\entry@\unmacro@
  \Sixtoc@
  \write\toc@{\noexpand\Page{\number\pageno}{\page@N}%
   {\page@P}{\page@Q}^^J}%
 \fi}
\def\mainfile#1{\def\mainfile@{#1}}
\def\checkmainfile@{\ifx\mainfile@\undefined
 \Err@{No \noexpand\mainfile specified}\fi}
\expandafter\newcount@\csname HL@C1\endcsname
\csname HL@C1\endcsname\z@
\expandafter\def\csname HL@S1\endcsname#1{#1\null.}
\expandafter\let\csname HL@N1\endcsname\arabic
\expandafter\let\csname HL@P1\endcsname\empty
\expandafter\let\csname HL@Q1\endcsname\empty
\expandafter\def\csname HL@F1\endcsname{\bf}
\expandafter\let\csname HL@W1\endcsname\empty
\expandafter\newcount@\csname hl@C1\endcsname
\csname hl@C1\endcsname\z@
\expandafter\def\csname hl@S1\endcsname#1{#1\/}
\expandafter\let\csname hl@N1\endcsname\arabic
\expandafter\let\csname hl@P1\endcsname\empty
\expandafter\let\csname hl@Q1\endcsname\empty
\expandafter\def\csname hl@F1\endcsname{\bf}
\expandafter\let\csname hl@W1\endcsname\empty
\expandafter\def\csname HL@1\endcsname#1\endHL{\bigbreak
 {\locallabel@
  \global\setbox\@ne\vbox{\Let@\tabskip\hss@
  \halign to\hsize{\bf\hfil\ignorespaces##\unskip\hfil\cr
  \expandafter\ifx\csname HL@W1\endcsname\empty\else
   \csname HL@W1\endcsname\space\fi
  {\HL@@F\ifx\thelabel@@\empty\else\thelabel@@\space\fi}%
  \ignorespaces#1\crcr}}%
  }%
 \unvbox\@ne\nobreak\medskip}
\expandafter\def\csname hl@1\endcsname#1{\medbreak\noindent@@
 {\locallabel@
 \bf{\hl@@F\ifx\thelabel@@\empty\else\thelabel@@\space\fi}%
 \ignorespaces#1\unskip\punct@{\null.}\addspace@\enspace}}
\expandafter\def\csname HL@I1\endcsname{\Reset\hl1{1}%
 \ifx\pref\empty\newpre\hl1{}\else\newpre\hl1{\pref.}\fi}
\def\NameHL#1#2{\define#2{}%
 \expandafter\ifx\csname HL@R#1\endcsname\relax
 \else
  \def\nextiv@{\let\nextiii@}%
  \expandafter\nextiv@\csname HL@R#1\endcsname
  \expandafter\let\nextiii@\undefined
  \expandafter\let\csname\exxx@\nextiii@ @C\endcsname\relax
  \expandafter\let\csname\exxx@\nextiii@ @P\endcsname\relax
  \expandafter\let\csname\exxx@\nextiii@ @Q\endcsname\relax
  \expandafter\let\csname\exxx@\nextiii@ @S\endcsname\relax
  \expandafter\let\csname\exxx@\nextiii@ @N\endcsname\relax
  \expandafter\let\csname\exxx@\nextiii@ @F\endcsname\relax
  \expandafter\let\csname\exxx@\nextiii@ @W\endcsname\relax
  \expandafter\let\csname end\exxx@\nextiii@\endcsname\undefined
 \fi
 \expandafter\gdef\csname HL@R#1\endcsname{#2}%
 \expandafter\gdef\csname\exstring@#2@R\endcsname{{HL}{#1}}%
 \iftoc@\write\toc@{\noexpand\NameHL#1\noexpand#2^^J}\fi
 \rightadd@#2\to\overlonglist@
 \edef\next@{\let\csname\exstring@#2@C\endcsname\expandafter\noexpand
  \csname HL@C#1\endcsname}\next@
 \edef\next@{\let\csname\exstring@#2@P\endcsname\expandafter\noexpand
  \csname HL@P#1\endcsname}\next@
 \edef\next@{\let\csname\exstring@#2@Q\endcsname\expandafter\noexpand
  \csname HL@Q#1\endcsname}\next@
 \edef\next@{\let\csname\exstring@#2@S\endcsname\expandafter\noexpand
  \csname HL@S#1\endcsname}\next@
 \edef\next@{\let\csname\exstring@#2@N\endcsname\expandafter\noexpand
  \csname HL@N#1\endcsname}\next@
 \edef\next@{\let\csname\exstring@#2@F\endcsname\expandafter\noexpand
  \csname HL@F#1\endcsname}\next@
 \edef\next@{\let\csname\exstring@#2@W\endcsname\expandafter\noexpand
  \csname HL@W#1\endcsname}\next@
 \edef\next@{\def\noexpand#2####1\expandafter\noexpand
  \csname end\exstring@#2\endcsname
  {\def\noexpand\HLtype@{\noexpand#2}%
   \def\noexpand\HLname@{\noexpand#2}%
   \gdef\noexpand\HLlevel@{#1}%
   \noexpand\FNSS@\noexpand\HL@####1\noexpand\endHL}}%
  \next@
 \edef\next@{\noexpand\Invalid@\expandafter\noexpand
  \csname end\exstring@#2\endcsname}%
 \next@}
\def\Namehl#1#2{\define#2{}%
 \expandafter\ifx\csname hl@R#1\endcsname\relax
 \else
  \def\nextiv@{\let\nextiii@}%
  \expandafter\nextiv@\csname hl@R#1\endcsname
  \expandafter\let\nextiii@\undefined
  \expandafter\let\csname\exxx@\nextiii@ @C\endcsname\relax
  \expandafter\let\csname\exxx@\nextiii@ @P\endcsname\relax
  \expandafter\let\csname\exxx@\nextiii@ @Q\endcsname\relax
  \expandafter\let\csname\exxx@\nextiii@ @S\endcsname\relax
  \expandafter\let\csname\exxx@\nextiii@ @N\endcsname\relax
  \expandafter\let\csname\exxx@\nextiii@ @F\endcsname\relax
  \expandafter\let\csname\exxx@\nextiii@ @W\endcsname\relax
 \fi
 \expandafter\gdef\csname hl@R#1\endcsname{#2}%
 \expandafter\gdef\csname\exstring@#2@R\endcsname{{hl}{#1}}%
 \iftoc@\write\toc@{\noexpand\Namehl#1\noexpand#2^^J}\fi
 \rightadd@#2\to\nofrillslist@%
 \edef\next@{\let\csname\exstring@#2@C\endcsname\expandafter\noexpand
  \csname hl@C#1\endcsname}\next@
 \edef\next@{\let\csname\exstring@#2@P\endcsname\expandafter\noexpand
  \csname hl@P#1\endcsname}\next@
 \edef\next@{\let\csname\exstring@#2@Q\endcsname\expandafter\noexpand
  \csname hl@Q#1\endcsname}\next@
 \edef\next@{\let\csname\exstring@#2@S\endcsname\expandafter\noexpand
  \csname hl@S#1\endcsname}\next@
 \edef\next@{\let\csname\exstring@#2@N\endcsname\expandafter\noexpand
  \csname hl@N#1\endcsname}\next@
 \edef\next@{\let\csname\exstring@#2@F\endcsname\expandafter\noexpand
  \csname hl@F#1\endcsname}\next@
 \edef\next@{\let\csname\exstring@#2@W\endcsname\expandafter\noexpand
  \csname hl@W#1\endcsname}\next@
 \edef\next@{\def\noexpand#2{%
  \def\noexpand\hltype@{\noexpand#2}%
  \def\noexpand\hlname@{\noexpand#2}%
  \gdef\noexpand\hllevel@{#1}%
  \noexpand\FNSS@\noexpand\hl@}}%
 \next@}%
\def\Initialize{\FN@\Init@}
\def\Init@{\ifx\next\HL\let\next@\InitH@\else\ifx\next\hl\let\next@\InitH@
  \else\let\next@\InitS@\fi\fi\next@}
\def\InitH@#1#2{\expandafter\ifx\csname\exstring@#1@C#2\endcsname\relax
 \DN@{\Err@{\noexpand#1level #2 not defined in this style}}\else
 \DN@{\expandafter\gdef\csname\exstring@#1@J#2\endcsname}\fi\next@}
\def\InitC@#1#2{\edef\nextii@{\expandafter\noexpand\csname#1\endcsname{#2}}}
\def\InitS@#1{\expandafter\ifx\csname\exstring@#1@R\endcsname\relax
 \Err@{\noexpand#1not defined in this style}\let\next@\relax\else
 \DN@{\let\next@}\expandafter\next@\csname\exstring@#1@R\endcsname
 \expandafter\InitC@\next@
 \DN@{\expandafter\InitH@\nextii@}\fi\next@}
\def\value#1{\expandafter
 \ifx\csname\exstring@#1@C\endcsname\relax
  \expandafter\ifx\csname\exstring@#1@C1\endcsname\relax
   \DN@{\Err@{\noexpand\value can't be used with \string#1}}%
  \else
   \DN@{\value@#1}%
  \fi
 \else
  \DN@{\number\csname\exstring@#1@C\endcsname\relax}%
 \fi
 \next@}
\def\value@#1#2{\expandafter
 \ifx\csname\exstring@#1@C#2\endcsname\relax
  \DN@{\Err@{\string\value\string#1 can't be followed by \string#2}}%
 \else
  \DN@{\number\csname\exstring@#1@C#2\endcsname\relax}%
 \fi
 \next@}
\newcount\Value
\def\Evaluate#1{\expandafter
 \ifx\csname\exstring@#1@C\endcsname\relax
  \expandafter\ifx\csname\exstring@#1@C1\endcsname\relax
   \DN@{\Err@{\noexpand\Evaluate can't be used with \string#1}}%
  \else
   \DN@{\Evaluate@#1}%
  \fi
 \else
  \DN@{\global\Value\csname\exstring@#1@C\endcsname}%
 \fi
 \next@}
\def\Evaluate@#1#2{\expandafter
 \ifx\csname\exstring@#1@C#2\endcsname\relax
  \DN@{\Err@{\string\Evaluate\string#1 can't be followed by \string#2}}%
 \else
  \DN@{\global\Value\csname\exstring@#1@C#2\endcsname}%
 \fi\next@}
\def\pre#1{\expandafter
 \ifx\csname\exstring@#1@P\endcsname\relax
  \expandafter\ifx\csname\exstring@#1@P1\endcsname\relax
   \DN@{\Err@{\noexpand\pre can't be used with \string#1}}%
  \else
   \DN@{\pre@#1}%
  \fi
 \else
  \DN@{{\csname\exstring@#1@P\endcsname}}%
 \fi
 \next@}
\def\pre@#1#2{\expandafter
 \ifx\csname\exstring@#1@P#2\endcsname\relax
  \DN@{\Err@{\string\pre\string#1 can't be followed by \string#2}}%
 \else
  \DN@{{\csname\exstring@#1@P#2\endcsname}}%
 \fi
 \next@}
\def\post#1{\expandafter
 \ifx\csname\exstring@#1@Q\endcsname\relax
  \expandafter\ifx\csname\exstring@#1@Q1\endcsname\relax
   \DN@{\Err@{\noexpand\post can't be used with \string#1}}%
  \else
   \DN@{\post@#1}%
  \fi
 \else
  \DN@{{\csname\exstring@#1@Q\endcsname}}%
 \fi
 \next@}
\def\post@#1#2{\expandafter
 \ifx\csname\exstring@#1@Q#2\endcsname\relax
  \DN@{\Err@{\string\post\string#1 can't be followed by \string#2}}%
 \else
  \DN@{{\csname\exstring@#1@Q#2\endcsname}}%
 \fi
 \next@}
\def\style#1{\expandafter
 \ifx\csname\exstring@#1@S\endcsname\relax
  \expandafter\ifx\csname\exstring@#1@S1\endcsname\relax
   \DN@{\Err@{\noexpand\style can't be used with \string#1}}%
  \else
   \DN@{\style@#1}%
  \fi
 \else
  \DN@{\csname\exstring@#1@S\endcsname}%
 \fi
 \next@}
\def\style@#1#2{\expandafter
 \ifx\csname\exstring@#1@S#2\endcsname\relax
  \DN@{\Err@{\string\style\string#1 can't be followed by \string#2}}%
 \else
  \DN@{\csname\exstring@#1@S#2\endcsname}%
 \fi
 \next@}
\def\fontstyle#1{\expandafter
 \ifx\csname\exstring@#1@F\endcsname\relax
  \expandafter\ifx\csname\exstring@#1@F1\endcsname\relax
   \DN@{\Err@{\noexpand\fontstyle can't be used with \string#1}}%
  \else
   \DN@{\fontstyle@#1}%
  \fi
 \else
  \DN@##1{{\csname\exstring@#1@F\endcsname##1}}%
 \fi
 \next@}
\def\fontstyle@#1#2{\expandafter
 \ifx\csname\exstring@#1@F#2\endcsname\relax
  \DN@{\Err@{\string\fontstyle\string#1 can't be followed by \string#2}}%
 \else
  \DN@##1{{\csname\exstring@#1@F#2\endcsname##1}}%
 \fi
 \next@}
\def\Reset#1{\expandafter
 \ifx\csname\exstring@#1@C\endcsname\relax
  \expandafter\ifx\csname\exstring@#1@C1\endcsname\relax
   \DN@{\Err@{\noexpand\Reset can't be used with \string#1}}%
  \else
   \DN@{\Reset@#1}%
  \fi
 \else
  \DN@##1{\count@##1\relax\ifx#1\page\else\advance\count@\m@ne\fi
   \global\csname\exstring@#1@C\endcsname\count@}%
 \fi
 \next@}
\def\Reset@#1#2{\expandafter
 \ifx\csname\exstring@#1@C#2\endcsname\relax
  \DN@{\Err@{\string\Reset\string#1 can't be followed by \string#2}}%
 \else
  \DN@##1{\count@##1\relax\advance\count@\m@ne
   \global\csname\exstring@#1@C#2\endcsname\count@}%
 \fi
 \next@}
\def\Offset#1{\expandafter
 \ifx\csname\exstring@#1@C\endcsname\relax
  \expandafter\ifx\csname\exstring@#1@C1\endcsname\relax
   \DN@{\Err@{\noexpand\Offset can't be used with \string#1}}%
  \else
   \DN@{\Offset@#1}%
  \fi
 \else
  \DN@##1{\count@##1\relax\advance\count@\m@ne\global\advance
   \csname\exstring@#1@C\endcsname\count@}%
 \fi
 \next@}
\def\Offset@#1#2{\expandafter
 \ifx\csname\exstring@#1@C#2\endcsname\relax
  \DN@{\Err@{\string\Offset\string#1 can't be followed by \string#2}}%
 \else
  \DN@##1{\count@##1\relax\advance\count@\m@ne
   \global\advance\csname\exstring@#1@C#2\endcsname\count@}%
 \fi
 \next@}
\def\getR@#1#2{\def\nextiv@{\let\nextiii@}\expandafter\nextiv@
 \csname\exstring@#1@R#2\endcsname}
\def\letR@#1#2#3{\expandafter\let\csname#1@#3#2\endcsname\Next@}
\def\letR@@#1#2{\expandafter\let\csname\exstring@#1@#2\endcsname\Next@}
\def\newpre#1{\expandafter
 \ifx\csname\exstring@#1@P\endcsname\relax
  \expandafter\ifx\csname\exstring@#1@P1\endcsname\relax
   \DN@{\Err@{\noexpand\newpre can't be used with \string#1}}%
  \else
   \DN@{\newpre@#1}%
  \fi
 \else
  \DN@{%
   \DNii@{%
    \endgroup
    \expandafter\let\csname\exstring@#1@P\endcsname\Next@
    \expandafter\ifx\csname\exstring@#1@R\endcsname\relax\else
    \getR@#1{}\expandafter\letR@\nextiii@ P\fi
    }%
   \begingroup\noexpands@\afterassignment\nextii@\xdef\Next@}%
 \fi
 \next@}
\def\newpre@#1#2{\expandafter
 \ifx\csname\exstring@#1@P#2\endcsname\relax
  \DN@{\Err@{\string\newpre\string#1 can't be followed by \string#2}}%
 \else
  \DN@{%
   \DNii@{%
    \endgroup
    \expandafter\let\csname\exstring@#1@P#2\endcsname\Next@
    \expandafter\ifx\csname\exstring@#1@R#2\endcsname\relax\else
    \getR@#1{#2}\expandafter\letR@@\nextiii@ P\fi
    }%
   \begingroup\noexpands@\afterassignment\nextii@\xdef\Next@}%
 \fi
 \next@}
\def\newpost#1{\expandafter
 \ifx\csname\exstring@#1@Q\endcsname\relax
  \expandafter\ifx\csname\exstring@#1@Q1\endcsname\relax
   \DN@{\Err@{\noexpand\newpost can't be used with \string#1}}%
  \else
   \DN@{\newpost@#1}%
  \fi
 \else
  \DN@{%
   \DNii@{%
    \endgroup
    \expandafter\let\csname\exstring@#1@Q\endcsname\Next@
    \expandafter\ifx\csname\exstring@#1@R\endcsname\relax\else
    \getR@#1{}\expandafter\letR@\nextiii@ Q\fi
    }%
   \begingroup\noexpands@\afterassignment\nextii@\xdef\Next@}%
 \fi
 \next@}
\def\newpost@#1#2{\expandafter
 \ifx\csname\exstring@#1@Q#2\endcsname\relax
  \DN@{\Err@{\string\newpost\string#1 can't be followed by \string#2}}%
 \else
  \DN@{%
   \DNii@{%
    \endgroup
    \expandafter\let\csname\exstring@#1@Q#2\endcsname\Next@
    \expandafter\ifx\csname\exstring@#1@R#2\endcsname\relax\else
    \getR@#1{#2}\expandafter\letR@@\nextiii@ Q\fi
    }%
   \begingroup\noexpands@\afterassignment\nextii@\xdef\Next@}%
 \fi
 \next@}
\def\newstyle#1{\expandafter
 \ifx\csname\exstring@#1@S\endcsname\relax
  \expandafter\ifx\csname\exstring@#1@S1\endcsname\relax
   \DN@{\Err@{\noexpand\newstyle can't be used
    with \string#1}}%
  \else
   \DN@{\newstyle@#1}%
  \fi
 \else
  \DN@{%
   \DNii@{%
    \expandafter\let\csname\exstring@#1@S\endcsname\Next@
    \expandafter\ifx\csname\exstring@#1@R\endcsname\relax\else
    \getR@#1{}\expandafter\letR@\nextiii@ S\fi
    }%
   \afterassignment\nextii@\gdef\Next@}%
 \fi
 \next@}
\def\newstyle@#1#2{\expandafter
 \ifx\csname\exstring@#1@S#2\endcsname\relax
  \DN@{\Err@{\string\newstyle\string#1 can't be followed by
   \string#2}}%
 \else
  \DN@{%
   \DNii@{%
    \expandafter\let\csname\exstring@#1@S#2\endcsname\Next@
    \expandafter\ifx\csname\exstring@#1@R#2\endcsname\relax\else
    \getR@#1{#2}\expandafter\letR@@\nextiii@ S\fi
    }%
   \afterassignment\nextii@\gdef\Next@}%
 \fi
 \next@}
\def\newnumstyle#1{\expandafter
 \ifx\csname\exstring@#1@N\endcsname\relax
  \expandafter\ifx\csname\exstring@#1@N1\endcsname\relax
   \DN@{\Err@{\noexpand\newnumstyle can't be used with
    \string#1}}%
  \else
   \DN@{\newnumstyle@#1}%
  \fi
 \else
  \DN@##1{%
   \gdef\Next@{##1}%
    \expandafter\let\csname\exstring@#1@N\endcsname\Next@
    \expandafter\ifx\csname\exstring@#1@R\endcsname\relax\else
    \getR@#1{}\expandafter\letR@\nextiii@ N\fi
    }%
 \fi
 \next@}
\def\newnumstyle@#1#2{\expandafter
 \ifx\csname\exstring@#1@N#2\endcsname\relax
  \DN@{\Err@{\string\newnumstyle\string#1 can't be followed by
   \string#2}}%
 \else
  \DN@##1{%
   \gdef\Next@{##1}%
    \expandafter\let\csname\exstring@#1@N#2\endcsname\Next@
    \expandafter\ifx\csname\exstring@#1@R#2\endcsname\relax\else
    \getR@#1{#2}\expandafter\letR@@\nextiii@ N\fi
    }%
  \fi
 \next@}
\def\newfontstyle#1{\expandafter
 \ifx\csname\exstring@#1@F\endcsname\relax
  \expandafter\ifx\csname\exstring@#1@F1\endcsname\relax
   \DN@{\Err@{\noexpand\newfontstyle can't be used with
    \string#1}}%
  \else
   \DN@{\newfontstyle@#1}%
  \fi
 \else
  \DN@##1{%
   \gdef\Next@{##1}%
    \expandafter\let\csname\exstring@#1@F\endcsname\Next@
    \expandafter\ifx\csname\exstring@#1@R\endcsname\relax\else
    \getR@#1{}\expandafter\letR@\nextiii@ F\fi
    }%
 \fi
 \next@}
\def\newfontstyle@#1#2{\expandafter
 \ifx\csname\exstring@#1@F#2\endcsname\relax
  \DN@{\Err@{\string\newfontstyle\string#1 can't be followed by
   \string#2}}%
 \else
  \DN@##1{%
   \gdef\Next@{##1}%
    \expandafter\let\csname\exstring@#1@F#2\endcsname\Next@
    \expandafter\ifx\csname\exstring@#1@R#2\endcsname\relax\else
    \getR@#1{#2}\expandafter\letR@@\nextiii@ F\fi
    }%
 \fi
 \next@}
\def\word#1{\expandafter
 \ifx\csname\exstring@#1@W\endcsname\relax
  \expandafter\ifx\csname\exstring@#1@W1\endcsname\relax
   \DN@{\Err@{\noexpand\word can't be used with \string#1}}%
  \else
   \DN@{\word@#1}%
  \fi
 \else
  \DN@{{\csname\exstring@#1@W\endcsname}}%
 \fi
 \next@}
\def\word@#1#2{\expandafter
 \ifx\csname\exstring@#1@W#2\endcsname\relax
  \DN@{\Err@{\string\word\noexpand#1can't be followed by \string#2}}%
 \else
  \DN@{{\csname\exstring@#1@W#2\endcsname}}%
 \fi
 \next@}
\def\newword#1{\expandafter
 \ifx\csname\exstring@#1@W\endcsname\relax
  \expandafter\ifx\csname\exstring@#1@W1\endcsname\relax
   \DN@{\Err@{\noexpand\newword can't be used  with \string#1}}%
  \else
   \DN@{\newword@#1}%
  \fi
 \else
  \DN@{%
   \DNii@{%
    \expandafter\let\csname\exstring@#1@W\endcsname\Next@
    \expandafter\ifx\csname\exstring@#1@R\endcsname\relax\else
     \getR@#1{}\expandafter\letR@\nextiii@ W\fi
    }%
   \afterassignment\nextii@\gdef\Next@}%
 \fi
 \next@}
\def\newword@#1#2{\expandafter
 \ifx\csname\exstring@#1@W#2\endcsname\relax
  \DN@{\Err@{\string\newword\noexpand#1can't be followed by \string#2}}%
 \else
  \DN@{%
   \DNii@{%
    \expandafter\let\csname\exstring@#1@W#2\endcsname\Next@
    \expandafter\ifx\csname\exstring@#1@R#2\endcsname\relax\else
     \getR@#1{#2}\expandafter\letR@@\nextiii@ W\fi
    }%
   \afterassignment\nextii@\gdef\Next@}%
 \fi
 \next@}
\newif\iffn@
\newcount\footmark@C
\footmark@C\z@
\def\footmark@S#1{$^{#1}$}
\let\footmark@N\arabic
\def\footmark@F{\rm}
\def\foottext@S#1{$^{#1}$}
\def\foottext@F{\rm}
\let\modifyfootnote@\relax
\def\modifyfootnote#1{\def\modifyfootnote@{#1}}
\def\vfootnote@#1{\insert\footins
 \bgroup
 \floatingpenalty\@MM\interlinepenalty\interfootnotelinepenalty
 \leftskip\z@\rightskip\z@\spaceskip\z@\xspaceskip\z@
 \rm\splittopskip\ht\strutbox\splitmaxdepth\dp\strutbox
 \locallabel@\noindent@@{\foottext@F#1}\modifyfootnote@
 \footstrut\FN@\fo@t}
\def\fo@t{\ifcat\bgroup\noexpand\next\expandafter\f@@t\else
 \expandafter\f@t\fi}
\def\f@t#1{#1\@foot}
\def\f@@t{\bgroup\aftergroup\@foot\afterassignment\FNSSP@\let\next@}
\def\@foot{\unskip\lower\dp\strutbox\vbox to\dp\strutbox{}\egroup
 \iffn@\expandafter\fn@false\else
 \expandafter\postvanish@\fi}
\newif\ifplainfn@
\plainfn@true
\def\fancyfootnotes{\plainfn@false}
\newcount\fancyfootmarkcount@
\fancyfootmarkcount@\z@
\newcount\lastfnpage@
\lastfnpage@-\@M
\let\justfootmarklist@\empty
\def\footmark{\let\@sf\empty
 \ifhmode\edef\@sf{\spacefactor\the\spacefactor}\/\fi
 \DN@{\ifx"\next\expandafter\nextii@\else\expandafter\footmark@\fi}%
 \DNii@"##1"{%
  \iffirstchoice@
   {\let\style\footmark@S\let\numstyle\footmark@N
   \footmark@F##1%
   \noexpands@
   \let\style\foottext@S
   \Qlabel@{##1}%
   }%
   \iffn@\else
    {\noexpands@
    \xdef\Next@{{\Thelabel@}{\Thelabel@@}{\Thelabel@@@}{\Thelabel@@@@}}%
    }%
    \expandafter\rightappend@\Next@\to\justfootmarklist@
   \fi
  \fi
  \@sf\relax}%
 \FN@\next@}
\def\footmark@{%
 \iffirstchoice@
  \global\advance\footmark@C\@ne
  \ifplainfn@
   \xdef\adjustedfootmark@{\number\footmark@C}%
  \else
   {\let\\\or\xdef\Next@{\ifcase\number\footmark@C\fnpages@\else
     -\@M\fi}}%
   \ifnum\Next@=-\@M
    \xdef\adjustedfootmark@{\number\footmark@C}%
   \else
    \ifnum\Next@=\lastfnpage@
     \global\advance\fancyfootmarkcount@\@ne
    \else
     \global\fancyfootmarkcount@\@ne
     \global\lastfnpage@\Next@
    \fi
    \xdef\adjustedfootmark@{\number\fancyfootmarkcount@}%
   \fi
  \fi
  {\noexpands@
  \xdef\Thelabel@@@{\adjustedfootmark@}%
  \xdefThelabel@\footmark@N
  \xdef\Thelabel@@@@{\Thelabel@}%
  \xdefThelabel@@\foottext@S
  }%
  \iffn@\else
   {\noexpands@
   \xdef\Next@{{\Thelabel@}{\Thelabel@@}{\Thelabel@@@}{\Thelabel@@@@}}%
   }%
   \expandafter\rightappend@\Next@\to\justfootmarklist@
  \fi
  \ifplainfn@
  \else
   \edef\next@{\write\laxwrite@{F\noexpand\the\pageno}}\next@
  \fi
 \fi
 \footmark@S{\footmark@N{\adjustedfootmark@}}%
 \@sf\relax}
\def\foottext{\prevanish@
 \ifx\justfootmarklist@\empty
  \Err@{There is no \noexpand\footmark for this \string\foottext}\fi
 \DN@\\##1##2\next@{\DN@{##1}\gdef\justfootmarklist@{##2}}%
 \expandafter\next@\justfootmarklist@\next@
 \expandafter\foottext@\next@}
\def\foottext@#1#2#3#4{{\noexpands@
  \xdef\Thelabel@{#1}\xdef\Thelabel@@{#2}%
  \xdef\Thelabel@@@{#3}\xdef\Thelabel@@@@{#4}}%
  \vfootnote@{\thelabel@@}}
\rightadd@\foottext\to\vanishlist@
\def\footnote{\fn@true
 \let\@sf\empty
 \ifhmode\edef\@sf{\spacefactor\the\spacefactor}\/\fi
 \DN@{\ifx"\next\expandafter\nextii@\else\expandafter\nextiii@\fi}%
 \DNii@"##1"{\footmark"##1"\vfootnote@{\let\style\foottext@S
  \let\numstyle\footmark@N##1}}%
 \def\nextiii@{\footmark\vfootnote@{\foottext@S{\footmark@N
  {\adjustedfootmark@}}}}%
 \FN@\next@}
\newdimen\litindent
\litindent20\p@
\newbox\litbox@
\newbox\Litbox@
\newcount\interlitpenalty@
\interlitpenalty@\@M
\newcount\litlines@
{\obeyspaces\gdef\defspace@{\def {\allowbreak\hskip.5emminus.15em}}}
{\obeylines\gdef\letM@{\let^^M\CtrlM@}}
\def\CtrlM@{\egroup
 \ifcase\litlines@\advance\litlines@\@ne\or
 \box\litbox@\advance\litlines@\@ne\else
 \penalty\interlitpenalty@\box\litbox@\fi
 \Lit@}
\def\Lit@{\setbox\litbox@\hbox\bgroup\litdefs@\hskip\litindent}
\newcount\littab@
\littab@8
\def\littab#1{\littab@#1\relax}
{\catcode`\^^I=\active\gdef\letTAB@{\let^^I\TAB@}}
\def\TAB@{\egroup
 \dimen@\wd\litbox@
 \advance\dimen@-\litindent
 \setboxz@h{\tt0}%
 \dimen@ii\littab@\wdz@
 \divide\dimen@\dimen@ii
 \multiply\dimen@\dimen@ii
 \advance\dimen@\littab@\wdz@
 \advance\dimen@\litindent
 \setbox\litbox@\hbox\bgroup\litdefs@\hbox to\dimen@{\unhbox\litbox@\hfil}}
{\catcode`\`=\active\gdef`{\relax\lq}}
\let\litbs@\relax
\let\litbs@@\relax
\def\litbackslash#1{%
 \edef\litbs@{\catcode`\string#1=\z@
 \def\noexpand\litbs@@{\def\expandafter\noexpand\csname\string#1\endcsname
  {\char`\string#1}}}}
\def\litcodes@{\catcode`\\=12
 \catcode`\{=12 \catcode`\}=12
 \catcode`\$=12 \catcode`\&=12
 \catcode`\#=12
 \catcode`\^=12 \catcode`\_=12
 \catcode`\@=12 \catcode`\~=12 \catcode`\"=12
 \catcode`\;=12 \catcode`\:=12 \catcode`\!=12 \catcode`\?=12
 \catcode`\%=12 \litbs@\catcode`\`=\active\obeyspaces\defspace@}
\def\activate@#1#2{{\lccode`\~=`#2%
 \lowercase{%
  \if0#1%
  \gdef\Next@{\def~{\egroup\endgroup\bigskip\vskip-\parskip
   \def\next@{\noindent@@\FN@\pretendspace@}\FNSS@\next@}}\else
  \gdef\Next@{\def~{\egroup\egroup\endgroup}}\fi
  }%
 }}
\def\litdefs@{\let\0\empty\let\1\litdelim@\def\ {\char32 }\litbs@@}%
\def\litdelimiter#1{%
 \edef\litdelim@{\char`#1}%
 \def\lit#1{\leavevmode\begingroup\litcodes@\litdefs@
  \tt\hyphenchar\tentt\m@ne\lit@}%
 \def\lit@##1#1{##1\endgroup\null}%
 \def\Lit#1{\ifhmode$$\abovedisplayskip\bigskipamount
  \abovedisplayshortskip\bigskipamount
  \belowdisplayskip\z@\belowdisplayshortskip\z@
  \postdisplaypenalty\@M
  $$\vskip-\baselineskip\else\bigskip\fi
  \begingroup\litlines@\z@
  \catcode`#1=\active\activate@0#1\Next@
  \def\displaybreak{\egroup\break\litlines@\z@\Lit@}%
  \def\allowdisplaybreak{\egroup\allowbreak\litlines@\z@\Lit@}%
  \def\allowdisplaybreaks{\egroup\allowbreak\interlitpenalty@\z@
   \litlines@\z@\Lit@}%
  \litcodes@\tt\catcode`\^^I=\active\letTAB@
  \obeylines\letM@\Lit@}%
 \def\Litbox##1=#1{\begingroup\ifodd##1\relax\aftergroup\global\fi
  \aftergroup\setbox\aftergroup##1\aftergroup\box\aftergroup\Litbox@
  \def\allowdisplaybreak{\egroup\allowbreak\litlines@\z@\Lit@}%
  \def\allowdisplaybreaks{\egroup\allowbreak\interlitpenalty@\z@
   \litlines@\z@\Lit@}%
  \catcode`#1=\active\activate@1#1\Next@
  \litcodes@\tt\catcode`\^^I=\active\letTAB@
  \obeylines\letM@\global\setbox\Litbox@\vbox\bgroup\litindent\z@%
  \litlines@\z@\Lit@}%
}
\newbox\titlebox@
\setbox\titlebox@\vbox{}
\rightadd@\title\to\overlonglist@
\def\title{\begingroup\Let@
 \global\setbox\titlebox@\vbox\bgroup\tabskip\hss@
 \halign to\hsize\bgroup\bf\hfil\ignorespaces##\unskip\hfil\cr}
\def\endtitle{\crcr\egroup\egroup\endgroup\overlong@false}
\newbox\authorbox@
\rightadd@\author\to\overlonglist@
\def\author{\begingroup\Let@
 \global\setbox\authorbox@\vbox\bgroup\tabskip\hss@
 \halign to\hsize\bgroup\rm\hfil\ignorespaces##\unskip\hfil\cr}
\def\endauthor{\crcr\egroup\egroup\endgroup\overlong@false}
\newbox\affilbox@
\def\affil{\begingroup\Let@
 \global\setbox\affilbox@\vbox\bgroup\tabskip\hss@
 \halign to\hsize\bgroup\rm\hfil\ignorespaces##\unskip\hfil\cr}%
\def\endaffil{\crcr\egroup\egroup\endgroup\overlong@false}
\let\date@\relax
\def\date#1{\gdef\date@{\ignorespaces#1\unskip}}
\def\today{\ifcase\month\or January\or February\or March\or April\or May\or
 June\or July\or August\or September\or October\or November\or December\fi
 \space\number\day, \number\year}
\def\maketitle{\hrule\height\z@\vskip-\topskip
 \vskip24\p@ plus12\p@ minus12\p@
 \unvbox\titlebox@
 \ifvoid\authorbox@\else\vskip12\p@ plus6\p@ minus3\p@\unvbox\authorbox@\fi
 \ifvoid\affilbox@\else\vskip10\p@ plus5\p@ minus2\p@\unvbox\affilbox@\fi
 \ifx\date@\relax\else\vskip6\p@ plus2\p@ minus\p@\centerline{\rm\date@}\fi
 \vskip18\p@ plus12\p@ minus6\p@}
\def\cite{%
 \DNii@(##1)##2{{\rm[}{##2}, {##1\/}{\rm]}}%
 \def\nextiii@##1{{\rm[}{##1\/}{\rm]}}%
 \DN@{\ifx\next(\expandafter\nextii@\else\expandafter\nextiii@\fi}%
 \FN@\next@}
\def\makebib@W{Bibliography}
\def\makebib{\begingroup\rm\bigbreak\centerline{\smc\makebib@W}%
 \nobreak\medskip
 \sfcode`\.=\@m\everypar{}\parindent\z@
 \def\nopunct{\nopunct@true}\def\nospace{\nospace@true}%
 \nopunct@false\nospace@false
 \def\lkerns@{\null\kern\m@ne sp\kern\@ne sp}%
 \def\nkerns@{\null\kern-\tw@ sp\kern\tw@ sp}%
}

\newif\ifnoprepunct@
\newif\ifnoprespace@
\newif\ifnoquotes@
\def\noprepunct{\noprepunct@true}
\def\noprespace{\noprespace@true}
\def\noquotes{\noquotes@true}
\newbox\nobox@
\newbox\keybox@
\newbox\bybox@
\newbox\paperbox@
\newbox\paperinfobox@
\newbox\jourbox@
\newbox\volbox@
\newbox\issuebox@
\newbox\yrbox@
\newbox\pgbox@
\newbox\ppbox@
\newbox\bookbox@
\newbox\inbookbox@
\newbox\bookinfobox@
\newbox\publbox@
\newbox\publaddrbox@
\newbox\edbox@
\newbox\edsbox@
\newbox\langbox@
\newbox\translbox@
\newbox\finalinfobox@
\def\setbibinfo@#1{\edef\next@{\ifnopunct@1\else0\fi
 \ifnospace@1\else0\fi\ifnoprepunct@1\else0\fi\ifnoprespace@1\else0\fi
 \ifnoquotes@1\else0\fi}%
 \DNii@{00000}%
 \ifx\next@\nextii@\else\xdef\bibinfo@{\bibinfo@\the#1,\next@}%
 \fi}
\def\getbibinfo@#1{\ifx\bibinfo@\empty
 \let\next@0\let\nextii@0\let\nextiii@0\let\nextiv@0\let\nextv@0\else
 \edef\next@{\def
  \noexpand\next@####1\the#1,####2####3####4####5####6####7\noexpand\next@
  {\let\noexpand\next@####2\let\noexpand\nextii@####3%
  \let\noexpand\nextiii@####4\let\noexpand\nextiv@####5%
  \let\noexpand\nextv@####6}%
  \noexpand\next@\bibinfo@\the#1,00000\noexpand\next@}\next@
 \fi}
\newif\ifbookinquotes@
\def\bookinquotes{\bookinquotes@true}
\newif\ifpaperinquotes@
\def\paperinquotes{\paperinquotes@true}
\newif\ifininbook@
\def\ininbook{\ininbook@true}
\newif\ifopenquotes@
\def\closequotes@{\ifopenquotes@''\openquotes@false\fi}
\newif\ifbeginbib@
\newif\ifendbib@
\newif\ifprevjour@
\newif\ifprevbook@
\newdimen\bibindent@
\bibindent@20\p@
\def\bib{\global\let\bibinfo@\empty\global\let\translinfo@\relax\beginbib@true
 \begingroup\noindent@
 \hangindent\bibindent@\hangafter\@ne\bib@}
\def\v@id#1{\setbox#1\box\voidb@x}
\def\bib@{\v@id\nobox@\v@id\keybox@\v@id\bybox@\v@id\paperbox@
 \v@id\paperinfobox@\v@id\jourbox@\v@id\volbox@\v@id\issuebox@
 \v@id\yrbox@\v@id\pgbox@\v@id\ppbox@\v@id\bookbox@\v@id\inbookbox@
 \v@id\bookinfobox@\v@id\publbox@\v@id\publaddrbox@\v@id\edbox@
 \v@id\edsbox@\v@id\langbox@\v@id\translbox@\v@id\finalinfobox@
 \bgroup}
\def\Setnonemptybox@#1#2{\unskip\setbibinfo@#1\egroup#2%
 \def\aftergroup@{\ifdim\wd#1=\z@\setbox#1\box\voidb@x\fi}%
 \setbox#1\vbox\bgroup\aftergroup\aftergroup@\hsize\maxdimen\leftskip\z@
 \rightskip\z@\hbadness\@M\hfuzz\maxdimen\noindent}
\def\setnonemptybox@#1{\Setnonemptybox@#1\relax}
\def\no{\setnonemptybox@\nobox@}
\def\key{\setnonemptybox@\keybox@\bf}
\def\by{\setnonemptybox@\bybox@}
\def\bysame{\setnonemptybox@\bybox@\leaders\hrule\hskip3em\null}
\def\paper{\setnonemptybox@\paperbox@
 \ifpaperinquotes@\getbibinfo@\paperbox@
 \if\nextv@1\else``\fi\else\it\fi}
\def\paperinfo{\setnonemptybox@\paperinfobox@}
\def\jour{\Setnonemptybox@\jourbox@\prevjour@true}
\def\vol{\setnonemptybox@\volbox@\bf}
\def\issue{\setnonemptybox@\issuebox@}
\def\yr{\setnonemptybox@\yrbox@}

\def\pg{\setnonemptybox@\pgbox@}
\def\pp{\setnonemptybox@\ppbox@}
\def\book{\Setnonemptybox@\bookbox@\prevbook@true
 \ifbookinquotes@\getbibinfo@\bookbox@
 \if\nextv@1\else``\fi\else\it\fi}
\def\inbook{\Setnonemptybox@\inbookbox@\prevbook@true
 \ifininbook@ in \fi\ifbookinquotes@\getbibinfo@\inbookbox@
 \if\nextv@1\else``\fi\fi}
\def\bookinfo{\setnonemptybox@\bookinfobox@}
\def\publ{\setnonemptybox@\publbox@}
\def\publaddr{\setnonemptybox@\publaddrbox@}
\def\ed{\setnonemptybox@\edbox@}
\def\eds{\setnonemptybox@\edsbox@}
\def\lang{\setnonemptybox@\langbox@}
\def\finalinfo{\setnonemptybox@\finalinfobox@}
\def\setboxzl@{\setbox\z@\lastbox}
\def\getbox@#1{\setbox\z@\vbox{\vskip-\@M\p@
 \unvbox#1%
 \setboxzl@
 \global\setbox\@ne\hbox{\unhbox\z@\unskip\unskip\unpenalty}%
 \ifdim\lastskip=-\@M\p@\else
 \loop\ifdim\lastskip=-\@M\p@
 \else\unskip\unpenalty\setboxzl@
 \global\setbox\@ne\hbox{\unhbox\z@\unhbox\@ne}%
 \repeat\fi}%
 \unhbox\@ne}
\def\adjustpunct@#1{\count@\lastkern
 \ifnum\count@=\z@#1\closequotes@\else
 \ifnum\count@>\tw@#1\closequotes@\else
 \ifnum\count@<-\tw@#1\closequotes@\else
  \unkern\unkern\setboxzl@
  \skip@\lastskip\unskip
  \count@@\lastpenalty\unpenalty
  \ifnum\count@=\tw@\unskip\setboxzl@\fi
  \ifdim\skip@=\z@\else\hskip\skip@\fi
  #1\closequotes@
  \ifnum\count@=\tw@\null\hfill\fi
  \penalty\count@@
 \fi\fi\fi}
\def\prepunct@#1#2{\getbibinfo@#2%
 \ifnopunct@
 \else
  \if\nextiii@0\adjustpunct@#1\fi
 \fi
 \closequotes@
 \ifnospace@
 \else
  \if\nextiv@0\space\else\fi
 \fi
 \nopunct@false\nospace@false
 \if\next@1\nopunct@true\fi
 \if\nextii@1\nospace@true\fi}
\def\ppunbox@#1#2{\prepunct@{#1}#2%
 \getbox@#2}
\let\semicolon@;
\def\endbib@{%
 \ifbeginbib@
  \ifvoid\nobox@
   \ifvoid\keybox@\else\hbox to\bibindent@{[\getbox@\keybox@]\hss}\fi
  \else\hbox to\bibindent@{\hss\getbox@\nobox@. }\fi
  \ifvoid\bybox@\else\getbox@\bybox@\fi
 \else
  \nopunct@true
  \ifvoid\bybox@\else\ppunbox@\relax\bybox@\fi
 \fi
 \ifvoid\translbox@\else\ppunbox@,\translbox@\fi
 \ifvoid\paperbox@\else\ppunbox@,\paperbox@\ifpaperinquotes@
  \if\nextv@1\else\openquotes@true\fi\fi
 \fi
 \ifvoid\paperinfobox@\else\ppunbox@,\paperinfobox@\fi
 \test@false
 \ifvoid\jourbox@\else\test@true\ppunbox@,\jourbox@\fi
 \ifprevjour@\test@true\fi
 \iftest@
  \ifvoid\volbox@\else\ppunbox@\relax\volbox@\fi
  \ifvoid\issuebox@
   \else\prepunct@\relax\issuebox@ no.~\getbox@\issuebox@\fi
  \ifvoid\yrbox@\else\prepunct@\relax\yrbox@(\getbox@\yrbox@)\fi
  \ifvoid\ppbox@\else\ppunbox@,\ppbox@\fi
  \ifvoid\pgbox@\else\prepunct@,\pgbox@ p.~\getbox@\pgbox@\fi
 \fi
 \test@false
 \ifvoid\bookbox@\else\test@true\ppunbox@,\bookbox@\ifbookinquotes@
  \if\nextv@1\else\openquotes@true\fi\fi\fi
 \ifvoid\inbookbox@\else\test@true\ppunbox@,\inbookbox@\ifbookinquotes@
  \if\nextv@1\else\openquotes@true\fi\fi\fi
 \ifprevbook@\test@true\fi
 \iftest@
  \ifvoid\edbox@\else\prepunct@\relax\edbox@(\getbox@\edbox@, ed.)\fi
  \ifvoid\edsbox@\else\prepunct@\relax\edsbox@(\getbox@\edsbox@, eds.)\fi
  \ifvoid\bookinfobox@\else\ppunbox@,\bookinfobox@\fi
  \ifvoid\publbox@\else\ppunbox@,\publbox@\fi
  \ifvoid\publaddrbox@\else\ppunbox@,\publaddrbox@\fi
  \ifvoid\yrbox@\else\ppunbox@,\yrbox@\fi
  \ifvoid\ppbox@\else\prepunct@,\ppbox@ pp.~\getbox@\ppbox@\fi
  \ifvoid\pgbox@\else\prepunct@,\pgbox@ p.~\getbox@\pgbox@\fi
 \fi
 \ifvoid\finalinfobox@
  \ifendbib@
   \ifnopunct@\else.\closequotes@\fi
  \else
  \ifvoid\langbox@\else\space(\getbox@\langbox@)\fi
   \/\semicolon@\closequotes@
  \fi
 \else
  \ifendbib@
   \ppunbox@{.\spacefactor3000\relax}\finalinfobox@
    \ifnopunct@\else.\fi
  \else
   \ppunbox@,\finalinfobox@\/\semicolon@\fi
 \fi
 \ifvoid\langbox@\else\space(\getbox@\langbox@)\fi
}
\def\endbib{\unskip\egroup\endbib@true\endbib@\par\endgroup}
\def\morebib{\unskip\egroup
 \endbib@false\endbib@
 \global\let\bibinfo@\empty\beginbib@false
 \bib@}
\def\anotherbib{\unskip\egroup
 \endbib@false\endbib@
 \global\let\bibinfo@\empty\beginbib@false
 \prevjour@false\prevbook@false\bib@}
\def\transl{\unskip
 \xdef\translinfo@{\the\translbox@,\ifnopunct@1\else0\fi
 \ifnospace@1\else0\fi\ifnoprepunct@1\else0\fi\ifnoprespace@1\else0\fi0}%
 \egroup\endbib@false\endbib@
 \global\let\bibinfo@\translinfo@\beginbib@false
 \bib@
 \egroup
 \def\aftergroup@{\ifdim\wd\translbox@=\z@\setbox\translbox@\box\voidb@x\fi}%
 \setbox\translbox@\vbox\bgroup\aftergroup\aftergroup@
 \hsize\maxdimen\leftskip\z@\rightskip\z@\hbadness\@M\hfuzz\maxdimen
 \noindent}
\newwrite\auxwrite@
\newread\bbl@
\def\UseBibTeX{\immediate\openout\auxwrite@=\jobname.aux
 \let\cite\BTcite@
 \def\nocite##1{\immediate\write\auxwrite@{\string\citation{##1}}}%
 \def\bibliographystyle##1{\immediate\write\auxwrite@{\string
  \bibstyle{##1}}}%
 \def\bibliography@W{Bibliography}%
 \def\bibliography##1{\immediate\write\auxwrite@{\string\bibdata{##1}}%
  \immediate\openin\bbl@=\jobname.bbl
  \ifeof\bbl@
   \W@{No .bbl file}%
  \else
   \immediate\closein\bbl@
   \begingroup\input bibtex \input\jobname.bbl \endgroup
  \fi}%
 }
\def\BTcite@{%
 \DNii@(##1)##2{{\rm[}\BTcite@@##2,\BTcite@@{\rm, }{##1\/}{\rm]}%
  \immediate\write\auxwrite@{\string\citation{##2}}}%
 \def\nextiii@##1{{\rm[}\BTcite@@##1,\BTcite@@\/{\rm]}%
  \immediate\write\auxwrite@{\string\citation{##1}}}%
 \DN@{\ifx\next(\expandafter\nextii@\else\expandafter\nextiii@\fi}%
 \FN@\next@}%
\def\BTcite@@#1,{\BTcite@@@{#1}\FN@\BTcite@@@@}
\def\BTcite@@@@{\ifx\next\BTcite@@
 \expandafter\eat@\else{\rm, }\expandafter\BTcite@@\fi}
\catcode`\~=11
\def\BTcite@@@#1{\nolabel@\cite{#1}\relax
 \DNii@##1~##2\nextii@{##1}%
 \csL@{#1}\expandafter\nextii@\Next@\nextii@\fi}
\catcode`\~=\active

\def\beginthebibliography@#1{\rm\setboxz@h{#1\ }\bibindent@\wdz@
 \bigbreak\centerline{\smc\bibliography@W}\nobreak\medskip
 \sfcode`\.=\@m\everypar{}\parindent\z@}

\newif\iffigproofing@
\def\Figureproofing{\figproofing@true}
\def\noFigureproofing{\figproofing@false}
\newif\ifHby@
\def\Hbyw#1{\global\Hby@true\hbyw\vsize{#1}}
\def\hbyw#1#2{%
 \hbox{%
  \ifHby@
  \else
   \iffigproofing@
    \setbox\z@\vbox{\hrule\width5\p@}\ht\z@\z@
    \vbox to#1{\hrule\height5\p@\width.4\p@\vfil\hrule\height5\p@\width.4\p@}%
    \kern-.4\p@\rlap{\copy\z@}\raise#1\hbox{\rlap{\copy\z@}}%
   \fi
  \fi
  \vbox to#1{\hbox to#2{}\vfil}%
  \ifHby@
  \else
   \iffigproofing@
    \vbox to#1{\hrule\height5\p@\width.4\p@\vfil\hrule\height5\p@\width.4\p@}%
    \kern-.4\p@\llap{\copy\z@}\raise#1\hbox{\llap{\boxz@}}%
   \fi
  \fi}}
\newcount\island@C
\let\island@P\empty
\let\island@Q\empty
\def\island@S#1{#1\null.}
\let\island@N\arabic
\def\island@F{\rm}
\def\island@@@P{\csname\exxx@\islandtype@ @P\endcsname}
\def\island@@@Q{\csname\exxx@\islandtype@ @Q\endcsname}
\def\island@@@S{\csname\exxx@\islandtype@ @S\endcsname}
\def\island@@@N{\csname\exxx@\islandtype@ @N\endcsname}
\def\island@@@F{\csname\exxx@\islandtype@ @F\endcsname}
\def\island@@@C{\csname island@C\islandclass@\endcsname}
\newif\ifplace@
\newif\ifisland@
\def\island{%
 \ifplace@
  \DN@{\let\islandclass@\empty\def\islandtype@{\island}\FN@\island@}%
 \else
  \long\DN@##1\endisland{\Err@{\noexpand\island must be used after some
   type of \string\...place}}%
 \fi
 \next@}
\def\island@{\ifx\next\c\let\next@\island@c\else
 \DN@{\FN@\island@@}\fi\next@}
\def\island@@{\ifcat\bgroup\noexpand\next\let\next@\island@@@\else
 \DN@{\Err@{\noexpand\island must be followed by a {prefix} for
 \string\caption's}}\fi\next@}
\newbox\islandbox@
\newcount\captioncount@
\def\island@@@#1{\def\captionprefix@{#1}\captioncount@\z@
 \global\setbox\islandbox@\vbox\bgroup}
\def\island@c\c#1{%
 \ifplace@
 \DN@{\def\islandclass@{#1}%
  \expandafter\ifx\csname island@C#1\endcsname\relax
  \expandafter\newcount@\csname island@C#1\endcsname
   \global\csname island@C#1\endcsname\z@\fi
  \FNSS@\island@c@}%
 \else
 \DN@{\edef\next@{\long\def\noexpand\next@########1\expandafter\noexpand
  \csname end\exxx@\islandtype@\endcsname{\noexpand\Err@{\noexpand\noexpand
  \expandafter\noexpand
  \islandtype@ must be used after some type of \noexpand\string
   \noexpand\...place}}}\next@\next@}%
 \fi
 \next@}
\def\island@c@{%
 \ifcat\bgroup\noexpand\next
  \let\next@\island@c@@
 \else
  \DN@{\Err@{\noexpand\island\string\c{\expandafter\string\islandclass@} must
   be followed by a {prefix} for \string\caption's}}%
 \fi\next@}
\def\island@c@@#1{\def\captionprefix@{#1}%
 \captioncount@\z@\global\setbox\islandbox@\vbox\bgroup}
\rightadd@\caption\to\nofrillslist@
\newbox\captionbox@
\newbox\Captionbox@
\def\caption{%
 \ifnum\captioncount@=\z@
  \ifnopunct@
   \DN@{\egroup\nopunct@true}%
  \else
   \let\next@\egroup
  \fi
 \else
  \let\next@\relax
 \fi
 \next@
 \advance\captioncount@\@ne
 \FN@\caption@}
\def\caption@{\ifx\next"\expandafter\caption@q\else\expandafter\caption@@\fi}
\def\caption@q"#1"{\quoted@true
 {\noexpands@
 \let\pre\island@@@P\let\post\island@@@Q
 \let\style\island@@@S\let\numstyle\island@@@N
 \Qlabel@{#1}\let\style\relax\xdef\Qlabel@@@@{#1}}%
 \finishcaption@}
\def\caption@@{\quoted@false
 \global\advance\island@@@C\@ne
 {\noexpands@
 \xdef\Thelabel@@@{\number\island@@@C}%
 \xdefThelabel@\island@@@N
 \xdef\Thelabel@@@@{\island@@@P\Thelabel@\island@@@Q}%
 \xdefThelabel@@\island@@@S
 \xdef\Thepref@{\Thelabel@@@@}}%
 \finishcaption@}
\long\def\captionformat@#1#2#3{\rm\strut#1 {\island@@@F#2} #3%
 \punct@.\strut}
\long\def\widerthanisland@#1#2#3{\test@true\setbox\z@\vbox{\hsize\maxdimen
 \noindent@@\captionformat@{#1}{#2}{#3}\par\setboxzl@}%
 \ifdim\wdz@=\z@
  \global\setbox\captionbox@\hbox{\noset@\unlabel@
   \captionformat@{#1}{#2}{#3}}%
  \ifdim\wd\captionbox@>\wd\islandbox@\else\test@false\fi
 \fi}
\long\def\captionformat@@#1#2#3{\widerthanisland@{#1}{#2}{#3}%
 \iftest@
  \global\setbox\captionbox@\vbox{\hsize\wd\islandbox@
   \vskip-\parskip\noindent@@\noset@\unlabel@
   \captionformat@{#1}{#2}{#3}\par}%
 \else
  \global\setbox\captionbox@
   \hbox to\wd\islandbox@{\hfil\box\captionbox@\hfil}%
 \fi}
\long\def\finishcaption@#1{\def\entry@{#1}%
 {\locallabel@
 \captionformat@@
  {\expandafter\ignorespaces\captionprefix@\unskip}%
  {\ifx\thelabel@@\empty\unskip\else\thelabel@@\fi}%
  {\ignorespaces#1\unskip}%
 \ifnum\captioncount@=\@ne
  \global\setbox\islandbox@\vbox{\ticwrite@\vbox{\box\islandbox@}}%
  \global\setbox\Captionbox@\vbox{\box\captionbox@}%
 \else
  \global\setbox\islandbox@\vbox{\unvbox\islandbox@\setboxzl@
   \ticwrite@\boxz@}%
  \global\setbox\Captionbox@\vbox{\unvbox\Captionbox@
   \smallskip\box\captionbox@}%
 \fi}%
 \nopunct@false\nospace@false\ignorespaces}
\def\Sixtic@{\ifx\macdef@\empty\else
 \DN@##1##2\next@{\def\macdef@{##1##2}}%
 \expandafter\next@\macdef@\next@
 \edef\next@
  {\noexpand\six@\tic@\macdef@
  \space\space\space\space\space\space\space\space\space\space\space\space
  \noexpand\six@}%
 \next@\let\macdef@\relax\fi}
\def\ticwrite@{%
 \iftoc@
  {\noexpands@\let\style\relax
  \DN@{\island}%
  \edef\next@{\write\tic@{%
   \ifnopunct@\noexpand\noexpand\noexpand\nopunct\fi
   \ifx\islandtype@\next@\noexpand\noexpand\noexpand\island
    \noexpand\string\noexpand\c{\islandclass@}{\captionprefix@}%
     {\QorThelabel@@@@}\else\noexpand\noexpand\expandafter\noexpand
     \islandtype@{\QorThelabel@@@@}}\fi}%
  \next@}%
  \expandafter\unmacro@\meaning\entry@\unmacro@
  \Sixtic@
  \write\tic@{\noexpand\Page{\number\pageno}{\page@N}{\page@P}{\page@Q}^^J}%
 \fi}
\def\Htrim@#1{%
 \ifHby@
  \dimen@\vsize
  \ifnum\captioncount@=\z@
  \else
   \advance\dimen@-\ht\Captionbox@
   \advance\dimen@-#1%
  \fi
  \global\Hby@false
  \dimen@ii\wd\islandbox@
  \global\setbox\islandbox@\vbox
   {\unvbox\islandbox@\setboxzl@
   \vbox to\z@{\vss\boxz@}\nointerlineskip\hbyw\dimen@\dimen@ii}%
  \global\Hby@true
 \fi}
\newif\ifdata@
\def\iclasstest@#1{\DN@{#1}\ifx\next@\islandclass@
 \test@true\else\test@false\fi}
\skipdef\skipi@=1
\def\endisland{\ifnum\captioncount@=\z@\expandafter\egroup\fi
 \ifdata@
 \else
  \iclasstest@{T}%
  \iftest@
   {\rm\global\skipi@-\dp\strutbox}\global\advance\skipi@\bigskipamount
   \Htrim@\skipi@
   \global\setbox\islandbox@\vbox
    {\ifnum\captioncount@=\z@\else
     \box\Captionbox@
     \nointerlineskip
     \vskip\skipi@\fi
     \box\islandbox@}%
  \else
   {\rm\global\skipi@\dp\strutbox}\global\advance\skipi@\medskipamount
   \Htrim@\skipi@
   \global\setbox\islandbox@\vbox
    {\box\islandbox@
     \ifnum\captioncount@=\z@\else
     \nointerlineskip
     \vskip\skipi@
     \box\Captionbox@
     \fi}%
  \fi
  \ifHby@
  \else
   \dimen@\ht\islandbox@\advance\dimen@\dp\islandbox@
   \ifdim\dimen@>\vsize
    \DN@{\island}%
    \Err@{%
     \ifx\islandtype@\next@\noexpand\island\else
      \expandafter\noexpand\islandtype@\fi
     \ifnum\captioncount@=\z@\else
       with \noexpand\caption\fi
      is larger than page}%
     \ht\islandbox@=\vsize
   \fi
  \fi
 \fi
 \global\Hby@false\island@true}
\def\newisland#1\c#2#3{\define#1{}%
 \iftoc@\immediate\write\tic@{\noexpand\newisland\noexpand#1%
  \string\c{#2}{#3}^^J}\fi
 \expandafter\def\csname\exstring@#1@S\endcsname{\island@S}%
 \expandafter\def\csname\exstring@#1@N\endcsname{\island@N}%
 \expandafter\def\csname\exstring@#1@P\endcsname{\island@P}%
 \expandafter\def\csname\exstring@#1@Q\endcsname{\island@Q}%
 \expandafter\def\csname\exstring@#1@F\endcsname{\island@F}%
 \expandafter\def\csname end\exstring@#1\endcsname{\endisland}%
 \expandafter
 \ifx\csname island@C#2\endcsname\relax
  \expandafter\newcount@\csname island@C#2\endcsname
  \global\csname island@C#2\endcsname\z@
 \fi
 \edef\next@{\noexpand\expandafter\noexpand\let\noexpand
  \csname\exstring@#1@C\noexpand\endcsname
  \csname island@C#2\endcsname}%
 \next@
 \def#1{\def\islandtype@{#1}\island@c\c{#2}{#3}}}
\newisland\Figure\c{F}{Figure}
\newisland\Table\c{T}{Table}
\newbox\islandboxi
\newbox\islandboxii
\newbox\islandboxiii
\newbox\captionboxi
\newbox\captionboxii
\newbox\captionboxiii
\long\def\islandpairdata#1#2{{\data@true
 \place@true
 #1%
 \global\setbox\islandboxi\box\islandbox@
 \global\setbox\captionboxi\box\Captionbox@
 #2%
 \global\setbox\islandboxii\box\islandbox@
 \global\setbox\captionboxii\box\Captionbox@
 }}
\long\def\islandpairbox#1#2{\islandpairdata{#1}{#2}%
 \dimen@\ht\captionboxi
 \ifdim\ht\captionboxii>\dimen@\dimen@\ht\captionboxii\fi
 \ifdim\dimen@>\z@
  \ifdim\ht\captionboxi<\dimen@
   \global\setbox\captionboxi\vbox to\dimen@{\unvbox\captionboxi\vfil}\fi
  \ifdim\ht\captionboxii<\dimen@
   \global\setbox\captionboxii\vbox to\dimen@{\unvbox\captionboxii\vfil}\fi
 \fi
 \global\setbox\islandbox@\vbox
 {\hbox to\hsize{\hfil\box\islandboxi\hfil\box\islandboxii\hfil}%
 \ifdim\dimen@>\z@\nointerlineskip
 {\rm\global\skipi@\dp\strutbox}\global\advance\skipi@\medskipamount
  \vskip\skipi@
  \hbox to\hsize{\hfil\box\captionboxi\hfil\box\captionboxii\hfil}\fi}}	
\long\def\islandpairboxa#1#2{\islandpairdata{#1}{#2}%
 \dimen@\ht\captionboxi
 \ifdim\ht\captionboxii>\dimen@\dimen@\ht\captionboxii\fi
 \ifdim\dimen@>\z@
  \ifdim\ht\captionboxi<\dimen@
   \global\setbox\captionboxi\vbox to\dimen@{\vfil\unvbox\captionboxi}\fi
  \ifdim\ht\captionboxii<\dimen@
   \global\setbox\captionboxii\vbox to\dimen@{\vfil\unvbox\captionboxii}\fi
 \fi
 \dimen@ii\ht\islandboxi
 \ifdim\ht\islandboxii>\dimen@ii \dimen@ii\ht\islandboxii\fi
 \ifdim\dimen@ii>\z@
  \ifdim\ht\islandboxi<\dimen@ii
   \global\setbox\islandboxi\vbox to\dimen@ii{\box\islandboxi\vfil}\fi
  \ifdim\ht\islandboxii<\dimen@ii
   \global\setbox\islandboxii\vbox to\dimen@ii{\box\islandboxii\vfil}\fi
 \fi
 \global\setbox\islandbox@\vbox{\ifdim\dimen@>\z@
  \hbox to\hsize{\hfil\box\captionboxi\hfil\box\captionboxii\hfil}%
  \nointerlineskip{\rm\global\skipi@-\dp\strutbox}%
  \global\advance\skipi@\bigskipamount\vskip\skipi@\fi
  \hbox to\hsize{\hfil\box\islandboxi\hfil\box\islandboxii\hfil}}}
\long\def\islandtripledata#1#2#3{{\data@true\place@true
 #1%
 \global\setbox\islandboxi\box\islandbox@
 \global\setbox\captionboxi\box\Captionbox@
 #2%
 \global\setbox\islandboxii\box\islandbox@
 \global\setbox\captionboxii\box\Captionbox@
 #3%
 \global\setbox\islandboxiii\box\islandbox@
 \global\setbox\captionboxiii\box\Captionbox@
 }}
\long\def\islandtriplebox#1#2#3{\islandtripledata{#1}{#2}{#3}%
 \dimen@\ht\captionboxi
 \ifdim\ht\captionboxii>\dimen@ \dimen@\ht\captionboxii\fi
 \ifdim\ht\captionboxiii>\dimen@ \dimen@\ht\captionboxiii\fi
 \ifdim\dimen@>\z@
  \ifdim\ht\captionboxi<\dimen@
   \global\setbox\captionboxi\vbox to\dimen@{\unvbox\captionboxi\vfil}\fi
  \ifdim\ht\captionboxii<\dimen@
   \global\setbox\captionboxii\vbox to\dimen@{\unvbox\captionboxii\vfil}\fi
  \ifdim\ht\captionboxiii<\dimen@
   \global\setbox\captionboxiii\vbox to\dimen@{\unvbox\captionboxiii\vfil}\fi
 \fi
 \global\setbox\islandbox@\vbox
  {\hbox to\hsize{\hfil\box\islandboxi\hfil\box\islandboxii\hfil
   \box\islandboxiii\hfil}%
 \ifdim\dimen@>\z@\nointerlineskip
  {\rm\global\skipi@\dp\strutbox}\global\advance\skipi@\medskipamount
  \vskip\skipi@
  \hbox to\hsize{\hfil\box\captionboxi\hfil\box\captionboxii\hfil
   \box\captionboxiii\hfil}\fi}}
\def\islandtripleboxa#1#2#3{\islandtripledata{#1}{#2}{#3}%
 \dimen@\ht\captionboxi
 \ifdim\ht\captionboxii>\dimen@ \dimen@\ht\captionboxii\fi
 \ifdim\ht\captionboxiii>\dimen@ \dimen@\ht\captionboxiii\fi
 \ifdim\dimen@>\z@
  \ifdim\ht\captionboxi<\dimen@
   \global\setbox\captionboxi\vbox to\dimen@{\vfil\unvbox\captionboxi}\fi
  \ifdim\ht\captionboxii<\dimen@
   \global\setbox\captionboxii\vbox to\dimen@{\vfil\unvbox\captionboxii}\fi
  \ifdim\ht\captionboxiii<\dimen@
   \global\setbox\captionboxiii\vbox to\dimen@{\vfil\unvbox\captionboxiii}\fi
 \fi
 \dimen@ii\ht\islandboxi
 \ifdim\ht\islandboxii>\dimen@ii \dimen@ii\ht\islandboxii\fi
 \ifdim\ht\islandboxiii>\dimen@ii \dimen@ii\ht\islandboxiii\fi
 \ifdim\dimen@ii>\z@
  \ifdim\ht\islandboxi<\dimen@ii
   \global\setbox\islandboxi\vbox to\dimen@ii{\box\islandboxi\vfil}\fi
  \ifdim\ht\islandboxii<\dimen@ii
   \global\setbox\islandboxii\vbox to\dimen@ii{\box\islandboxii\vfil}\fi
  \ifdim\ht\islandboxiii<\dimen@ii
   \global\setbox\islandboxiii\vbox to\dimen@ii{\box\islandboxiii\vfil}\fi
 \fi
 \global\setbox\islandbox@\vbox
  {\ifdim\dimen@>\z@
  \hbox to\hsize{\hfil\box\captionboxi\hfil\box\captionboxii\hfil
   \box\captionboxiii\hfil}%
  \nointerlineskip{\rm\global\skipi@-\dp\strutbox}%
  \global\advance\skipi@\bigskipamount\vskip\skipi@\fi
  \hbox to\hsize{\hfil\box\islandboxi\hfil\box\islandboxii\hfil
   \box\islandboxiii\hfil}}}
\def\Figurepair#1\and#2\endFigurepair{\island@true
 \islandpairbox{\Figure#1\endFigure}{\Figure#2\endFigure}}
\def\Figuretriple#1\and#2\and#3\endFiguretriple{\island@true
 \islandtriplebox{\Figure#1\endFigure}{\Figure#2\endFigure}%
  {\Figure#3\endFigure}}
\def\Tablepair#1\and#2\endTablepair{\island@true
 \islandpairboxa{\Table#1\endTable}{\Table#2\endTable}}
\def\Tabletriple#1\and#2\and#3\endTabletriple{\island@true
 \islandtripleboxa{\Table#1\endTable}{\Table#2\endTable}%
 {\Table#3\endTable}}
\def\place#1{\place@true\island@false
 #1%
 \ifisland@
  \box\islandbox@
 \else
  \Err@{Whoa ... there's no \string\Figure, \string\Table,
   etc., here}%
 \fi
 \place@false}
\newskip\belowtopfigskip
\belowtopfigskip 15\p@ plus 5\p@ minus5\p@
\newskip\abovebotfigskip
\abovebotfigskip 18\p@ plus 6\p@ minus6\p@
\newdimen\minpagesize
\minpagesize 5pc
\dimen@\belowtopfigskip
\advance\dimen@-\abovebotfigskip
\skip\topins\dimen@
\dimen\topins\z@
\newcount\topinscount@
\newbox\topinsdims@
\def\storedim@{\global\setbox\topinsdims@
 \vbox{\hbox to\dimen@{}\unvbox\topinsdims@}}
\def\advancedimtopins@{%
 \ifnum\pageno=\@ne
 \else
   \advance\dimen@\dimen\topins
   \global\dimen\topins\dimen@
 \fi}
\newcount\flipcount@
\def\fliptopins@{%
 \global\flipcount@\z@
 \ifvoid\topins\else
 \setbox\z@\vbox
  {\vskip\p@
   \unvbox\topins
   \global\setbox\topins\vbox{}%
   \loop
    \test@false
    \ifdim\lastskip=\z@\unskip
     \ifdim\lastskip=\z@
      \test@true\fi\fi
    \iftest@
    \global\advance\flipcount@\@ne
    \setboxzl@
    \global\setbox\topins\vbox{\unvbox\topins\boxz@}%
    \unpenalty
   \repeat}\fi}
\newif\ifPar@
\newcount\Parcount@
\newbox\Parbox@
\expandafter\newbox\csname Parfigbox1\endcsname
\expandafter\newbox\csname Parfigbox2\endcsname
\expandafter\newbox\csname Parfigbox3\endcsname
\expandafter\newbox\csname Parfigbox4\endcsname
\expandafter\newbox\csname Parfigbox5\endcsname
\expandafter\newdimen\csname Parprev1\endcsname
\expandafter\newdimen\csname Parprev2\endcsname
\expandafter\newdimen\csname Parprev3\endcsname
\expandafter\newdimen\csname Parprev4\endcsname
\expandafter\newdimen\csname Parprev5\endcsname
\expandafter\newdimen\csname Parprev6\endcsname
\def\Par{\par\global\csname Parprev1\endcsname\prevdepth
 \global\Parcount@\@ne
 \global\Par@true\global\let\Parlist@\empty
 \global\setbox\Parbox@\vbox\bgroup\break}
\def\place@#1#2{%
 \ifisland@
  \ifhmode
   \ifPar@
    \ifnum\Parcount@>5
     \Err@{Only 5 \string\place's allowed per
      \string\Par...\noexpand\endPar paragraph}%
    \else
     \expandafter\expandafter\expandafter
      \global\expandafter\setbox
       \csname Parfigbox\number\Parcount@\endcsname\box\islandbox@
     \global\advance\Parcount@\@ne
     \xdef\Parlist@{\Parlist@#1}%
    \fi
   \else
    \vadjust{#2}%
   \fi
  \else
   #2%
  \fi
 \else
  \Err@{Whoa ... there's no \string\Figure,
   \string\Table, etc., here}%
 \fi
 \place@false}
\long\def\Aplace#1{\prevanish@
 \place@true\island@false
 #1%
 \place@ a\Aplace@
 \postvanish@}
\long\def\AAplace#1{\prevanish@\place@true\island@false
 #1%
 \place@ A\AAplace@
 \postvanish@}
\newif\ifAA@
\def\AAplace@{\AA@true\Aplace@\AA@false}
\let\AAlist@\empty
\def\Aplace@{\allowbreak
 \dimen@=\ht\islandbox@
 \advance\dimen@\abovebotfigskip
 \ht\islandbox@\dimen@
 \advance\dimen@\dp\islandbox@
 \storedim@
 \ifAA@
  \xdef\AAlist@{\AAlist@1}%
  \advancedimtopins@
 \else
  \xdef\AAlist@{\AAlist@0}%
  \ifnum\topinscount@>\@ne\else\advancedimtopins@\fi
 \fi
 \insert\topins{\penalty\z@\splittopskip\z@\floatingpenalty\z@
  \box\islandbox@}%
 \global\advance\topinscount@\@ne}
\long\def\Bplace#1{\prevanish@\place@true\island@false
 #1%
 \place@ b\Bplace@
 \postvanish@}
\def\Bplace@{\allowbreak
 \ifnum\topinscount@=\z@
  \setbox\z@\vbox{\vbox to-\belowtopfigskip{}}%
  \dimen@-\skip\topins
  \ht\z@\dimen@
  \storedim@
  \advancedimtopins@
  \insert\topins{\boxz@}%
  \global\advance\topinscount@\@ne
  \xdef\AAlist@{\AAlist@0}%
 \fi
 \dimen@\ht\islandbox@
 \advance\dimen@\abovebotfigskip
 \ht\islandbox@\dimen@
 \advance\dimen@\dp\islandbox@
 \storedim@
 \xdef\AAlist@{\AAlist@0}%
 \ifnum\topinscount@>\@ne\else\advancedimtopins@\fi
 \insert\topins{\penalty\z@\splittopskip\z@
  \floatingpenalty\z@
  \box\islandbox@}%
 \global\advance\topinscount@\@ne}
\def\breakisland@{\global\setbox\@ne\lastbox\global\skipi@\lastskip\unskip
 \global\setbox\thr@@\lastbox}%
\def\printisland@{\centerline{\box\thr@@}\nobreak\nointerlineskip
 \vskip\skipi@
 \ifdim\ht\@ne<\z@\box\@ne\else\centerline{\box\@ne}\fi}
\def\bottomfigs@{%
 \count@\@ne
 \loop
  \ifnum\count@<\flipcount@
  \nointerlineskip
  \vskip\abovebotfigskip
  \global\setbox\topins\vbox{\unvbox\topins\setboxzl@
   \unvbox\z@
   \breakisland@}%
  \printisland@
  \advance\count@\@ne
  \repeat}
\def\resetdimtopins@{%
 \global\advance\topinscount@-\flipcount@
 \global\setbox\topinsdims@\vbox
  {\unvbox\topinsdims@
   \count@\z@
   \DN@##1##2\next@{\gdef\AAlist@{##2}}%
   \loop
    \ifnum\count@<\flipcount@\setboxzl@
    \expandafter\next@\AAlist@\next@
    \advance\count@\@ne
    \repeat
   \dimen@\z@
   \count@\z@
   \setbox\tw@\vbox{}%
   \edef\nextiii@{\AAlist@}%
   \DN@##1##2\next@{\DNii@{##1}\def\nextiii@{##2}}%
   \loop
    \test@false
    \ifnum\count@<\topinscount@
    \expandafter\next@\nextiii@\next@
     \ifnum\count@<\tw@
      \test@true
     \else
      \if\nextii@ 1\test@true\fi
     \fi
    \fi
    \iftest@
     \setboxzl@
     \advance\dimen@\wdz@
     \setbox\tw@\vbox{\boxz@\unvbox\tw@}%
     \advance\count@\@ne
    \repeat
    \unvbox\tw@
    \global\dimen\topins\dimen@}}
\def\Place@#1#2{%
 \ifisland@
  \ifhmode
   \ifPar@
    \ifnum\Parcount@>5
     \Err@{Only 5 \string\place's allowed per
       \string\Par...\noexpand\endPar paragraph}%
    \else
     \expandafter\expandafter\expandafter\global\expandafter\setbox
      \csname Parfigbox\number\Parcount@\endcsname\box\islandbox@
     \global\advance\Parcount@\@ne
     \xdef\Parlist@{\Parlist@#1}%
     \vadjust{\break}%
    \fi
   \else
    \Err@{\noexpand#2allowed only in a \string\Par...\noexpand\endPar
     paragraph}%
   \fi
  \else
   #2%
  \fi
 \else
  \Err@{Who ... there's no \string\Figure, \string\Table,
   etc., here}%
 \fi
 \place@false}
\newif\ifC@
\newdimen\Cdim@
\long\def\Cplace#1{\prevanish@\place@true\island@false
 #1%
 \Place@ c\Cplace@
 \postvanish@}
\def\Cplace@{\allowbreak
 \ifnum\topinscount@>\z@\else
  \global\C@true\global\Cdim@\pagetotal\fi
 \Aplace@}
\long\def\Mplace#1{\prevanish@\place@true\island@false
 #1%
 \Place@ m\Mplace@
 \postvanish@}
\long\def\MXplace#1{\prevanish@\place@true\island@false
 #1%
 \Place@ M\MXplace@
 \postvanish@}
\newif\ifMX@
\def\MXplace@{\MX@true\Mplace@\MX@false}
\def\Mplace@{\allowbreak
 \dimen@\ht\islandbox@\advance\dimen@\dp\islandbox@
 \ifdim\pagetotal=\z@\else
  \ifdim\lastskip<\abovebotfigskip\advance\dimen@\abovebotfigskip
  \advance\dimen@-\lastskip\fi
 \fi
 \advance\dimen@\pagetotal
 \ifdim\dimen@>\pagegoal
  \Aplace@
 \else
  \nointerlineskip
  \ifdim\lastskip<\abovebotfigskip\removelastskip\vskip\abovebotfigskip\fi
  \setbox\z@\vbox{\unvbox\islandbox@
   \breakisland@}%
  \printisland@
  \ifnum\topinscount@=\z@
   \setbox\z@\vbox{\vbox to-\belowtopfigskip{}}%
   \dimen@-\skip\topins
   \ht\z@\dimen@
   \storedim@
   \advancedimtopins@
   \insert\topins{\boxz@}%
   \global\advance\topinscount@\@ne
   \xdef\AAlist@{\AAlist@0}%
  \fi
  \ifMX@
   \ifnum\topinscount@=\@ne
    \setbox\z@\vbox{\vbox to-\abovebotfigskip{}}%
    \ht\z@\z@
    \dimen@\z@
    \storedim@
    \advancedimtopins@
    \insert\topins{\boxz@}%
    \global\advance\topinscount@\@ne
    \xdef\AAlist@{\AAlist@0}%
   \fi
  \fi
  \nointerlineskip
  \vskip\belowtopfigskip
 \fi}
\expandafter\newbox\csname Parbox1\endcsname
\expandafter\newbox\csname Parbox2\endcsname
\expandafter\newbox\csname Parbox3\endcsname
\expandafter\newbox\csname Parbox4\endcsname
\expandafter\newbox\csname Parbox5\endcsname
\def\endPar{\egroup
 \count@\@ne
 {\vbadness\@M\vfuzz\maxdimen\splitmaxdepth\maxdimen\splittopskip\ht\strutbox
 \setbox\z@\vsplit\Parbox@ to\ht\Parbox@
 \loop
  \ifnum\count@<\Parcount@
  \expandafter\expandafter\expandafter\global\expandafter\setbox
   \csname Parbox\number\count@\endcsname\vsplit\Parbox@ to\ht\Parbox@
  \count@@\count@\advance\count@@\@ne
  \global\csname Parprev\number\count@@\endcsname
   \dp\csname Parbox\number\count@\endcsname
  \advance\count@\@ne
  \repeat}%
 \vskip\parskip
 \count@\@ne
 \def\nextv@##1##2\nextv@{\DN@{##1}\gdef\Parlist@{##2}}%
 \loop
  \ifnum\count@<\Parcount@
   \dimen@\csname Parprev\number\count@\endcsname
   \advance\dimen@\ht\strutbox
   \ifdim\dimen@<\baselineskip
    \advance\dimen@-\baselineskip\vskip-\dimen@
   \else
    \vskip\lineskip
   \fi
   \unvbox\csname Parbox\number\count@\endcsname
   \global\setbox\islandbox@\box\csname Parfigbox\number\count@\endcsname
   \expandafter\nextv@\Parlist@\nextv@
   \if a\next@\Aplace@\else
   \if A\next@\AAplace@\else
   \if b\next@\Bplace@\else
   \if c\next@\Cplace@\else
   \if m\next@\Mplace@\else
   \if M\next@\MXplace@\fi\fi\fi\fi\fi\fi
  \advance\count@\@ne
  \repeat
 \global\Par@false
 \ifvoid\Parbox@
  \prevdepth\csname Parprev\number\count@\endcsname
 \else
  \dimen@\csname Parprev\number\count@\endcsname\advance\dimen@\ht\strutbox
  \ifdim\dimen@<\baselineskip
    \advance\dimen@-\baselineskip\vskip-\dimen@
  \else
    \vskip\lineskip
  \fi
  \dimen@\dp\Parbox@
  \unvbox\Parbox@
  \prevdepth\dimen@
 \fi}
\def\folio{{\page@F\page@S{\page@P\page@N{\number\page@C}\page@Q}}}
\def\advancepageno{\global\advance\pageno\@ne}
\newif\ifspecialsplit@
\newbox\outbox@
\let\shipout@\shipout
\def\plainoutput{\specialsplit@false\ifvoid\topins\else\ifdim\ht\topins=\z@
 \specialsplit@true\advance\minpagesize-\skip\topins\fi\fi
 \fliptopins@
 \setbox\outbox@\vbox{\makeheadline\pagebody\makefootline}%
 {\noexpands@\let\style\relax
 \shipout@\box\outbox@}%
 \advancepageno
 \resetdimtopins@
 \ifvoid\@cclv\else\unvbox\@cclv\penalty\outputpenalty\fi
 \ifnum\outputpenalty>-\@MM\else\dosupereject\fi}
\def\pagebody{\vbox to\vsize{\boxmaxdepth\maxdepth
 \ifvoid\margin@\else
 \rlap{\kern\hsize\vbox to\z@{\kern4\p@\box\margin@\vss}}\fi
 \pagecontents}}
\newif\ifonlytop@
\def\pagecontents{%
 \onlytop@false
 \ifdim\ht\@cclv<\minpagesize\ifnum\flipcount@<\tw@\ifvoid\footins
  \onlytop@true\fi\fi\fi
 \test@false
 \ifC@
  \ifnum\flipcount@=\@ne
   \global\multiply\Cdim@\tw@
   \ifdim\Cdim@>\ht\@cclv
    \test@true
   \fi
  \fi
 \fi
 \global\C@false
 \iftest@
  \dimen@\ht\@cclv
  \advance\dimen@\skip\topins
  {\vfuzz\maxdimen\vbadness\@M
  \splitmaxdepth\maxdepth\splittopskip\topskip
  \setbox\z@\vsplit\@cclv to\dimen@
  \unvbox\z@}%
  \global\setbox\topins\vbox{\unvbox\topins
   \global\setbox\@ne\lastbox}%
  \setbox\z@\vbox{\unvbox\@ne
   \breakisland@}%
  \nointerlineskip
  \vskip\abovebotfigskip
  \printisland@
 \else
  \ifnum\flipcount@>\z@
   \global\setbox\topins\vbox{\unvbox\topins\global\setbox\@ne\lastbox}%
   \setbox\z@\vbox{\unvbox\@ne
    \breakisland@}%
   \printisland@
   \ifonlytop@\kern-\prevdepth\vfill\else\vskip\belowtopfigskip\fi
  \fi
 \fi
 \ifdim\ht\@cclv<\minpagesize
  \ifonlytop@\else\vfill\fi
 \else
  \ifspecialsplit@
   {\vfuzz\maxdimen\vbadness\@M
   \splitmaxdepth\maxdepth\splittopskip\topskip
   \dimen@ii\ht\@cclv \advance\dimen@ii\skip\topins
   \setbox\z@\vsplit\@cclv to\dimen@ii
   \unvbox\z@}%
  \else
   \unvbox\@cclv
  \fi
 \fi
 \bottomfigs@
 \ifvoid\footins\else\vskip\skip\footins\footnoterule\unvbox\footins\fi}
\newread\readdata@
\def\readthedata@#1{\expandafter
 \ifx\csname#1@D\endcsname\relax
  \immediate\openin\readdata@=#1.dat
  \ifeof\readdata@
   \Err@{No file #1.dat}%
  \else
   {\endlinechar\m@ne\gdef\Next@{}%
   \DNii@##1 ##2 ##3pt{\global\data@ht##1\global\data@dp##2%
    \global\data@wd##3pt}%
   \loop
    \ifeof\readdata@
    \else
    \read\readdata@ to\next@
    \ifx\next@\empty\else
     \edef\next@{\expandafter\nextii@\next@}%
     \expandafter\rightadd@\next@\to\Next@
    \fi
    \repeat}%
   \immediate\closein\readdata@
   \expandafter\expandafter\expandafter\global\expandafter
    \let\csname#1@D\endcsname\Next@\global\let\Next@\relax
  \fi
 \fi}
\newdimen\data@ht
\newdimen\data@dp
\newdimen\data@wd
\newif\ifgetdata@
\def\getdata@#1#2{\global\getdata@true\count@#2\relax
 {\let\\\or\xdef\Next@{\ifcase\number\count@#1\else
 \global\noexpand\getdata@false\fi}}\Next@}
\def\paste#1#2{\readthedata@{#1}%
 \getdata@{\csname#1@D\endcsname}{#2}%
 \ifgetdata@
 \dimen@\data@ht \advance\dimen@\data@dp
  \hbox{\special{dvipaste: #1 #2}%
   \lower\data@dp\vbox to\dimen@{\hbox to\data@wd{}\vfil}}%
 \else
  {\lccode`\Z=`\#\lccode`\N=`\N\lccode`\F=`\F%
   \lowercase{\Err@{No data for File [#1], Z#2}}}%
 \fi}
\newdimen\httable
\newdimen\dptable
\newdimen\wdtable
\def\measuretable#1#2{\readthedata@{#1}%
 \getdata@{\csname#1@D\endcsname}{#2}%
 \ifgetdata@
  \httable\data@ht \dptable\data@dp \wdtable\data@wd
 \else
  {\lccode`\Z=`\#\lccode`\N=`\N\lccode`\F=`\F%
  \lowercase{\Err@{No data for File [#1], Z#2}}}%
 \fi}
\def\East#1#2{\setboxz@h{$\m@th\ssize\;{#1}\;\;$}%
 \setbox\tw@\hbox{$\m@th\ssize\;{#2}\;\;$}\setbox4=\hbox{$\m@th#2$}%
 \dimen@\minaw@
 \ifdim\wdz@>\dimen@\dimen@\wdz@\fi\ifdim\wd\tw@>\dimen@\dimen@\wd\tw@\fi
 \ifdim\wd4 >\z@
  \mathrel{\mathop{\hbox to\dimen@{\rightarrowfill}}\limits^{#1}_{#2}}%
 \else
  \mathrel{\mathop{\hbox to\dimen@{\rightarrowfill}}\limits^{#1}}%
 \fi}
\def\West#1#2{\setboxz@h{$\m@th\ssize\;\;{#1}\;$}%
 \setbox\tw@\hbox{$\m@th\ssize\;\;{#2}\;$}\setbox4=\hbox{$\m@th#2$}%
 \dimen@\minaw@
 \ifdim\wdz@>\dimen@\dimen@\wdz@\fi\ifdim\wd\tw@>\dimen@\dimen@\wd\tw@\fi
 \ifdim\wd4 >\z@
  \mathrel{\mathop{\hbox to\dimen@{\leftarrowfill}}\limits^{#1}_{#2}}%
 \else
  \mathrel{\mathop{\hbox to\dimen@{\leftarrowfill}}\limits^{#1}}%
 \fi}
\font\arrow@i=lams1
\font\arrow@ii=lams2
\font\arrow@iii=lams3
\font\arrow@iv=lams4
\font\arrow@v=lams5
\newdimen\standardcgap
\standardcgap40\p@
\newdimen\hunit
\hunit\tw@\p@
\newdimen\standardrgap
\standardrgap32\p@
\newdimen\vunit
\vunit1.6\p@
\def\Cgaps#1{\RIfM@
 \standardcgap#1\standardcgap\relax\hunit#1\hunit\relax
 \else\nonmatherr@\Cgaps\fi}
\def\Rgaps#1{\RIfM@
 \standardrgap#1\standardrgap\relax\vunit#1\vunit\relax
 \else\nonmatherr@\Rgaps\fi}
\newdimen\getdim@
\def\getcgap@#1{\ifcase#1\or\getdim@\z@\else\getdim@\standardcgap\fi}
\def\getrgap@#1{\ifcase#1\getdim@\z@\else\getdim@\standardrgap\fi}
\def\cgaps{\RIfM@\expandafter\cgaps@\else\expandafter\nonmatherr@
 \expandafter\cgaps\fi}
\def\cgaps@{\ifnum\catcode`\;=\active\expandafter\cgapsA@\else
 \expandafter\cgapsO@\fi}
\def\cgapsO@#1{\toks@{\ifcase\i@\or\getdim@=\z@}%
 \gaps@@\standardcgap#1;\gaps@@\gaps@@
 \edef\next@{\the\toks@\noexpand\else\noexpand\getdim@\noexpand\standardcgap
  \noexpand\fi}%
 \toks@=\expandafter{\next@}%
 \edef\getcgap@##1{\i@##1\relax\the\toks@}\toks@{}}
{\catcode`\;=\active
 \gdef\cgapsA@#1{\toks@{\ifcase\i@\or\getdim@=\z@}%
 \gaps@@\standardcgap#1;\gaps@@\gaps@@
 \edef\next@{\the\toks@\noexpand\else\noexpand\getdim@\noexpand\standardcgap
  \noexpand\fi}%
 \toks@=\expandafter{\next@}%
 \edef\getcgap@##1{\i@##1\relax\the\toks@}\toks@{}}
}
\def\Gaps@@{\gaps@@}
\def\gaps@@#1#2;#3{\mgaps@#1#2\mgaps@
 \edef\next@{\the\toks@\noexpand\or\noexpand\getdim@
  \noexpand#1\the\mgapstoks@@}%
 \toks@\expandafter{\next@}%
 \DN@{#3}%
 \ifx\next@\Gaps@@\def\next@##1\gaps@@{}\else
  \def\next@{\gaps@@#1#3}\fi\next@}
{\catcode`\;=\active
 \gdef\rgaps#1{\RIfM@{\ifnum\catcode`\;=\active\def;{\string;}\fi
   \xdef\Next@{\noexpand\rgaps@{#1}}}%
  \Next@\edef\getrgap@##1{\i@##1\relax\the\toks@}\toks@{}\else
  \nonmatherr@\rgaps\fi}
}
\def\rgaps@#1{\toks@{\ifcase\i@\getdim@=\z@}%
 \gaps@@\standardrgap#1;\gaps@@\gaps@@
 \edef\next@{\the\toks@\noexpand\else\noexpand\getdim@\noexpand\standardrgap
  \noexpand\fi}%
 \toks@=\expandafter{\next@}}
\newbox\ZER@
\def\mgaps@#1{\let\mgapsnext@#1\FNSS@\mgaps@@}
\def\mgaps@@{\ifx\next\w\expandafter\mgaps@@@\else
 \expandafter\mgaps@@@@\fi}
\newtoks\mgapstoks@@
\def\mgaps@@@@#1\mgaps@{\getdim@\mgapsnext@\getdim@#1\getdim@
 \edef\next@{\noexpand\getdim@\the\getdim@}%
 \mgapstoks@@\expandafter{\next@}}
\def\mgaps@@@\w#1#2\mgaps@{\mgaps@@@@#2\mgaps@
 \setbox\ZER@\hbox{$\m@th\hskip15\p@\tsize@#1$}%
 \dimen@\wd\ZER@
 \ifdim\dimen@>\getdim@\getdim@\dimen@\fi
 \edef\next@{\noexpand\getdim@\the\getdim@}%
 \mgapstoks@@\expandafter{\next@}}
\def\changewidth#1#2{\setbox\ZER@{$\m@th#2}%
 \hbox to\wd\ZER@{\hss$\m@th#1$\hss}}
\atdef@({\FN@\ARROW@}
\def\ARROW@{\ifx\next)\let\next@\OPTIONS@\else
 \DN@{\csname\string @(\endcsname}\fi\next@}
\newif\ifoptions@
\def\OPTIONS@){\ifoptions@\let\next@\relax\else
 \DN@{\global\options@true\begingroup\optioncodes@}\fi\next@}
\newif\ifN@
\newif\ifE@
\newif\ifNESW@
\newif\ifH@
\newif\ifV@
\newif\ifHshort@
\expandafter\def\csname\string @(\endcsname #1,#2){%
 \ifoptions@\expandafter\endgroup\fi
 \N@false\E@false\H@false\V@false\Hshort@false
 \ifnum#1>\z@\E@true\fi
 \ifnum#1=\z@\V@true\global\tX@false\global\tY@false\global\a@false\fi
 \ifnum#2>\z@\N@true\fi
 \ifnum#2=\z@\H@true\global\tX@false\global\tY@false\global\a@false
  \ifshort@\Hshort@true\fi\fi
 \NESW@false
 \ifN@\ifE@\NESW@true\fi\else\ifE@\else\NESW@true\fi\fi
 \arrow@{#1}{#2}%
 \global\options@false
 \global\scount@\z@\global\tcount@\z@\global\arrcount@\z@
 \global\s@false\global\sxdimen@\z@\global\sydimen@\z@
 \global\tX@false\global\tXdimen@i\z@\global\tXdimen@ii\z@
 \global\tY@false\global\tYdimen@i\z@\global\tYdimen@ii\z@
 \global\a@false\global\exacount@\z@
 \global\x@false\global\xdimen@\z@
 \global\X@false\global\Xdimen@\z@
 \global\y@false\global\ydimen@\z@
 \global\Y@false\global\Ydimen@\z@
 \global\p@false\global\pdimen@\z@
 \global\label@ifalse\global\label@iifalse
 \global\dl@ifalse\global\ldimen@i\z@
 \global\dl@iifalse\global\ldimen@ii\z@
 \global\short@false\global\unshort@false}
\newif\iflabel@i
\newif\iflabel@ii
\newcount\scount@
\newcount\tcount@
\newcount\arrcount@
\newif\ifs@
\newdimen\sxdimen@
\newdimen\sydimen@
\newif\iftX@
\newdimen\tXdimen@i
\newdimen\tXdimen@ii
\newif\iftY@
\newdimen\tYdimen@i
\newdimen\tYdimen@ii
\newif\ifa@
\newcount\exacount@
\newif\ifx@
\newdimen\xdimen@
\newif\ifX@
\newdimen\Xdimen@
\newif\ify@
\newdimen\ydimen@
\newif\ifY@
\newdimen\Ydimen@
\newif\ifp@
\newdimen\pdimen@
\newif\ifdl@i
\newif\ifdl@ii
\newdimen\ldimen@i
\newdimen\ldimen@ii
\newif\ifshort@
\newif\ifunshort@
\def\zero@#1{\ifnum\scount@=\z@
 \if#1e\global\scount@\m@ne\else
 \if#1t\global\scount@\tw@\else
 \if#1h\global\scount@\thr@@\else
 \if#1'\global\scount@6 \else
 \if#1`\global\scount@7 \else
 \if#1(\global\scount@8 \else
 \if#1)\global\scount@9 \else
 \if#1s\global\scount@12 \else
 \if#1H\global\scount@13 \else
 \Err@{\Invalid@@ option \string\0}\fi\fi\fi\fi\fi\fi\fi\fi\fi
 \fi}
\def\one@#1{\ifnum\tcount@=\z@
 \if#1e\global\tcount@\m@ne\else
 \if#1h\global\tcount@\tw@\else
 \if#1t\global\tcount@\thr@@\else
 \if#1'\global\tcount@4 \else
 \if#1`\global\tcount@5 \else
 \if#1(\global\tcount@\ten@ \else
 \if#1)\global\tcount@11 \else
 \if#1s\global\tcount@12 \else
 \if#1H\global\tcount@13 \else
 \Err@{\Invalid@@ option \string\1}\fi\fi\fi\fi\fi\fi\fi\fi\fi
 \fi}
\def\a@#1{\ifnum\arrcount@=\z@
 \if#10\global\arrcount@\m@ne\else
 \if#1+\global\arrcount@\@ne\else
 \if#1-\global\arrcount@\tw@\else
 \if#1=\global\arrcount@\thr@@\else
 \Err@{\Invalid@@ option \string\a}\fi\fi\fi\fi
 \fi}
\def\ds@{\ifnum\catcode`\;=\active\expandafter\dsA@\else
 \expandafter\dsO@\fi}
\def\dsO@(#1;#2){\ds@@{#1}{#2}}
\def\ds@@#1#2{\ifs@\else
 \global\s@true
 \global\sxdimen@\hunit\global\sxdimen@#1\sxdimen@\relax
 \global\sydimen@\vunit\global\sydimen@#2\sydimen@\relax
 \fi}
\def\dtX@{\ifnum\catcode`\;=\active\expandafter\dtXA@\else
 \expandafter\dtXO@\fi}
\def\dtXO@(#1;#2){\dtX@@{#1}{#2}}
\def\dtX@@#1#2{\iftX@\else
 \global\tX@true
 \global\tXdimen@i\hunit\global\tXdimen@i#1\tXdimen@i\relax
 \global\tXdimen@ii\vunit\global\tXdimen@ii#2\tXdimen@ii\relax
 \fi}
\def\dtY@{\ifnum\catcode`\;=\active\expandafter\dtYA@\else
 \expandafter\dtYO@\fi}
\def\dtYO@(#1;#2){\dtY@@{#1}{#2}}
\def\dtY@@#1#2{\iftY@\else
 \global\tY@true
 \global\tYdimen@i\hunit\global\tYdimen@i#1\tYdimen@i\relax
 \global\tYdimen@ii\vunit\global\tYdimen@ii#2\tYdimen@ii\relax
 \fi}
{\catcode`\;=\active
 \gdef\dsA@(#1;#2){\ds@@{#1}{#2}}
 \gdef\dtXA@(#1;#2){\dtX@@{#1}{#2}}
 \gdef\dtYA@(#1;#2){\dtY@@{#1}{#2}}
}
\def\da@#1{\ifa@\else\global\a@true\global\exacount@#1\relax\fi}
\def\dx@#1{\ifx@\else
 \global\x@true
 \global\xdimen@\hunit\global\xdimen@#1\xdimen@\relax
 \fi}
\def\dX@#1{\ifX@\else
 \global\X@true
 \global\Xdimen@\hunit\global\Xdimen@#1\Xdimen@\relax
 \fi}
\def\dy@#1{\ify@\else
 \global\y@true
 \global\ydimen@\vunit\global\ydimen@#1\ydimen@\relax
 \fi}
\def\dY@#1{\ifY@\else
 \global\Y@true
 \global\Ydimen@\vunit\global\Ydimen@#1\Ydimen@\relax
 \fi}
\def\p@@#1{\ifp@\else
 \global\p@true
 \global\pdimen@\hunit\global\divide\pdimen@\tw@
 \global\pdimen@#1\pdimen@\relax
 \fi}
\def\L@#1{\iflabel@i\else
 \global\label@itrue\gdef\label@i{#1}%
 \fi}
\def\l@#1{\iflabel@ii\else
 \global\label@iitrue\gdef\label@ii{#1}%
 \fi}
\def\dL@#1{\ifdl@i\else
 \global\dl@itrue\global\ldimen@i\hunit\global\ldimen@i#1\ldimen@i\relax
 \fi}
\def\dl@#1{\ifdl@ii\else
 \global\dl@iitrue\global\ldimen@ii\hunit\global\ldimen@ii#1\ldimen@ii\relax
 \fi}
\def\s@{\ifunshort@\else\global\short@true\fi}
\def\uns@{\ifshort@\else\global\unshort@true\global\short@false\fi}
\def\optioncodes@{\let\0\zero@\let\1\one@\let\a\a@\let\ds\ds@\let\dtX\dtX@
 \let\dtY\dtY@\let\da\da@\let\dx\dx@\let\dX\dX@\let\dY\dY@\let\dy\dy@
 \let\p\p@@\let\L\L@\let\l\l@\let\dL\dL@\let\dl\dl@\let\s\s@\let\uns\uns@}
\def\slopes@{\\161\\152\\143\\134\\255\\126\\357\\238\\349\\45{10}\\56{11}%
 \\11{12}\\65{13}\\54{14}\\43{15}\\32{16}\\53{17}\\21{18}\\52{19}\\31{20}%
 \\41{21}\\51{22}\\61{23}}
\newcount\tan@i
\newcount\tan@ip
\newcount\tan@ii
\newcount\tan@iip
\newdimen\slope@i
\newdimen\slope@ip
\newdimen\slope@ii
\newdimen\slope@iip
\newcount\angcount@
\newcount\extracount@
\def\slope@{{\slope@i\secondy@\advance\slope@i-\firsty@
 \ifN@\else\multiply\slope@i\m@ne\fi
 \slope@ii\secondx@\advance\slope@ii-\firstx@
 \ifE@\else\multiply\slope@ii\m@ne\fi
 \ifdim\slope@ii<\z@
  \global\tan@i6 \global\tan@ii\@ne\global\angcount@23
 \else
  \dimen@\slope@i\multiply\dimen@6
  \ifdim\dimen@<\slope@ii
   \global\tan@i\@ne\global\tan@ii6 \global\angcount@\@ne
  \else
   \dimen@\slope@ii\multiply\dimen@6
   \ifdim\dimen@<\slope@i
    \global\tan@i6 \global\tan@ii\@ne\global\angcount@23
   \else
    \global\tan@ip\z@\global\tan@iip\@ne
    \def\\##1##2##3{\global\angcount@##3\relax
     \slope@ip\slope@i\slope@iip\slope@ii
     \multiply\slope@iip##1\relax\multiply\slope@ip##2\relax
     \ifdim\slope@iip<\slope@ip
      \global\tan@ip##1\relax\global\tan@iip##2\relax
     \else
      \global\tan@i##1\relax\global\tan@ii##2\relax
      \def\\####1####2####3{}%
     \fi}%
    \slopes@
    \slope@i\secondy@\advance\slope@i-\firsty@
    \ifN@\else\multiply\slope@i\m@ne\fi
    \multiply\slope@i\tan@ii\multiply\slope@i\tan@iip\multiply\slope@i\tw@
    \count@\tan@i\multiply\count@\tan@iip
    \extracount@\tan@ip\multiply\extracount@\tan@ii
    \advance\count@\extracount@
    \slope@ii\secondx@\advance\slope@ii-\firstx@
    \ifE@\else\multiply\slope@ii\m@ne\fi
    \multiply\slope@ii\count@
    \ifdim\slope@i<\slope@ii
     \global\tan@i\tan@ip\global\tan@ii\tan@iip
     \global\advance\angcount@\m@ne
    \fi
   \fi
  \fi
 \fi}%
}
\def\slope@a#1{{\def\\##1##2##3{\ifnum##3=#1\global\tan@i##1\relax
 \global\tan@ii##2\relax\fi}\slopes@}}
\newcount\i@
\newcount\j@
\newcount\colcount@
\newcount\Colcount@
\newcount\tcolcount@
\newdimen\rowht@
\newdimen\rowdp@
\newcount\rowcount@
\newcount\Rowcount@
\newcount\maxcolrow@
\newtoks\colwidthtoks@
\newtoks\Rowheighttoks@
\newtoks\Rowdepthtoks@
\newtoks\widthtoks@
\newtoks\Widthtoks@
\newtoks\heighttoks@
\newtoks\Heighttoks@
\newtoks\depthtoks@
\newtoks\Depthtoks@
\newif\iffirstCDcr@
\def\dotoks@i{%
 \global\widthtoks@\expandafter{\the\widthtoks@\else\getdim@\z@\fi}%
 \global\heighttoks@\expandafter{\the\heighttoks@\else\getdim@\z@\fi}%
 \global\depthtoks@\expandafter{\the\depthtoks@\else\getdim@\z@\fi}}
\def\dotoks@ii{%
 \global\widthtoks@{\ifcase\j@}%
 \global\heighttoks@{\ifcase\j@}%
 \global\depthtoks@{\ifcase\j@}}
\def\preCD@#1\endCD{\setbox\ZER@
 \vbox{%
  \def\arrow@##1##2{{}}%
  \global\rowcount@\m@ne\global\colcount@\z@\global\Colcount@\z@
  \global\firstCDcr@true\toks@{}%
  \global\widthtoks@{\ifcase\j@}%
  \global\Widthtoks@{\ifcase\i@}%
  \global\heighttoks@{\ifcase\j@}%
  \global\Heighttoks@{\ifcase\i@}%
  \global\depthtoks@{\ifcase\j@}%
  \global\Depthtoks@{\ifcase\i@}%
  \global\Rowheighttoks@{\ifcase\i@}%
  \global\Rowdepthtoks@{\ifcase\i@}%
  \Let@
  \everycr{%
   \noalign{%
    \global\advance\rowcount@\@ne
    \ifnum\colcount@<\Colcount@
    \else
     \global\Colcount@\colcount@\global\maxcolrow@\rowcount@
    \fi
    \global\colcount@\z@
    \iffirstCDcr@
     \global\firstCDcr@false
    \else
     \edef\next@{\the\Rowheighttoks@\noexpand\or\noexpand\getdim@\the\rowht@}%
      \global\Rowheighttoks@\expandafter{\next@}%
     \edef\next@{\the\Rowdepthtoks@\noexpand\or\noexpand\getdim@\the\rowdp@}%
      \global\Rowdepthtoks@\expandafter{\next@}%
     \global\rowht@\z@\global\rowdp@\z@
     \dotoks@i
     \edef\next@{\the\Widthtoks@\noexpand\or\the\widthtoks@}%
      \global\Widthtoks@\expandafter{\next@}%
     \edef\next@{\the\Heighttoks@\noexpand\or\the\heighttoks@}%
      \global\Heighttoks@\expandafter{\next@}%
     \edef\next@{\the\Depthtoks@\noexpand\or\the\depthtoks@}%
      \global\Depthtoks@\expandafter{\next@}%
     \dotoks@ii
    \fi}}%
  \tabskip\z@
  \halign{&\setbox\ZER@\hbox{\vrule\height\ten@\p@\width\z@\depth\z@     
   $\m@th\displaystyle{##}$}\copy\ZER@
   \ifdim\ht\ZER@>\rowht@\global\rowht@\ht\ZER@\fi
   \ifdim\dp\ZER@>\rowdp@\global\rowdp@\dp\ZER@\fi
   \global\advance\colcount@\@ne
   \edef\next@{\the\widthtoks@\noexpand\or\noexpand\getdim@\the\wd\ZER@}%
    \global\widthtoks@\expandafter{\next@}%
   \edef\next@{\the\heighttoks@\noexpand\or\noexpand\getdim@\the\ht\ZER@}%
    \global\heighttoks@\expandafter{\next@}%
   \edef\next@{\the\depthtoks@\noexpand\or\noexpand\getdim@\the\dp\ZER@}%
    \global\depthtoks@\expandafter{\next@}%
   \cr#1\crcr}}%
 \Rowcount@\rowcount@
 \global\Widthtoks@\expandafter{\the\Widthtoks@\fi\relax}%
 \edef\Width@##1##2{\i@##1\relax\j@##2\relax\the\Widthtoks@}%
 \global\Heighttoks@\expandafter{\the\Heighttoks@\fi\relax}%
 \edef\Height@##1##2{\i@##1\relax\j@##2\relax\the\Heighttoks@}%
 \global\Depthtoks@\expandafter{\the\Depthtoks@\fi\relax}%
 \edef\Depth@##1##2{\i@##1\relax\j@##2\relax\the\Depthtoks@}%
 \edef\next@{\the\Rowheighttoks@\noexpand\fi\relax}%
 \global\Rowheighttoks@\expandafter{\next@}%
 \edef\Rowheight@##1{\i@##1\relax\the\Rowheighttoks@}%
 \edef\next@{\the\Rowdepthtoks@\noexpand\fi\relax}%
 \global\Rowdepthtoks@\expandafter{\next@}%
 \edef\Rowdepth@##1{\i@##1\relax\the\Rowdepthtoks@}%
 \global\colwidthtoks@{\fi}%
 \setbox\ZER@\vbox{%
  \unvbox\ZER@
  \count@\rowcount@
  \loop
   \unskip\unpenalty
   \setbox\ZER@\lastbox
   \ifnum\count@>\maxcolrow@\advance\count@\m@ne
   \repeat
  \hbox{%
   \unhbox\ZER@
   \count@\z@
   \loop
    \unskip
    \setbox\ZER@\lastbox
    \edef\next@{\noexpand\or\noexpand\getdim@\the\wd\ZER@\the\colwidthtoks@}%
     \global\colwidthtoks@\expandafter{\next@}%
    \advance\count@\@ne
    \ifnum\count@<\Colcount@
    \repeat}}%
 \edef\next@{\noexpand\ifcase\noexpand\i@\the\colwidthtoks@}%
  \global\colwidthtoks@\expandafter{\next@}%
 \edef\Colwidth@##1{\i@##1\relax\the\colwidthtoks@}%
 \global\colwidthtoks@{}\global\Rowheighttoks@{}\global\Rowdepthtoks@{}%
 \global\widthtoks@{}\global\Widthtoks@{}\global\heighttoks@{}%
 \global\Heighttoks@{}\global\depthtoks@{}\global\Depthtoks@{}%
}
\newcount\xoff@
\newcount\yoff@
\newcount\endcount@
\newcount\rcount@
\newdimen\firstx@
\newdimen\firsty@
\newdimen\secondx@
\newdimen\secondy@
\newdimen\tocenter@
\newdimen\charht@
\newdimen\charwd@
\def\outside@{\Err@{This arrow points outside the \string\CD}}
\newif\ifsvertex@
\newif\iftvertex@
\def\arrow@#1#2{\global\xoff@#1\relax\global\yoff@#2\relax
 \count@\rowcount@\advance\count@-\yoff@
 \ifnum\count@<\@ne\outside@\else\ifnum\count@>\Rowcount@\outside@\fi\fi
 \count@\colcount@\advance\count@\xoff@
 \ifnum\count@<\@ne\outside@\else\ifnum\count@>\Colcount@\outside@\fi\fi
 \tcolcount@\colcount@\advance\tcolcount@\xoff@
 \Width@\rowcount@\colcount@\divide\getdim@\tw@\tocenter@-\getdim@
 \ifdim\getdim@=\z@
  \firstx@\z@\firsty@\mathaxis@\svertex@true
 \else
  \svertex@false
  \ifHshort@
   \Colwidth@\colcount@\divide\getdim@\tw@
   \ifE@ \firstx@\getdim@ \else \firstx@-\getdim@ \fi
  \else
   \ifE@ \firstx@\getdim@ \else \firstx@-\getdim@ \fi
  \fi
  \ifE@
   \ifH@ \advance\firstx@\thr@@\p@ \else \advance\firstx@-\thr@@\p@ \fi  
  \else
   \ifH@ \advance\firstx@-\thr@@\p@ \else \advance\firstx@\thr@@\p@ \fi  
  \fi
  \ifN@
   \Height@\rowcount@\colcount@ \firsty@\getdim@                         
   \ifV@ \advance\firsty@\thr@@\p@ \fi                                   
  \else
   \ifV@
    \Depth@\rowcount@\colcount@ \firsty@-\getdim@                        
    \advance\firsty@-\thr@@\p@                                           
   \else
    \firsty@\z@                                                          
   \fi
  \fi
 \fi
 \ifV@
 \else
  \Colwidth@\colcount@\divide\getdim@\tw@
  \ifE@\secondx@\getdim@\else\secondx@-\getdim@\fi
  \ifE@\else\getcgap@\colcount@\advance\secondx@-\getdim@\fi
  \endcount@\colcount@\advance\endcount@\xoff@
  \count@\colcount@
  \ifE@
   \advance\count@\@ne
   \loop
    \ifnum\count@<\endcount@
    \Colwidth@\count@\advance\secondx@\getdim@
    \getcgap@\count@\advance\secondx@\getdim@
    \advance\count@\@ne
    \repeat
  \else
   \advance\count@\m@ne
   \loop
    \ifnum\count@>\endcount@
    \Colwidth@\count@\advance\secondx@-\getdim@
    \getcgap@\count@\advance\secondx@-\getdim@
    \advance\count@\m@ne
    \repeat
  \fi
  \Colwidth@\count@\divide\getdim@\tw@
  \ifHshort@
  \else
   \ifE@\advance\secondx@\getdim@\else\advance\secondx@-\getdim@\fi
  \fi
  \ifE@\getcgap@\count@\advance\secondx@\getdim@\fi
  \rcount@\rowcount@\advance\rcount@-\yoff@
  \Width@\rcount@\count@\divide\getdim@\tw@
  \tvertex@false
  \ifH@\ifdim\getdim@=\z@\tvertex@true\Hshort@false\fi\fi
  \ifHshort@
  \else
   \ifE@\advance\secondx@-\getdim@\else\advance\secondx@\getdim@\fi
  \fi
  \iftvertex@
   \advance\secondx@.4\p@
  \else
   \ifE@\advance\secondx@-\thr@@\p@\else\advance\secondx@\thr@@\p@\fi    
  \fi
 \fi
 \ifH@
 \else
  \ifN@
   \Rowheight@\rowcount@\secondy@\getdim@
  \else
   \Rowdepth@\rowcount@\secondy@-\getdim@
   \getrgap@\rowcount@\advance\secondy@-\getdim@
  \fi
  \endcount@\rowcount@\advance\endcount@-\yoff@
  \count@\rowcount@
  \ifN@
   \advance\count@\m@ne
   \loop
    \ifnum\count@>\endcount@
    \Rowheight@\count@\advance\secondy@\getdim@
    \Rowdepth@\count@\advance\secondy@\getdim@
    \getrgap@\count@\advance\secondy@\getdim@
    \advance\count@\m@ne
    \repeat
  \else
   \advance\count@\@ne
   \loop
    \ifnum\count@<\endcount@
    \Rowheight@\count@\advance\secondy@-\getdim@
    \Rowdepth@\count@\advance\secondy@-\getdim@
    \getrgap@\count@\advance\secondy@-\getdim@
    \advance\count@\@ne
    \repeat
  \fi
  \tvertex@false
  \ifV@\Width@\count@\colcount@\ifdim\getdim@=\z@\tvertex@true\fi\fi
  \ifN@
   \getrgap@\count@\advance\secondy@\getdim@
   \Rowdepth@\count@\advance\secondy@\getdim@
   \iftvertex@
    \advance\secondy@\mathaxis@
   \else
    \Depth@\count@\tcolcount@\advance\secondy@-\getdim@
    \advance\secondy@-\thr@@\p@                                          
   \fi
  \else
   \Rowheight@\count@\advance\secondy@-\getdim@
   \iftvertex@
    \advance\secondy@\mathaxis@
   \else
    \Height@\count@\tcolcount@\advance\secondy@\getdim@
    \advance\secondy@\thr@@\p@                                           
   \fi
  \fi
 \fi
 \ifV@\else\advance\firstx@\sxdimen@\fi
 \ifH@\else\advance\firsty@\sydimen@\fi
 \iftX@
  \advance\secondy@\tXdimen@ii
  \advance\secondx@\tXdimen@i
  \slope@
 \else
  \iftY@
   \advance\secondy@\tYdimen@ii
   \advance\secondx@\tYdimen@i
   \slope@
   \secondy@\secondx@\advance\secondy@-\firstx@
   \ifNESW@\else\multiply\secondy@\m@ne\fi
   \multiply\secondy@\tan@i\divide\secondy@\tan@ii\advance\secondy@\firsty@
  \else
   \ifa@
    \slope@
    \ifNESW@\global\advance\angcount@\exacount@\else
     \global\advance\angcount@-\exacount@\fi
    \ifnum\angcount@>23 \global\angcount@23 \fi
    \ifnum\angcount@<\@ne\global\angcount@\@ne\fi
    \slope@a\angcount@
    \ifY@
     \advance\secondy@\Ydimen@
    \else
     \ifX@
      \advance\secondx@\Xdimen@
      \dimen@\secondx@\advance\dimen@-\firstx@
      \ifNESW@\else\multiply\dimen@\m@ne\fi
      \multiply\dimen@\tan@i\divide\dimen@\tan@ii
      \advance\dimen@\firsty@\secondy@\dimen@
     \fi
    \fi
   \else
    \ifH@\else\ifV@\else\slope@\fi\fi
   \fi
  \fi
 \fi
 \ifH@\else\ifV@\else\ifsvertex@\else
  \dimen@6\p@\multiply\dimen@\tan@ii
  \count@\tan@i\advance\count@\tan@ii\divide\dimen@\count@
  \ifE@\advance\firstx@\dimen@\else\advance\firstx@-\dimen@\fi
  \multiply\dimen@\tan@i\divide\dimen@\tan@ii
  \ifN@\advance\firsty@\dimen@\else\advance\firsty@-\dimen@\fi
 \fi\fi\fi
 \ifp@
  \ifH@\else\ifV@\else
   \getcos@\pdimen@\advance\firsty@\dimen@\advance\secondy@\dimen@
   \ifNESW@\advance\firstx@-\dimen@ii\else\advance\firstx@\dimen@ii\fi
  \fi\fi
 \fi
 \ifH@\else\ifV@\else
  \ifnum\tan@i>\tan@ii
   \charht@\ten@\p@\charwd@\ten@\p@
   \multiply\charwd@\tan@ii\divide\charwd@\tan@i
  \else
   \charwd@\ten@\p@\charht@\ten@\p@
   \divide\charht@\tan@ii\multiply\charht@\tan@i
  \fi
  \ifnum\tcount@=\thr@@
   \ifN@\advance\secondy@-.3\charht@\else\advance\secondy@.3\charht@\fi
  \fi
  \ifnum\scount@=\tw@
   \ifE@\advance\firstx@.3\charht@\else\advance\firstx@-.3\charht@\fi
  \fi
  \ifnum\tcount@=12
   \ifN@\advance\secondy@-\charht@\else\advance\secondy@\charht@\fi
  \fi
  \iftY@
  \else
   \ifa@
    \ifX@
    \else
     \secondx@\secondy@\advance\secondx@-\firsty@
     \ifNESW@\else\multiply\secondx@\m@ne\fi
     \multiply\secondx@\tan@ii\divide\secondx@\tan@i
     \advance\secondx@\firstx@
    \fi
   \fi
  \fi
 \fi\fi
 \ifH@\harrow@\else\ifV@\varrow@\else\arrow@@\fi\fi}
\newdimen\mathaxis@
\mathaxis@90\p@\divide\mathaxis@36
\def\harrow@b{\ifE@\hskip\tocenter@\hskip\firstx@\fi}
\def\harrow@bb{\ifE@\hskip\xdimen@\else\hskip\Xdimen@\fi}
\def\harrow@e{\ifE@\else\hskip-\firstx@\hskip-\tocenter@\fi}
\def\harrow@ee{\ifE@\hskip-\Xdimen@\else\hskip-\xdimen@\fi}
\def\harrow@{\dimen@\secondx@\advance\dimen@-\firstx@
 \ifE@\let\next@\rlap\else\multiply\dimen@\m@ne\let\next@\llap\fi
 \next@{%
  \harrow@b
  \smash{\raise\pdimen@\hbox to\dimen@
   {\harrow@bb\arrow@ii
    \ifnum\arrcount@=\m@ne\else\ifnum\arrcount@=\thr@@\else
     \ifE@
      \ifnum\scount@=\m@ne
      \else
       \ifcase\scount@\or\or\char118 \or\char117 \or\or\or\char119 \or
       \char120 \or\char121 \or\char122 \or\or\or\arrow@i\char125 \or
       \char117 \hskip\thr@@\p@\char117 \hskip-\thr@@\p@\fi
      \fi
     \else
      \ifnum\tcount@=\m@ne
      \else
       \ifcase\tcount@\char117 \or\or\char117 \or\char118 \or\char119 \or
       \char120 \or\or\or\or\or\char121 \or\char122 \or\arrow@i\char125
       \or\char117 \hskip\thr@@\p@\char117 \hskip-\thr@@\p@\fi
      \fi
     \fi
    \fi\fi
    \dimen@\mathaxis@\advance\dimen@.2\p@
    \dimen@ii\mathaxis@\advance\dimen@ii-.2\p@
    \ifnum\arrcount@=\m@ne
     \let\leads@\null
    \else
     \ifcase\arrcount@
      \def\leads@{\hrule\height\dimen@\depth-\dimen@ii}\or
      \def\leads@{\hrule\height\dimen@\depth-\dimen@ii}\or
      \def\leads@{\hbox to\ten@\p@{%
       \leaders\hrule\height\dimen@\depth-\dimen@ii\hfil
       \hfil
      \leaders\hrule\height\dimen@\depth-\dimen@ii\hskip\z@ plus2fil\relax
       \hfil
       \leaders\hrule\height\dimen@\depth-\dimen@ii\hfil}}\or
     \def\leads@{\hbox{\hbox to\ten@\p@{\dimen@\mathaxis@\advance\dimen@1.2\p@
       \dimen@ii\dimen@\advance\dimen@ii-.4\p@
       \leaders\hrule\height\dimen@\depth-\dimen@ii\hfil}%
       \kern-\ten@\p@
       \hbox to\ten@\p@{\dimen@\mathaxis@\advance\dimen@-1.2\p@
       \dimen@ii\dimen@\advance\dimen@ii-.4\p@
       \leaders\hrule\height\dimen@\depth-\dimen@ii\hfil}}}\fi
    \fi
    \cleaders\leads@\hfil
    \ifnum\arrcount@=\m@ne\else\ifnum\arrcount@=\thr@@\else
     \arrow@i
     \ifE@
      \ifnum\tcount@=\m@ne
      \else
       \ifcase\tcount@\char119 \or\or\char119 \or\char120 \or\char121 \or
       \char122 \or \or\or\or\or\char123 \or\char124 \or
       \char125 \or\char119 \hskip-\thr@@\p@\char119 \hskip\thr@@\p@\fi
      \fi
     \else
      \ifcase\scount@\or\or\char120 \or\char119 \or\or\or\char121 \or\char122
      \or\char123 \or\char124 \or\or\or\char125 \or
      \char119 \hskip-\thr@@\p@\char119 \hskip\thr@@\p@\fi
     \fi
    \fi\fi
    \harrow@ee}}%
  \harrow@e}%
 \iflabel@i
  \dimen@ii\z@\setbox\ZER@\hbox{$\m@th\tsize@@\label@i$}%
  \ifnum\arrcount@=\m@ne
  \else
   \advance\dimen@ii\mathaxis@
   \advance\dimen@ii\dp\ZER@\advance\dimen@ii\tw@\p@
   \ifnum\arrcount@=\thr@@\advance\dimen@ii\tw@\p@\fi
  \fi
  \advance\dimen@ii\pdimen@
  \next@{\harrow@b\smash{\raise\dimen@ii\hbox to\dimen@
   {\harrow@bb\hskip\tw@\ldimen@i\hfil\box\ZER@\hfil\harrow@ee}}\harrow@e}%
 \fi
 \iflabel@ii
  \ifnum\arrcount@=\m@ne
  \else
   \setbox\ZER@\hbox{$\m@th\tsize@\label@ii$}%
   \dimen@ii-\ht\ZER@\advance\dimen@ii-\tw@\p@
   \ifnum\arrcount@=\thr@@\advance\dimen@ii-\tw@\p@\fi
   \advance\dimen@ii\mathaxis@\advance\dimen@ii\pdimen@
   \next@{\harrow@b\smash{\raise\dimen@ii\hbox to\dimen@
    {\harrow@bb\hskip\tw@\ldimen@ii\hfil\box\ZER@\hfil\harrow@ee}}\harrow@e}%
  \fi
 \fi}
\let\tsize@\tsize
\def\tsizeCDlabels{\let\tsize@\tsize}
\def\ssizeCDlabels{\let\tsize@\ssize}
\def\tsize@@{\ifnum\arrcount@=\m@ne\else\tsize@\fi}
\def\varrow@{\dimen@\secondy@\advance\dimen@-\firsty@
 \ifN@\else\multiply\dimen@\m@ne\fi
 \setbox\ZER@\vbox to\dimen@
  {\ifN@\vskip-\Ydimen@\else\vskip\ydimen@\fi
   \ifnum\arrcount@=\m@ne\else\ifnum\arrcount@=\thr@@\else
    \hbox{\arrow@iii
     \ifN@
      \ifnum\tcount@=\m@ne
      \else
       \ifcase\tcount@\char117 \or\or\char117 \or\char118 \or\char119 \or
       \char120 \or\or\or\or\or\char121 \or\char122 \or\char123 \or
       \vbox{\hbox{\char117}\nointerlineskip\vskip\thr@@\p@
       \hbox{\char117}\vskip-\thr@@\p@}\fi
      \fi
     \else
      \ifcase\scount@\or\or\char118 \or\char117 \or\or\or\char119 \or
      \char120 \or\char121 \or\char122 \or\or\or\char123 \or
      \vbox{\hbox{\char117}\nointerlineskip\vskip\thr@@\p@
      \hbox{\char117}\vskip-\thr@@\p@}\fi
     \fi}%
    \nointerlineskip
   \fi\fi
   \ifnum\arrcount@=\m@ne
    \let\leads@\null
   \else
    \ifcase\arrcount@\let\leads@\vrule\or\let\leads@\vrule\or
    \def\leads@{\vbox to\ten@\p@{%
     \hrule\height1.67\p@\depth\z@\width.4\p@
     \vfil
     \hrule\height3.33\p@\depth\z@\width.4\p@
     \vfil
     \hrule\height1.67\p@\depth\z@\width.4\p@}}\or
    \def\leads@{\hbox{\vrule\height\p@\hskip\tw@\p@\vrule}}\fi
   \fi
  \cleaders\leads@\vfill\nointerlineskip
   \ifnum\arrcount@=\m@ne\else\ifnum\arrcount@=\thr@@\else
    \hbox{\arrow@iv
     \ifN@
      \ifcase\scount@\or\or\char118 \or\char117 \or\or\or\char119 \or
      \char120 \or\char121 \or\char122 \or\or\or\arrow@iii\char123 \or
      \vbox{\hbox{\char117}\nointerlineskip\vskip-\thr@@\p@
      \hbox{\char117}\vskip\thr@@\p@}\fi
     \else
      \ifnum\tcount@=\m@ne
      \else
       \ifcase\tcount@\char117 \or\or\char117 \or\char118 \or\char119 \or
       \char120 \or\or\or\or\or\char121 \or\char122 \or\arrow@iii\char123 \or
       \vbox{\hbox{\char117}\nointerlineskip\vskip-\thr@@\p@
       \hbox{\char117}\vskip\thr@@\p@}\fi
      \fi
     \fi}%
   \fi\fi
   \ifN@\vskip\ydimen@\else\vskip-\Ydimen@\fi}%
 \ifN@
  \dimen@ii\firsty@
 \else
  \dimen@ii-\firsty@\advance\dimen@ii\ht\ZER@\multiply\dimen@ii\m@ne
 \fi
 \rlap{\smash{\hskip\tocenter@\hskip\pdimen@\raise\dimen@ii\box\ZER@}}%
 \iflabel@i
  \setbox\ZER@\vbox to\dimen@{\vfil
   \hbox{$\m@th\tsize@@\label@i$}\vskip\tw@\ldimen@i\vfil}%
  \rlap{\smash{\hskip\tocenter@\hskip\pdimen@
  \ifnum\arrcount@=\m@ne\let\next@\relax\else\let\next@\llap\fi
  \next@{\raise\dimen@ii\hbox{\ifnum\arrcount@=\m@ne\hskip-.5\wd\ZER@\fi
   \box\ZER@\ifnum\arrcount@=\m@ne\else\hskip\tw@\p@\fi}}}}%
 \fi
 \iflabel@ii
  \ifnum\arrcount@=\m@ne
  \else
   \setbox\ZER@\vbox to\dimen@{\vfil
    \hbox{$\m@th\tsize@\label@ii$}\vskip\tw@\ldimen@ii\vfil}%
   \rlap{\smash{\hskip\tocenter@\hskip\pdimen@
   \rlap{\raise\dimen@ii\hbox{\ifnum\arrcount@=\thr@@\hskip4.5\p@\else
    \hskip2.5\p@\fi\box\ZER@}}}}%
  \fi
 \fi
}
\newdimen\goal@
\newdimen\shifted@
\newcount\Tcount@
\newcount\Scount@
\newbox\shaft@
\newcount\slcount@
\def\getcos@#1{%
 \ifnum\tan@i<\tan@ii
  \dimen@#1%
  \ifnum\slcount@<8 \count@9 \else \ifnum\slcount@<12 \count@8 \else
   \count@7 \fi\fi
  \multiply\dimen@\count@\divide\dimen@\ten@
  \dimen@ii\dimen@\multiply\dimen@ii\tan@i\divide\dimen@ii\tan@ii
 \else
  \dimen@ii#1%
  \count@-\slcount@\advance\count@24
  \ifnum\count@<8 \count@9 \else \ifnum\count@<12 \count@8
   \else\count@7 \fi\fi
  \multiply\dimen@ii\count@\divide\dimen@ii\ten@
  \dimen@\dimen@ii\multiply\dimen@\tan@ii\divide\dimen@\tan@i
 \fi}
\newdimen\adjust@
\def\Nnext@{\ifN@\let\next@\raise\else\let\next@\lower\fi}
\def\arrow@@{\slcount@\angcount@
 \ifNESW@
  \ifnum\angcount@<\ten@
   \let\arrowfont@\arrow@i\global\advance\angcount@\m@ne
   \global\multiply\angcount@13
  \else
   \ifnum\angcount@<19
    \let\arrowfont@\arrow@ii\global\advance\angcount@-\ten@
    \global\multiply\angcount@13
   \else
    \let\arrowfont@\arrow@iii\global\advance\angcount@-19
    \global\multiply\angcount@13
  \fi\fi
  \Tcount@\angcount@
 \else
  \ifnum\angcount@<5
   \let\arrowfont@\arrow@iii\global\advance\angcount@\m@ne
   \global\multiply\angcount@13 \global\advance\angcount@65
  \else
   \ifnum\angcount@<14
    \let\arrowfont@\arrow@iv\global\advance\angcount@-5
    \global\multiply\angcount@13
   \else
    \ifnum\angcount@<23
     \let\arrowfont@\arrow@v\global\advance\angcount@-14
     \global\multiply\angcount@13
    \else
     \let\arrowfont@\arrow@i\global\angcount@117
  \fi\fi\fi
  \ifnum\angcount@=117 \Tcount@115 \else\Tcount@\angcount@\fi
 \fi
 \Scount@\Tcount@
 \ifE@
  \ifnum\tcount@=\z@\advance\Tcount@\tw@\else\ifnum\tcount@=13
   \advance\Tcount@\tw@\else\advance\Tcount@\tcount@\fi\fi
  \ifnum\scount@=\z@\else\ifnum\scount@=13 \advance\Scount@\thr@@\else
   \advance\Scount@\scount@\fi\fi
 \else
  \ifcase\tcount@\advance\Tcount@\thr@@\or\or\advance\Tcount@\thr@@\or
  \advance\Tcount@\tw@\or\advance\Tcount@6 \or\advance\Tcount@7
  \or\or\or\or\or\advance\Tcount@8 \or\advance\Tcount@9 \or
  \advance\Tcount@12 \or\advance\Tcount@\thr@@\fi
  \ifcase\scount@\or\or\advance\Scount@\thr@@\or\advance\Scount@\tw@\or
  \or\or\advance\Scount@4 \or\advance\Scount@5 \or\advance\Scount@\ten@
  \or\advance\Scount@11 \or\or\or\advance\Scount@12 \or\advance
  \Scount@\tw@\fi
 \fi
 \ifcase\arrcount@\or\or\global\advance\angcount@\@ne\else\fi
 \ifN@\shifted@\firsty@\else\shifted@-\firsty@\fi
 \ifE@\else\advance\shifted@\charht@\fi
 \goal@\secondy@\advance\goal@-\firsty@
 \ifN@\else\multiply\goal@\m@ne\fi
 \setbox\shaft@\hbox{\arrowfont@\char\angcount@}%
 \ifnum\arrcount@=\thr@@
  \getcos@{1.5\p@}%
  \setbox\shaft@\hbox to\wd\shaft@{\arrowfont@
   \rlap{\hskip\dimen@ii
    \smash{\ifNESW@\let\next@\lower\else\let\next@\raise\fi
     \next@\dimen@\hbox{\arrowfont@\char\angcount@}}}%
   \rlap{\hskip-\dimen@ii
    \smash{\ifNESW@\let\next@\raise\else\let\next@\lower\fi
      \next@\dimen@\hbox{\arrowfont@\char\angcount@}}}\hfil}%
 \fi
 \rlap{\smash{\hskip\tocenter@\hskip\firstx@
  \ifnum\arrcount@=\m@ne
  \else
   \ifnum\arrcount@=\thr@@
   \else
    \ifnum\scount@=\m@ne
    \else
     \ifnum\scount@=\z@
     \else
      \setbox\ZER@\hbox{\ifnum\angcount@=117 \arrow@v\else\arrowfont@\fi
       \char\Scount@}%
      \ifNESW@
       \ifnum\scount@=\tw@
        \dimen@\shifted@\advance\dimen@-\charht@
        \ifN@\hskip-\wd\ZER@\fi
        \Nnext@
        \next@\dimen@\copy\ZER@
        \ifN@\else\hskip-\wd\ZER@\fi
       \else
        \Nnext@
        \ifN@\else\hskip-\wd\ZER@\fi
        \next@\shifted@\copy\ZER@
        \ifN@\hskip-\wd\ZER@\fi
       \fi
       \ifnum\scount@=12
        \advance\shifted@\charht@\advance\goal@-\charht@
        \ifN@\hskip\wd\ZER@\else\hskip-\wd\ZER@\fi
       \fi
       \ifnum\scount@=13
        \getcos@{\thr@@\p@}%
        \ifN@\hskip\dimen@\else\hskip-\wd\ZER@\hskip-\dimen@\fi
        \adjust@\shifted@\advance\adjust@\dimen@ii
        \Nnext@
        \next@\adjust@\copy\ZER@
        \ifN@\hskip-\dimen@\hskip-\wd\ZER@\else\hskip\dimen@\fi
       \fi
      \else
       \ifN@\hskip-\wd\ZER@\fi
       \ifnum\scount@=\tw@
        \ifN@\hskip\wd\ZER@\else\hskip-\wd\ZER@\fi
        \dimen@\shifted@\advance\dimen@-\charht@
        \Nnext@
        \next@\dimen@\copy\ZER@
        \ifN@\hskip-\wd\ZER@\fi
       \else
        \Nnext@
        \next@\shifted@\copy\ZER@
        \ifN@\else\hskip-\wd\ZER@\fi
       \fi
       \ifnum\scount@=12
        \advance\shifted@\charht@\advance\goal@-\charht@
        \ifN@\hskip-\wd\ZER@\else\hskip\wd\ZER@\fi
       \fi
       \ifnum\scount@=13
        \getcos@{\thr@@\p@}%
        \ifN@\hskip-\wd\ZER@\hskip-\dimen@\else\hskip\dimen@\fi
        \adjust@\shifted@\advance\adjust@\dimen@ii
        \Nnext@
        \next@\adjust@\copy\ZER@
        \ifN@\hskip\dimen@\else\hskip-\dimen@\hskip-\wd\ZER@\fi
       \fi	
      \fi
  \fi\fi\fi\fi
  \ifnum\arrcount@=\m@ne
  \else
   \loop
    \ifdim\goal@>\charht@
    \ifE@\else\hskip-\charwd@\fi
    \Nnext@
    \next@\shifted@\copy\shaft@
    \ifE@\else\hskip-\charwd@\fi
    \advance\shifted@\charht@\advance\goal@-\charht@
    \repeat
   \ifdim\goal@>\z@
    \dimen@\charht@\advance\dimen@-\goal@
    \divide\dimen@\tan@i\multiply\dimen@\tan@ii
    \ifE@\hskip-\dimen@\else\hskip-\charwd@\hskip\dimen@\fi
    \adjust@\shifted@\advance\adjust@-\charht@\advance\adjust@\goal@
    \Nnext@
    \next@\adjust@\copy\shaft@
    \ifE@\else\hskip-\charwd@\fi
   \else
    \adjust@\shifted@\advance\adjust@-\charht@
   \fi
  \fi
  \ifnum\arrcount@=\m@ne
  \else
   \ifnum\arrcount@=\thr@@
   \else
    \ifnum\tcount@=\m@ne
    \else
     \setbox\ZER@
      \hbox{\ifnum\angcount@=117 \arrow@v\else\arrowfont@\fi\char\Tcount@}%
     \ifnum\tcount@=\thr@@
      \advance\adjust@\charht@
      \ifE@\else\ifN@\hskip-\charwd@\else\hskip-\wd\ZER@\fi\fi
     \else
      \ifnum\tcount@=12
       \advance\adjust@\charht@
       \ifE@\else\ifN@\hskip-\charwd@\else\hskip-\wd\ZER@\fi\fi
      \else
       \ifE@\hskip-\wd\ZER@\fi
     \fi\fi
     \Nnext@
     \next@\adjust@\copy\ZER@
     \ifnum\tcount@=13
      \hskip-\wd\ZER@
      \getcos@{\thr@@\p@}%
      \ifE@\hskip-\dimen@\else\hskip\dimen@\fi
      \advance\adjust@-\dimen@ii
      \Nnext@
      \next@\adjust@\box\ZER@
     \fi
  \fi\fi\fi}}%
 \iflabel@i
  \rlap{\hskip\tocenter@
  \dimen@\firstx@\advance\dimen@\secondx@\divide\dimen@\tw@
  \advance\dimen@\ldimen@i
  \dimen@ii\firsty@\advance\dimen@ii\secondy@\divide\dimen@ii\tw@
  \global\multiply\ldimen@i\tan@i\global\divide\ldimen@i\tan@ii
  \ifNESW@\advance\dimen@ii\ldimen@i\else\advance\dimen@ii-\ldimen@i\fi
  \setbox\ZER@\hbox{\ifNESW@\else\ifnum\arrcount@=\thr@@\hskip4\p@\else
   \hskip\tw@\p@\fi\fi
   $\m@th\tsize@@\label@i$\ifNESW@\ifnum\arrcount@=\thr@@\hskip4\p@\else
   \hskip\tw@\p@\fi\fi}%
  \ifnum\arrcount@=\m@ne
   \ifNESW@\advance\dimen@.5\wd\ZER@\advance\dimen@\p@\else
    \advance\dimen@-.5\wd\ZER@\advance\dimen@-\p@\fi
   \advance\dimen@ii-.5\ht\ZER@
  \else
   \advance\dimen@ii\dp\ZER@
   \ifnum\slcount@<6 \advance\dimen@ii\tw@\p@\fi
  \fi
  \hskip\dimen@
  \ifNESW@\let\next@\llap\else\let\next@\rlap\fi
  \next@{\smash{\raise\dimen@ii\box\ZER@}}}%
 \fi
 \iflabel@ii
  \ifnum\arrcount@=\m@ne
  \else
   \rlap{\hskip\tocenter@
   \dimen@\firstx@\advance\dimen@\secondx@\divide\dimen@\tw@
   \ifNESW@\advance\dimen@\ldimen@ii\else\advance\dimen@-\ldimen@ii\fi
   \dimen@ii\firsty@\advance\dimen@ii\secondy@\divide\dimen@ii\tw@
   \global\multiply\ldimen@ii\tan@i\global\divide\ldimen@ii\tan@ii
   \advance\dimen@ii\ldimen@ii
   \setbox\ZER@\hbox{\ifNESW@\ifnum\arrcount@=\thr@@\hskip4\p@\else
    \hskip\tw@\p@\fi\fi
    $\m@th\tsize@\label@ii$\ifNESW@\else\ifnum\arrcount@=\thr@@\hskip4\p@
    \else\hskip\tw@\p@\fi\fi}%
   \advance\dimen@ii-\ht\ZER@
   \ifnum\slcount@<9 \advance\dimen@ii-\thr@@\p@\fi
   \ifNESW@\let\next@\rlap\else\let\next@\llap\fi
   \hskip\dimen@\next@{\smash{\raise\dimen@ii\box\ZER@}}}%
  \fi
 \fi
}
\def\outCD@#1{\def#1{\Err@{\noexpand#1must not be used within \string\CD}}}
\newskip\preCDskip@
\newskip\postCDskip@
\preCDskip@\z@
\postCDskip@\z@
\def\preCDspace#1{\RIfMIfI@
 \onlydmatherr@\preCDspace\else\advance\preCDskip@#1\relax\fi\else
 \onlydmatherr@\preCDspace\fi}
\def\postCDspace#1{\RIfMIfI@
 \onlydmatherr@\postCDspace\else\advance\postCDskip@#1\relax\fi\else
 \onlydmatherr@\postCDspace\fi}
\def\predisplayspace#1{\RIfMIfI@
 \onlydmatherr@\predisplayspace\else
 \advance\abovedisplayskip#1\relax
 \advance\abovedisplayshortskip#1\relax\fi
 \else\onlydmatherr@\preCDspace\fi}
\def\postdisplayspace#1{\RIfMIfI@
 \onlydmatherr@\postdisplayspace\else
 \advance\belowdisplayskip#1\relax
 \advance\belowdisplayshortskip#1\relax\fi
 \else\onlydmatherr@\postdisplayspace\fi}
\def\PreCDSpace#1{\global\preCDskip@#1\relax}
\def\PostCDSpace#1{\global\postCDskip@#1\relax}
\def\CD#1\endCD{%
 \outCD@\cgaps\outCD@\rgaps\outCD@\Cgaps\outCD@\Rgaps
 \preCD@#1\endCD
 \advance\abovedisplayskip\preCDskip@
 \advance\abovedisplayshortskip\preCDskip@
 \advance\belowdisplayskip\postCDskip@
 \advance\belowdisplayshortskip\postCDskip@
 \vcenter{\offinterlineskip
  \vskip\preCDskip@\Let@\global\colcount@\@ne\global\rowcount@\z@
  \everycr{%
   \noalign{%
    \ifnum\rowcount@=\Rowcount@
    \else
     \getrgap@\rowcount@\vskip\getdim@
     \global\advance\rowcount@\@ne\global\colcount@\@ne
    \fi}}%
  \tabskip\z@
  \halign{&\global\xoff@\z@\global\yoff@\z@
   \getcgap@\colcount@\hskip\getdim@
   \hfil\vrule\height\ten@\p@\width\z@\depth\z@
   $\m@th\displaystyle{##}$\hfil
   \global\advance\colcount@\@ne\cr
   #1\crcr}\vskip\postCDskip@}%
 \preCDskip@\z@\postCDskip@\z@
 \def\getcgap@##1{\ifcase##1\or\getdim@\z@\else\getdim@\standardcgap\fi}%
 \def\getrgap@##1{\ifcase##1\getdim@\z@\else\getdim@\standardrgap\fi}%
 \let\Width@\relax\let\Height@\relax\let\Depth@\relax\let\Rowheight@\relax
 \let\Rowdepth@\relax\let\Colwidth@\relax
}
\let\enddocument\bye
\def\alloc@#1#2#3#4#5{\global\advance\count1#1by\@ne
  \ch@ck#1#4#2%
  \allocationnumber=\count1#1%
  \global#3#5=\allocationnumber
  \wlog{\string#5=\string#2\the\allocationnumber}}
\catcode`\@=\active

\catcode`\"=12
\font\black=cmbx10
\font\sblack=cmbx7
\font\ssblack=cmbx5
\font\blackital=cmmib10  \skewchar\blackital='177
\font\sblackital=cmmib7  \skewchar\sblackital='177
\font\ssblackital=cmmib5  \skewchar\ssblackital='177
\font\sanss=cmss10
\font\ssanss=cmss8 scaled 900
\font\sssanss=cmss8 scaled 600
\font\blackboard=msbm10
\font\sblackboard=msbm7
\font\ssblackboard=msbm5
\font\caligr=eusm10
\font\scaligr=eusm7
\font\sscaligr=eusm5

\font\fraktur=eufm10
\font\sfraktur=eufm7
\font\ssfraktur=eufm5

\font\bsymb=cmsy10 scaled\magstep2
\def\all#1{\setbox0=\hbox{\lower1.5pt\hbox{\bsymb
       \char"38}}\setbox1=\hbox{$_{#1}$} \box0\lower2pt\box1\;}
\def\exi#1{\setbox0=\hbox{\lower1.5pt\hbox{\bsymb \char"39}}
       \setbox1=\hbox{$_{#1}$} \box0\lower2pt\box1\;}

\def\tx#1{{\fam0\relax#1}}

\newfam\bifam
\textfont\bifam=\blackital
\scriptfont\bifam=\sblackital
\scriptscriptfont\bifam=\ssblackital
\def\bi#1{{\fam\bifam\relax#1}}

\newfam\blfam
\textfont\blfam=\black
\scriptfont\blfam=\sblack
\scriptscriptfont\blfam=\ssblack

\newfam\bbfam
\textfont\bbfam=\blackboard
\scriptfont\bbfam=\sblackboard
\scriptscriptfont\bbfam=\ssblackboard
\def\bb#1{{\fam\bbfam\relax#1}}

\newfam\ssfam
\textfont\ssfam=\sanss
\scriptfont\ssfam=\ssanss
\scriptscriptfont\ssfam=\sssanss
\def\ss#1{{\fam\ssfam\relax#1}}

\newfam\clfam
\textfont\clfam=\caligr
\scriptfont\clfam=\scaligr
\scriptscriptfont\clfam=\sscaligr
\def\cl#1{{\fam\clfam\relax#1}}

\newfam\frfam
\textfont\frfam=\fraktur
\scriptfont\frfam=\sfraktur
\scriptscriptfont\frfam=\ssfraktur

\def\hpb#1{\setbox0=\hbox{${#1}$}
    \copy0 \kern-\wd0 \kern.2pt \box0}
\def\vpb#1{\setbox0=\hbox{${#1}$}
    \copy0 \kern-\wd0 \raise.08pt \box0}

\def\pmb#1{\setbox0\hbox{${#1}$} \copy0 \kern-\wd0 \kern.2pt \box0}
\def\pmbb#1{\setbox0\hbox{${#1}$} \copy0 \kern-\wd0
      \kern.2pt \copy0 \kern-\wd0 \kern.2pt \box0}
\def\pmbbb#1{\setbox0\hbox{${#1}$} \copy0 \kern-\wd0
      \kern.2pt \copy0 \kern-\wd0 \kern.2pt
    \copy0 \kern-\wd0 \kern.2pt \box0}
\def\pmxb#1{\setbox0\hbox{${#1}$} \copy0 \kern-\wd0
      \kern.2pt \copy0 \kern-\wd0 \kern.2pt
      \copy0 \kern-\wd0 \kern.2pt \copy0 \kern-\wd0 \kern.2pt \box0}
\def\pmxbb#1{\setbox0\hbox{${#1}$} \copy0 \kern-\wd0 \kern.2pt
      \copy0 \kern-\wd0 \kern.2pt
      \copy0 \kern-\wd0 \kern.2pt \copy0 \kern-\wd0 \kern.2pt
      \copy0 \kern-\wd0 \kern.2pt \box0}

\mathchardef\za="710B  
\mathchardef\zb="710C  
\mathchardef\zg="710D  
\mathchardef\zd="710E  
\mathchardef\zve="710F 
\mathchardef\zz="7110  
\mathchardef\zh="7111  
\mathchardef\zvy="7112 
\mathchardef\zi="7113  
\mathchardef\zk="7114  
\mathchardef\zl="7115  
\mathchardef\zm="7116  
\mathchardef\zn="7117  
\mathchardef\zx="7118  
\mathchardef\zp="7119  
\mathchardef\zr="711A  
\mathchardef\zs="711B  
\mathchardef\zt="711C  
\mathchardef\zu="711D  
\mathchardef\zvf="711E 
\mathchardef\zq="711F  
\mathchardef\zc="7120  
\mathchardef\zw="7121  
\mathchardef\ze="7122  
\mathchardef\zy="7123  
\mathchardef\zvp="7124 
\mathchardef\zvr="7125 
\mathchardef\zvs="7126 
\mathchardef\zf="7127  
\mathchardef\zG="7000  
\mathchardef\zD="7001  
\mathchardef\zY="7002  
\mathchardef\zL="7003  
\mathchardef\zX="7004  
\mathchardef\zP="7005  
\mathchardef\zS="7006  
\mathchardef\zU="7007  
\mathchardef\zF="7008  
\mathchardef\zC="7009  
\mathchardef\zW="700A  

\catcode`\"=\active

\loadmsam
\newsymbol\leqslant 1336
\newsymbol\geqslant 133E
\newsymbol\centerdot 1205

\def\fpr#1{\underset{{#1}}\to\times}

\newcounter\secno
\newcounter\ssecno

\define\sect#1{\Reset\ssecno1\bigpagebreak
	\flushpar {\secno.}\,{\bf #1}\vskip1.2mm}
\newfontstyle\secno{\bf}

\define\ssca#1{\Offset\secno0\bigpagebreak\vskip-4mm
	\flushpar {\secno.\ssecno.}\,{\bf #1}\vskip1.2mm}
\newfontstyle\ssecno{\bf}

\define\sscx#1{\Offset\secno0\bigpagebreak
	\flushpar {\secno.\ssecno.}\,{\bf #1}\vskip1.2mm}
\newfontstyle\ssecno{\bf}

\def\*{{\textstyle *}}
\def\ast{{\scriptstyle *}}

\newsymbol\blacktriangle 104E

\def\proof{\demo{Proof}}
\def\endproof{\hfill \vrule height4pt width6pt depth2pt \enddemo}

\def\N{{\bb N}}
\def\R{{\bb R}}

\def\ssT{{\scriptscriptstyle {\ss T}}}

\def\by{{\bi y}}

\def\sT{{\ss T}}
\def\sV{{\ss V}}

\def\st{{\ss t}}

\def\xa{\tx{a}}
\def\xd{\tx{d}}
\def\xi{\tx{i}}
\def\xD{\tx{D}}

\def\im{\operatorname{im}}

\def\sign{\operatorname{sign}}

\def\cD{{\cl D}}
\def\cE{{\cl E}}
\def\cF{{\cl F}}
\def\cG{{\cl G}}
\def\cH{{\cl H}}

\def\cL{{\cl L}}
\def\cP{{\cl P}}

\input paper.st\relax
\hsize=37pc
\hoffset=-10pt
\vsize=53pc
\voffset=6pt
\TagsOnRight
\document
\line{PLBEP.TEX \hfill \today}
\input xy
\xyoption{all}

        \title
                Dynamics of autonomous systems with external forces
        \endtitle

        \author
        Giuseppe Marmo \\
        Dipartimento di Scienze Fisiche \\
                Universit\`a Federico II di Napoli \\
                Complesso Universitario di Monte Sant'Angelo \\
        Via Cinthia, 80126 Napoli, Italy \\
        Istituto Nazionale di Fisica Nucleare,
        Sezione di Napoli, Italy \\
                {\tt Giuseppe.Marmo\@na.infn.it} \\
                        \\
        W\l odzimierz M. Tulczyjew \\
        Dipartimento di Matematica e Fisica \\
        Universit\`a di Camerino \\
        62032 Camerino, Italy \\
        Istituto Nazionale di Fisica Nucleare,
        Sezione di Napoli, Italy \\
                {\tt tulczy\@campus.unicam.it} \\
                        \\
                Pawe\l\ Urba\'nski \\
                Division of Mathematical Methods in Physics \\
                University of Warsaw \\
                Ho\D{z}a 74, 00-682 Warszawa \\
                {\tt urbanski\@fuw.edu.pl}
        \endauthor

                \thanks{Supported by PRIN SINTESI and KBN, grant No 2 PO3A 041 18}

\abstract

We consider a geometric framework for analytical mechanics with external
forces. Four versions of this framework are considered. A variational
principle with boundary therms and external forces.The second and the third
versions are the Lagrangian and Hamiltonian formulations,respectively. The
last one is the Poisson formulation. An extensive introductory section
presents some well known and some little known geometric constructions to
put our formulation in the appropriate setting to make the comparison of
the different formulations more easy.

\endabstract

        \maketitle

        \sect{Introduction.}

    We distinguish three types of situations to which methods of
analytical mechanics can be applied.  A mechanical system can be studied
as an isolated system not subject to external interference.  The study of
planetary motion is an example. The motion of a mechanical system in a
finite time interval can be studied.  The external interaction with the
system is limited to setting up or observing the initial and final
conditions without interfering with the system during its motion. Such
situations are studied in old-time ballistics.  The motion of system in a
finite time interval can be considered with both the boundary conditions
and the motion itself during the time interval are subject to control.
Such situations are studied in modern ballistics of guided missiles.  The
flight of an airplane or the motion of a car are also examples of such
situations.  Early formulations of analytical mechanics were applicable to
all three types of situations.  Recent geometrical formulations left out
the possibility of analyzing the external interaction with a mechanical
system during its motion.  The external interaction can take different
forms.  One possibility is to influence the motion of a mechanical object
by subjecting the object to constraints varying in time.  Driving a car is
an example of this type of control.  A geometric study of this type of
control was initiated by Marle [1], [2], [3]. Another possibility is to
control a mechanical system by applying external forces.  This happens
when trajectory of a space vehicle or the orbit of a satellite are
corrected by remotely activating jet engines mounted on the vehicle.

    A geometric framework for analytical mechanics with external forces is
the subject of the present paper.  Four versions of this framework are
considered.  The first version is a variational formulation.  Variational
principles found in current literature are almost exclusively versions of
the Hamilton principle.  A recent paper by Gr\` acia, Mar\`\i n-Solano and
Mu\~noz-Lecanda [4] gives a clear geometric description of the Hamilton
principle in presence of constraints.  Hamilton's principle does not take
into account momenta at the boundary or external forces.  A variational
principle with boundary momenta was used by Schwinger [5]. The variational
principle with boundary terms appearing in [6] is not related to
Schwinger's principle.  We use a variational principle with boundary terms
and external forces. The second version of the framework is the Lagrangian
formulation of mechanics and the third is the Hamiltonian formulation. The
Lagrangian formulation usually presented is completely equivalent to the
Hamilton principle and does not include momenta or external forces.
Momenta but not external forces are present in the usual Hamiltonian
formulation.  In our interpretation a Lagrangian system and the
corresponding Hamiltonian system are the same object described in terms of
two different generating functions -- the Lagrangian and the Hamiltonian.
The last version is the Poisson formulation different from the Hamiltonian
formulation only in the use of the Poisson structure of the phase space in
place of the equivalent symplectic structure.

    Our formulations are based on some well known and some little known
geometric constructions.  These are presented in an extensive introductory
section.

    Work related to our formulations can be found in the references [7],
[8], [9], [10] and [11].

        \sect{Geometric constructions.} \vskip3mm
        \ssca{Tangent functors.}
    Let $M$ be a differential manifold.  A local chart $(x^\zk) \colon M
\rightarrow \R^m$ of $M$ with coordinates $x^\zk \colon M \rightarrow \R$
will be treated as defined on all of $M$.  Simple modifications have to be
applied to constructions involving charts if truly local charts are used.

    In the set $C^\infty(M|\R)$ of differentiable mappings from $\R$ to
$M$ we introduce an equivalence relation.  Two mappings $\zg$ and $\zg'$
are equivalent if $\zg'(0) = \zg(0)$ and $\xD (f \circ \zg')(0) = \xD (f
\circ \zg)(0)$ for each differentiable function $f \colon M \rightarrow
\R$.

    The set of equivalence classes will be denoted by $\sT M$.  The
equivalence class of a curve $\zg \colon \R \rightarrow M$ will be denoted
by $\st\zg(0)$.  The set $\sT M$ is the set of {\it tangent vectors} in
$M$ or the {\it tangent bundle} of $M$.

    We have the surjective mapping
        $$\zt_M \colon \sT M \rightarrow M \colon \st\zg(0) \mapsto \zg(0).
                                                        \tag \label{Fplb1}$$
    From a function $f \colon M \rightarrow \R$ we construct functions
        $$f_{1;0} \colon \sT M \rightarrow \R \colon \st\zg(0) \mapsto (f
\circ \zg)(0)
                                                        \tag \label{Fplb2}$$
    and
        $$f_{1;1} \colon \sT M \rightarrow \R \colon \st\zg(0) \mapsto
\xD(f \circ \zg)(0).
                                                        \tag \label{Fplb3}$$
    These constructions can be applied to functions defined on open
subsets of $M$. The functions $f_{1;0}$ and $f_{1;1}$ constructed from a
function $f$ on $U \subset M$ are defined on $\zt_M{}^{-1}(U)$.  The
function $f_{1;0}$ is the composition $f \circ \zt_M$.

    The set $\sT M$ is a differential manifold.  A local chart $(x^\zk)
\colon M \rightarrow \R^m$ induces a local chart
        $$(x^\zk,\dot x^\zl) \colon \sT M \rightarrow \R^{2m}
                                                        \tag \label{Fplb4}$$
     of $\sT M$.  The local coordinates $x^\zk$ and $\dot x^\zl$ are the
functions $x^\zk{}_{1;0}$ and $x^\zl{}_{1;1}$ constructed from the
coordinates of $M$.  Note that we are using the symbol $x^\zk$ to denote
local coordinates of $M$ and also of $\sT M$.  The diagram
    \vskip1mm
        $$\xymatrix@R+0mm{{\sT M} \ar[d]_*{\zt_M} \\ M}
                                                        \tag \label{Fplb5}$$
    \vskip2mm
    \noindent is a differential fibration.

    Each curve $\zg \colon \R \rightarrow M$ has a {\it tangent
prolongation}
        $$\st\zg \colon \R \rightarrow \sT M \colon t \mapsto \st\zg(t +
\cdot)(0).
                                                        \tag \label{Fplb6}$$
    The value $\st\zg(t)$ of the prolongation $\st\zg$ depends only on the
definition of the curve $\zg$ in the immediate neighbourhood of $t \in
\R$.  It follows that a local curve $\zg \colon I \rightarrow M$ defined
on an open subset $I \subset \R$ has a well defined prolongation $\st\zg
\colon I \rightarrow \sT M$.  The curve $\zg$ is obtained from $\st\zg$ as
the projection $\zt_M \circ \st\zg$.

    Let $M$ and $N$ be differential manifolds and let $\za \colon M
\rightarrow N$ be a differentiable mapping.  We have the mapping
        $$\sT\za \colon \sT M \rightarrow \sT N \colon \st\zg(0) \mapsto
\st(\za \circ \zg)(0).
                                                        \tag \label{Fplb7}$$
    The diagram
    \vskip1mm
        $$\xymatrix@R+0mm @C+6mm{{\sT M} \ar[d]_*{\zt_M} \ar[r]^*{\sT\za} &
            \sT N \ar[d]_*{\zt_N} \\
            M \ar[r]^*{\za} & N}
                                                        \tag \label{Fplb8}$$
    \vskip2mm
    \noindent is a differential fibration morphism.  If $M$, $N$, and $O$
are differential manifolds and $\za \colon M \rightarrow N$ and $\zb
\colon N \rightarrow O$ are differentiable mappings, then
        $$\sT(\zb \circ \za) = \sT\zb \circ \sT\za.
                                                        \tag \label{Fplb9}$$

    We have introduced a covariant functor $\sT$ in the category of
differential manifolds and differentiable mappings.

    We introduce a second equivalence relation in the set $C^\infty(M|\R)$
of differentiable mappings from $\R$ to $M$. Two mappings $\zg$ and $\zg'$
are equivalent if $\zg'(0) = \zg(0)$, $\xD(f \circ \zg')(0) = \xD(f \circ
\zg)(0)$, and $\xD^2(f \circ \zg')(0) = \xD^2(f \circ \zg)(0)$ for each
differentiable function $f \colon M \rightarrow \R$.  The set of
equivalence classes will be denoted by $\sT^2 M$ and the equivalence class
of a curve $\zg \colon \R \rightarrow M$ will be denoted by $\st^2\zg(0)$.
The set $\sT^2 M$ is the set of {\it second tangent vectors} in $M$ or the
{\it second tangent bundle} of $M$.

    We have surjective mappings
        $$\zt_2{}_M \colon \sT^2 M \rightarrow M \colon \st^2\zg(0)
\mapsto \zg(0)
                                                        \tag \label{Fplb10}$$
    and
        $$\zt^1{}_2{}_M \colon \sT^2 M \rightarrow \sT M \colon
\st^2\zg(0) \mapsto \st\zg(0)
                                                        \tag \label{Fplb11}$$
    satisfying $\zt{}_M \circ \zt^1{}_2{}_M = \zt_2{}_M$.

    From a function $f \colon M \rightarrow \R$ we construct functions
        $$f_{2;0} \colon \sT^2 M \rightarrow \R \colon \st^2\zg(0) \mapsto
(f \circ \zg)(0),
                                                        \tag \label{Fplb12}$$
        $$f_{2;1} \colon \sT^2 M \rightarrow \R \colon \st^2\zg(0) \mapsto
\xD(f \circ \zg)(0),
                                                        \tag \label{Fplb13}$$
    and
        $$f_{2;2} \colon \sT^2 M \rightarrow \R \colon \st^2\zg(0) \mapsto
\xD^2(f \circ \zg)(0).
                                                        \tag \label{Fplb14}$$
    These constructions can be applied to functions defined on open
subsets of $M$.

    The set $\sT^2 M$ is a differential manifold.  A local chart $(x^\zk)
\colon M \rightarrow \R^m$ induces a local chart
        $$(x^\zk,\dot x^\zl,\ddot x^\zm) \colon \sT^2 M \rightarrow \R^{3m}
                                                        \tag \label{Fplb15}$$
     of $\sT M$.  The local coordinates $x^\zk$, $\dot x^\zl$, and $\ddot
x^\zm$ are the functions $x^\zk{}_{2;0}$, $x^\zl{}_{2;1}$, and
$x^\zm{}_{2;2}$ constructed from the coordinates of $M$.  Note that
$x^\zk$ denote local coordinates of $M$ and also of $\sT^2 M$.  Diagrams
    \vskip1mm
        $$\xymatrix@R+3mm{{\sT^2 M} \ar[d]_*{\zt_2{}_M} \\ M} \hskip20mm
\xymatrix@R+4mm{{\sT^2 M} \ar[d]_*{\zt^1{}_2{}_M} \\ \sT M}
                                                        \tag \label{Fplb16}$$
    \vskip2mm
    \noindent are differential fibrations.

    Each curve $\zg \colon \R \rightarrow M$ has a {\it second tangent
prolongation}
        $$\st^2\zg \colon \R \rightarrow \sT^2 M \colon t \mapsto
\st^2\zg(t + \cdot)(0).
                                                        \tag \label{Fplb17}$$
    Second tangent prolongations of local curves can be constructed.
Relations $\zt_2{}_M \circ \st^2\zg = \zg$ and $\zt^1{}_2{}_M \circ
\st^2\zg = \st\zg$ hold.

    Let $M$ and $N$ be differential manifolds and let $\za \colon M
\rightarrow N$ be a differentiable mapping.  We have the mapping
        $$\sT^2\za \colon \sT^2 M \rightarrow \sT^2 N \colon \st^2\zg(0)
\mapsto \st^2(\za \circ \zg)(0).
                                                        \tag \label{Fplb18}$$
    Diagrams
    \vskip1mm
        $$\xymatrix@R+3mm @C+10mm{{\sT^2 M} \ar[d]_*{\zt_2{}_M}
\ar[r]^*{\sT^2\za} &
            \sT^2 N \ar[d]_*{\zt_2{}_N} \\
            M \ar[r]^*{\za} & N}
                                                        \tag \label{Fplb19}$$
    \vskip1mm
    \noindent and
    \vskip1mm
        $$\xymatrix@R+3mm @C+10mm{{\sT^2 M} \ar[d]_*{\zt^1{}_2{}_M}
\ar[r]^*{\sT^2\za} &
            \sT^2 N \ar[d]_*{\zt^1{}_2{}_N} \\
            \sT M \ar[r]^*{\sT\za} & \sT N}
                                                        \tag \label{Fplb20}$$
    \vskip2mm
    \noindent are morphisms of differential fibrations.  If $M$, $N$, and
$O$ are differential manifolds and $\za \colon M \rightarrow N$ and $\zb
\colon N \rightarrow O$ are differentiable mappings, then
        $$\sT^2(\zb \circ \za) = \sT^2\zb \circ \sT^2\za.
                                                        \tag \label{Fplb21}$$

    We have introduced a covariant functor $\sT^2$ in the category of
differential manifolds and differentiable mappings.

        \sscx{Tangent and cotangent vectors.}

    The fibration
    \vskip1mm
        $$\xymatrix@R+3mm{{\sT M} \ar[d]_*{\zt_M} \\ M}
                                                        \tag \label{Fplb22}$$
    \vskip2mm
    \noindent is a vector fibration.  It is called the {\it tangent
fibration}. Since representatives of vectors (curves in $M$) can not be
added the construction of linear operations in fibres of $\zt_M$ is
somewhat indirect.  Let $v $, $v_1$, and $v_2$ be elements of the same
fibre $\sT_x M = \zt_M{}^{-1}(x)$.  We write
        $$v = v_1 + v_2
                                                        \tag \label{Fplb23}$$
    if
        $$f_{1,1}(v) = f_{1,1}(v_1) + f_{1,1}(v_2)
                                                        \tag \label{Fplb24}$$
    for each function $f$ on $M$.  Note that
        $$f_{1,1}(v) = \langle \xd f, v\rangle.
                                                        \tag \label{Fplb25}$$
    We have defined a relation between three elements of a fibre $\sT_x
M$.  This relation will turn into a binary operation if we show that for
each pair $(v_1,v_2) \in \sT_x M \times \sT_x M$ there is an unique vector
$v \in \sT_x M$ such that $v = v_1 + v_2$.  The coordinate construction
        $$(x^\zk \circ \zg)(t) = x^\zk(v_1) + (\dot x^\zk(v_1) + \dot
x^\zk(v_2))t
                                                        \tag \label{Fplb26}$$
    of a representative $\zg$ of $v$ proves existence.  Let $v$ and $v'$
be in relations $v = v_1 + v_2$ and $v' = v_1 + v_2$ with $v_1$ and $v_2$.
Then
        $$f_{1,1}(v') = f_{1,1}(v_1) + f_{1,1}(v_2) = f_{1,1}(v)
                                                        \tag \label{Fplb27}$$
    for each function $f$.  It follows that $v' = v$.  This proves
uniqueness.

    Let $v$ and $u$ be elements of $\sT_x M$ and let $k$ be a number.  We
write
        $$v = ku
                                                        \tag \label{Fplb28}$$
    if
        $$f_{1,1}(v) = kf_{1,1}(u)
                                                        \tag \label{Fplb29}$$
    for each function $f$ on $M$.  The coordinate construction
        $$(x^\zk \circ \zg)(t) = x^\zk(u) + k\dot x^\zk(u)t
                                                        \tag \label{Fplb30}$$
    shows that for each $k \in \R$ and $u \in \sT_x M$ there is a vector
$v \in \sT_x M$ such that $v = ku$.  If $v$ and $v'$ are two such vectors,
then
        $$f_{1,1}(v') = kf_{1,1}(u) = f_{1,1}(v).
                                                        \tag \label{Fplb31}$$
    It follows that the vector $v$ is unique.

    We have defined operations
        $$+ \colon \sT M \fpr{(\zt_M,\zt_M)} \sT M \rightarrow \sT M
                                                        \tag \label{Fplb32}$$
    and
        $$\cdot\, \colon \R \times \sT M \rightarrow \sT M.
                                                        \tag \label{Fplb33}$$
    The symbol $\sT M \fpr{(\zt_M,\zt_M)} \sT M$ denotes the {\it fibre
product}
        $$\{(v_1,v_2) \in \sT M \times \sT M ;\; \zt_M(v_1) = \zt_M(v_2)\}.
                                                        \tag \label{Fplb34}$$

    A section $X \colon M \rightarrow \sT M$ of the tangent fibration
$\zt_M$ is a {\it vector field}.  The {\it zero section} will be denoted
by $O_{\zt_M}$.

    Let $C^\infty(\R|M)$ denote the algebra of differentiable functions on
a differential manifold $M$.  In the set $C^\infty(\R|M) \times M$ we
introduce an equivalence relation.  Two pairs $(f,x)$ and $(f',x')$ are
equivalent if $x' = x$ and
        $$\xD(f' \circ \zg)(0) = \xD(f \circ \zg)(0)
                                                        \tag \label{Fplb35}$$
    for each differentiable curve $\zg \colon \R \rightarrow M$ such that
$\zg(0) = x$.  The set of equivalence classes denoted by $\sT^\*M$ is
called the {\it cotangent bundle} of $M$.  Elements of $\sT^\*M$ are
called {\it cotangent vectors}.  The equivalence class of $(f,x)$ denoted
by $\xd f(x)$ is called the {\it differential} of $f$ at $x$.  The mapping
        $$\zp_M \colon \sT^\*M \rightarrow M \colon \xd f(x) \mapsto x
                                                        \tag \label{Fplb36}$$
    is called the {\it cotangent bundle projection}.

    The cotangent bundle $\sT^\*M$ is a differential manifold.  A local
chart $(x^\zk) \colon M \rightarrow \R^m$ induces a local chart
$(x^\zk,p_\zl) \colon \sT^\*M \rightarrow \R^{2m}$.  The local coordinates
$x^\zk$ and $p_\zl$ are the functions
        $$x^\zk \colon \sT^\*M \rightarrow \R \colon \xd f(x) \mapsto
x^\zk(x)
                                                        \tag \label{Fplb37}$$
    and
        $$p_\zl \colon \sT^\*M \rightarrow \R \colon \xd f(x) \mapsto
\xD(f \circ \zg_{\zl,x})(0),
                                                        \tag \label{Fplb38}$$
    where $\zg_{\zl,x}$ are curves in $M$ characterized by
        $$x^\zk(\zg_{\zl,x}(t)) = x^\zk(x) + \zd^\zk{}_\zl t
                                                        \tag \label{Fplb39}$$
    for $t$ sufficiently close to $0 \in \R$.  Note that $x^\zk$ is used
to denote local coordinates in $M$ and also in $\sT^\*M$.

    The diagram
    \vskip1mm
        $$\xymatrix@R+3mm{{\sT^\* M} \ar[d]_*{\zp_M} \\ M}
                                                        \tag \label{Fplb40}$$
    \vskip1mm \noindent is a differential vector fibration.  It is called
the {\it cotangent fibration}.  The linear operations in fibres of the
cotangent fibration have natural definitions
        $$+ \colon \sT^\* M \fpr{(\zp_M,\zp_M)} \sT^\* M \rightarrow
\sT^\* M \colon (\xd f_1(x),\xd f_2(x)) \mapsto \xd(f_1 + f_2)(x)
                                                        \tag \label{Fplb41}$$
    and
        $$\cdot\, \colon \R \times \sT^\* M \rightarrow \sT^\* M \colon
(k,\xd f(x)) \mapsto \xd(kf)(x).
                                                        \tag \label{Fplb42}$$

    The mapping
        $$\langle \,\; ,\; \rangle \colon \sT^\*M \fpr{(\zp_M,\zt_M)} \sT
M \rightarrow \R \colon (\xd f(a),\st\zg(0)) \mapsto \xD(f \circ \zg)(0)
                                                        \tag \label{Fplb43}$$
    is a differentiable, bilinear, and nondegenerate pairing.  The symbol
$\sT^\*M \fpr{(\zp_M,\zt_M)} \sT M$ is again the fibre product defined as
the equalizer
        $$\sT^\*M \fpr{(\zp_M,\zt_M)} \sT M =  \{(p,v) \in \sT^\*M \times
\sT M ;\; \zp_M(p) = \zt_M(v)\}
                                                        \tag \label{Fplb44}$$
    of the projections $\zp_M$ and $\zt_M$.  The tangent fibration and the
cotangent fibration are a dual pair of vector fibrations.

    The {\it Liouville form} is a 1-form $\zy_M$ on $\sT^\*M$ defined as
        $$\zy_M \colon \sT(\sT^\*M) \rightarrow \R \colon w \mapsto
\langle \zt_{\sT^\*M}(w), \sT\zp_M(w) \rangle.
                                                        \tag \label{Fplb45}$$

    A 1-form on $M$ can interpreted as a section of the cotangent
projection $\zp_M$.  If $\zm \colon M \rightarrow \sT^\*M$, then
        $$\align
    \langle \zm^\*\zy_M, v\rangle &= \langle \zy_M, \sT\zm(v)\rangle \\
            &= \langle \zt_{\sT^\*M}(\sT\zm(v)),
\sT\zp_M(\sT\zm(v))\rangle \\
            &= \langle \zm(\zt_M(v)), \sT(\zp_M \circ \zm)(v)\rangle \\
            &= \langle \zm, v\rangle
                                                \tag \label{Fplb46}\endalign$$
    for each $v \in \sT M$.  Hence, $\zm^\*\zy_M = \zm$.  The {\it zero
section} of the cotangent fibration will be denoted by $O_{\zp_M}$.

    The cotangent bundle $\sT^\*M$ together with the 2-form $\zw_M =
\xd\zy_M$ form a symplectic manifold $(\sT^\*M,\zw_M)$.  It follows from
the local expressions
        $$\zy_M = p_\zk \xd x^\zk
                                                        \tag \label{Fplb47}$$
    and
        $$\zw_M = \xd p_\zk \wedge \xd x^\zk
                                                        \tag \label{Fplb48}$$
    that the form $\zw_M$ is non degenerate.

        \sscx{Special symplectic structures.}
    Let $(P,\zw)$ be a symplectic manifold.  A {\it special symplectic
structure} for $(P,\zw)$ is a diagram
    \vskip1mm
        $$\xymatrix@R+0mm @C+10mm{{(P,\zy)} \ar[d]_*{\zp} \\ M}
                                                        \tag \label{Fplb49}$$
    \vskip2mm
    \noindent where $\zp \colon P \rightarrow Q$ is a vector fibration and
$\zy$ is a vertical one-form on $P$ such that $\xd\zy = \zw$.  An
additional requirement is the existence of a vector fibration morphism
    \vskip1mm
        $$\xymatrix@R+3mm @C+10mm{{P} \ar[d]_*{\zp} \ar[r]^*{\za} &
            \sT^\*M \ar[d]_*{\zp{}_M} \\
            M \ar@{=}[r] & M}
                                                        \tag \label{Fplb50}$$
    \vskip2mm
    \noindent such that $\zy = \za^\* \zy_M$.  This morphism is
necessarily an isomorphism.  For each $w \in \sT P$ we have
        $$\align
    \langle \zy, w \rangle &= \langle \za^\*\zy_M, w \rangle \\
            &= \langle \zy_M, \sT\za(w) \rangle \\
            &= \langle \zt_{\sT^\*M}(\sT\za(w)), \sT\zp{}_M(\sT\za(w))
\rangle \\
            &= \langle \za(\zt{}_P(w)), \sT(\zp{}_M \circ \za)(w) \rangle
\\
            &= \langle \za(\zt{}_P(w)), \sT\zp(w) \rangle.
                                                \tag \label{Fplb51}\endalign$$
    It follows that the mapping $\za \colon P \rightarrow \sT^\*M$ is
completely characterized by
        $$\langle \za(p), v \rangle = \langle \zy, w \rangle,
                                                        \tag \label{Fplb52}$$
    where $v \in \sT_{\zp(p)}M$ and $w$ is any vector in $\sT_p P$ such
that $\sT\zp(w) = v$.  We conclude that if the morphism \Ref{Fplb50}
associated with a special symplectic structure exists, then it is unique.
It can be shown that if the 1-form $\zy$ interpreted as a mapping $\zy
\colon \sT P \rightarrow \R$ is linear on fibres of the vector fibration
$\sT\zp \colon \sT P \rightarrow \sT M$, then the morphism \Ref{Fplb50}
exists.  We will usually present a special symplectic structure together
with the associated vector fibration isomorphism.

        \sscx{Generating functions and Legendre transformations.}
    Let
    \vskip2mm
        $$\xymatrix@R+3mm @C+10mm{{(P,\zy)} \ar[d]_*{\zp} \\ M}
            \hskip20mm \xymatrix@R+3mm @C+10mm{{P} \ar[d]_*{\zp}
\ar[r]^*{\za} &
            \sT^\*M \ar[d]_*{\zp{}_M} \\
            M \ar@{=}[r] & M}
                                                        \tag \label{Fplb53}$$
    \vskip2mm
    \noindent be a special symplectic structure for a symplectic manifold
$(P,\zw)$ with its associated vector fibre morphism.  Let $S \subset P$ be
the image $\im(\zs)$ of a section $\zs \colon M \rightarrow P$ of $\zp$.
If $S$ is a Lagrangian submanifold of $(P,\zw)$, then the 1-form
$\zs^\*\zy$ is closed since $\xd\zs^\*\zy = \zs^\*\zw = 0$.  Let this form
be exact.  A function $F \colon M \rightarrow \R$ such that $\zs^\*\zy =
\xd F$ is called a {\it generating function} of $S$ relative to the
special symplectic structure \Ref{Fplb53}.  From
        $$\zs^\*\zy = \zs^\*\za^\*\zy_M = (\za \circ \zs)^\*\zy_M = \za
\circ \zs
                                                        \tag \label{Fplb54}$$
    it follows that $\zs = \za^{-1} \circ \xd F$.  From
        $$\xd F = (\zp \circ \zs)^\*\xd F = \zs^\*\zp^\*\xd F =
\zs^\*\xd(F \circ \zp)
                                                        \tag \label{Fplb55}$$
    it follows that $\zs^\*\zy = \zs^\*\xd(F \circ \zp)$.  Consequently
$\zr^\*\zy = \zr^\*\xd(F \circ \zp)$ for any mapping $\zr \colon R
\rightarrow P$ such that $\im(\zr) \subset S$.  This is in particular true
for the canonical injection $\zi_S \colon S \rightarrow P$.  Since the
forms $\zy$ and $\xd(F \circ \zp)$ are both vertical we have $\zy \circ
\zi_S = \xd(F \circ \zp) \circ \zi_S$ or $\zy|S = \xd(F \circ \zp)|S$. The
set $S$ can be obtained from its generating function $F$ as the equalizer
        $$S = \left\{p \in P;\;\zy(p) = \xd(F \circ \zp)(p) \right\}
                                                        \tag \label{Fplb56}$$
    of the two forms.  If $S = \im(\zs)$ is not Lagrangian we define its
{\it generating form} relative to the special symplectic structure
\Ref{Fplb53} as the 1-form $\zf$ on $M$ such that $\zs^\*\zy = \zf$.
Relations $\zs = \za^{-1} \circ \zf$ and
        $$S = \left\{p \in P;\;\zy(p) = (\zp^\*\zf)(p) \right\}
                                                        \tag \label{Fplb57}$$
    replace the corresponding relations of the Lagrangian case.

    Let
    \vskip2mm
        $$\xymatrix@R+6mm @C+10mm{{(P,\zy')} \ar[d]_*{\zp'} \\ M'}
            \hskip20mm \xymatrix@R+6mm @C+10mm{{P} \ar[d]_*{\zp'}
\ar[r]^*{\za'} &
            \sT^\*M' \ar[d]_*{\zp{}_M'} \\
            M' \ar@{=}[r] & M'}
                                                        \tag \label{Fplb58}$$
    \vskip2mm
    \noindent be a second special symplectic structure for $(P,\zw)$.  The
difference $\zy' - \zy$ is a closed form. Let it be exact and let $G$ be a
function on $P$ such that $\zy' - \zy = \xd G$.  Let $S$ be the image
$\im(\zs')$ of a section $\zs' \colon M' \rightarrow P$ of $\zp'$.  The
function
        $$F' = (F \circ \zp + G) \circ \zs'
                                                        \tag \label{Fplb59}$$
    is a generating function of $S$ relative to the new special symplectic
structure since
        $$\xd F' = \zs'{}^\*\xd(F \circ \zp + G) = \zs'{}^\*\xd(F \circ
\zp) + \zs'{}^\*(\zy' - \zy) = \zs'{}^\*\xd(F \circ \zp) - \zs'{}^\*\zy +
\zs'{}^\*\zy' = \zs'{}^\*\zy'.
                                                        \tag \label{Fplb60}$$
    The passage from $F$ to $F'$ is a version of the Legendre
transformation.  If $S = \im(\zs) = \im(\zs')$ is generated by a
generating form $\zf$ with respect to the special symplectic structure
\Ref{Fplb53}, then the form
        $$\zf' = \zs'{}^\*(\zp^\*\zf + \xd G)
                                                        \tag \label{Fplb61}$$
    is the generating form of $S$ relative to the special symplectic
structure \Ref{Fplb58}.

        \sscx{Iterated tangent bundles $\sT\sT M$, $\sT\sT^2 M$, and
$\sT^2\sT M$.}

    The sets $\sT\sT M$, $\sT\sT^2 M$, and $\sT^2\sT M$ obtained by
repeated application of tangent functors are differential manifolds.  From
a function $f \colon M \rightarrow \R$ we construct sets of functions
        $$\{f_{1,0;1,0}, f_{1,0;1,1}, f_{1,1;1,0}, f_{1,1;1,1}\}
                                                        \tag \label{Fplb62}$$
    on $\sT\sT M$,
        $$\{f_{1,0;2,0}, f_{1,0;2,1}, f_{1,0;2,2}, f_{1,1;2,0},
f_{1,1;2,1}, f_{1,1;2,2}\}
                                                        \tag \label{Fplb63}$$
     on $\sT\sT^2 M$, and
        $$\{f_{2,0;1,0}, f_{2,0;1,1}, f_{2,1;1,0}, f_{2,1;1,1},
f_{2,2;1,0}, f_{2,2;1,1}\}
                                                        \tag \label{Fplb64}$$
    on $\sT^2\sT M$.  The functions are obtained by repeated application
of the constructions introduced in formulae \Ref{Fplb2} and \Ref{Fplb3}.
The general definition is
        $$f_{k',k;i,j} = (f_{k;j})_{k';i}
                                                        \tag \label{Fplb65}$$
    with suitable values of the indices.  These constructions apply to
local functions as well.  By applying these constructions to the local
coordinates for a chart $(x^\zk) \colon M \rightarrow \R^m$ we construct
charts
        $$(x^\zk{}_{1,0;1,0}, x^\zl{}_{1,0;1,1}, x^\zm{}_{1,1;1,0},
x^\zn{}_{1,1;1,1}) \colon \sT\sT M \rightarrow \R^{4m},
                                                        \tag \label{Fplb66}$$
        $$(x^\zk{}_{1,0;2,0}, x^\zl{}_{1,0;2,1}, x^\zm{}_{1,0;2,2},
x^\zn{}_{1,1;2,0}, x^\zw{}_{1,1;2,1}, x^\zr{}_{1,1;2,2}) \colon \sT\sT^2 M
\rightarrow \R^{6m},
                                                        \tag \label{Fplb67}$$
    and
        $$(x^\zk{}_{2,0;1,0}, x^\zl{}_{2,0;1,1}, x^\zm{}_{2,1;1,0},
x^\zn{}_{2,1;1,1}, x^\zw{}_{2,2;1,0}, x^\zr{}_{2,2;1,1}) \colon \sT^2\sT M
\rightarrow \R^{6m}.
                                                        \tag \label{Fplb68}$$
    Coordinates in $\sT\sT M$ are usually denoted by $(x^\zk,\dot
x^\zl,\zd x^\zm,\zd\dot x^\zn)$, coordinates in $\sT\sT^2 M$ are denoted
by $(x^\zk,\dot x^\zl,\ddot x^\zm,\zd x^\zn,\zd\dot x^\zw,\zd\ddot
x^\zr)$, and coordinates in $\sT^2\sT M$ could be denoted by $(x^\zk,\dot
x^\zl,x'{}^\zm,\dot x'{}^\zn,x''{}^\zw,\dot x''{}^\zr)$.

    Each element of one of the spaces $\sT\sT M$, $\sT\sT^2 M$, and
$\sT^2\sT M$ is conveniently represented by a mapping $\zq \colon \R^2
\rightarrow M$.  Mappings
        $$\zh \colon \R \rightarrow \sT M \colon s \mapsto
\st\zq(s,\cdot)(0)
                                                        \tag \label{Fplb69}$$
    and
        $$\zz \colon \R \rightarrow \sT^2 M \colon s \mapsto
\st^2\zq(s,\cdot)(0)
                                                        \tag \label{Fplb70}$$
    derived from the mapping $\zq$ serve as representatives of elements
$\st\zh(0) \in \sT\sT M$, $\st\zz(0) \in \sT\sT^2 M$, and $\st^2\zh(0) \in
\sT^2\sT M$.  We denote these elements by $\st^{1,1}\zq(0,0)$,
$\st^{1,2}\zq(0,0)$, and $\st^{2,1}\zq(0,0)$ respectively.  The mapping
$\zq$ characterized by
        $$x^\zk(\zq(s,t)) = x^\zk(x) + t\dot x^\zk(x) + s\zd x^\zk(x) +
st\zd\dot x^\zk(x)
                                                        \tag \label{Fplb71}$$
    for $(s,t)$ sufficiently close to $(0,0) \in \R^2$ is a representative
of an element $x$ of $\sT\sT M$.  This coordinate construction proves the
existence of representatives.  The corresponding coordinate constructions
of representatives of elements $y \in \sT\sT^2 M$ and $z \in \sT^2\sT M$
are provided by
        $$x^\zk(\zq(s,t)) = x^\zk(y) + t\dot x^\zk(y) +
\frac{1}{2}t^2\ddot x^\zk(y) + s\zd x^\zk(x) + st\zd\dot x^\zk(y) +
\frac{1}{2}s^2t\zd\dot x^\zk(y)
                                                        \tag \label{Fplb72}$$
    and
        $$x^\zk(\zq(s,t)) = x^\zk(z) + t\dot x^\zk(z) + sx'{}^\zk(z) +
st\dot x'{}^\zk(z) + \frac{1}{2}s^2 x''{}^\zk(z) + \frac{1}{2}s^2t\dot
x''{}^\zk(z)
                                                        \tag \label{Fplb73}$$
    respectively.  Relations
        $$\zt^{k''}{}_{k'}{}_{\sT^k M}(\st^{k',k}\zq(0,0)) =
\st^{k'',k}\zq(0,0)
                                                        \tag \label{Fplb74}$$
    and
        $$\sT^{k'}\zt^{k''}{}_k{}_M(\st^{k',k}\zq(0,0)) =
\st^{k',k''}\zq(0,0)
                                                        \tag \label{Fplb75}$$
    are easily verified.  The definition \Ref{Fplb65} is equivalent to
        $$f_{k',k;i,j}(\st^{k',k}\zq(0,0)) = \xD^{(i,j)}(f \circ \zq)(0,0).
                                                        \tag \label{Fplb76}$$
    The equalities
        $$\langle \xd f_{k;i}, w\rangle = f_{1,k;1,i}(w)
                                                        \tag \label{Fplb77}$$
    for $k = 1$ or $k = 2$, $i \leqslant k$, and $w \in \sT\sT^k$ follow
from
        $$\align
    \langle \xd f_{k;i}, \st^{1,k}\zq(0,0)\rangle &= \xD(f_{k;i} \circ
\zz)(0) \\
            &= \frac{\xd}{\xd s}(f_{k;i}(\st^k\zq(s,\cdot)(0)))_{|s=0} \\
            &= \frac{\partial^{i+1}}{\partial s\partial
t^i}(f(\zq(s,t)(0)))_{|s=0,t=0} \\
            &= \xD^{(1,i)(f \circ \zq)(0,0)} \\
            &= f_{1,k;1,i}(\st^{1,k}\zq(0,0))
                                                \tag \label{Fplb78}\endalign$$
    with 
        $$\zz \colon \R \rightarrow \sT^k M \colon s \mapsto
\st^k\zq(s,\cdot)(0).
                                                        \tag \label{Fplb79}$$

    From a mapping $\zq \colon \R^2 \rightarrow M$ we derive the mapping
$\widetilde\zq \colon \R^2 \rightarrow M$ by setting $\widetilde\zq(t,s) =
\zq(s,t)$. This construction is used in the definitions of mappings
        $$\zk^{1,1}{}_M \colon \sT\sT M \rightarrow \sT\sT M \colon
\st^{1,1}\zq(0,0) \mapsto \st^{1,1}\widetilde\zq(0,0),
                                                        \tag \label{Fplb80}$$
        $$\zk^{1,2}{}_M \colon \sT\sT^2 M \rightarrow \sT^2\sT M \colon
\st^{1,2}\zq(0,0) \mapsto \st^{2,1}\widetilde\zq(0,0),
                                                        \tag \label{Fplb81}$$
    and
        $$\zk^{2,1}{}_M \colon \sT^2\sT M \rightarrow \sT\sT^2 M \colon
\st^{2,1}\zq(0,0) \mapsto \st^{1,2}\widetilde\zq(0,0).
                                                        \tag \label{Fplb82}$$
    Diagrams
    \vskip1mm
        $$\xymatrix@R+9mm @C+17mm{{\sT^{k'}\sT^k M}
\ar[d]_*{\zt^{k''}{}_{k'}{}_{\sT^k M}} \ar[r]^*{\zk^{k',k}{}_M} &
            \sT^k\sT^{k'}M \ar[d]_*{\sT^k\zt^{k''}{}_{k'}{}_M} \\
            \sT^{k''}\sT^k M \ar[r]^*{\zk^{k'',k}{}_M} & \sT^k\sT^{k''}M}
                                                        \tag \label{Fplb83}$$
    \vskip2mm
    \noindent for $k'' \leqslant k'$ are commutative and relations
        $$\zk^{k,k'}{}_M \circ \zk^{k',k}{}_M = 1_{\sT^{k'}\sT^k M}
                                                        \tag \label{Fplb84}$$
    are satisfied for all applicable values of $k$, $k'$, and $k''$.  The
special case $\zk{}_M = \zk^{1,1}{}_M$ is the most frequently used.  It is
known as the canonical involution in $\sT\sT M$.

        \sscx{Derivations.}
    Let $\zW(M)$ be the exterior algebra of differential forms on a
differential manifold $M$.  A linear operator $\xa \colon \zW(M)
\rightarrow \zW(M)$ is called a {\it derivation} of $\zW(M)$ of degree $p$
if $\xa\zm$ is a form of degree $q+p$ and
        $$\xa(\zm \wedge \zn) = \xa\zm \wedge \zn + (-1)^{pq}\zm \wedge
\xa\zn
                                                        \tag \label{Fplb85}$$
    when $\zm$ is a form of degree $q$ and $\zn$ is any form on $M$.  The
exterior differential $\xd \colon \zW(M) \rightarrow \zW(M)$ is a
derivation of degree 1. The {\it commutator}
        $$[\xa,\xa'] = \xa\xa' - (-1)^{pp'}\xa'\xa
                                                        \tag \label{Fplb86}$$
    of derivations $\xa$ and $\xa'$ of degrees $p$ and $p'$ respectively
is a derivation of degree $p+p'$.  A derivation $\xa$ is said to be of
{\it type} $\xi_\*$ if $\xa f = 0$ for each function $f$ on $M$.  A
derivation $\xa$ is said to be of {\it type} $\xd_\*$ if $[\xa,\xd] = 0$.
If $\xi_A$ is a derivation of type $\xi_\*$, then $\xd_A = [\xi_A,\xd]$ is
a derivation of type $\xd_\*$.  Derivations are local operators: if $\xa$
is a derivation and $\zm$ is a differential form on $M$ vanishing on an
open subset $U \subset M$, then $\xa\zm$ vanishes on $U$.  A derivation is
fully characterized by its action on functions and differentials of
functions since each differential form is locally represented as a sum of
exterior products of differentials of functions multiplied by functions. A
derivation of type $\xd_\*$ is fully characterized by its action on
functions.

    A {\it vector-valued} $p$-{\it form} is a linear mapping
        $$A\, \colon \wedge^p \sT M \rightarrow \sT M.
                                                        \tag \label{Fplb87}$$
    If $w \in \wedge^p \sT_a M$, then $A(w) \in \sT_a M$.  Following
Fr\"olicher and Nijenhuis [FN] we associate with a vector-valued $p$-form
$A$ a derivation $\xi_A$ of type $\xi_\*$ and degree $p-1$ and the
derivation $\xd_A = [\xi_A,\xd]$.  The derivation $\xi_A$ is characterized
by its action on 1-forms.  If $\zm$ is a 1-form, then $\xi_A\zm$ is a
$p$-form and
        $$\langle \xi_A\zm, w\rangle = \langle \zm, A(w)\rangle
                                                        \tag \label{Fplb88}$$
    for each $w \in \wedge^p \sT M$.

    For $k = 1$ or $k = 2$, and each $n \in \N$ we define a linear mapping
        $$F(k;n) \colon \sT\sT^k M \rightarrow \sT\sT^k M \colon
\st^{1,k}\zq(0,0) \rightarrow \st^{1,k}\zq^n(0,0),
                                                        \tag \label{Fplb89}$$
    where $\zq$ is a mapping from $\R^2$ to $M$ and
        $$\zq^n \colon \R^2 \rightarrow M \colon (s,t) \mapsto \zq(st^n,t).
                                                        \tag \label{Fplb90}$$
    Relations
        $$F(k;0) = 1_{\sT\sT^k M},
                                                        \tag \label{Fplb91}$$
        $$F(k;n') \circ F(k;n) = F(k;n' + n),
                                                        \tag \label{Fplb92}$$
    and
        $$F(k;n) = 0 \hskip5mm \text{if } n \geqslant k
                                                        \tag \label{Fplb93}$$
    are easily established.  It follows that $F(1;1)$, $F(2;1)$, and
$F(2;2)$ are the only non trivial cases.  The diagrams
    \vskip1mm
        $$\xymatrix@R+6mm @C+10mm{{\sT\sT^k M} \ar[d]_*{\zt_{\sT^k M}}
\ar[r]^*{F(k;n)} &
            \sT\sT^k M \ar[d]_*{\zt_{\sT^k M}} \\
            \sT^k M \ar@{=}[r] & \sT^k M}
                                                        \tag \label{Fplb94}$$
    \vskip2mm
    \noindent are commutative since $\zq^n(0,\cdot) = \zq(0,\cdot)$ and
the diagrams
    \vskip1mm
        $$\xymatrix@R+6mm @C+10mm{{\sT\sT^2 M} \ar[d]_*{\sT\zt^1{}_2{}_M}
\ar[r]^*{F(2;n)} &
            \sT\sT^2 M \ar[d]_*{\sT\zt^1{}_2{}_M} \\
            \sT\sT M \ar[r]^*{F(1;n)} & \sT\sT M}
                                                        \tag \label{Fplb95}$$
    \vskip2mm
    \noindent are obviously commutative.  The mappings $F(k;n)$ are
vector-valued 1-forms.

        \claim \c{p}{Proposition}{}
\label{Cplb1}

        $$\langle \xd f_{k;i}, F(k;n)(w)\rangle = \frac{i!}{(i-n)!}
\langle \xd f_{k;i-n}, w\rangle
                                                        \tag \label{Fplb96}$$
    if $i \geqslant n$ and
        $$\langle \xd f_{k;i}, F(k;n)(w)\rangle = 0
                                                        \tag \label{Fplb97}$$
    if $i < n$.
        \endclaim
        \proof
    The proof is established by the calculations
        $$\align
    \langle \xd f_{k;i}, F(k;n)(\st^{1,k}\zq(0,0))\rangle &=
f_{1,k;1,i}(F(k;n)(\st^{1,k}\zq(0,0))) \\
            &= \xD^{(1,i)}(f \circ \zq^n)(0,0) \\
            &= \frac{\partial^{i+1}}{\partial s\partial
t^i}(f(\zq(st^n,t)))_{|s=0,t=0} \\
            &= \frac{\partial^i}{\partial t^i}(t^n\frac{\partial}{\partial
u} f(\zq(u,t)))_{|u=0,t=0} \\
            &= \frac{i!}{(i-n)!}\frac{\partial^{i-n+1}}{\partial u\partial
t^{i-n}}(f(\zq(u,t)))_{|u=0,t=0} \\
            &= \frac{i!}{(i-n)!}\xD^{(1,i-n)}(f \circ \zq)(0,0) \\
            &= \frac{i!}{(i-n)!}\langle \xd f_{k;i-n},
\st^{1,k}\zq(0,0)\rangle
                                                \tag \label{Fplb98}\endalign$$
    if $i \geqslant n$ and
        $$\langle \xd f_{k;i}, F(k;n)(\st^{1,k}\zq(0,0))\rangle =
\frac{\partial^i}{\partial t^i}\left(t^n\frac{\partial}{\partial u}
f(\zq(u,t))\right)_{|u=0,t=0} = 0
                                                        \tag \label{Fplb99}$$
    if $i < n$.
        \endproof

    Here are the non trivial cases of formulae \Ref{Fplb96} and
\Ref{Fplb97}:
        $$\langle \xd f_{1;0}, F(1;1)(w)\rangle = 0,
                                                        \tag \label{Fplb100}$$
        $$\langle \xd f_{1;1}, F(1;1)(w)\rangle = \langle \xd f_{1;0},
w\rangle,
                                                        \tag \label{Fplb101}$$
        $$\langle \xd f_{2;0}, F(2;1)(w)\rangle = 0,
                                                        \tag \label{Fplb102}$$
        $$\langle \xd f_{2;1}, F(2;1)(w)\rangle = \langle \xd f_{2;0},
w\rangle,
                                                        \tag \label{Fplb103}$$
        $$\langle \xd f_{2;2}, F(2;1)(w)\rangle = 2\langle \xd f_{2;1},
w\rangle,
                                                        \tag \label{Fplb104}$$
        $$\langle \xd f_{2;0}, F(2;2)(w)\rangle = 0,
                                                        \tag \label{Fplb105}$$
        $$\langle \xd f_{2;1}, F(2;2)(w)\rangle = 0,
                                                        \tag \label{Fplb106}$$
        $$\langle \xd f_{2;2}, F(2;2)(w)\rangle = 2\langle \xd f_{2;0},
w\rangle.
                                                        \tag \label{Fplb107}$$

    It follows from formulae \Ref{Fplb100} and \Ref{Fplb102} that if $w
\in \im(F(1;1))$, then $\langle \xd f_{1;0}, w\rangle = 0$ and if $w \in
\im(F(2;1))$, then $\langle \xd f_{2;0}, w\rangle = 0$ for each function
$f$ on $M$.

        \claim \c{p}{Proposition}{}
\label{Cplb2}
    If $w \in \sT\sT M$ and $\langle \xd f_{1;0}, w\rangle = 0$ for each
function $f$, then $w \in \im(F(1;1))$ and if $w \in \sT\sT^2 M$ and
$\langle \xd f_{2;0}, w\rangle = 0$ for each function $f$, then $w \in
\im(F(2;1))$.
        \endclaim
        \proof
    Let $(x^\zk,\dot x^\zl,\zd x^\zm,\zd\dot x^\zn) \colon \sT\sT M
\rightarrow \R^{4m}$ be a chart of $\sT\sT M$ derived from a chart
$(x^\zk) \colon M \rightarrow \R^m$.  If $w \in \sT\sT M$ and $\langle \xd
f_{1;0}, w\rangle = 0$ for each function $f$, then $\zd x^\zk{}(w) =
\langle \xd x^\zk, w\rangle = 0$.  A representative $\zq$ of $w$ such that
        $$(x^\zk \circ \zq)(s,t) = x^\zk(w) + \dot x^\zk(w)t + \zd\dot
x^\zk(w)st
                                                        \tag \label{Fplb108}$$
    can be chosen.  If $\zz$ is the mapping
        $$\zz \colon \R^2 \rightarrow M \colon (s,t) \mapsto \lim_{u
\rightarrow t} \zq(su^{-n},u),
                                                        \tag \label{Fplb109}$$
  then $\zq = \zz^1$ and $w = F(1;1)(\st^{1,1}\zz(0,0))$.

    We use in $\sT\sT^2 M$ coordinates $(x^\zk,\dot x^\zl,\ddot x^\zm,\zd
x^\zn,\zd\dot x^\zw,\zd\ddot x^\zr)$ derived from a chart $(x^\zk) \colon
M \rightarrow \R^m$.  If $w \in \sT\sT^2 M$ and $\langle \xd f_{2;0},
w\rangle = 0$ for each function $f$, then $\zd x^\zk{}(w) = 0$.  We choose
a representative $\zq$ of $w$ such that
        $$(x^\zk \circ \zq)(s,t) = x^\zk(w) + \dot x^\zk(w)t +
\frac{1}{2}\ddot x^\zk(w)t^2 + \zd\dot x^\zk(w)st + \frac{1}{2}\zd\ddot
x^\zk(w)st^2.
                                                        \tag \label{Fplb110}$$
    If $\zz$ is the mapping defined in formula \Ref{Fplb109}, then $\zq =
\zz^1$ and $w = F(2;1)(\st^{2,1}\zz(0,0))$.
        \endproof

        \claim \c{p}{Proposition}{}
\label{Cplb3}
    $\im(F(1;1)) = \ker(F(1;1))$ and $\im(F(2;1)) = \ker(F(2;2))$.
        \endclaim
        \proof
    From $F(1;1) \circ F(1;1) = F(1;2) = 0$ and $F(2;1) \circ F(2;2) =
F(2;3) = 0$ we derive the inclusions $\im(F(1;1)) \subset \ker(F(1;1))$
and $\im(F(2;1)) \subset \ker(F(2;2))$. If $F(1;1)(w) = 0$, then
        $$\langle \xd f_{1;0}, w\rangle =  \langle \xd f_{1;1},
F(1;1)(w)\rangle = 0.
                                                        \tag \label{Fplb111}$$
    Hence $w \in \im(F(1;1))$.  If $F(2;2)(w) = 0$, then
        $$\langle \xd f_{2;0}, w\rangle = \frac{1}{2}\langle \xd f_{2;2},
F(2;2)(w)\rangle = 0.
                                                        \tag \label{Fplb112}$$
    Hence $w \in \im(F(2;1))$.
        \endproof

        \claim \c{p}{Proposition}{}
\label{Cplb4}
    $\ker(\sT\zt_M) = \ker(F(1;1))$ and $\ker(\sT\zt_2{}_M) = \ker(F(2;2))$
        \endclaim
        \proof
    The results follow directly from the identities
        $$\langle \xd f_{1;1}, F(1;1)(w)\rangle =  \langle \xd f_{1;0},
w\rangle = \langle \xd f, \sT\zt_M(w)\rangle
                                                        \tag \label{Fplb113}$$
    for $w \in \sT\sT M$ and
        $$\langle \xd f_{2;2}, F(2;2)(w)\rangle =  2\langle \xd f_{2;0},
w\rangle = 2\langle \xd f, \sT\zt_2{}_M(w)\rangle
                                                        \tag \label{Fplb114}$$
    for $w \in \sT\sT^2 M$.
        \endproof

    The relations $\ker(\sT\zt_M) = \im(F(1;1))$ and $\ker(\sT\zt_2{}_M) =
\im(F(2;1))$ follow from the two above propositions.

    Let $\zW_1(M)$ and $\zW_2(M)$ denote the exterior algebras of
differential forms on the tangent bundles $\sT M$ and $\sT^2 M$
respectively.  We will denote by $\zs_2{}^1{}_M$ the homomorphism
        $$\zt^1{}_2{}_M{}^\* \colon \zW_1(M) \rightarrow \zW_2(M).
                                                        \tag \label{Fplb115}$$

    Derivations $\xi_{F(k;n)}$ and $\xd_{F(k;n)}$ are associated with the
vector-valued 1-forms $F(k;n)$.  The diagram
    \vskip1mm
        $$\xymatrix@R+6mm @C+15mm{{\zW_1(M)} \ar[d]_*{\zs_2{}^1{}_M}
\ar[r]^*{\xi_{F(1;1)}} &
           \zW_1(M) \ar[d]_*{\zs_2{}^1{}_M} \\
            \zW_2(M) \ar[r]^*{\xi_{F(2;1)}} & \zW_2(M)}
                                                        \tag \label{Fplb116}$$
    \vskip2mm
    \noindent is commutative.

    The article [12] offers a generalization of the Fr\"olicher and
Nijenhuis theory.  Let $\zf \colon N \rightarrow M$ be a differentiable
mapping.  The mapping $\zf^\* \colon \zF(M) \rightarrow \zF(N)$ is a
homomorphism of the exterior algebras.  A {\it derivation of degree} $p$
{\it relative to} $\zf^\*$ is a linear operator $\xa \colon \zF(M)
\rightarrow \zF(N)$ such that $\xa\zm$ is a form on $N$ of degree $q+p$ and
        $$\xa(\zm \wedge \zn) = \xa\zm \wedge \zf^\*\zn +
(-1)^{pq}\zf^\*\zm \wedge \xa\zn
                                                        \tag \label{Fplb117}$$
    if $\zm$ is a form on $M$ of degree $q$ and $\zn$ is any form on $M$.
A derivation of the algebra $\zF(M)$ is a derivation relative to the
identity mapping $1_M$.  A derivation $\xa$ relative to $\zf$ is said to
be of {\it type} $\xi_\*$ if $\xa f = 0$ for each function $f$ on $M$.  A
relative derivation $\xa$ of degree $p$ is said to be of {\it type}
$\xd_\*$ if $\xa\xd - (-1)^p \xd\xa = 0$.  If $\xi_A$ is a derivation of
type $\xi_\*$ relative to $\zf$, then $\xd_A = \xi_A\xd - (-1)^p\xd\xi_A$
is a derivation of type $\xd_\*$ relative to $\zf$.  Note that the
expressions $\xa\xd - (-1)^p \xd\xa$ and $\xi_A\xd - (-1)^p\xd\xi_A$ are
not commutators since each of these expressions involves two different
exterior differentials $\xd$.  If $\xa$ is a derivation of degree $p$
relative to $\zf^\*$ and $\zc \colon O \rightarrow N$ is a differentiable
mapping, then the operator $\zc^\*\xa \colon \zF(M) \rightarrow \zF(O)$ is
a derivation of degree $p$ relative to $(\zf \circ \zc)^\*$ since
        $$\align
    \zc^\*\xa(\zm \wedge \zn) &= \zc^\*\xa\zm \wedge \zc^\*\zf^\*\zn +
(-1)^{pq}\zc^\*\zf^\*\zm \wedge \zc^\*\xa\zn \\
        &= \zc^\*\xa\zm \wedge (\zf \circ \zc)^\*\zn + (-1)^{pq}(\zf \circ
\zc)^\*\zm \wedge \zc^\*\xa\zn
                                            \tag \label{Fplb118}\endalign$$
    if $\zm$ is a form on $M$ of degree $q$ and $\zn$ is any form on $M$.
If $\xa$ is a derivation of type $\xi_\*$ or $\xd_\*$, then $\zc^\*\xa$ is
a derivation of the same type.  Relative derivations are again local
operators and are completely characterized by their action on functions
and differentials of functions.

    A {\it vector-valued} $p$-{\it form relative to} $\zf \colon N
\rightarrow M$ is a linear mapping
        $$A\, \colon \wedge^p \sT N \rightarrow \sT M
                                                        \tag \label{Fplb119}$$
    such that if $w \in \wedge^p \sT_b N$, then $A(w) \in \sT_{\zf(b)}M$.
We associate with a vector-valued $p$-form $A$ relative to $\zf$ a
derivation $\xi_A$ relative to $\zf^\*$ of type $\xi_\*$ and degree $p-1$
and the relative derivation $\xd_A = \xi_A\xd - (-1)^p\xd\xi_A$.  If $\zm$
is a 1-form on $M$, then $\xi_A\zm$ is a $p$-form on $N$ and
        $$\langle \xi_A\zm, w\rangle = \langle \zm, A(w)\rangle
                                                        \tag \label{Fplb120}$$
     for each $w \in \wedge^p \sT N$.

    Let $T(0) \colon \sT M \rightarrow \sT M$ be the identity mapping
interpreted as a deformation of the tangent projection $\zt_M \colon \sT M
\rightarrow M$.  We associate with $T(0)$ derivations $\xi_{T(0)} \colon
\zW(M) \rightarrow \zW_1(M)$ and $\xd_{T(0)} \colon \zW(M) \rightarrow
\zW_1(M)$ relative to $\zs_M = \zt_M{}^\*$.  The derivation $\xi_{T(0)}$
is a derivation of degree -1.  If $\zm$ is a $(q+1)$-form on $M$, then
$\xi_{T(0)}\zm$ is a $q$-form on $\sT M$ and if $w_1,\ldots,w_q$ are
elements of $\sT\sT M$ such that $\zt_{\sT M}(w_1) = \ldots = \zt_{\sT
M}(w_q)$, then
        $$\langle \xi_{T(0)}\zm, w_1 \wedge \ldots \wedge w_q\rangle =
\langle \zm, \zt_{\sT M}(w_1) \wedge \sT\zt_M(w_1) \wedge \ldots \wedge
\sT\zt_M(w_q)\rangle.
                                                        \tag \label{Fplb121}$$

    Let $X \colon M \rightarrow \sT M$ be a vector field.  The operator
$X^\*\xi_{T(0)}$ is a derivation of $\zW(M)$ of type $\xi_\*$ and degree
-1.  For each 1-form $\zm$ on $M$ we have
        $$X^\*\xi_{T(0)}\zm = \xi_{T(0)}\zm \circ X = \langle \zm, T(0)
\circ X\rangle = \langle \zm, X\rangle = \xi_X\zm.
                                                        \tag \label{Fplb122}$$
    Hence, $X^\*\xi_{T(0)} = \xi_X$.  The relation $X^\*\xd_{T(0)} =
\xd_X$ is established by
        $$X^\*\xd_{T(0)} = X^\*(\xi_{T(0)}\xd + \xd\xi_{T(0)}) =
X^\*\xi_{T(0)}\xd + \xd X^\*\xi_{T(0)} = \xi_X\xd + \xd\xi_X = \xd_X.
                                                        \tag \label{Fplb123}$$

    Let $f$ be a function on $M$.  For each vector $v = \dot\zx(0) =
\st\zx(0) \in \sT M$ we find
        $$\align
    \xd_{T(0)}f(v) &= \xi_{T(0)}\xd f(v) \\
            &= \langle \xd f, T(0)(v)\rangle \\
            &= \langle \xd f, v\rangle \\
            &= \xD(f \circ \zx)(0).
                                                \tag \label{Fplb124}\endalign$$
    Hence, $\xi_{T(0)}\xd f = f_{1;1}$, $\xd_{T(0)}f = f_{1;1}$, and
$\xd_{T(0)}\xd f = \xd f_{1;1}$.

    For each vector $w \in \sT\sT M$ there is a mapping $\zd\zx \colon \R
\rightarrow \sT M$ such that $w = \zk^{1,1}(\st\zd\zx(0))$.  Let
$w_1,\ldots,w_q$ be elements of $\sT\sT M$ such that
        $$\zt_{\sT M}(w_1) = \ldots = \zt_{\sT M}(w_q)
                                                        \tag \label{Fplb125}$$
    and let
        $$\zd\zx_1 \colon \R \rightarrow \sT M, \ldots , \zd\zx_q \colon
\R \rightarrow \sT M
                                                        \tag \label{Fplb126}$$
    be the mappings such that
        $$w_1 = \zk^{1,1}(\st\zd\zx_1(0)), \ldots , w_q =
\zk^{1,1}(\st\zd\zx_q(0)).
                                                        \tag \label{Fplb127}$$
    We will require that these mappings satisfy the condition
        $$\zt_M \circ \zd\zx_1 = \cdots = \zt_M \circ \zd\zx_q.
                                                        \tag \label{Fplb128}$$
    The following construction proves the existence of such mappings.  Let
$(x^\zk,\dot x^\zl) \colon \sT M \rightarrow \R^{2m}$ be a chart of $\sT
M$ and $(x^\zk,\dot x^\zl,\zd x^\zm,\zd\dot x^\zn) \colon \sT\sT M
\rightarrow \R^{4m}$ a chart of $\sT\sT M$ derived from a chart $(x^\zk)
\colon M \rightarrow \R^m$. Mappings $\zd\zx_1,\ldots,\zd\zx_q$
characterized by
        $$\align
    (x^\zk,\dot x^\zl)(\zd\zx_1(t)) &= (x^\zk(w_1) + t\dot x^\zk(w_1),\zd
x^\zl(w_1) + t\zd\dot x^\zl(w_1)) \\
        &\hskip1cm ................................ \\
    (x^\zk,\dot x^\zl)(\zd\zx_q(t)) &= (x^\zk(w_q) + t\dot x^\zk(w_q),\zd
x^\zl(w_q) + t\zd\dot x^\zl(w_1))
                                                \tag \label{Fplb129}\endalign$$
    for $t$ close to $0 \in \R$ have the required property since
$x^\zk(w_1) = \ldots = x^\zk(w_q)$ and $\dot x^\zk(w_1) = \ldots = \dot
x^\zk(w_q)$.  We denote by $\zx$ the mapping $\zt_M \circ \zd\zx_1 =
\ldots = \zt_M \circ \zd\zx_q$.  The following proposition is stated in
terms of the mappings $\zx$ and $\zd\zx_1, \ldots , \zd\zx_q$.

        \claim \c{p}{Proposition}{}
\label{Cplb5}
    If $q > 0$ and $\zm$ is a $q$-form on $M$, then $\xd_{T(0)}\zm$ is a
$q$-form on $\sT M$ and
        $$\langle \xd_{T(0)}\zm, w_1 \wedge \ldots \wedge w_q\rangle =
\xD\langle \zm, \zd\zx_1 \wedge \ldots \wedge \zd\zx_q\rangle(0)
                                                        \tag \label{Fplb130}$$
    where $w_1,\ldots,w_q$ are vectors in $\sT\sT M$ such that $\zt_{\sT
M}(w_1) = \ldots = \zt_{\sT M}(w_q)$.
        \endclaim
        \proof
    Let an operator $\xa \colon \zW(M) \rightarrow \zW_1(M)$ of degree 0
be defined by
        $$\xa f = \xd_{T(0)} f
                                                        \tag \label{Fplb131}$$
    for each function $f$ on $M$ and
        $$\langle \xa\zm, w_1 \wedge \ldots \wedge w_q \rangle =
\xD\langle \zm, \zd\zx_1 \wedge \ldots \wedge \zd\zx_q\rangle(0),
                                                        \tag \label{Fplb132}$$
    if $q > 0$, $\zm$ is a $q$-form on $M$, and $w_1,\ldots,w_q$ are
elements of $\sT\sT M$ such that $\zt_{\sT M}(w_1) = \ldots = \zt_{\sT
M}(w_q)$.

    We show that $\xa$ is a derivation relative to $\zs_M$.  If $f_1$ and
$f_2$ are functions on $M$, then
        $$\xa(fg) = \xd_{T(0)}(f g \circ \zt_M) = \xd_{T(0)}f g \circ
\zt_M + f \circ \zt_M\xd_{T(0)}g = \xa f g \circ \zt_M + f \circ \zt_M\xa
g.
                                                        \tag \label{Fplb133}$$
    If $f$ is a function on $M$ and $\zm$ is a $q$-form on $M$ with $q >
0$, then
        $$\align
    \langle \xa(f\zm), w_1 \wedge \ldots \wedge w_q\rangle &= \xD\langle
f\zm, \zd\zx_1 \wedge \ldots \wedge \zd\zx_q\rangle(0) \\
            &= \xD((f \circ \zx)\langle \zm, \zd\zx_1 \wedge \ldots \wedge
\zd\zx_q\rangle)(0) \\
            &= \langle \xd f, \st\zx(0)\rangle\langle \zm, \sT\zt_M(w_1)
\wedge \ldots \wedge \sT\zt_M(w_q)\rangle \\
    &\hskip40mm + f(\zx(0))\xD\langle \zm, \zd\zx_1 \wedge \ldots \wedge
\zd\zx_q\rangle(0) \\
            &= \xa f(\st\zx(0))\langle \zs_M\zm, w_1 \wedge \ldots \wedge
w_q \rangle + \zs_M f(\st\zx(0))\langle \xa\zm, w_1 \wedge \ldots \wedge
w_q \rangle \\
            &= \langle \xa f\zs_M\zm + \zs_M f \xa\zm, w_1 \wedge \ldots
\wedge w_q \rangle.
                                            \tag \label{Fplb134}\endalign$$
    If $\zm_1$ and $\zm_2$ are forms on $M$ of degrees $q_1 > 0$ and $q_2
> 0$ respectively and $q = q_1 + q_2$, then
        $$\align
    \langle \xa(\zm_1 \wedge &\zm_2), w_1 \wedge \ldots \wedge w_q \rangle
= \xD\langle \zm_1 \wedge \zm_2, \zd\zx_1 \wedge \ldots \wedge
\zd\zx_q\rangle(0) \\
            &= \frac{1}{q_1!q_2!}\sum_{\zs \in
S_q}\sign(\zs)\xD\left(\left\langle \zm_1, \zd\zx_{\zs(1)} \wedge \ldots
\wedge \zd\zx_{\zs(q_1)}\right\rangle \left\langle \zm_2,
\zd\zx_{\zs(q_1+1)} \wedge \ldots \wedge
\zd\zx_{\zs(q)}\right\rangle\right)(0) \\
            &= \frac{1}{q_1!q_2!}\sum_{\zs \in
S_q}\sign(\zs)\left(\xD\left\langle \zm_1, \zd\zx_{\zs(1)} \wedge \ldots
\wedge \zd\zx_{\zs(q_1)}\right\rangle(0) \left\langle \zm_2,
\zd\zx_{\zs(q_1+1)} \wedge \ldots \wedge \zd\zx_{\zs(q)}
\right\rangle(0)\right. \\
        &\hskip15mm + \left.\xD\left\langle \zm_1, \zd\zx_{\zs(1)} \wedge
\ldots \wedge \zd\zx_{\zs(q_1)}\right\rangle(0) \left\langle \zm_2,
\zd\zx_{\zs(q_1+1)} \wedge \ldots \wedge \zd\zx_{\zs(q)}
\right\rangle(0)\right)\\
            &= \frac{1}{q_1!q_2!}\sum_{\zs \in
S_q}\sign(\zs)\left(\left\langle \xa\zm_1, w_{\zs(1)} \wedge \ldots \wedge
w_{\zs(q_1)}\right\rangle \left\langle \zm_2, \sT\zt_M(w_{\zs(q_1+1)})
\wedge \ldots \wedge \sT\zt_M(w_{\zs(q)}) \right\rangle\right. \\
        &\hskip15mm + \left.\left\langle \zm_1, \sT\zt_M(w_{\zs(1)})
\wedge \ldots \wedge \sT\zt_M(w_{\zs(q_1)})\right\rangle \left\langle
\xa\zm_2, w_{\zs(q_1+1)} \wedge \ldots \wedge w_{\zs(q)}
\right\rangle\right)\\
            &= \frac{1}{q_1!q_2!}\sum_{\zs \in
S_q}\sign(\zs)\left(\left\langle \xa\zm_1, w_{\zs(1)} \wedge \ldots \wedge
w_{\zs(q_1)}\right\rangle \left\langle \zs_M\zm_2, w_{\zs(q_1+1)} \wedge
\ldots \wedge w_{\zs(q)} \right\rangle\right. \\
        &\hskip15mm + \left.\left\langle \zs_M\zm_1, w_{\zs(1)} \wedge
\ldots \wedge w_{\zs(q_1)}\right\rangle \left\langle \xa\zm_2,
w_{\zs(q_1+1)} \wedge \ldots \wedge w_{\zs(q)} \right\rangle\right)\\
            &= \langle \xa\zm_1 \wedge \zs_M\zm_2 + \zs_M\zm_1 \wedge
\xa\zm_2, w_1 \wedge \ldots \wedge w_q\rangle.
                                                \tag \label{Fplb135}\endalign$$
    This completes the proof that $\xa$ is a derivation relative to
$\zs_M$.

    Let $w$ be an element of $\sT\sT M$.  We associate the mapping
        $$\zd\zx \colon \R \rightarrow \sT M \colon t \mapsto
\st\zq(\cdot,t)(0)
                                                        \tag \label{Fplb136}$$
    with a representative $\zq \colon \R^2 \rightarrow M$ of $w$.  If $f$
is a function on $M$, then
        $$\align
    \langle \xa\xd f, w\rangle &= \xD\langle \xd f, \zd\zx\rangle(0) \\
            &= \frac{\xd}{\xd t}\langle \xd f, \zd\zx(t)\rangle_{|t=0} \\
            &= \frac{\xd}{\xd t}\langle \xd f,
\st\zq(\cdot,t)(0)\rangle_{|t=0} \\
            &= \xD^{(1,1)}(f \circ \zq)(0,0) \\
            &= f_{1,1;1,1}(\st^{1,1}\zq(0,0)) \\
            &= \langle \xd f_{1;1}, \st^{1,1}\zq(0,0)\rangle \\
            &= \langle \xd_{T(0)}\xd f, w\rangle.
                                                \tag \label{Fplb137}\endalign$$
    The equality $\xa\xd f = \xd_{T(0)}\xd f$ together with $\xd_{T(0)} f
= \xa f$ for each function $f$ imply the equality $\xd_{T(0)} = \xa$.
        \endproof

    The mapping
        $$T(1) \colon \sT^2 M \rightarrow \sT\sT M \colon \st^2\zx(0)
\mapsto \st\st\zx(0)
                                                        \tag \label{Fplb138}$$
    is a vector valued 0-form relative to $\zt^1{}_2{}M$.  We associate
with $T(1)$ derivations $\xi_{T(1)} \colon \zW_1(M) \rightarrow \zW_2(M)$
and $\xd_{T(1)} \colon \zW_1(M) \rightarrow \zW_2(M)$ relative to
$\zs_2{}^1{}_M$. Derivations $\xi_{T(1)}$ and $\xd_{T(1)}$ have properties
analogous to those of derivations $\xi_{T(0)}$ and $\xd_{T(0)}$.  If $\zm$
is a $(q+1)$-form on $\sT M$, then $\xi_{T(1)}\zm$ is a $q$-form on $\sT^2
M$ and if $w_1,\ldots,w_q$ are elements of $\sT\sT^2 M$ such that
$\zt_{\sT^2 M}(w_1) = \ldots = \zt_{\sT^2 M}(w_q)$, then
        $$\langle \xi_{T(1)}\zm, w_1 \wedge \ldots \wedge w_q\rangle =
\langle \zm, \zt_{\sT^2 M}(w_1) \wedge \sT\zt^1{}_2{}_M(w_1) \wedge \ldots
\wedge \sT\zt^1{}_2{}_M(w_q)\rangle.
                                                        \tag \label{Fplb139}$$

    Let $F$ be a function on $\sT M$.  For each element $a = \st^2\zx(0)
\in \sT^\2 M$ we have
        $$\align
    \xd_{T(1)}F(a) &= \xi_{T(1)}\xd F(a) \\
            &= \langle \xd F, T(1)(\st^2\zx(0))\rangle \\
            &= \langle \xd F, \st\st\zx(0)\rangle \\
            &= \xD(F \circ \st\zx)(0).
                                                \tag \label{Fplb140}\endalign$$

    If $w_1,\ldots,w_q$ are elements of $\sT\sT^2 M$ such that
        $$\zt_{\sT^2 M}(w_1) = \ldots = \zt_{\sT^2 M}(w_q),
                                                        \tag \label{Fplb141}$$
    then it is possible to choose mappings
        $$\zd\zx_1 \colon \R \rightarrow \sT M, \ldots , \zd\zx_q \colon
\R \rightarrow \sT M
                                                        \tag \label{Fplb142}$$
    such that
        $$w_1 = \zk^{2,1}(\st^2\zd\zx_1(0)), \ldots , w_q =
\zk^{2,1}(\st^2\zd\zx_q(0))
                                                        \tag \label{Fplb143}$$
    and
        $$\zt_M \circ \zd\zx_1 = \cdots = \zt_M \circ \zd\zx_q.
                                                        \tag \label{Fplb144}$$
    Let $(x^\zk,\dot x^\zl) \colon \sT M \rightarrow \R^{2m}$ be a chart
of $\sT M$ and $(x^\zk,\dot x^\zl,\ddot x^\zm,\zd x^\zn,\zd\dot
x^\zw,\zd\ddot x^\zp) \colon \sT\sT^2 M \rightarrow \R^{6m}$ a chart of
$\sT\sT^2 M$ derived from a chart $(x^\zk) \colon M \rightarrow \R^m$.
Mappings $\zd\zx_1,\ldots,\zd\zx_q$ such that
        $$\align
    (x^\zk,\dot x^\zl)(\zd\zx_1(t)) &= \left(x^\zk(w_1) + t\dot x^\zk(w_1)
+ \frac{t^2}{2}\ddot x^\zk(w_1),\zd x^\zl(w_1) + t\zd\dot x^\zl(w_1) +
\frac{t^2}{2}\zd\ddot x^\zl(w_1)\right) \\
        &\hskip1cm ................................ \\
    (x^\zk,\dot x^\zl)(\zd\zx_q(t)) &= \left(x^\zk(w_q) + t\dot x^\zk(w_q)
+ \frac{t^2}{2}\ddot x^\zk(w_q),\zd x^\zl(w_q) + t\zd\dot x^\zl(w_q) +
\frac{t^2}{2}\zd\ddot x^\zl(w_q)\right) \\
                                                \tag \label{Fplb145}\endalign$$
    for $t$ close to $0 \in \R$ are a correct choice since $x^\zk(w_1) =
\ldots = x^\zk(w_q)$, $\dot x^\zk(w_1) = \ldots = \dot x^\zk(w_q)$, and
$\ddot x^\zk(w_1) = \ldots = \ddot x^\zk(w_q)$.  We introduce mappings
        $$\zd\dot\zx_1 = \zk^{1,1} \circ \st\zd\zx_q, \ldots ,
\zd\dot\zx_q = \zk^{1,1} \circ \st\zd\zx_q.
                                                        \tag \label{Fplb146}$$
    The following proposition is stated in terms of these mappings.

        \claim \c{p}{Proposition}{}
\label{Cplb6}
    If $q > 0$ and $\zm$ is a $q$-form on $\sT M$, then $\xd_{T(1)}\zm$ is
a $q$-form on $\sT^2 M$ and
        $$\langle \xd_{T(1)}\zm, w_1 \wedge \ldots \wedge w_q\rangle =
\xD\langle \zm, \zd\dot\zx_1 \wedge \ldots \wedge \zd\dot\zx_q\rangle(0)
                                                        \tag \label{Fplb147}$$
    where $w_1,\ldots,w_q$ are vectors in $\sT\sT^2 M$ such that
$\zt_{\sT^2 M}(w_1) = \ldots = \zt_{\sT^2 M}(w_q)$.
        \endclaim
        \proof
    The proof of this proposition is analogous to that of Proposition
\Ref{Cplb5}. An operator $\xa \colon \zW_1(M) \rightarrow \zW_2(M)$ of
degree 0 is defined by
        $$\xa g = \xd_{T(1)}g
                                                        \tag \label{Fplb148}$$
    for each function $g$ on $\sT M$ and
        $$\langle \xa\zm, w_1 \wedge \ldots \wedge w_q \rangle =
\xD\langle \zm, \zd\dot\zx_1 \wedge \ldots \wedge \zd\dot\zx_q\rangle(0),
                                                        \tag \label{Fplb149}$$
    if $q > 0$, $\zm$ is a $q$-form on $\sT M$, and $w_1,\ldots,w_q$ are
elements of $\sT\sT^2 M$ such that $\zt_{\sT M}(w_1) = \ldots = \zt_{\sT
M}(w_q)$.  It is shown that $\xa$ is a derivation relative to
$\zs_2{}^1{}_M$ by performing essentially the same calculations as in the
proof of Proposition \Ref{Cplb5}.

    With a representative $\zq \colon \R^2 \rightarrow M$ of an element $w
= \st^{1,2}\zq(0,0)$ of $\sT\sT^2 M$ we associate the mapping
        $$\zd\dot\zx \colon \R \rightarrow \sT\sT M \colon t \mapsto
\st^{1,1}\zq(\cdot,t).
                                                        \tag \label{Fplb150}$$
    If $f$ is a function on $M$, then
        $$\align
    \langle \xa\xd f_{1;0}, w\rangle &= \xD\langle \xd f_{1;0},
\zd\dot\zx\rangle(0) \\
            &= \frac{\xd}{\xd t}\langle \xd f_{1;0},
\zd\zx(t)\rangle_{|t=0} \\
            &= \frac{\xd}{\xd t}\langle \xd f_{1;0},
\st^{1,1}\zq(\cdot,t)(0)\rangle_{|t=0} \\
            &= \frac{\xd}{\xd t}\langle f_{1,1;1,0},
\st^{1,1}\zq(\cdot,t)(0)\rangle_{|t=0} \\
            &= \frac{\xd}{\xd t}\left(\xD^{(1,0)}(f \circ
\zq)(0,t)\right)_{|t=0} \\
            &= \xD^{(1,1)}(f \circ \zq)(0,0) \\
            &= f_{1,2;1,1}(\st^{1,2}\zq(0,0)) \\
            &= f_{1,2;1,1}(w)) \\
            &= \langle \xd f_{2;1}, w\rangle \\
            &= \langle \xd_{T(1)}\xd f_{1;0}, w\rangle
                                                \tag \label{Fplb151}\endalign$$
    and
        $$\align
    \langle \xa\xd f_{1;1}, w\rangle &= \xD\langle \xd f_{1;1},
\zd\dot\zx\rangle(0) \\
            &= \frac{\xd}{\xd t}\langle \xd f_{1;1},
\zd\dot\zx(t)\rangle_{|t=0} \\
            &= \frac{\xd}{\xd t}\langle \xd f_{1;1},
\st^{1,1}\zq(\cdot,t)(0)\rangle_{|t=0} \\
            &= \frac{\xd}{\xd t}\langle f_{1,1;1,1},
\st^{1,1}\zq(\cdot,t)(0)\rangle_{|t=0} \\
            &= \frac{\xd}{\xd t}\left(\xD^{(1,1)}(f \circ
\zq)(0,t)\right)_{|t=0} \\
            &= \xD^{(1,2)}(f \circ \zq)(0,0) \\
            &= f_{1,2;1,2}(\st^{1,2}\zq(0,0)) \\
            &= f_{1,2;1,2}(w)) \\
            &= \langle \xd f_{2;2}, w\rangle \\
            &= \langle \xd_{T(1)}\xd f_{1;1}, w\rangle.
                                                \tag \label{Fplb152}\endalign$$
    Equalities $\xd_{T(1)} f_{1;0} = \xa f_{1;0}$, $\xd_{T(1)} f_{1;1} =
\xa f_{1;1}$, $\xd_{T(1)} \xd f_{1;0} = \xa \xd f_{1;0}$, and $\xd_{T(1)}
\xd f_{1;1} = \xa \xd f_{1;1}$ for each function $f$ imply the equality
$\xd_{T(1)} = \xa$.
        \endproof

        \claim \c{p}{Proposition}{}
\label{Cplb7}
    The relation
        $$\xi_{F(2;1)}\xd_{T(1)} - \xd_{T(1)}\xi_{F(1;1)} =
\zs_2{}^1{}_M\xi_{F(1;0)}
                                                        \tag \label{Fplb153}$$
    holds.
        \endclaim
        \proof
    We show that the operator $\xi_{F(2;1)}\xd_{T(1)} -
\xd_{T(1)}\xi_{F(1;1)}$ is a derivation of type $\xi_\*$ and degree 0
relative to $\zs_2{}^1{}_M$.  For each function $g$ on $\sT M$ we have
        $$(\xi_{F(2;1)}\xd_{T(1)} - \xd_{T(1)}\xi_{F(1;1)})g = 0.
                                                        \tag \label{Fplb154}$$
    For any two forms $\zm$ and $\zn$ on $\sT M$ we have
        $$\align
    (\xi_{F(2;1)}\xd_{T(1)} - \xd_{T(1)}\xi_{F(1;1)})(\zm \wedge \zn) &=
\xi_{F(2;1)}(\xd_{T(1)}\zm \wedge \zs_2{}^1{}_M\zn + \zs_2{}^1{}_M\zm
\wedge \xd_{T(1)}\zn) \\
            &\hskip7mm - \xd_{T(1)}(\xi_{F(1;1)}\zm \wedge \zn + \zm
\wedge \xi_{F(1;1)}\zn) \\
            &= \xi_{F(2;1)}\xd_{T(1)}\zm \wedge \zs_2{}^1{}_M\zn +
\xd_{T(1)}\zm \wedge \xi_{F(2;1)}\zs_2{}^1{}_M\zn \\
            &\hskip7mm + \xi_{F(2;1)}\zs_2{}^1{}_M\zm \wedge \xd_{T(1)}\zn
+ \zs_2{}^1{}_M\zm \wedge \xi_{F(2;1)}\xd_{T(1)}\zn \\
            &\hskip7mm - \xd_{T(1)}\xi_{F(1;1)}\zm \wedge \zs_2{}^1{}_M\zn
- \zs_2{}^1{}_M\xi_{F(1;1)}\zm \wedge \xd_{T(1)}\zn \\
            &\hskip7mm - \xd_{T(1)}\zm \wedge \zs_2{}^1{}_M\xi_{F(1;1)}\zn
- \zs_2{}^1{}_M\zm \wedge \xd_{T(1)}\xi_{F(1;1)}\zn \\
            &= \xi_{F(2;1)}\xd_{T(1)}\zm \wedge \zs_2{}^1{}_M\zn +
\zs_2{}^1{}_M\zm \wedge \xi_{F(2;1)}\xd_{T(1)}\zn \\
            &\hskip7mm - \xd_{T(1)}\xi_{F(1;1)}\zm \wedge \zs_2{}^1{}_M\zn
- \zs_2{}^1{}_M\zm \wedge \xd_{T(1)}\xi_{F(1;1)}\zn\\
            &= (\xi_{F(2;1)}\xd_{T(1)} - \xd_{T(1)}\xi_{F(1;1)})\zm \wedge
\zs_2{}^1{}_M\zn \\
            &\hskip7mm + \zs_2{}^1{}_M\zm \wedge (\xi_{F(2;1)}\xd_{T(1)} -
\xd_{T(1)}\xi_{F(1;1)})\zn.
                                                \tag \label{Fplb155}\endalign$$
    This proves that the operator under consideration is a derivation of
the stated type.  The equalities
        $$(\xi_{F(2;1)}\xd_{T(1)} - \xd_{T(1)}\xi_{F(1;1)})\xd f_{1;0} =
\xi_{F(2;1)}\xd f_{2;1} = \xd f_{2;0} = \zs_2{}^1{}_M\xi_{F(1;0)}\xd
f_{1;0}
                                                        \tag \label{Fplb156}$$
    and
        $$\align
    (\xi_{F(2;1)}\xd_{T(1)} - \xd_{T(1)}\xi_{F(1;1)})\xd f_{1;1} &=
\xi_{F(2;1)}\xd f_{2;2} - \xd_{T(1)}\xd f_{1;0} \\
            &= 2\xd f_{2;1} - \xd f_{2;1} \\
            &= \zs_2{}^1{}_M\xi_{F(1;0)} \xd f_{1;1}
                                                \tag \label{Fplb157}\endalign$$
    complete the proof.
        \endproof

        \sscx{The Lagrange differential.}

    We define a linear operator $\bi E \colon \zW_1(M) \rightarrow
\zW_2(M)$ by the formula
        $$\bi E = \zs_2{}^1{}_M - \xd_{T(1)}\xi_{F(1;1)}.
                                                        \tag \label{Fplb158}$$

        \claim \c{p}{Proposition}{}
\label{Cplb8}
    For each 1-form $\zm$ on $\sT M$ the 1-form $\bi E\zm$ on $\sT^2 M$ is
vertical with respect to the projection $\zt_2{}_M \colon \sT^2 M
\rightarrow M$.
        \endclaim
        \proof
    Verticality means that $\langle \bi E\zm, w\rangle = 0$ for each $w
\in \ker(\sT\zt_2{}_M)$.  Verticality is established by showing that
$\xi_{F(2;1)}\bi E\zm = 0$ since $\ker(\sT\zt_2{}_M) = \im(F(2;1))$ and
$\langle \xi_{F(2;1)}\bi E\zm, v\rangle = \langle \bi E\zm,
F(2;1)(v)\rangle$. The equality
        $$\align
    \xi_{F(2;1)}\bi E\zm &= \xi_{F(2;1)}(\zs_2{}^1{}_M -
\xd_{T(1)}\xi_{F(1;1)})\zm\\
        &= (\zs_2{}^1{}_M\xi_{F(1;1)} - \xd_{T(1)}\xi_{F(1;1)}\xi_{F(1;1)}
- \zs_2{}^1{}_M\xi_{F(1;1)})\zm = 0
                                                \tag \label{Fplb159}\endalign$$
    follows from $\xi_{F(1;1)}\xi_{F(1;1)}\zm = \xi_{F(1;2)}\zm = 0$.  We
have used formulae \Ref{Fplb92} and \Ref{Fplb153} and the commutativity of
the diagram \Ref{Fplb116}.
        \endproof

    The operator $\bi P = \xi_{F(1;1)} \colon \zW_1 \rightarrow \zW_1$
appears in the decomposition $\zs_2{}^1{}_M = \bi E + \xd_{T(1)}\bi P$
used in the calculus of variations.  The decomposition $\zs_2{}^1{}_M\zm =
\bi E\zm + \xd_{T(1)}\bi P\zm$ for a 1-form $\zm$ on $\sT M$ is usually
obtained by using local coordinates and integrating by parts.  For each
1-form $\zm$ on $\sT M$ the 1-form $\bi P\zm$ is vertical with respect to
the tangent projection $\zt_M \colon \sT M \rightarrow M$. This property
follows from $\xi_{F(1;1)}\bi P\zm = \xi_{F(1;1)}\xi_{F(1;1)}\zm =
\xi_{F(1;2)}\zm = 0$.

    Let $L$ be a function on $\sT M$.  Verticality of the form $\bi E\xd
L$ makes it possible to construct a mapping $\cE L \colon \sT^2 M
\rightarrow \sT^\* M$ such that $\zp_M \circ \cE L = \zt_2{}_M$.  This
mapping is characterized by $\langle \bi E\xd L, w\rangle = -\langle \cE
L(\zt_{\sT^2M}(w)), \sT\zt_2{}_M(w)\rangle$ for each $w \in \sT\sT^2 M$.
Verticality of the form $\bi P\xd L$ implies the existence of a mapping
$\cP L \colon \sT M \rightarrow \sT^\*M$ such that $\zp_M \circ \cP L =
\zt_M$.  This mapping is characterized by $\langle \bi P\xd L, w\rangle =
\langle \cP L(\zt_{\sT M}(w)), \sT\zt_M(w)\rangle$ for each $w \in \sT\sT
M$.

    The equation $\cE L \circ \st^2\zg = 0$ is a second order differential
equation for a curve $\zg \colon I \rightarrow M$ known as the {\it
Euler-Lagrange equation} derived from the {\it Lagrangian} $L \colon \sT M
\rightarrow \R$.  The mapping $\cP L$ is called the {\it Legendre mapping}.

        \sscx{The tangent of a vector fibration and its dual.}
    Let $\ze \colon E \rightarrow M$ be a differential fibration.  Local
triviality implies the existence of {\it adapted charts}.  An adapted
chart $(x^\zk,e^i) \colon E \rightarrow \R^{m+k}$ associated with a chart
$(x^\zk) \colon M \rightarrow \R^m$ is characterized by the equality
$x^\zk \circ \ze = x^\zk$.  Note that the symbol $x^\zk$ is used to denote
a coordinate of $M$ and also of $E$.  The chart $(x^\zk,\dot x^\zl) \colon
\sT M \rightarrow \R^{2m}$ is an adapted chart for the tangent fibration
$\zt_M$.  There are two fibrations $\zt_E \colon \sT E \rightarrow E$ and
$\sT\ze \colon \sT E \rightarrow \sT M$ for the tangent bundle $\sT E$.
The tangent chart $(x^\zk,e^i,\dot x^\zl,\dot e^j)$ for $\sT E$ induced by
an adapted chart $(x^\zk,e^i)$ is adapted to both fibrations since
$(x^\zk,e^i) \circ \zt_E = (x^\zk,e^i)$ and $(x^\zk,\dot x^\zl) \circ
\sT\ze = (x^\zk,\dot x^\zl)$.

    Let $\ze \colon E \rightarrow M$ and $\zf \colon F \rightarrow M$ be
differential fibrations.  The equalizer
        $$F \fpr{(\zf,\ze)} E = \{(f,e) \in F \times E ;\; \zf(f) =
\ze(e)\}
                                                        \tag \label{Fplb160}$$
    of the projections $\zf$ and $\ze$ is called the {\it fibre product}
of $F$ and $E$.  A curve $(\zs,\zr) \colon \R \rightarrow F
\fpr{(\zf,\ze)} E$ consists of two curves $\zr \colon \R \rightarrow E$
and $\zs \colon \R \rightarrow F$ such that $\ze \circ \zr = \zf \circ
\zs$.  The mapping
        $$\zq \colon \sT(F \fpr{(\zf,\ze)} E) \rightarrow \sT F
\fpr{(\sT\zf,\sT\ze)} \sT E \colon \st(\zs,\zr)(0) \mapsto
(\st\zs(0),\st\zr(0))
                                                        \tag \label{Fplb161}$$
    is obviously injective.  Let $(x^\zk,e^i,\dot x^\zl,\dot e^j)$ and
$(x^\zk,f^A,\dot x^\zl,\dot f^B)$ be tangent charts for $\sT E$ and $\sT
F$ respectively induced by adapted charts $(x^\zk,e^i)$ and
$(x^\zk,f^A)$.  If $(w,v) \in \sT F \fpr{(\sT\zf,\sT\ze)} \sT E$, then
$(x^\zk,\dot x^\zl)(w) = (x^\zk,\dot x^\zl)(v)$ since $\sT\zf(w) =
\sT\ze(v)$. Curves $\zr \colon \R \rightarrow E$ and $\zs \colon \R
\rightarrow F$ characterized locally by
        $$(x^\zk,e^i) \circ \zr \colon \R \rightarrow \R^{m+k} \colon s
\mapsto (x^\zk(v) + s\dot x^\zk(v),e^i(v) + s\dot e^i(v))
                                                        \tag \label{Fplb162}$$
    and
        $$(x^\zk,f^A) \circ \zr \colon \R \rightarrow \R^{m+l} \colon s
\mapsto (x^\zk(w) + s\dot x^\zk(w),f^A(w) + s\dot f^A(w))
                                                        \tag \label{Fplb163}$$
    define a curve $(\zs,\zr) \colon \R \rightarrow F \fpr{(\zf,\ze)} E$
such that $\zq(\st(\zs,\zr)) = (w,v)$.  It follows that $\zq$ is
surjective.  We will identify the space $\sT(F \fpr{(\zf,\ze)} E)$ with
$\sT F \fpr{(\sT\zf,\sT\ze)} \sT E$.  The diagram
    \vskip1mm
        $$\xymatrix@R+4mm @C+10mm{{\sT F \fpr{(\sT\zf,\sT\ze)} \sT E}
\ar[d]_*{(\zt_F,\zt_E)} \\ F \fpr{(\zf,\ze)} E}
                                                        \tag \label{Fplb164}$$
    \vskip2mm
    \noindent with
        $$(\zt_F,\zt_E) \colon \sT F \fpr{(\sT\zf,\sT\ze)} \sT E
\rightarrow F \fpr{(\zf,\ze)} E \colon (w,v) \mapsto (\zt_F(w),\zt_E(v))
                                                        \tag \label{Fplb165}$$
    is a vector fibration.  If $(w_1,v_1)$ and $(w_2,v_2)$ are elements of
$\sT F \fpr{(\sT\zf,\sT\ze)} \sT E$ such that $\zt_F(w_1) = \zt_F(w_2)$
and $\zt_E(v_1) = \zt_E(v_2)$, then $(w_1,v_1) + (w_2,v_2) = (w_1 +
w_2,v_1 + v_2)$.  If $(w,v) \in \sT F \fpr{(\sT\zf,\sT\ze)} \sT E$ and $k
\in \R$, then $k(w,v) = (kw,kv)$.  The diagram
    \vskip1mm
        $$\xymatrix@R+6mm @C+10mm{{\sT(F \fpr{(\zf,\ze)} E)}
\ar[d]_*{\zt_{F \fpr{(\zf,\ze)} E}} \ar[r]^*{\zq} &
            \sT F \fpr{(\sT\zf,\sT\ze)} \sT E \ar[d]_*{(\zt_F,\zt_E)}\\
            F \fpr{(\zf,\ze)} E \ar@{=}[r] & F \fpr{(\zf,\ze)} E}
                                                        \tag \label{Fplb166}$$
    \vskip2mm
    \noindent is an isomorphism of vector fibrations.

    Let
    \vskip1mm
        $$\xymatrix@R+0mm @C+10mm{{E} \ar[d]_*{\ze} \\ M}
                                                        \tag \label{Fplb167}$$
    \vskip2mm
    \noindent be a vector fibration with operations
        $$+ \colon E \fpr{(\ze,\ze)} E \rightarrow E
                                                        \tag \label{Fplb168}$$
    and
        $$\cdot\, \colon \R \times E \rightarrow E.
                                                        \tag \label{Fplb169}$$
    Let
        $$O_\ze \colon M \rightarrow E
                                                        \tag \label{Fplb170}$$
    be the zero section.

    The tangent fibration
    \vskip1mm
        $$\xymatrix@R+0mm @C+10mm{{\sT E} \ar[d]_*{\zt_E} \\ E}
                                                        \tag \label{Fplb171}$$
    \vskip2mm
    \noindent is a vector fibration with operations
        $$+ \colon \sT E \fpr{(\zt_E,\zt_E)} \sT E \rightarrow \sT E
                                                        \tag \label{Fplb172}$$
    and
        $$\cdot\, \colon \R \times \sT E \rightarrow \sT E
                                                        \tag \label{Fplb173}$$
    and the zero section
        $$O_{\zt_E} \colon E \rightarrow \sT E.
                                                        \tag \label{Fplb174}$$

    The diagram
    \vskip1mm
        $$\xymatrix@R+0mm @C+10mm{{\sT E} \ar[d]_*{\sT\ze} \\ \sT M}
                                                        \tag \label{Fplb175}$$
    \vskip2mm
    \noindent is again a vector fibration with operations
        $$+^\ssT \colon \sT E \fpr{(\sT\ze,\sT\ze)} \sT E \rightarrow \sT E
                                                        \tag \label{Fplb176}$$
    and
        $$\cdot^\ssT\, \colon \R \times \sT E \rightarrow \sT E
                                                        \tag \label{Fplb177}$$
    and the zero section
        $$O_{\sT\ze} \colon \sT M \rightarrow \sT E.
                                                        \tag \label{Fplb178}$$
    The operation $+^\ssT$ is obtained from the tangent mapping
        $$\sT+ \colon \sT(E \fpr{(\ze,\ze)} E) \rightarrow \sT E
                                                        \tag \label{Fplb179}$$
    by identifying the space $\sT(E \fpr{(\ze,\ze)} E)$ with $\sT E
\fpr{(\sT\ze,\sT\ze)} \sT E$.  The operation $\cdot ^\ssT$ is constructed
from the tangent mapping
        $$\sT\cdot\, \colon \sT(\R \times E) \rightarrow \sT E.
                                                        \tag \label{Fplb180}$$
    The space $\sT(\R \times E)$ is identified with $\sT\R \times \sT E =
\R^2 \times \sT E$ and the operation $\cdot ^\ssT$ is defined as the
mapping
        $$\cdot^\ssT\, \colon \R \times \sT E \rightarrow \sT E \colon
(k,v) \mapsto (k,0)\, \sT\!\cdot \,v.
                                                        \tag \label{Fplb181}$$

    In the diagrams
    \vskip1mm
        $$\xymatrix@R+3mm @C+10mm{{\sT E} \ar[d]_*{\zt_E} \ar[r]^*{\sT\ze}
&
            \sT M \ar[d]_*{\zt_M} \\
            E \ar[r]^*{\ze} & M} \hskip10mm
    \xymatrix@R+3mm @C+10mm{{\sT E} \ar[r]^*{\zt_E} \ar[d]_*{\sT\ze} &
            E \ar[d]_*{\ze} \\
            \sT M \ar[r]^*{\zt_M} & M}
                                                        \tag \label{Fplb182}$$
    \vskip2mm
    \noindent vertical arrows are vector fibrations and horizontal arrows
define vector fibration morphisms.  The space $\sT E$ with its two vector
bundle structures forms a {\it double vector bundle}.

    Let
    \vskip1mm
        $$\xymatrix@R+0mm @C+10mm{{E} \ar[d]_*{\ze} \\ M} \hskip35mm
            \xymatrix@R+0mm @C+10mm{{F} \ar[d]_*{\zf} \\ M}
                                                        \tag \label{Fplb183}$$
    \vskip2mm
    \noindent represent a vector fibration $\ze$ and its dual vector
fibration $\zf$. Let
        $$\langle \,\;,\;\rangle \colon F \fpr{(\zf,\ze)} E \rightarrow \R
                                                        \tag \label{Fplb184}$$
    be the canonical pairing.  We have the double vector bundle structures
for $\sT E$ and $\sT F$.  The tangent fibrations
    \vskip1mm
        $$\xymatrix@R+3mm @C+10mm{{\sT E} \ar[d]_*{\sT\ze} \\ \sT M}
\hskip30mm
            \xymatrix@R+3mm @C+10mm{{\sT F} \ar[d]_*{\sT\zf} \\ \sT M}
                                                        \tag \label{Fplb185}$$
    \vskip2mm
    \noindent are a dual pair of vector fibrations.  The {\it tangent
pairing}
        $$\langle \,\;,\;\rangle^\ssT \colon \sT F \fpr{(\sT\zf,\sT\ze)}
\sT E \rightarrow \R
                                                        \tag \label{Fplb186}$$
    is constructed from the tangent mapping
        $$\sT\langle \,\;,\;\rangle \colon \sT(F \fpr{(\zf,\ze)} E)
\rightarrow \sT\R
                                                        \tag \label{Fplb187}$$
    by identifying the space $\sT(F \fpr{(\zf,\ze)} E)$ with $\sT F
\fpr{(\sT\zf,\sT\ze)} \sT E$ and composing the tangent mapping with the
second projection of $\sT\R = \R^2$ onto $\R$. If $(w,v)$ is an element of
$\sT F \fpr{(\sT\zf,\sT\ze)} \sT E$ and $(\zs,\zr)$ is a curve in $F
\fpr{(\zf,\ze)} E$ such that $(w,v) = (\st\zs(0),\st\zr(0))$, then
        $$\langle w, v\rangle^\ssT = \langle \st\zs(0),
\st\zr(0)\rangle^\ssT = \xD\langle \zs, \zr\rangle(0).
                                                        \tag \label{Fplb188}$$
    If $(\zs,\zr)$ is a curve in $F \fpr{(\zf,\ze)} E$, then
        $$\langle \st\zs, \st\zr\rangle^\ssT = \xD\langle \zs, \zr\rangle.
                                                        \tag \label{Fplb189}$$

    Linearity of the tangent mapping \Ref{Fplb187} implies linearity of
the tangent pairing.  If $(w_1,v_1)$ and $(w_2,v_2)$ are elements of $\sT
F \fpr{(\sT\zf,\sT\ze)} \sT E$ such that $\zt_F(w_1) = \zt_F(w_2)$ and
$\zt_E(v_1) = \zt_E(v_2)$, then
        $$\langle w_1 + w_2, v_1 + v_2 \rangle^\ssT = \langle w_1, v_1
\rangle^\ssT + \langle w_2, v_2 \rangle^\ssT.
                                                        \tag \label{Fplb190}$$
    If $(w,v) \in \sT F \fpr{(\sT\zf,\sT\ze)} \sT E$ and $k \in \R$, then
        $$\langle kw, kv\rangle^\ssT = k\langle w, v\rangle^\ssT.
                                                        \tag \label{Fplb191}$$

    There are mappings
        $$\zm_\ze \colon E \fpr{(\ze,\ze)} E \rightarrow \sT E \colon
(e,e') \mapsto \st\zg(0)
                                                        \tag \label{Fplb192}$$
    with
        $$\zg \colon \R \rightarrow E \colon s \mapsto e + se'
                                                        \tag \label{Fplb193}$$
    and
        $$\zc_\ze \colon E \fpr{(\ze,\ze\circ\zt_E)} \sT E \rightarrow \sT
E \colon (e,\dot e) \mapsto \dot e - \zm_\ze(\zt_E(\dot e),e).
                                                        \tag \label{Fplb194}$$
    The image of $\zm_\ze$ is the subbundle
        $$\sV E = \{v \in \sT E;\; \st\ze(v) = 0\}
                                                        \tag \label{Fplb195}$$
     of vertical vectors.

    Let $e$ be an element of a vector bundle $E$ and let $f$ and $f'$ be
elements of the dual bundle $F$ such that $\zf(f) = \zf(f') = \ze(e)$.
The vector $\zm_\zf(f,f') \in \sT F$ is the tangent vector of the curve
$\zs \colon \R \rightarrow F \colon s \mapsto f + sf'$ and $O_{\zt_E}(e)
\in \sT E$ is the tangent vector of the constant curve $\zr \colon \R
\rightarrow E \colon s \mapsto e$.  We have $\xD\langle \zs, \zr\rangle(0)
= \langle f', e\rangle$ since $\langle \zs, \zr\rangle(s) = \langle f,
e\rangle + s\langle f', e\rangle$.  It follows that
        $$\langle \zm_\zf(f,f'), O_{\zt_E}(e)\rangle = \langle f',
e\rangle.
                                                        \tag \label{Fplb196}$$

        \sscx{The structure of the tangent bundle of the cotangent bundle.}

    As was stated earlier the fibrations
    \vskip1mm
        $$\xymatrix@R+6mm{{\sT^\* M} \ar[d]_*{\zp_M} \\ M}
                \hskip35mm
        \xymatrix@R+6mm{{\sT M} \ar[d]_*{\zt_M} \\ M}
                                                        \tag \label{Fplb197}$$
    \vskip2mm
    \noindent for each differential manifold $M$ are a dual pair of vector
fibrations with the canonical pairing
        $$\langle \,\;,\; \rangle \colon \sT^\* M \fpr{(\zp_M,\zt_M)} \sT
M \rightarrow \R.
                                                        \tag \label{Fplb198}$$

    The fibrations
    \vskip1mm
        $$\xymatrix@R+6mm{{\sT^\*\sT M} \ar[d]_*{\zp_{\sT M}} \\ \sT M}
                \hskip35mm
        \xymatrix@R+6mm{{\sT\sT M} \ar[d]_*{\zt_{\sT M}} \\ \sT M}
                                                        \tag \label{Fplb199}$$
    \vskip1mm \noindent are again a dual pair with the pairing
        $$\langle \,\;,\; \rangle \colon \sT^\*\sT M \fpr{(\zp_{\sT
M},\zt_{\sT M})} \sT\sT M \rightarrow \R.
                                                        \tag \label{Fplb200}$$

    By applying the tangent functor to fibrations \Ref{Fplb197} we obtain
a dual pair of vector fibrations
    \vskip1mm
        $$\xymatrix@R+6mm{{\sT\sT^\* M} \ar[d]_*{\sT\zp_M} \\ \sT M}
                \hskip35mm
            \xymatrix@R+6mm{{\sT\sT M} \ar[d]_*{\sT\zt_M} \\ \sT M}
                                                        \tag \label{Fplb201}$$
    \vskip1mm \noindent with the pairing
        $$\langle \,\;,\; \rangle^\ssT \colon \sT\sT^\* M
\fpr{(\sT\zp_M,\sT\zt_M)} \sT\sT M \rightarrow \R.
                                                        \tag \label{Fplb202}$$
    If $\zd\zx \colon \R \rightarrow \sT M$ and $\zh \colon \R \rightarrow
\sT^\* M$ are curves such that $\zp_M \circ \zh = \zt_M \circ \zd\zx$ and
if $w = \st\zd\zx(0)$ and $z = \st\zh(0)$, then
        $$\langle z, w\rangle^\ssT = \xD\langle \zh, \zd\zx\rangle(0).
                                                        \tag \label{Fplb203}$$

    We have two vector bundle structures for the manifold $\sT\sT M$ and
the diagram
    \vskip1mm
        $$\xymatrix@R+6mm @C+15mm{{\sT\sT M} \ar[d]_*{\sT\zt_M} &
            \sT\sT M \ar[d]_*{\zt_{\sT M}} \ar[l]_*{\zk_M} \\
            \sT M \ar@{=}[r] & \sT M}
                                                        \tag \label{Fplb204}$$
    \vskip2mm
    \noindent represents an isomorphism of vector fibrations.  This is the
diagram \Ref{Fplb83} with $k' = k = 1$ and $k'' = 0$.

    Pairings \Ref{Fplb200} and \Ref{Fplb202} permit the introduction of
the vector fibration isomorphism
    \vskip1mm
        $$\xymatrix@R+8mm @C+15mm{{\sT\sT^\* M} \ar[d]_*{\sT\zp_M}
\ar[r]^*{\za_M} &
            \sT^\*\sT M \ar[d]_*{\zp_{\sT M}} \\
            \sT M \ar@{=}[r] & \sT M}
                                                        \tag \label{Fplb205}$$
    \vskip2mm
    \noindent dual to the vector fibration isomorphism \Ref{Fplb204}.  If
$w \in \sT\sT M$, $z \in \sT\sT^\*M$ and $\zt_{\sT M}(w) = \sT\zp_M(z)$,
then
        $$\langle \za_M(z), w\rangle = \langle z, \zk_M(w)\rangle^\ssT.
                                                        \tag \label{Fplb206}$$

    Let $\xd_T$ and $\xi_T$ denote the derivations $\xd_{T(0)} \colon
\zW(\sT^\*M) \rightarrow \zW_1(\sT^\*M)$ and $\xi_{T(0)} \colon
\zW(\sT^\*M) \rightarrow \zW_1(\sT^\*M)$ respectively. The manifold
$\sT\sT^\*M$ with the 2-form $\xd_T\zw_M = \xd\xd_T\zy_M$ form a
symplectic manifold $(\sT\sT^\*M,\xd_T\zw_M)$.  We construct two natural
special symplectic structures for this symplectic manifold.

    The following proposition implies that the diagrams
    \vskip1mm
        $$\xymatrix@R+8mm @C+10mm{{(\sT\sT^\*M,\xd_T\zy_M)}
\ar[d]_*{\sT\zp{}_M} \\ \sT M}
            \hskip20mm \xymatrix@R+8mm @C+15mm{{\sT\sT^\*M}
\ar[d]_*{\sT\zp{}_M} \ar[r]^*{\za{}_M} &
            \sT^\*\sT M \ar[d]_*{\zp{}_{\sT M}} \\
            \sT M \ar@{=}[r] & \sT M}
                                                        \tag \label{Fplb207}$$
    \vskip2mm
    \noindent constitute a special symplectic structure for the symplectic
manifold $(\sT\sT^\*M,\xd_T\zw_M)$.
        \claim \c{p}{Proposition}{}
\label{Cplb9}
    If $z \in \sT\sT^\*M$, $w \in \sT\sT\sT^\*M$, and $v \in \sT\sT M$
satisfy relations $\zt_{\sT\sT^\*M}(w) = z$ and $\sT\sT\zp_M(w) = v$, then
        $$\langle \za_M(z), v\rangle = \langle \xd_T\zy_M, w\rangle.
                                                        \tag \label{Fplb208}$$
        \endclaim
        \proof
    Let $\zq \colon \R^2 \rightarrow \sT^\*M$ be a representative of $w$.
The mapping $\zf = \zp_M \circ \zq$ is a representative of $v$ and $\zz =
\zq(0,\cdot)$ is a representative of $z$. The vector $\zk_M(v) =
\st^{1,1}\widetilde\zf(0,0)$ is represented by the curve
        $$\zh \colon \R \rightarrow \sT M \colon t \mapsto
\st\widetilde\zf(t,\cdot)(0) = \st\zf(\cdot,t)(0).
                                                        \tag \label{Fplb209}$$
    The mapping
        $$\zd\zz \colon \R \rightarrow \sT\sT^\* M \colon t \mapsto
\st\zq(\cdot,t)(0)
                                                        \tag \label{Fplb210}$$
    appears in the formula
        $$\langle \xd_T\zy_M, w\rangle = \xD\langle \zy_M, \zd\zz\rangle(0)
                                                        \tag \label{Fplb211}$$
    derived in Proposition \Ref{Cplb5}.  Relations $\zh = \sT\zp_M \circ
\zd\zz$ and $\zz = \zt_{\sT^\*M} \circ \zd\zz$ follow from
        $$(\sT\zp_M \circ \zd\zz)(t) = \sT\zp_M(\st\zq(\cdot,t)(0)) =
\st(\zp_M \circ \zq)(\cdot,t)(0) = \st\zf(\cdot,t)(0)
                                                        \tag \label{Fplb212}$$
    and
        $$(\zt_{\sT^\*M} \circ \zd\zz)(t) =
\zt_{\sT^\*M}(\st\zq(\cdot,t)(0)) = \zq(0,t).
                                                        \tag \label{Fplb213}$$
    The calculation
        $$\align
    \langle \za_M(z), v\rangle &= \langle z, \zk_M(v)\rangle^\ssT \\
            &= \langle \st\zz(0), \st\zh(0)\rangle^\ssT \\
            &= \xD\langle \zz, \zh\rangle(0) \\
            &= \xD\langle \zy_M, \zd\zz\rangle(0) \\
            &= \langle \xd_T\zy_M, w\rangle
                                                \tag \label{Fplb214}\endalign$$
    completes the proof.
        \endproof

    The mapping $\zb_{(\sT^\*M,\zw_M)} \colon \sT\sT^\*M \rightarrow
\sT^\*\sT^\*M$ characterized by
        $$\langle \zb_{(\sT^\*M,\zw_M)}(u),v \rangle = \langle \zw_M, u
\wedge v \rangle
                                                        \tag \label{Fplb215}$$
    with $u \in \sT\sT^\*M$ and $v \in \sT\sT^\*M$ such that
$\zt_{\sT^\*M}(u) = \zt_{\sT^\*M}(v)$ establishes an isomorphism
    \vskip1mm
        $$\xymatrix@R+8mm @C+20mm{{\sT\sT^\*M} \ar[d]_*{\zt_{\sT^\*M}}
\ar[r]^*{\zb_{(\sT^\*M,\zw_M)}} &
            \sT^\*\sT^\*M \ar[d]_*{\zp_{\sT^\*M}} \\
            \sT^\*M \ar@{=}[r] & \sT^\*M}
                                                        \tag \label{Fplb216}$$
    \vskip2mm
    \noindent of vector fibrations.

        \claim \c{p}{Proposition}{}
\label{Cplb10}
    A special symplectic structure for the symplectic manifold
$(\sT\sT^\*M,\xd_T\zw_M)$ is introduced by the diagrams
    \vskip1mm
        $$\xymatrix@R+8mm @C+10mm{{(\sT\sT^\*M,\xi_T\zw_M)}
\ar[d]_*{\zt{}_{\sT^\*M}} \\ \sT^\*M}
            \hskip15mm \xymatrix@R+8mm @C+20mm{{\sT\sT^\*M}
\ar[d]_*{\zt{}_{\sT^\*M}} \ar[r]^*{\zb_{(\sT^\*M,\zw_M)}} &
            \sT^\*\sT^\*M \ar[d]_*{\zp{}_{\sT^\*M}} \\
            \sT^\*M \ar@{=}[r] & \sT^\*M}.
                                                        \tag \label{Fplb217}$$
    \vskip2mm
        \endclaim
        \proof
    We have    $\xd_T\zw_M = \xd\xi_T\zw_M$ since $\xd\zw_M = 0$.  Further
        $$\align
    \langle \zb_{(\sT^\*M,\zw_M)}{}^\*\zy_{\sT^\*M}, w\rangle &= \langle
\zy_{\sT^\*M}, \sT\zb_{(\sT^\*M,\zw_M)}(w)\rangle \\
            &= \langle \zt_{\sT^\*\sT^\*M}(\sT\zb_{(\sT^\*M,\zw_M)}(w)),
\sT\zp_{\sT^\*M}(\sT\zb_{(\sT^\*M,\zw_M)}(w))\rangle \\
            &= \langle \zb_{(\sT^\*M,\zw_M)}(\zt_{\sT\sT^\*M}(w)),
\sT(\zp_{\sT^\*M} \circ \zb_{(\sT^\*M,\zw_M)})(w) \rangle \\
            &= \langle \zb_{(\sT^\*M,\zw_M)}(\zt_{\sT\sT^\*M}(w)),
\sT\zt_{\sT^\*M}(w) \rangle \\
            &= \langle \zw_M, \zt_{\sT\sT^\*M}(w) \wedge
\sT\zt_{\sT^\*M}(w) \rangle \\
            &= \langle \xi_T\zw_M, w\rangle
                                                \tag \label{Fplb218}\endalign$$
    for each $w \in \sT\sT\sT^\*M$.  Hence,
$\zb_{(\sT^\*M,\zw_M)}^\*\zy_{\sT^\*M} = \xi_T\zw_M$.
        \endproof

    The symplectic form $\xd_T\zw_M$ can be obtained as the pull back
$\za_M^\*\zw_{\sT M}$ of the symplectic form $\zw_{\sT M}$ on $\sT^\*\sT
M$.  The pull back $\zb_{(\sT^\*M,\zw_M)}^\* \zw_{\sT^\*M}$ is again the
symplectic form $\xd_T\zw_M$.

    Comparing the Hamiltonian special symplectic structure with the
Lagrangian structure we observe that $\xi_T\zw_M - \xd_T\zy_M =
-\xd\xi_T\zy_M$.  The function $G_M = -\xi_T\zy_M$ on $\sT\sT^\*M$ plays
the role of the function $G$ introduced in Section 2.4.

    An alternative analysis of the structure of $\sT\sT^\*M$ is offered in
a recent {\it Springer-Verlag} text [mr] in Section 6.8 on page 161.  We
are not in total agreement with this analysis.  In particular, we failed
to identify the second symplectic structure whose existence is claimed in
this publication.  We suspect that the claim may be false. \goodbreak

        \sect{Dynamics of autonomous systems.} \vskip3mm
        \ssca{Motions and variations.}
    Let $M$ be the {\it configuration manifold} of an autonomous
mechanical system. A {\it configuration} is a point $x \in M$ and a {\it
motion} of the system is a differentiable curve $\zx \colon I \rightarrow
M$ defined on an open interval $I \subset \R$.  The first and second
prolongations of a motion $\zx$ denoted by $\dot\zx$ and $\ddot\zx$
represent the {\it velocity} and the {\it acceleration} along the motion.

    An {\it infinitesimal variation} of a motion $\zx \colon I \rightarrow
M$ is a differentiable mapping $\zd\zx \colon I \rightarrow \sT M$ such
that $\zt_M \circ \zd\zx = \zx$. Mappings $\zd\dot\zx = \zk^{1,1}{}_M
\circ \st\zd\zx$ and $\zd\ddot\zx = \zk^{2,1}{}_M \circ \st^2\zd\zx$ are
the infinitesimal variations of the velocity $\dot\zx$ and the
acceleration $\ddot\zx$.  Relations $\zt{}_M \circ \zd\zx = \zx$,
$\zt{}_{\sT M} \circ \zd\dot\zx = \dot\zx$, and $\zt{}_{\sT^2 M} \circ
\zd\ddot\zx = \ddot\zx$ are satisfied.  The derivations of additional
relations
        $$\align
    \sT\zt{}_M \circ \zd\dot\zx &= \sT\zt{}_M \circ \zk{}_M \circ
\st\zd\zx \\
            &= \zt{}_{\sT M} \circ \st\zd\zx \\
            &= \zd\zx
                                                \tag \label{Fplb219}\endalign$$
    and
        $$\align
    \sT\zt^1{}_2{}_M \circ \zd\ddot\zx &= \sT\zt^1{}_2{}_M \circ
\zk^{2,1}{}_M \circ \st^2\zd\zx \\
            &= \zk{}_M \circ \zt^1{}_2{}_{\sT M} \circ \st^2\zd\zx \\
            &= \zk{}_M \circ \st\zd\zx \\
            &= \zd\dot\zx
                                                \tag \label{Fplb220}\endalign$$
    are based on the commutativity of the diagram \Ref{Fplb212} with $k' =
k = 1$ and $k'' = 0$ and the same diagram with $k = 1$, $k' = 2$, and $k''
= 1$.

        \sscx{Force-momentum trajectories.}
    The fibre product
        $$\sT^\*M \fpr{(\zp{}_M,\zp{}_M)} \sT^\*M
                                                        \tag \label{Fplb221}$$
     is the {\it force-momentum phase space} $Ph$ of the system.  A pair
$(f,p) \in Ph$ consists of the {\it external force} $f$ and the {\it
momentum} $p$ at $x = \zp_M(f) = \zp_M(p)$.  A {\it force-momentum
trajectory} of an autonomous system is a curve
        $$(\zz,\zh) \colon I \rightarrow Ph.
                                                        \tag \label{Fplb222}$$
    The two curves $\zz \colon I \rightarrow \sT^\* M$ and $\zh \colon I
\rightarrow \sT^\* M$ represent the external force and the momentum along
the motion $\zx = \zp{}_M \circ \zz = \zp{}_M \circ \zh$.

        \sscx{The variational principle of dynamics.}
    Let $L \colon \sT M \rightarrow \R$ be the Lagrangian of the
mechanical system. The {\it action} is a function $A$ which associates the
integral
        $$A(\zx,[a,b]) = \int_a^b L \circ \dot\zx
                                                        \tag \label{Fplb223}$$
    with a motion $\zx \colon I \rightarrow M$ and an interval $[a,b]
\subset I$. The {\it variation} of the action is the integral
        $$\zd A(\zx,[a,b]) = \int_a^b \langle \xd L, \zd\dot\zx \rangle
                                                        \tag \label{Fplb224}$$
    associated with an infinitesimal variation $\zd\zx \colon I
\rightarrow \sT M$.

    The {\it dynamics} of the system is a set $\cD$ of force-momentum
trajectories satisfying the following variational principle.  A trajectory
$(\zz,\zh) \colon I \rightarrow Ph$ is in $\cD$ if
        $$\zd A(\zx,[a,b]) = -\int_a^b \langle \zz, \zd\zx \rangle +
\langle \zh(b), \zd\zx(b) \rangle - \langle \zh(a), \zd\zx(a) \rangle
                                                        \tag \label{Fplb225}$$
    for each $[a,b] \subset I$ and each variation $\zd\zx$ such that
$\zt{}_M \circ \zd\zx = \zp{}_M \circ \zz = \zp{}_M \circ \zh$.

    The variation of the action can be converted to an equivalent
expression:
        $$\align
    \zd A(\zx,[a,b]) &= \int_a^b \langle \xd L, \zd\dot\zx \rangle \\
            &= \int_a^b \langle \xd L, \sT\zt^1{}_2{}_M \circ \zd\ddot\zx
\rangle \\
            &= \int_a^b \langle \zs_2{}^1{}_M\xd L, \zd\ddot\zx \rangle \\
            &= \int_a^b \langle \zs_2{}^1{}_M\xd L -
\xd_{T(1)}\xi_{F(1;1)}\xd L, \zd\ddot\zx \rangle + \int_a^b \langle
\xd_{T(1)}\xi_{F(1;1)}\xd L, \zd\ddot\zx \rangle \\
            &= - \int_a^b \langle \cE L \circ \ddot\zx, \zd\zx \rangle +
\int_a^b \xD\langle \cP L \circ \dot\zx, \zd\zx \rangle \\
            &= - \int_a^b \langle \cE L \circ \ddot\zx, \zd\zx \rangle +
\langle (\cP L \circ \dot\zx)(b), \zd\zx(b) \rangle - \langle (\cP L \circ
\dot\zx)(a), \zd\zx(a) \rangle.
                                                \tag \label{Fplb226}\endalign$$
    By using variations $\zd\zx$ with $\zd\zx(a) = 0$ and $\zd\zx(b) = 0$
we first derive from the variational principle the Euler-Lagrange equations
        $$\cE L \circ \ddot\zx = \zz
                                                        \tag \label{Fplb227}$$
    in $[a,b]$.  Equations
        $$(\cP L \circ \dot\zx)(a) = \zh(a)
                                                        \tag \label{Fplb228}$$
    and
        $$(\cP L \circ \dot\zx)(b) = \zh(b)
                                                        \tag \label{Fplb229}$$
    follow.  These equations are satisfied for each interval $[a,b]
\subset I$.  It follows that a force-momentum trajectory $(\zz,\zh)$ is in
$\cD$ if and only if equations
        $$\cE L \circ \ddot\zx = \zz
                                                        \tag \label{Fplb230}$$
    and
        $$\cP L \circ \dot\zx = \zh
                                                        \tag \label{Fplb231}$$
    are satisfied in $I$.

    Sets
        $$E = \left\{(f,a) \in \sT^\*M \fpr{(\zp{}_M,\zt{}_{2\,M})} \sT^2
M ;\; f = \cE L(a)\right\}
                                                        \tag \label{Fplb232}$$
    and
        $$P = \left\{(p,v) \in \sT^\*M \fpr{(\zp{}_M,\zt{}_M)} \sT M ;\; p
= \cP L(v)\right\}
                                                        \tag \label{Fplb233}$$
    are graphs of the Euler-Lagrange and the Legendre maps respectively.
The dynamics can be stated in terms of these sets treated as differential
equations. Equation \Ref{Fplb230} means that the mapping $(\zz,\zx)$ is a
solution curve of the differential equation $E$ and equation \Ref{Fplb231}
means that the mapping $(\zh,\zx)$ is a solution curve of the differential
equation $P$.  The relation $\zx = \zp{}_M \circ \zz = \zp{}_M \circ \zh$
is always imposed.  The Euler-Lagrange equation alone does not provide a
complete characterization of dynamics.  The equation \Ref{Fplb231} could
be called the {\it velocity-momentum} relation.  It is an essential part
of dynamics.  The set
        $$E_0 = \left\{a \in \sT^2 M ;\; 0 = \cE L(a)\right\}
                                                        \tag \label{Fplb234}$$
    is a version of the Euler-Lagrange equations without external forces.
Solution curves are motions of the system with zero external forces.

        \sscx{Lagrangian formulation of dynamics.}
    The version
        $$\int_a^b \langle \xd L, \zd\dot\zx \rangle = -\int_a^b \langle
\zz, \zd\zx \rangle + \int_a^b \xD\langle \zh, \zd\zx \rangle
                                                        \tag \label{Fplb235}$$
    of the variational principle is suitable for deriving the
infinitesimal limits. Infinitesimal limits are obtained by dividing both
sides of the equality by $b - a$ and passing to the limit of $b = a = t
\in I$.  The resulting equality
        $$\langle \xd L, \zd\dot\zx \rangle = - \langle \zz, \zd\zx
\rangle + \xD\langle \zh, \zd\zx \rangle
                                                        \tag \label{Fplb236}$$
    satisfied by the force-momentum trajectory $(\zz,\zh) \colon I
\rightarrow Ph$ for each variation $\zd\zx \colon I \rightarrow \sT\sT M$
such that $\zt{}_M \circ \zd\zx = \zp{}_M \circ \zz$ is a characterization
of the dynamics $\cD$ equivalent to the original variational principle.
The equality
        $$\langle \zz, \zd\zx \rangle = \langle \zm_{\zp_M} \circ
(\zh,\zz), O_{\zt{}_{\sT M}} \circ \zd\zx \rangle^\ssT
                                                        \tag \label{Fplb237}$$
    is derived from formula \Ref{Fplb196} and the equality
        $$\xD\langle \zh, \zd\zx \rangle = \langle \st\zh, \st\zd\zx
\rangle^\ssT = \langle \dot\zh, \zk{}_M \circ \zd\dot\zx \rangle^\ssT
                                                        \tag \label{Fplb238}$$
    is a version of formula \Ref{Fplb189}.  By combining the two
equalities we obtain the formula
        $$- \langle \zz, \zd\zx \rangle + \xD\langle \zh, \zd\zx \rangle =
- \langle \zm_{\zp_M} \circ (\zh,\zz), O_{\zt{}_{\sT M}} \circ \zd\zx
\rangle^\ssT + \langle \dot\zh, \zk{}_M \circ \zd\dot\zx \rangle^\ssT.
                                                        \tag \label{Fplb239}$$
    Relations
        $$\zt_{\sT^\*M} \circ \zm_{\zp_M} \circ (\zh,\zz) = \zt_{\sT^\*M}
\circ \dot\zh = \zh
                                                        \tag \label{Fplb240}$$
    and
        $$\zt_{\sT M} \circ O_{\zt{}_{\sT M}} \circ \zd\zx = \zt_{\sT M}
\circ \zk{}_M \circ \zd\dot\zx
                                                        \tag \label{Fplb241}$$
    permit the use of formula \Ref{Fplb190}.  The result is
        $$\align
    - \langle \zz, \zd\zx \rangle + \xD\langle \zh, \zd\zx \rangle &=
\langle \dot\zh - \zm_{\zp_M} \circ (\zh,\zz), \zk{}_M \circ \zd\dot\zx
\rangle^\ssT \\
        &= \langle \zc{}_M \circ (\zz,\dot\zh), \zk{}_M \circ \zd\dot\zx
\rangle^\ssT \\
        &= \langle \za{}_M \circ \zc{}_M \circ (\zz,\dot\zh), \zd\dot\zx
\rangle.
                                                \tag \label{Fplb242}\endalign$$
    We have obtained a characterization of the dynamics $\cD$ in terms of
a first order differential equation
        $$\zc{}_M \circ (\zz,\dot\zh) = \za_M{}^{-1} \circ \xd L \circ
\dot\zx
                                                        \tag \label{Fplb243}$$
    with $\zx = \zp{}_M \circ \zz = \zp{}_M \circ \zh$.  The codomain of
$(\zz,\dot\zh)$ is the fibre product $\sT^\*M \fpr{(\zp{}_M,\zp{}_M \circ
\zt{}_{\sT^\ast M})} \sT\sT^\*M$.

    Starting with the identity
        $$\int_a^b \langle \xd L, \zd\dot\zx \rangle = - \int_a^b \langle
\cE L \circ \ddot\zx, \zd\zx \rangle + \int_a^b \xD\langle \cP L \circ
\dot\zx, \zd\zx \rangle
                                                        \tag \label{Fplb244}$$
    instead of equation \Ref{Fplb236} and performing the operations
leading from \Ref{Fplb236} to \Ref{Fplb243} with $\zh$ and $\zz$ replaced
by $\cP L \circ \dot\zx$ and $\cE L \circ \ddot\zx$ respectively we obtain
the identity
        $$\zc{}_M \circ (\cE L \circ \ddot\zx,\st(\cP L \circ \dot\zx)) =
\za_M{}^{-1} \circ \xd L \circ \dot\zx.
                                                        \tag \label{Fplb245}$$
    A useful characterization
        $$\cP L = \zt{}_{\sT^\*M} \circ \za_M{}^{-1} \circ \xd L
                                                        \tag \label{Fplb246}$$
    of the Legendre mapping follows from this identity.

    Equation \Ref{Fplb243} means that the curve $(\zz,\zh)$ satisfies the
differential equation
        $$D = \left\{(f,w) \in \sT^\*M \fpr{(\zp{}_M,\zp{}_M \circ
\zt{}_{\sT^\ast M})} \sT\sT^\*M ;\; \zc_M(f,w) \in D_0 \right\},
                                                        \tag \label{Fplb247}$$
    where
        $$D_0 = \left\{w \in \sT\sT^\*M ;\; \za_M(w) = \xd
L(\zt_{\sT^\*M}(w))\right\}.
                                                        \tag \label{Fplb248}$$
    The set $D_0$ is a version of the Lagrange equation without external
forces. The image of the differential $\xd L \colon \sT M \rightarrow
\sT^\*\sT M$ is a Lagrangian submanifold of $(\sT^\*\sT M,\zw_{\sT M})$.
Consequently the set $D_0 = \za_M{}^{-1}(\im(\xd L))= \im(\za_M{}^{-1}
\circ \xd L)$ is a Lagrangian submanifold of $(\sT\sT^\*M,\xd_T\zw_M)$ and
the Lagrangian is its generating function relative to the Lagrangian
special symplectic structure
    \vskip2mm
        $$\xymatrix@R+3mm @C+10mm{{(\sT\sT^\*M,\xd_T\zy_M)}
\ar[d]_*{\sT\zp{}_M} \\ \sT M}
            \hskip20mm \xymatrix@R+3mm @C+13mm{{\sT\sT^\*M}
\ar[d]_*{\sT\zp{}_M} \ar[r]^*{\za{}_M} &
            \sT^\*\sT M \ar[d]_*{\zp{}_{\sT M}} \\
            \sT M \ar@{=}[r] & \sT M}
                                                        \tag \label{Fplb249}$$
    \vskip2mm
    \noindent for the symplectic manifold $(\sT\sT^\*M,\xd_T\zw_M)$.

        \sscx{Hamiltonian formulation of dynamics.}

    A Lagrangian is said to be {\it hyperregular} if the Legendre mapping
$\cP L = \zt{}_{\sT^\*M} \circ \za_M{}^{-1} \circ \xd L$ is a
diffeomorphism.  We will denote by $\zL$ the inverse of the Legendre
mapping for a hyperregular Lagrangian.

    For a hyperregular Lagrangian the set $D_0$ is the image $\im(Z)$ of
the mapping $Z = \za_M{}^{-1} \circ \xd L \circ \zL$.  This mapping is a
vector field on $\sT^\*M$ since $\zt{}_{\sT^\*M} \circ Z =\zt{}_{\sT^\*M}
\circ \za_M{}^{-1} \circ \xd L \circ \zL = 1_{\sT^\*M}$.  Let a function
$H \colon \sT^\*M \rightarrow \R$ be defined by $H(p) = \langle p, \zL(p)
\rangle - L(\zL(p))$.  We show that the function $-H$ is a generating
function of $D_0$ relative to the Hamiltonian special symplectic structure
    \vskip2mm
        $$\xymatrix@R+3mm @C+10mm{{(\sT\sT^\*M,\xi_T\zw_M)}
\ar[d]_*{\zt{}_{\sT^\*M}} \\ \sT^\*M}
            \hskip15mm \xymatrix@R+3mm @C+20mm{{\sT\sT^\*M}
\ar[d]_*{\zt{}_{\sT^\*M}} \ar[r]^*{\zb_{(\sT^\*M,\zw_M)}} &
            \sT^\*\sT^\*M \ar[d]_*{\zp{}_{\sT^\*M}} \\
            \sT^\*M \ar@{=}[r] & \sT^\*M}
                                                        \tag \label{Fplb250}$$
    \vskip2mm
    \noindent for $(\sT\sT^\*M,\xd_T\zw_M)$.  The generating function of
$D_0$ relative to the Hamiltonian special symplectic structure is the
function
        $$\left(L \circ \sT\zp{}_M + G_M \right) \circ Z,
                                                        \tag \label{Fplb251}$$
    where $G_M = -\xi_T\zy_M$.  From
        $$\sT\zp{}_M \circ Z = \sT\zp{}_M \circ \za_M{}^{-1} \circ \xd L
\circ \zL = \zp{}_{\sT M} \circ \xd L \circ \zL = \zL
                                                        \tag \label{Fplb252}$$
    and
        $$\align
    -G_M(Z(p)) &= \xi_T\zy_M((\za_M{}^{-1} \circ \xd L \circ \zL)(p)) \\
            &= \langle \zy_M, (\za_M{}^{-1} \circ \xd L \circ \zL)(p)
\rangle \\
            &= \langle (\zt{}_{\sT^\*M} \circ \za_M{}^{-1} \circ \xd L
\circ \zL)(p), (\sT\zp{}_M \circ \za_M{}^{-1} \circ \xd L \circ \zL)(p)
\rangle \\
            &= \langle p, (\zp{}_{\sT M} \circ \xd L \circ \zL)(p)\rangle
\\
            &= \langle p, \zL(p)\rangle
                                                \tag \label{Fplb253}\endalign$$
    it follows that
        $$\left(L \circ \sT\zp{}_M + G_M \right) \circ Z = - H.
                                                        \tag \label{Fplb254}$$
    The field $Z$ is a Hamiltonian vector field and the function $H$ is a
Hamiltonian for this field since
        $$\xi_Z\zw_M = Z^\*\xi_T\zw_M = - \xd H.
                                                        \tag \label{Fplb255}$$
    The formula
        $$Z = - \zb_{(\sT^\*M,\zw_M)}^{-1} \circ \xd H
                                                        \tag \label{Fplb256}$$
    is an equivalent expression of the field $Z$ in terms of the
Hamiltonian.  A force-momentum trajectory $(\zz,\zh)$ is in $\cD$ if and
only if
        $$\dot\zh(t) = Z(\zh(t)) + \zm_{\zp_M}(\zz(t),\zh(t))
                                                        \tag \label{Fplb257}$$
    for each $t \in I$.

        \sscx{Poisson formulation of dynamics.}

    We define the {\it Poisson tensor} $W_M \colon \sT^\*\sT^\*M
\rightarrow \sT\sT^\*M$ by $W_M = - \zb_{(\sT^\*M,\zw_M)}^{-1}$.  The {\it
Poisson bracket} of two functions $F$ and $G$ on $\sT^\*M$ is the function
$\{F,G\} = \langle \xd F, W_M \circ \xd G \rangle$.  The vector field $Z$
is expressed as $Z = W_M \circ \xd H$ and the Lie derivative $\xd_Z F =
\langle \xd F, Z \rangle$ of a function $F$ on $\sT^\*M$ is expressed as
the Poisson bracket $\{F,H\}$.  A force-momentum trajectory $(\zz,\zh)
\colon I \rightarrow Ph$ is in $\cD$ if and only if
        $$\xD(F \circ \zh)(t) = \{F,H\}(\zh(t)) + \langle \xd F,
\zm_{\zp_M}(\zz(t),\zh(t))\rangle
                                                        \tag \label{Fplb258}$$
    for each function $F$ on $\sT^\*M$ and each $t \in I$.

        \sect{Dynamics in the presence of non potential forces.}
    If non potential internal forces are present, then the dynamics is no
longer represented by a Lagrangian. Formulations similar to those for the
potential case are still possible if the differential $\xd L$ of the
Lagrangian is replaced by a 1-form $\zl$ on $\sT M$.  This form is
typically the difference $\xd L - \zr$ of a potential part $\xd L$ and a
1-form $\zr$ on $\sT M$ vertical with respect to the tangent projection
$\zt_M$ representing velocity dependent forces.  With the help of
operators $\bi E$ and $\bi P$ we construct mappings $\cE\zl \colon \sT^2 M
\rightarrow \sT^\* M$ and $\cP\zl \colon \sT M \rightarrow \sT^\*M$
characterized by $\langle \bi E\zl, w\rangle = -\langle \cE\zl(\zt_{\sT^2
M}(w)), \sT\zt_2{}_M(w)\rangle$ for each $w \in \sT\sT^2 M$ and $\langle
\bi P\zl, w\rangle = \langle \cP\zl(\zt_{\sT M}(w)), \sT\zt_M(w)\rangle$
for each $w \in \sT\sT M$. These constructions are possible due to
verticality properties of forms $\bi E\zl$ and $\bi P\zl$ established in
Section 2.7.

        \sscx{The variational principle.}
    The dynamics can be derived from a variational principle even if the
action and its variation are no longer defined.  The {\it dynamics} of the
system is a set $\cD$ of force-momentum trajectories satisfying this
variational principle.  A trajectory $(\zz,\zh) \colon I \rightarrow Ph$
is in $\cD$ if
        $$\int_a^b \langle \zl, \zd\dot\zx \rangle = -\int_a^b \langle
\zz, \zd\zx \rangle + \langle \zh(b), \zd\zx(b) \rangle - \langle \zh(a),
\zd\zx(a) \rangle
                                                        \tag \label{Fplb259}$$
    for each $[a,b] \subset I$ and each variation $\zd\zx$ such that
$\zt{}_M \circ \zd\zx = \zp{}_M \circ \zz = \zp{}_M \circ \zh$.

    From the equivalent form
        $$\int_a^b \langle \zl, \zd\dot\zx \rangle = - \int_a^b \langle
\cE\zl \circ \ddot\zx, \zd\zx \rangle + \langle (\cP\zl \circ \dot\zx)(b),
\zd\zx(b) \rangle - \langle (\cP\zl \circ \dot\zx)(a), \zd\zx(a) \rangle
                                                        \tag \label{Fplb260}$$
    of the variational principle equations
        $$\cE\zl \circ \ddot\zx = \zz,
                                                        \tag \label{Fplb261}$$
        $$(\cP\zl \circ \dot\zx)(a) = \zh(a),
                                                        \tag \label{Fplb262}$$
    and
        $$(\cP\zl \circ \dot\zx)(b) = \zh(b)
                                                        \tag \label{Fplb263}$$
    are derived.  These equations are satisfied for each interval $[a,b]
\subset I$. It follows that a force-momentum trajectory $(\zz,\zh)$ is in
$\cD$ if and only if equations
        $$\cE\zl \circ \ddot\zx = \zz
                                                        \tag \label{Fplb264}$$
    and
        $$\cP\zl \circ \dot\zx = \zh
                                                        \tag \label{Fplb265}$$
    are satisfied in $I$.

        \sscx{Lagrangian formulation of dynamics.}
    The Lagrangian formulation is the infinitesimal limit derived from the
version
        $$\int_a^b \langle \zl, \zd\dot\zx \rangle = -\int_a^b \langle
\zz, \zd\zx \rangle + \int_a^b \xD\langle \zh, \zd\zx \rangle
                                                        \tag \label{Fplb266}$$
    of the variational principle.  The first order differential equation
        $$\zc{}_M \circ (\zz,\dot\zh) = \za_M{}^{-1} \circ \zl \circ
\dot\zx
                                                        \tag \label{Fplb267}$$
    follow from this principle.

    The identity \Ref{Fplb245} is replaced by
        $$\zc{}_M \circ (\cE L \circ \ddot\zx,\st(\cP L \circ \dot\zx)) =
\za_M{}^{-1} \circ \xd L \circ \dot\zx.
                                                        \tag \label{Fplb268}$$
    The formula
        $$\cP\zl = \zt{}_{\sT^\*M} \circ \za_M{}^{-1} \circ \zl
                                                        \tag \label{Fplb269}$$
    follows.

    The differential equation
        $$D = \left\{(f,w) \in \sT^\*M \fpr{(\zp{}_M,\zp{}_M \circ
\zt{}_{\sT^\ast M})} \sT\sT^\*M ;\; \zc_M(f,w) \in D_0 \right\},
                                                        \tag \label{Fplb270}$$
    with
        $$D_0 = \left\{w \in \sT\sT^\*M ;\; \za_M(w) = \zl
(\zt_{\sT^\*M}(w))\right\}
                                                        \tag \label{Fplb271}$$
    can be introduced.  The set $D_0$ is submanifold of
$(\sT\sT^\*M,\xd_T\zw_M)$ but not a Lagrangian submanifold unless $\zl$ is
closed.  The form $\zl$ can be considered a {\it generating form} of $D_0$
relative to the Lagrangian special symplectic structure since $D_0 =
\za_M{}^{-1}(\im(\zl))$.

        \sscx{Hamiltonian formulation of dynamics.}
    As in the potential case we say that the Lagrangian form $\zl$ is {\it
hyperregular} if the mapping $\cP\zl$ is a diffeomorphism.  In the
hyperregular case we denote by $\zL$ the inverse of the mapping $\cP\zl$.
The set $D_0$ is the image of the vector field $Z = \za_M{}^{-1} \circ \zl
\circ \zL$.  This field is not necessarily a Hamiltonian vector field.
Let a 1-form $\zq$ on $\sT^\*M$ be defined by
        $$\langle \zq, z\rangle = \langle z, \sT\zL(z)\rangle^\ssT -
\langle \zl, \sT\zL(z)\rangle.
                                                        \tag \label{Fplb272}$$
    We will show that $-\zq$ is the generating form of $D_0$ relative to
the Hamiltonian special symplectic structure.

    According to formula \Ref{Fplb61} adapted to the present case the
generating form of $D_0$ relative to the Hamiltonian special symplectic
structure is the form
        $$Z^\*((\sT\zp_M)^\*\zl + \xd G_M).
                                                        \tag \label{Fplb273}$$
    The formula $Z^\*G_M(p) = - \langle p, \zL(p)\rangle$ derived for the
potential case is still valid in the non potential case with $\xd L$
replaced by $\zl$.  The equality
        $$\langle Z^\*\xd G_M, z\rangle = \langle \xd Z^\*G_M, z\rangle =
- \langle v, \sT\zL(z) \rangle^\ssT
                                                        \tag \label{Fplb274}$$
    follows from this formula.  We have
        $$\langle Z^\*((\sT\zp_M)^\*\zl + \xd G_M), z\rangle = \langle
\zl, \sT\zL(z)\rangle - \langle z, \sT\zL(z) \rangle^\ssT = - \langle \zq,
z\rangle
                                                        \tag \label{Fplb275}$$
    since
        $$\sT\zp{}_M \circ Z = \sT\zp{}_M \circ \za_M{}^{-1} \circ \zl
\circ \zL = \zp{}_{\sT M} \circ \zl \circ \zL = \zL.
                                                        \tag \label{Fplb276}$$
    Hence, the form $-\zq$ defined in \Ref{Fplb272} is the generating form
of $D_0$ relative to the Hamiltonian special symplectic structure.  The
formula
        $$Z = - \zb_{(\sT^\*M,\zw_M)}^{-1} \circ \zq
                                                        \tag \label{Fplb277}$$
    follows.

    In the special case of $\zl = \xd L - \zr$ we have $\cP\zl = \cP L$
since $\xi_{F(1;1)}\zr = 0$ due to verticality of $\zr$.  Let $H \colon
\sT^\*M \rightarrow \R$ be the Hamiltonian corresponding to $L$.  This
Hamiltonian is defined by $H(p) = \langle p, \zL(p) \rangle - L(\zL(p))$
and satisfies the relation
        $$\langle \xd H, z\rangle = \langle z, \sT\zL(z)\rangle^\ssT -
\langle \xd L, \sT\zL(z)\rangle.
                                                        \tag \label{Fplb278}$$
    By comparing this relation with
        $$\langle \zq, z\rangle = \langle z, \sT\zL(z)\rangle^\ssT -
\langle \xd L, \sT\zL(z)\rangle + \langle \zr, \sT\zL(z)\rangle
                                                        \tag \label{Fplb279}$$
    we derive the formula
        $$\zq = \xd H + \zL^\*\zr.
                                                        \tag \label{Fplb280}$$

    As in the potential case a force-momentum trajectory $(\zz,\zh)$ is in
$\cD$ if and only if
        $$\dot\zh(t) = Z(\zh(t)) + \zm_{\zp_M}(\zz(t),\zh(t))
                                                        \tag \label{Fplb281}$$
    for each $t \in I$.

        \sscx{Poisson formulation of dynamics.}
    The vector field $Z$ of the Hamiltonian formulation is expressed as $Z
= W_M \circ \zq$.  A force-momentum trajectory $(\zz,\zh) \colon I
\rightarrow Ph$ is in $\cD$ if and only if
        $$\xD(F \circ \zh)(t) = \langle \xd F, W_M \circ \zq\rangle +
\langle \xd F, \zm_{\zp_M}(\zz(t),\zh(t))\rangle
                                                        \tag \label{Fplb282}$$
    for each function $F$ on $\sT^\*M$ and each $t \in I$.

        \sect{Local expressions.}

    Coordinate definitions of objects add nothing to the clarity of the
conceptual structure of a theory.  Covariance of a definition with respect
to coordinate transformations guarantees the existence of an intrinsic
interpretation of the object being defined without providing an
interpretation.  We have provided intrinsic definitions.  Now we give
local expressions of most objects introduced earlier in order to
facilitate comparison with traditional formulations of mechanics.  Local
expressions are also used in calculations and appear in examples.
    \vskip6mm

        \ssca{The tangent and the cotangent fibrations.}

    Coordinates
        $$(x^\zk,\zd x^\zl) \colon \sT M \rightarrow \R^{2m}
                                                        \tag \label{Fplb283}$$
    and
        $$(x^\zk,f_\zl) \colon \sT^\* M \rightarrow \R^{2m}
                                                        \tag \label{Fplb284}$$
    induced by coordinates
        $$(x^\zk) \colon M \rightarrow \R^m.
                                                        \tag \label{Fplb285}$$

    Projections:
        $$(x^\zk) \circ \zt_M = (x^\zk),
                                                        \tag \label{Fplb286}$$
        $$(x^\zk) \circ \zf_M = (x^\zk).
                                                        \tag \label{Fplb287}$$

    Zero sections and linear operations:
        $$(x^\zk,\zd x^\zl) \circ O_{\zt_M} = (x^\zk,0),
                                                        \tag \label{Fplb288}$$
        $$(x^\zk,\zd x^\zl)(v + v') = (x^\zk(v),\zd x^\zl(v) + \zd
x^\zl(v'))
                                                        \tag \label{Fplb289}$$
    defined if
        $$(x^\zk(v')) = (x^\zk(v)),
                                                        \tag \label{Fplb290}$$
        $$(x^\zk,\zd x^\zl)(kv) = (x^\zk(v),k\zd x^\zl(v)),
                                                        \tag \label{Fplb291}$$
        $$(x^\zk,f_\zl) \circ O_{\zp_M} = (x^\zk,0),
                                                        \tag \label{Fplb292}$$
        $$(x^\zk,f_\zl)(f + f') = (x^\zk(f), f_\zl(f) + f_\zl(f'))
                                                        \tag \label{Fplb293}$$
    defined if
        $$(x^\zk(f')) = (x^\zk(f)),
                                                        \tag \label{Fplb294}$$
        $$(x^\zk,f_\zl)(kf) = (x^\zk(f),kf_\zl(f)).
                                                        \tag \label{Fplb295}$$

    The canonical pairing:
        $$\langle f, v\rangle = f_\zk(f) \zd x^\zk(v)
                                                        \tag \label{Fplb296}$$
    defined if
        $$x^\zk(f) = x^\zk(v).
                                                        \tag \label{Fplb297}$$

    Coordinates
        $$(x^\zk,\dot x^\zl) \colon \sT M \rightarrow \R^{2m}
                                                        \tag \label{Fplb298}$$
    and
        $$(x^\zk,p_\zl) \colon \sT^\* M \rightarrow \R^{2m}
                                                        \tag \label{Fplb299}$$
    are also used.

        \sscx{The dual pair $\sT\sT M$ and $\sT^\*\sT M$.}

    Coordinates:
        $$(x^\zk,\dot x^\zl,\zd x^\zm,\zd\dot x^\zn) \colon \sT\sT M
\rightarrow \R^{4m},
                                                        \tag \label{Fplb300}$$
        $$(x^\zk,\dot x^\zl,a_\zm,b_\zn) \colon \sT^\*\sT M \rightarrow
\R^{4m}.
                                                        \tag \label{Fplb301}$$
    Projections:
        $$(x^\zk,\dot x^\zl) \circ \zt_{\sT M} = (x^\zk,\dot x^\zl),
                                                        \tag \label{Fplb302}$$
        $$(x^\zk,\dot x^\zl) \circ \zp_{\sT M} = (x^\zk,\dot x^\zl).
                                                        \tag \label{Fplb303}$$

    Zero sections and linear operations:
        $$(x^\zk,\dot x^\zl,\zd x^\zm,\dot \zd x^\zn) \circ O_{\zt_{\sT
M}} = (x^\zk,\dot x^\zl,0,0),
                                                        \tag \label{Fplb304}$$
        $$(x^\zk,\dot x^\zl,\zd x^\zm,\zd\dot x^\zn)(w + w') =
(x^\zk(w),\dot x^\zl(w),\zd x^\zm(w) + \zd x^\zm(w'),\zd\dot x^\zn(w) +
\zd\dot x^\zn(w'))
                                                        \tag \label{Fplb305}$$
    defined if
        $$(x^\zk,\dot x^\zl)(w') = (x^\zk,\dot x^\zl)(w),
                                                        \tag \label{Fplb306}$$
        $$(x^\zk,\dot x^\zl,\zd x^\zm,\zd\dot x^\zn)(kw) = (x^\zk(w),\dot
x^\zl(w),k\zd x^\zm(w),k\zd\dot x^\zn(w)),
                                                        \tag \label{Fplb307}$$
        $$(x^\zk,\dot x^\zl,a_\zm,b_\zn) \circ O_{\zp_{\sT M}} =
(x^\zk,\dot x^\zl,0,0),
                                                        \tag \label{Fplb308}$$
        $$(x^\zk,\dot x^\zl,a_\zm,b_\zn)(z + z') = (x^\zk(z),\dot
x^\zl(z),a_\zm(z) + a_\zm(z'),b_\zn(z) + b_\zn(z'))
                                                        \tag \label{Fplb309}$$
    defined if
        $$(x^\zk,\dot x^\zl)(z') = (x^\zk,\dot x^\zl)(z),
                                                        \tag \label{Fplb310}$$
        $$(x^\zk,\dot x^\zl,a_\zm,b_\zn)(kz) = (x^\zk(z),\dot
x^\zl(z),ka_\zm(z),kb_\zn(z)).
                                                        \tag \label{Fplb311}$$

    The canonical pairing
        $$\langle z, w\rangle = a_\zk(z) \zd x^\zk(w) + b_\zk(z) \zd\dot
x^\zk(w)
                                                        \tag \label{Fplb312}$$
    defined if
        $$(x^\zk,\dot x^\zl)(z) = (x^\zk,\dot x^\zl)(w).
                                                        \tag \label{Fplb313}$$

        \sscx{The dual pair $\sT\sT M$ and $\sT\sT^\* M$.}

    Coordinates:
        $$(x^\zk,\zd x^\zl,\dot x^\zm,\dot \zd x^\zn) \colon \sT\sT M
\rightarrow \R^{4m},
                                                        \tag \label{Fplb314}$$
        $$(x^\zk,p_\zl,\dot x^\zm,\dot p_\zn) \colon \sT\sT^\* M
\rightarrow \R^{4m}.
                                                        \tag \label{Fplb315}$$
    Coordinates $(x^\zk,\dot x^\zl)$ in $\sT M$ are used.

    Projections:
        $$(x^\zk,\dot x^\zl) \circ \sT\zt_M = (x^\zk,\dot x^\zl),
                                                        \tag \label{Fplb316}$$
        $$(x^\zk,\dot x^\zl) \circ \sT\zp_M = (x^\zk,\dot x^\zl).
                                                        \tag \label{Fplb317}$$

    Zero sections and tangent linear operations:
        $$(x^\zk,\zd x^\zl,\dot x^\zm,\dot \zd x^\zn) \circ O_{\sT\zt_M} =
(x^\zk,0,\dot x^\zm,0),
                                                        \tag \label{Fplb318}$$
        $$(x^\zk,\zd x^\zl,\dot x^\zm,\dot \zd x^\zn)(w +^\ssT w') =
(x^\zk(w),\zd x^\zl(w) + \zd x^\zl(w'),\dot x^\zm(w),\zd\dot x^\zn(w) +
\zd\dot x^\zn(w'))
                                                        \tag \label{Fplb319}$$
    defined if
        $$(x^\zk,\dot x^\zl)(w') = (x^\zk,\dot x^\zl)(w),
                                                        \tag \label{Fplb320}$$
        $$(x^\zk,\zd x^\zl,\dot x^\zm,\dot \zd x^\zn)(k \cdot^\ssT w) =
(x^\zk(w),k\zd x^\zl(w),\dot x^\zm(w),k\zd\dot x^\zn(w)),
                                                        \tag \label{Fplb321}$$
        $$(x^\zk,p_\zl,\dot x^\zm,\dot p_\zn) \circ O_{\sT\zp_M} =
(x^\zk,0,\dot x^\zm,0),
                                                        \tag \label{Fplb322}$$
        $$(x^\zk,p_\zl,\dot x^\zm,\dot p_\zn)(z +^\ssT z') =
(x^\zk(z),p_\zl(z) + p_\zl(z'),\dot x^\zm(z),\dot p_\zn(z) + \dot
p_\zn(z'))
                                                        \tag \label{Fplb323}$$
    defined if
        $$(x^\zk,\dot x^\zl)(z') = (x^\zk,\dot x^\zl)(z),
                                                        \tag \label{Fplb324}$$
        $$(x^\zk,p_\zl,\dot x^\zm,\dot p_\zn)(k \cdot^\ssT z) =
(x^\zk(z),kp_\zl(z),\dot x^\zm(z),k\dot p_\zn(z)).
                                                        \tag \label{Fplb325}$$

    The tangent pairing:
        $$\langle z, w\rangle^\ssT = \dot p_\zk(z) \zd x^\zk(w) + p_\zk(z)
\zd\dot x^\zk(w)
                                                        \tag \label{Fplb326}$$
    defined if
        $$(x^\zk,\dot x^\zl)(z) = (x^\zk,\dot x^\zl)(w).
                                                        \tag \label{Fplb327}$$
    Let $\zx \colon \R \rightarrow M$, $\zd\zx \colon \R \rightarrow \sT
M$, and $\zh \colon \R \rightarrow \sT^\* M$ be curves such that $\zp_M
\circ \zh = \zt_M \circ \zd\zx = \zx$.  Let $\dot\zx \colon \R \rightarrow
\sT M$, $\zd\dot\zx \colon \R \rightarrow \sT\sT M$, and $\dot\zh \colon
\R \rightarrow \sT\sT^\* M$ be prolongations of these curves.
        $$\langle \dot\zh(0), \zd\dot\zx(0)\rangle^\ssT = \frac{\xd}{\xd
t}(\zh_\zk(t)\zd\zx^\zk(t))_{|t=0} = \dot\zh_\zk(0)\zd\zx^\zk(0)
+\zh_\zk(0)\zd\dot\zx^\zk(0).
                                                        \tag \label{Fplb328}$$

        \sscx{Relations between $\sT\sT^\* M$ and $\sT^\*\sT M$.}

    The local expression
        $$(x^\zk,\dot x^\zl,a_\zm,b_\zn) \circ \za_M = (x^\zk,\dot
x^\zl,\dot p_\zm,p_\zn)
                                                        \tag \label{Fplb329}$$
    of the isomorphism $\za_M \colon \sT\sT^\* M \rightarrow \sT^\*\sT M$
dual to the canonical involution $\zk^{1,1} \colon \sT\sT M \rightarrow
\sT\sT M$ defined locally by
        $$(x^\zk,\zd x^\zl,\dot x^\zm,\zd\dot x^\zn) \circ \zk^{1,1} =
(x^\zk, \dot x^\zl,\zd x^\zm,\zd\dot x^\zn).
                                                        \tag \label{Fplb330}$$
    Coordinate systems \Ref{Fplb301}, \Ref{Fplb314}, and \Ref{Fplb315} are
used in the local expressions.

        \sscx{The dual pair $\sT\sT^\* M$ and $\sT^\*\sT^\* M$.}

    Coordinates:
        $$(x^\zk,p_\zl,\dot x^\zm,\dot p_\zn) \colon \sT\sT^\* M
\rightarrow \R^{4m},
                                                        \tag \label{Fplb331}$$
        $$(x^\zk,p_\zl,y_\zm,z^\zn) \colon \sT^\*\sT^\* M \rightarrow
\R^{4m}.
                                                        \tag \label{Fplb332}$$

    Projections:
        $$(x^\zk,p_\zl) \circ \zt_{\sT^\*M} = (x^\zk,p_\zl)
                                                        \tag \label{Fplb333}$$
        $$(x^\zk,p_\zl) \circ \zp_{\sT^\*M} = (x^\zk,p_\zl)
                                                        \tag \label{Fplb334}$$

    Zero sections and linear operations:
        $$(x^\zk,p_\zl,\dot x^\zm,\dot p_\zn) \circ O_{\zt_{\sT^\ast M}} =
(x^\zk,p_\zl,0,0),
                                                        \tag \label{Fplb335}$$
        $$(x^\zk,p_\zl,\dot x^\zm,\dot p_\zn)(z + z') =
(x^\zk(z),p_\zl(z),\dot x^\zm(z) + \dot x^\zm(z'),\dot p_\zn(z) + \dot
p_\zn)(z'))
                                                        \tag \label{Fplb336}$$
    defined if
        $$(x^\zk,p_\zl)(z') = (x^\zk,p_\zl)(z),
                                                        \tag \label{Fplb337}$$
        $$(x^\zk,p_\zl,\dot x^\zm,\dot p_\zn)(kz) =
(x^\zk(z),p_\zl(z),k\dot x^\zm(z),k\dot p_\zn(z)),
                                                        \tag \label{Fplb338}$$
        $$(x^\zk,p_\zl,y_\zm,z^\zn) \circ O_{\zp_{\sT^\ast M}} =
(x^\zk,p_\zl,0,0),
                                                        \tag \label{Fplb339}$$
        $$(x^\zk,p_\zl,y_\zm,z^\zn)(b + b') = (x^\zk(b),p_\zl(b),y_\zm(b)
+ y_\zm(b'),z^\zn(b) + z^\zn(b'))
                                                        \tag \label{Fplb340}$$
    defined if
        $$(x^\zk,p_\zl)(b') = (x^\zk,p_\zl)(b),
                                                        \tag \label{Fplb341}$$
        $$(x^\zk,p_\zl,y_\zm,z^\zn)(kb) =
(x^\zk(b),p_\zl(b),ky_\zm(b),kz^\zn(b)).
                                                        \tag \label{Fplb342}$$

    The canonical pairing
        $$\langle b, z\rangle = y_\zk(b) \dot x^\zk(z) + z^\zk(b) \dot
p^\zk(z)
                                                        \tag \label{Fplb343}$$
    defined if
        $$(x^\zk,p_\zl)(b) = (x^\zk,p_\zl)(z).
                                                        \tag \label{Fplb344}$$

    The isomorphism $\zb_{(\sT^\*M,\zw_M)} \colon \sT\sT^\* M \rightarrow
\sT^\*\sT^\* M$ has a local expression
        $$(x^\zk,p_\zl,y_\zm,z^\zn) \circ \zb_{(\sT^\*M,\zw_M)} =
(x^\zk,p_\zl,\dot p_\zm,- \dot x^\zn).
                                                        \tag \label{Fplb345}$$

        \sscx{The bundles $\sT^2 M$, $\sT\sT^2 M$, and $\sT^2\sT M$.}

    Coordinates:
        $$(x^\zk,\dot x^\zl,\ddot x^\zm) \colon \sT^2 M \rightarrow \R^{3m}
                                                        \tag \label{Fplb346}$$

        $$(x^\zk,\dot x^\zl,\ddot x^\zm,x'{}^\zn,\dot x'{}^\zw,\ddot
x'{}^\zp) \colon \sT\sT^2 M \rightarrow \R^{6m}
                                                        \tag \label{Fplb347}$$

        $$(x^\zk,\dot x^\zl,x'{}^\zm,\dot x'{}^\zn,x''{}^\zw,\dot
x''{}^\zp) \colon \sT^2\sT M \rightarrow \R^{6m}
                                                        \tag \label{Fplb348}$$

    Projections:
        $$(x^\zk) \circ \zt_2{}_M = (z^\zk),
                                                        \tag \label{Fplb349}$$
        $$(x^\zk,\dot x^\zl) \circ \zt^1{}_2{}_M = (x^\zk,\dot x^\zl),
                                                        \tag \label{Fplb350}$$
        $$(x^\zk,\dot x^\zl,\ddot x^\zm) \circ \zt_{\sT^2 M} = (x^\zk,\dot
x^\zl,\ddot x^\zm),
                                                        \tag \label{Fplb351}$$
        $$(x^\zk,\dot x^\zl) \circ \sT\zt_2{}_M = (x^\zk,x'{}^\zl),
                                                        \tag \label{Fplb352}$$
        $$(x^\zk,\dot x^\zl,x'{}^\zm,\dot x'{}^\zn) \circ \sT\zt^1{}_2{}_M
= (x^\zk,\dot x^\zl,x'{}^\zm,\dot x'{}^\zn).
                                                        \tag \label{Fplb353}$$

    Mappings:
        $$(x^\zk,\dot x^\zl,x'{}^\zm,\dot x'{}^\zn,x''{}^\zw,\dot
x''{}^\zp) \circ \zk^{2,1} = (x^\zk,x'{}^\zl,\dot x^\zm,\dot
x'{}^\zn,\ddot x^\zw,\ddot x'{}^\zp),
                                                        \tag \label{Fplb354}$$
        $$(x^\zk,\dot x^\zl,\ddot x^\zm,x'{}^\zn,\dot x'{}^\zw,\ddot
x'{}^\zp) \circ \zk^{1,2} = (x^\zk,x'{}^\zl,x''{}^\zm,\dot x^\zn,\dot
x'{}^\zw,\dot x''{}^\zp).
                                                        \tag \label{Fplb355}$$

        \sscx{Tangent prolongations and tangent mappings.}

    If $\dot\zx \colon \R \rightarrow \sT M$ and $\ddot\zx \colon \R
\rightarrow \sT^2 M$ are prolongations of a curve $\zx \colon \R
\rightarrow M$ and $x^\zk \circ \zx = \zx^\zk$, then
        $$(x^\zk,\dot x^\zl) \circ \dot\zx = (\zx^\zk,\dot\zx^\zl) =
(\zx^\zk,\xD\zx^\zl).
                                                        \tag \label{Fplb356}$$
    and
        $$(x^\zk,\dot x^\zl,\ddot x^\zm) \circ \ddot\zx =
(\zx^\zk,\dot\zx^\zl,\ddot\zx^\zm) = (\zx^\zk,\xD\zx^\zl,\xD^2\zx^\zm).
                                                        \tag \label{Fplb357}$$

    Let $(x^\zk)$ and $(y^i)$ be charts of manifolds $M$ and $N$, let
$(x^\zk\dot x^\zl)$ and $(y^i,\dot y^j)$ be induced charts of $\sT M$ and
$\sT N$, and let $(x^\zk\dot x^\zl,\ddot x^\zm)$ and $(y^i,\dot y^j,\ddot
y^k)$ be induced charts of $\sT^2 M$ and $\sT^2 N$.  A differentiable
mapping $\za \colon M \rightarrow N$ is represented locally by a set of
functions $\za^i \colon \R^m \rightarrow \R$ defined by $\za^i = y^i \circ
\za \circ (x^\zk)^{-1}$. The tangent mapping $\sT\za$ and the second
tangent mapping $\sT^2\za$ have local representations
        $$(y^i,\dot y^j) \circ \sT\za = \left(\za^i(x^\zm),\frac{\partial
\za^j}{\partial x^\zk}(x^\zm)\dot x^\zk\right)
                                                        \tag \label{Fplb358}$$
    and
        $$(y^i,\dot y^j,\ddot y^k) \circ \sT^2\za =
\left(\za^i(x^\zm),\frac{\partial \za^j}{\partial x^\zk}(x^\zm)\dot
x^\zk,\frac{\partial^2 \za^k}{\partial x^\zk \partial x^\zl}(x^\zm)\dot
x^\zk \dot x^\zl + \frac{\partial \za^k}{\partial x^\zk}(x^\zm)\ddot
x^\zk\right).
                                                        \tag \label{Fplb359}$$

        \sscx{Vector valued forms and derivations.}

    Local expressions
        $$(x^\zk,\dot x^\zl) \circ T(0) = (x^\zk,\dot x^\zl),
                                                        \tag \label{Fplb360}$$
        $$(x^\zk,p_\zl,\dot x^\zm,\dot p_\zn) \circ T = (x^\zk,p_\zl,\dot
x^\zm,\dot p_\zn),
                                                        \tag \label{Fplb361}$$
        $$(x^\zk,\dot x^\zl,x'{}^\zm,\dot x'{}^\zn) \circ T(1) =
(x^\zk,\dot x^\zl,\dot x^\zm,\ddot x^\zn),
                                                        \tag \label{Fplb362}$$
        $$(x^\zk,\dot x^\zl,x'{}^\zm,\dot x'{}^\zn) \circ F(1;1) =
(x^\zk,\dot x^\zl,0,\dot x^\zn),
                                                        \tag \label{Fplb363}$$
        $$(x^\zk,\dot x^\zl,\ddot x^\zm,x'{}^\zn,\dot x'{}^\zw,\ddot
x'{}^\zp) \circ F(2;1) = (x^\zk,\dot x^\zl,\ddot x^\zm,0,x'{}^\zw,\dot
x'{}^\zp),
                                                        \tag \label{Fplb364}$$
        $$(x^\zk,\dot x^\zl,\ddot x^\zm,x'{}^\zn,\dot x'{}^\zw,\ddot
x'{}^\zp) \circ F(2;2) = (x^\zk,\dot x^\zl,\ddot x^\zm,0,0,2x'{}^\zp)
                                                        \tag \label{Fplb365}$$
    of vector valued forms
        $$T(0) \colon \sT M \rightarrow \sT M,
                                                        \tag \label{Fplb366}$$
        $$T \colon \sT\sT^\* M \rightarrow \sT\sT^\* M,
                                                        \tag \label{Fplb367}$$
        $$T(1) \colon \sT^2 M \rightarrow \sT\sT M,
                                                        \tag \label{Fplb368}$$
        $$F(1;1) \colon \sT\sT M \rightarrow \sT\sT M,
                                                        \tag \label{Fplb369}$$
        $$F(2;1) \colon \sT\sT^2 M \rightarrow \sT\sT^2 M,
                                                        \tag \label{Fplb370}$$
    and
        $$F(2;2) \colon \sT\sT^2 M \rightarrow \sT\sT^2 M.
                                                        \tag \label{Fplb371}$$

    Corresponding derivations:
        $$\xd_{T(0)}x^\zk = \xi_{T(0)}\xd x^\zk = \dot x^\zk,
                                                        \tag \label{Fplb372}$$
        $$\xd_T x^\zk = \xi_T \xd x^\zk = \dot x^\zk, \; \xd_T p_\zk =
\xi_T \xd p_\zk = \dot p_\zk,
                                                        \tag \label{Fplb373}$$
        $$\xd_{T(1)}x^\zk = \xi_{T(1)}\xd x^\zk = \dot x^\zk, \;
\xd_{T(1)}\dot x^\zk = \xi_{T(1)}\xd\dot x^\zk = \ddot x^\zk,
                                                        \tag \label{Fplb374}$$
        $$\xi_{F(1;1)}\xd x^\zk = 0, \; \xi_{F(1;1)}\xd\dot x^\zk = \xd
x^\zk,
                                                        \tag \label{Fplb375}$$
        $$\xi_{F(2;1)}\xd x^\zk = 0, \; \xi_{F(2;1)}\xd\dot x^\zk = \xd
x^\zk, \; \xi_{F(2;1)}\xd\ddot x^\zk = \xd\dot x^\zk,
                                                        \tag \label{Fplb376}$$
    and
        $$\xi_{F(2;2)}\xd x^\zk = 0, \; \xi_{F(2;2)}\xd\dot x^\zk = 0, \;
\xi_{F(2;2)}\xd\ddot x^\zk = 2\xd x^\zk.
                                                        \tag \label{Fplb377}$$

        \sscx{Louville, symplecting and Poisson structures.}

    On $\sT^\* M$:
        $$\zy_M = p_\zk \xd x^\zk,
                                                        \tag \label{Fplb378}$$
        $$\zw_M = \xd p_\zk \wedge \xd x^\zk.
                                                        \tag \label{Fplb379}$$

    On $\sT^\*\sT M$:
        $$\zy_{\sT M} = a_\zk \xd x^\zk + b_\zk \xd\dot x^\zk,
                                                        \tag \label{Fplb380}$$
        $$\zw_{\sT M} = \xd a_\zk \wedge \xd x^\zk + \xd b_\zk \wedge
\xd\dot x^\zk.
                                                        \tag \label{Fplb381}$$

    On $\sT^\*\sT^\* M$:
        $$\zy_{\sT^\* M} = y_\zk \xd x^\zk + z^\zk \xd p_\zk,
                                                        \tag \label{Fplb382}$$
        $$\zy_{\sT^\* M} = \xd y_\zk \wedge \xd x^\zk + \xd z^\zk \wedge
\xd p_\zk.
                                                        \tag \label{Fplb383}$$

    On $\sT\sT^\* M$:
        $$\xd_T\zy_M = \dot p_\zk \xd x^\zk + p_\zk \xd\dot x^\zk,
                                                        \tag \label{Fplb384}$$
        $$\xi_T\zw_M = \dot p_\zk \xd x^\zk - \dot x^\zk \xd p_\zk,
                                                        \tag \label{Fplb385}$$
        $$\xd_T\zw_M = \xd\dot p_\zk \wedge \xd x^\zk + \xd p_\zk \wedge
\xd\dot x^\zk.
                                                        \tag \label{Fplb386}$$

    The function $G_M$ on $\sT\sT^\* M$:
        $$G_M = \xi_T\zy_M = p_\zk \dot x^\zk.
                                                        \tag \label{Fplb387}$$

    Local expression
        $$\{F,G\} \circ (x^\zk,p_\zl)^{-1} = \frac{\partial \cF}{\partial
x^\zk}\frac{\partial \cG}{\partial p_\zk} - \frac{\partial \cG}{\partial
x^\zk}\frac{\partial \cF}{\partial p_\zk}
                                                        \tag \label{Fplb388}$$
    of the Poisson bracket of functions $F$ and $G$ on $\sT^\*M$ with
local expressions
        $$\cF = F \circ (x^\zk,p_\zl)^{-1}
                                                        \tag \label{Fplb389}$$
    and
        $$\cG = G \circ (x^\zk,p_\zl)^{-1}
                                                        \tag \label{Fplb390}$$

        \sscx{Other constructions.}

    The mapping
        $$\zm_{\zp_M} \colon \sT^\*M \fpr{(\zp{}_M,\zp{}_M)} \sT^\*M
\rightarrow \sT\sT^\*M
                                                        \tag \label{Fplb391}$$
    has a local expression
        $$(x^\zk,p_\zl,\dot x^\zm,\dot p_\zn) \circ \zm_{\zp_M} =
(x^\zk,p_\zl,0,f_\zn)
                                                        \tag \label{Fplb392}$$
    in terms of coordinates
        $$(x^\zk,f_\zl,p_\zm) \colon \sT^\*M \fpr{(\zp{}_M,\zp{}_M)}
\sT^\*M \rightarrow \R^{3m}
                                                        \tag \label{Fplb393}$$
    and \Ref{Fplb315}.

    The mapping
        $$\zc_{\zp_M} \colon \sT^\*M \fpr{(\zp_M,\zp_M \circ \zt_{\sT^\ast
M})} \sT\sT^\* M \rightarrow \sT\sT^\*M
                                                        \tag \label{Fplb394}$$
    is defined locally by

        $$(x^\zk,p_\zl,\dot x^\zm,\dot p_\zn) \circ \zc_{\zp_M} =
(x^\zk,p_\zl,\dot x^\zm,\dot p_\zn - f_\zn)
                                                        \tag \label{Fplb395}$$
    in terms of coordinates
        $$(x^\zk,f_\zl,p_\zm,\dot x^\zn,\dot p_\zw) \colon \sT^\*M
\fpr{(\zp_M,\zp_M \circ \zt_{\sT^\ast M})} \sT\sT^\* M \rightarrow \R^{5m}
                                                        \tag \label{Fplb396}$$
    and \Ref{Fplb315}.

        \sect{Local description of dynamics.} \vskip3mm
        \ssca{The variational principle.}

    We associate mappings
        $$\zx^\zk = x^\zk \circ \zd\dot\zx
                                                        \tag \label{Fplb397}$$
        $$\dot\zx^\zk = \dot x^\zk \circ \zd\dot\zx
                                                        \tag \label{Fplb398}$$
        $$\zd\zx^\zk = \zd x^\zk \circ \zd\dot\zx
                                                        \tag \label{Fplb399}$$
        $$\zd\dot\zx^\zk = \zd\dot x^\zk \circ \zd\dot\zx
                                                        \tag \label{Fplb400}$$
    defined in terms of coordinates
        $$(x^\zk,\dot x^\zl,\zd x^\zm,\zd\dot x^\zn) \colon \sT\sT M
\rightarrow \R^{4m},
                                                        \tag \label{Fplb401}$$
    with a variation
        $$\zd\dot\zx \colon I \rightarrow \sT\sT M.
                                                        \tag \label{Fplb402}$$
    We will denote by $\cL$ the local expression
        $$L \circ (x^\zk,\dot x^\zl)^{-1} \colon \R^{2m} \rightarrow \R 
                                                        \tag \label{Fplb403}$$
    of the Lagrangian.

    The variation of the action
        $$\int_a^b \cL(\zx^\zm(t),\dot\zx^\zn(t))\xd t
                                                        \tag \label{Fplb404}$$
    is expressed as the integral
        $$\int_a^b \left(\frac{\partial \cL}{\partial
x^\zk}(\zx^\zm(t),\dot\zx^\zn(t))\zd\zx^\zk(t) + \frac{\partial
\cL}{\partial\dot x^\zk}(\zx^\zm(t),\dot\zx^\zn(t))\zd\dot\zx^\zk(t)
\right)\xd t.
                                                        \tag \label{Fplb405}$$
    If non potential internal forces represented by the form
        $$\zr = \zr_\zk(x^\zm,\dot x^\zn)\xd x^\zk
                                                        \tag \label{Fplb406}$$
    are present, then the differential of the Lagrangian is replaced by
        $$\zl = \frac{\partial \cL}{\partial x^\zk}(x^\zm,\dot x^\zn)\xd
x^\zk
    + \frac{\partial \cL}{\partial\dot x^\zk}(x^\zm,\dot x^\zn)\xd\dot
x^\zk - \zr_\zk(x^\zm,\dot x^\zn)\xd x^\zk
                                                        \tag \label{Fplb407}$$
    and the variation takes the form
        $$\int_a^b \left(\frac{\partial \cL}{\partial
x^\zk}(\zx^\zm(t),\dot\zx^\zn(t))\zd\zx^\zk(t)
    + \frac{\partial \cL}{\partial\dot
x^\zk}(\zx^\zm(t),\dot\zx^\zn(t))\zd\dot\zx^\zk(t) -
\zr_\zk(\zx^\zm(t),\dot\zx^\zn(t))\zd\zx^\zk(t) \right)\xd t.
                                                        \tag \label{Fplb408}$$
    Integration by parts results in
        $$\align
    \int_a^b &\left(\frac{\partial \cL}{\partial
x^\zk}(\zx^\zm(t),\dot\zx^\zn(t)) - \frac{\xd}{\xd t}\frac{\partial
\cL}{\partial\dot x^\zk}(\zx^\zm(t),\dot\zx^\zn(t),\ddot\zx^\zw(t)) -
\zr_\zk(\zx^\zm(t),\dot\zx^\zn(t)) \right)\zd\zx^\zk(t)\xd t \\
    &\hskip20mm + \frac{\partial \cL}{\partial\dot
x^\zk}(\zx^\zm(b),\dot\zx^\zn(b))\zd\zx^\zk(b) - \frac{\partial
\cL}{\partial\dot x^\zk}(\zx^\zm(a),\dot\zx^\zn(a))\zd\zx^\zk(a).
                                                \tag \label{Fplb409}\endalign$$
    The expression
        $$\frac{\xd}{\xd t}\frac{\partial \cL}{\partial\dot
x^\zk}(\zx^\zm(t),\dot\zx^\zn(t),\ddot\zx^\zw(t))
                                                        \tag \label{Fplb410}$$
    stands for
        $$\frac{\xd}{\xd t}\left(\frac{\partial \cL}{\partial\dot
x^\zk}(\zx^\zm(t),\dot\zx^\zn(t))\right) = \frac{\partial^2
\cL}{\partial\dot x^\zk\partial
x^\zl}(\zx^\zm(t),\dot\zx^\zn(t))\dot\zx^\zl(t) + \frac{\partial^2
\cL}{\partial\dot x^\zk\partial\dot
x^\zl}(\zx^\zm(t),\dot\zx^\zn(t))\ddot\zx^\zl(t)
                                                        \tag \label{Fplb411}$$
    The variational principle requires that the variation \Ref{Fplb409} be
equal to
        $$-\int_a^b \zz_\zk(t)\zd\zx^\zk(t)\xd t + \zh_\zk(b)\zd\zx^\zk(b)
- \zh_\zk(a)\zd\zx^\zk(a),
                                                        \tag \label{Fplb412}$$
    where
        $$\zh_\zk = p_\zk \circ \zh
                                                        \tag \label{Fplb413}$$
    and
        $$\zz_\zk = f_\zk \circ \zz
                                                        \tag \label{Fplb414}$$
    are mappings derived from a force-momentum trajectory
        $$(\zz,\zh) \colon I \rightarrow Ph.
                                                        \tag \label{Fplb415}$$
    The Euler-Lagrange equation
        $$\frac{\xd}{\xd t} \frac{\partial \cL}{\partial\dot
x^\zk}(\zx^\zm(t),\dot\zx^\zn(t),\ddot\zx^\zw(t)) - \frac{\partial
\cL}{\partial x^\zk}(\zx^\zm(t),\dot\zx^\zn(t)) +
\zr_\zk(\zx^\zm(t),\dot\zx^\zn(t)) = \zz_\zk(t)
                                                        \tag \label{Fplb416}$$
    in $[a,b]$ and the momentum-velocity relations
        $$\frac{\partial \cL}{\partial\dot
x^\zk}(\zx^\zm(t),\dot\zx^\zn(a)) = \zh_\zk(a),
                                                        \tag \label{Fplb417}$$
        $$\frac{\partial \cL}{\partial\dot
x^\zk}(\zx^\zm(t),\dot\zx^\zn(b)) = \zh_\zk(b)
                                                        \tag \label{Fplb418}$$
    are derived from this variational principle.  The variational
principle is to be satisfied in each interval $[a,b] \subset I$.  It
follows that the Euler-Lagrange equations and the relation
        $$\frac{\partial \cL}{\partial\dot
x^\zk}(\zx^\zm(t),\dot\zx^\zn(t)) = \zh_\zk(t)
                                                        \tag \label{Fplb419}$$
    are satisfied in $I$.

        \sscx{The Lagrangian formulation.}

    Lagrange equations
        $$\zh_\zk(t) = \frac{\partial \cL}{\partial\dot
x^\zk}(\zx^\zm(t),\dot\zx^\zn(t))
                                                        \tag \label{Fplb420}$$
    and
        $$\dot\zh_\zk(t) - \zz_\zk(t) = \frac{\partial \cL}{\partial
x^\zk}(\zx^\zm(t),\dot\zx^\zn(t)) - \zr_\zk(\zx^\zm(t),\dot\zx^\zn(t))
                                                        \tag \label{Fplb421}$$
    are derived from the infinitesimal form
        $$\align
    \frac{\partial \cL}{\partial
x^\zk}(\zx^\zm(t),\dot\zx^\zn(t))&\zd\zx^\zk(t) + \frac{\partial
\cL}{\partial\dot x^\zk}(\zx^\zm(t),\dot\zx^\zn(t))\zd\dot\zx^\zk(t) -
\zr_\zk(\zx^\zm(t),\dot\zx^\zn(t))\zd\zx^\zk(t) \\
    & = - \zz_\zk(t)\zd\zx^\zk(t) + \dot\zh_\zk(t)\zd\zx^\zk(t) +
\zh_\zk(t)\zd\dot\zx^\zk(t)
                                                \tag \label{Fplb422}\endalign$$
    of the variational principle.

    Equations describing the dynamics in the potential case are obtained
by setting $\zr = 0$.

        \sscx{The Hamiltonian formulation.}

    The Legendre mapping $\cP\zl$ is represented locally by
        $$(x^\zk,p_\zl) \circ \cP\zl = \left(x^\zk,\frac{\partial
\cL}{\partial\dot x^\zl}(x^\zm,\dot x^\zn)\right).
                                                        \tag \label{Fplb423}$$
    It is convenient to introduce functions
        $$\zP_\zl = p_\zl \circ \cP\zl \circ (x^\zm,x^\zn)^{-1} =
\frac{\partial \cL}{\partial\dot x^\zl}.
                                                        \tag \label{Fplb424}$$
    If the Legendre mapping is a diffeomorphism, then the inverse
diffeomorphism $\zL$ is represented locally by
        $$(x^\zk,\dot x^\zl) \circ \zL = (x^\zk,\zL^\zl(x^\zm,p_\zn)),
                                                        \tag \label{Fplb425}$$
    where $\zL^\zk$ are the functions
        $$\zL^\zk = \dot x^\zk \circ \zL \circ (x^\zm,p_\zn)^{-1}.
                                                        \tag \label{Fplb426}$$
    Relations
        $$\zP_\zl\left(x^\zm(p),\zL^\zn(x^\zr(p),p_\zs(p))\right) =
p_\zl(p)
                                                        \tag \label{Fplb427}$$
    and
        $$\zL^\zl\left(x^\zm(v),\zP_\zn(x^\zr(v),\dot x^\zs(v))\right) =
\dot x^\zl(v)
                                                        \tag \label{Fplb428}$$
    are satisfied.

    In the potential case we have a Hamiltonian $H \colon \sT^\*
\rightarrow \R$ represented locally by the function
        $$\cH = H \circ (x^\zk,p_\zl)^{-1}.
                                                        \tag \label{Fplb429}$$
    This function is obtained from the formula
        $$\cH(x^\zk,p_\zl) = p_zk\zL^\zk(x^\zm,p_\zn) -
\cL(x^\zk,\zL^\zl(x^\zm,p_\zn)).
                                                        \tag \label{Fplb430}$$
    In the non potential case there is the Hamiltonian form
        $$\align
    \zq &= \zq_\zk(x^\zm,p_\zn)\xd x^\zk + \zq^\zk(x^\zm,p_\zn)\xd p_\zk \\
            &= \frac{\partial \cH}{\partial p_\zk}(x^\zm,p_\zn)\xd p_\zk +
\frac{\partial \cH}{\partial x^\zk}(x^\zm,p_\zn)\xd x^\zk +
\zr_\zk(x^\zm,\zL^\zn(x^\zr,p_\zs))\xd x^\zk \\
            &= \zL^\zk(x^\zm,p_\zn)\xd p_\zk - \frac{\partial
\cL}{\partial x^\zk}(x^\zm,\zL^\zn(x^\zr,p_\zs))\xd x^\zk +
\zr_\zk(x^\zm,\zL^\zn(x^\zr,p_\zs))\xd x^\zk
                                                \tag \label{Fplb431}\endalign$$
    obtained from formula \Ref{Fplb280}.  The function $\cH$ is the local
expression of the Hamiltonian $H$ associated with $L$.

    The vector field $Z = \za_M{}^{-1} \circ \zl \circ \zL = -
\zb_{(\sT^\*M,\zw_M)}^{-1} \circ \zq$ is expressed by
        $$\align
    (x^\zk,p_\zl,\dot x^\zm,\dot p_\zn) \circ Z &=
(x^\zk,p_\zl,\zq^\zm(x^\zr,p_\zs),-\zq_\zn(x^\zr,p_\zs)) \\
            &= \left(x^\zk,p_\zl,\frac{\partial \cH}{\partial
p_\zm}(x^\zr,p_\zs), - \frac{\partial \cH}{\partial x^\zn}(x^\zr,p_\zs) -
\zr_\zn(x^\zr,\zL^\zs(x^\zw,p_\zp))\right) \\
            &= \left(x^\zk,p_\zl,\zL^\zm(x^\zr,p_\zs),\frac{\partial
\cL}{\partial x^\zn}(x^\zr,\zL^\zs(x^\zw,p_\zp)) -
\zr_\zn(x^\zr,\zL^\zs(x^\zw,p_\zp))\right).
                                                \tag \label{Fplb432}\endalign$$

    A force-momentum trajectory $(\zz,\zh)$ satisfies Hamilton's equations
        $$\dot\zx^\zk(t) = \frac{\partial \cH}{\partial
p_\zk}(\zx^\zm(t),\zh_\zn(t))
                                                        \tag \label{Fplb433}$$
    and
        $$\dot\zh_\zk(t) - \zz_\zk(t) = - \frac{\partial \cH}{\partial
x^\zk}(\zx^\zm(t),\zh_\zn(t)) - \zr_\zk(\zx^\zm(t),\dot\zx^\zn(t)).
                                                        \tag \label{Fplb434}$$

        \sscx{The Poisson formulation.}

    Hamilton's equations are equivalent to equations
        $$\align
    \frac{\partial \cF}{\partial x^\zk}&(\zx^\zm(t),\zh_\zn(t))\dot\zx^\zk
+ \frac{\partial \cF}{\partial p_\zk}(\zx^\zm(t),\zh_\zn(t))\dot\zh_\zk \\
            &= \left(\frac{\partial \cF}{\partial x^\zk}\frac{\partial
\cH}{\partial p_\zk} - \frac{\partial \cH}{\partial x^\zk}\frac{\partial
\cF}{\partial p_\zk}\right)(\zx^\zm(t),\zh_\zn(t)) + \frac{\partial
\cF}{\partial p_\zk}(\zx^\zm(t),\zh_\zn(t))(\zz_\zk(t) -
\zr_\zk(\zx^\zm(t),\zh_\zn(t)))
                                                \tag \label{Fplb435}\endalign$$
    satisfied for each function $F$ on $\sT^\*M$ with local expression
        $$\cF = F \circ (x^\zk,p_\zl)^{-1}.
                                                        \tag \label{Fplb436}$$

        \sect{An example.}

    An aircraft is travelling in a vertical plane $M$.  The force of
gravity and the force due to air viscosity are the internal forces.  The
jet propulsion force and the aerodynamic forces acting on the wings, the
rudder, and the elevator are controlled by the pilot and are considered
external forces.

    An Euclidean affine chart
        $$(x^h,x^v) \colon M \rightarrow \R^2
                                                        \tag \label{Fplb437}$$
    induces charts
        $$(x^h,x^v,\dot x^h,\dot x^v) \colon \sT M \rightarrow \R^4,
                                                        \tag \label{Fplb438}$$
        $$(x^h,x^v,\dot x^h,\dot x^v,\ddot x^h,\ddot x^v) \colon \sT M
\rightarrow \R^6,
                                                        \tag \label{Fplb439}$$
    and
        $$(x^h,x^v,f_h,f_v,p_h,p_v) \colon \sT^\*M \fpr{(\zp_M,\zp_M)}
\sT^\* M \rightarrow \R^6.
                                                        \tag \label{Fplb440}$$

    A force-momentum trajectory $(\zz,\zh)$ has a local representation
        $$(\zx^h,\zx^v,\zz_h,\zz_v,\zh_h,\zh_v) =
(x^h,x^v,f_h,f_v,p_h,p_v) \circ (\zz,\zh).
                                                        \tag \label{Fplb441}$$
    The mapping $\zx = \zp{}_M \circ \zz = \zp{}_M \circ \zh$ and its
prolongations $\dot\zx$ and $\ddot\zx$ have local representations
        $$(\zx^h,\zx^v) = (x^h,x^v)\circ \zx,
                                                        \tag \label{Fplb442}$$
        $$(\zx^h,\zx^v,\dot\zx^h,\dot\zx^v) = (x^h,x^v,\dot x^h,\dot x^v)
\circ \dot\zx,
                                                        \tag \label{Fplb443}$$
    and
        $$(\zx^h,\zx^v,\dot\zx^h,\dot\zx^v,\ddot\zx^h,\ddot\zx^v) =
(x^h,x^v,\dot x^h,\dot x^v,\ddot x^h,\ddot x^v) \circ \ddot\zx.
                                                        \tag \label{Fplb444}$$

    The form
        $$\zl = \xd L - \zr = m\dot x^h \xd\dot x^h + m\dot x^v \xd\dot
x^v - mg\xd x^v - \zg_h \dot x^h\xd x^h - \zg_v \dot x^v\xd x^v
                                                        \tag \label{Fplb445}$$
    is constructed from the Lagrangian
        $$L = \frac{m}{2} ((\dot x^h)^2 + (\dot x^v)^2) - mgx^v
                                                        \tag \label{Fplb446}$$
    and the form
        $$\zr =  \zg_h \dot x^h\xd x^h + \zg_v \dot x^v\xd x^v
                                                        \tag \label{Fplb447}$$
    representing the non potential force of viscosity.

    The dynamics in a time interval $[0,T]$ is governed by the
Euler-Lagrange equations
        $$m\ddot\zx^h(t) + \zg_h\dot\zx^h(t) = \zz_h(t),
                                                        \tag \label{Fplb448}$$
        $$m\ddot\zx^v(t) + \zg_v\dot\zx^v(t) + mg = \zz_v(t),
                                                        \tag \label{Fplb449}$$
    and the momentum-velocity relations
        $$\zh_h(0) = m\dot\zx^h(0), \;\;\zh_v(0) = m\dot\zx^v(0),
\;\;\zh_h(T) = m\dot\zx^h(T), \;\;\zh_v(T) = m\dot\zx^v(T)
                                                        \tag \label{Fplb450}$$
    at the boundary.  In the absence of external forces and with initial
conditions
        $$(\zx^h,\zx^v,\dot\zx^h,\dot\zx^v)(0) = (0,0,v_0,0)
                                                        \tag \label{Fplb451}$$
    we obtain the trajectory
        $$\zx^h(t) = \frac{mv_0}{\zg_h}\left(1 -
\exp\left(-\frac{\zg_h}{m}t\right)\right),
                                                        \tag \label{Fplb452}$$
        $$\zx^v(t) = \frac{m^2 g}{\zg_v{}^2}\left(1 -
\exp\left(-\frac{\zg_v}{m}t\right)\right) - \frac{mg}{\zg_v}t,
                                                        \tag \label{Fplb453}$$
    and the momenta
        $$\zh_h(0) = mv_0, \;\;\zh_v(0) = 0, \;\;\zh_h(T) =
\frac{mv_0}{\zg_h}\exp\left(-\frac{\zg_h}{m}T\right) , \;\;\zh_v(T) = -
\frac{m^2 g}{\zg_v}\left(1 - \exp\left(-\frac{\zg_h}{m}T\right)\right)
                                                        \tag \label{Fplb454}$$
    at the boundary.  In order to maintain a horizontal trajectory
        $$\zx^h(t) = v_0 t, \;\; \zx^v(t) = 0
                                                        \tag \label{Fplb455}$$
    with constant velocity it is necessary to supply external forces
        $$\zz_h(t) = \zg_h v_0,\;\; \zz_v(t) = mg.
                                                        \tag \label{Fplb456}$$
    The external forces are true forces of control.  We have described a
very simple situation.  In reality these forces can not be chosen in
advance since they may have to compensate the effects of varying weather
conditions and allow changes of the trajectory necessary due to
unforeseeable circumstances.

        \sect{References.}

    [1] C.-M. Marle, {\it Kinematic and Geometric Constraints,
Servomechanism and Control of Mechanical Systems}, Rend. Sem. Mat. Univ.
Pol. Torino {\bf 54} (1996)

    [2] C.-M. Marle, {\it Sur la g\'eom\'etrie des syst\`emes m\'ecaniques
\`a liasons actives}, C. R. Acad. Sci. Paris {\bf 311} (1990)

    [3] C.-M. Marle, {\it Reduction of constrained mechanical systems and
stability of relative equilibria}, Commun. Math. Phys. {\bf 174} (1995)

    [4] X. Gr\`acia, J. Marin-Solano, M.-C. M\~unoz-Lecanda, {\it Some
Geometric Aspects of Variational Calculus in Constrained Systems},
arXiv:math-ph/0004019 (2000)

    [5] J. Schwinger, {\it Lectures on Quantum Dynamics}, Les Houches
Summer School (1955)

    [6] J. E. Marsden, T. S. Ratiu, J. Scheurle, {\it Reduction theory and
the Lagrange-Routh equations}, J. Math. Phys., {\bf 41} (2000)

    [7] J. F. Cari\~nena, {\it Sections Along Maps in Geometry and
Physics}, Rend. Sem. Mat. Univ. Pol. Torino {\bf 54} (1996)

    [8] J. F. Cari\~nena, C. L\'opez, {\it Geometric study of Hamilton's
variational principle}, Rev. Math. Phys. {\bf 3} (1991)

    [9] A. Ibort, M. de Le\'on, E. A. Lacomba, de Diego, P. Pitanga {\it
Mechanical systems subjected to impulsive constraints}, J. Phys. A: Math.
Gen. {\bf 30} (1997)

    [10] M. de Le\'on, D. M. de Diego, {\it On the geometry of
non-holonomic Lagrangian systems}, J. Math. Phys. {\bf 37} (1996)

    [11] M. de Le\'on, J. C. Marrero, D. M. de Diego, {\it Non-holonomic
Lagrangian systems in jet manifolds}, J. Phys. A: Math. Gen. {\bf 30}
(1997)

    [12] G. Pidello and W. M. Tulczyjew, {\it Derivations of differential
forms on jet bundles}, Ann. Mat. Pura Appl., {\bf 147} (1987)

\enddocument